\documentclass[10pt,a4paper,twoside,dutch,english]{book}                

\usepackage{pseudocode}
\usepackage{times}      
\usepackage[times]{quotchap}   
\usepackage{fancyhdr}
\usepackage[dutch,english]{babel}
\usepackage[a4paper,verbose, asymmetric, centering]{geometry}  
\usepackage{enumerate}  
\usepackage{verbatim}   
\usepackage[bf]{caption}     
\usepackage{subfigure}  
\usepackage{cite}
\usepackage{enumerate}  
\usepackage{verbatim}   

\usepackage{float}      
\usepackage{multirow}   

\usepackage[pdftex]{graphicx}

\usepackage{chappg}     
\usepackage{url}        
\usepackage{expdlist}   
\usepackage{acronym_expdlist}   
\usepackage{hhline}     
\usepackage{afterpage}  
\usepackage{amsmath,amsfonts,amsthm}
\usepackage{listings}
\usepackage[squaren]{SIunits}
\usepackage{ulem}
\usepackage{bbm}

\fancypagestyle{plain}{
\fancyhf{}

}

\renewcommand{\chaptermark}[1]{\markboth{#1}{}}

\newcommand\oddpageleftmark{}
\newcommand\evenpagerightmark{}

\def\PARstart#1#2{\begingroup\def\par{\endgraf\endgroup\lineskiplimit=0pt}
    \setbox2=\hbox{\uppercase{#2} }\newdimen\tmpht \tmpht \ht2
    \advance\tmpht by \baselineskip\font\hhuge=cmr10 at \tmpht
    \setbox1=\hbox{{\hhuge #1}}
    \count7=\tmpht \count8=\ht1\divide\count8 by 1000 \divide\count7 by\count8
    \tmpht=.001\tmpht\multiply\tmpht by \count7\font\hhuge=cmr10 at \tmpht
    \setbox1=\hbox{{\hhuge #1}} \noindent \hangindent1.05\wd1
    \hangafter=-2 {\hskip-\hangindent \lower1\ht1\hbox{\raise1.0\ht2\copy1}%
    \kern-0\wd1}\copy2\lineskiplimit=-1000pt}

\AtBeginDocument{%

   \renewcommand{\bibname}{References}%
}

\begin{document}
\graphicspath{{fig/}}

\restylefloat{figure}
\restylefloat{table}
\newfloat{algorithm}{ht}{alg}

\newcommand{\etal}{\textit{et al.\ }}
\newcommand{\mic}{\unit{}{\micro\meter}}


\newcommand{\dd}{\partial}
\newcommand{\nn}{\nonumber}
\newcommand{\eps}{\epsilon}
\newcommand{\OM}{\Omega}
\newcommand{\tens}[1]{{\mathbf #1}}
\newcommand{\bpsi}{{\boldsymbol \psi}}
\newcommand{\bY}{{\mathbf Y}}
\newcommand{\bPhi}{\mathbf{F}}
\newcommand{\bF}{\mathbf{F}}
\newcommand{\bP}{\mathbf{P}}
\newcommand{\bR}{\mathbf{R}}
\newcommand{\bs}{{\boldsymbol \sigma}}
\newcommand{\beps}{{\boldsymbol \epsilon}}
\newcommand{\veps}{\varepsilon}

\newcommand{\eval}[2]{   \left. #1\right|_{#2} }
\newcommand{\bsub}{\begin{subequations}}
\newcommand{\esub}{\end{subequations}}
\newcommand{\bea}{\begin{eqnarray}}
\newcommand{\eea}{\end{eqnarray}}
\newcommand{\bal}{\begin{align}}
\newcommand{\eal}{\end{align}}
\newcommand{\hs}{\hspace{6pt}}

\newcommand{\uA}{{\underline{A}}}
\newcommand{\uB}{{\underline{B}}}
\newcommand{\uC}{{\underline{C}}}
\newcommand{\uD}{{\underline{D}}}
\newcommand{\uE}{{\underline{E}}}
\newcommand{\uF}{{\underline{F}}}

\newcommand{\DD}{\mathcal{D}}
\newcommand{\uu}{\mathbf{u}}
\newcommand{\tuu}{\tilde{\mathbf{u}}}
\newcommand{\buu}{\bar{\mathbf{u}}}
\newcommand{\vv}{\mathbf{v}}
\newcommand{\ww}{\mathbf{w}}
\newcommand{\DT}{\mathbf{D}}
\newcommand{\qq}{\mathbf{q}}
\newcommand{\rr}{\mathbf{r}}
\newcommand{\FF}{\mathcal{F}}
\newcommand{\gambar}{\bar{\gamma}}
\newcommand{\mum}{\,\mu\mathrm{m}}
\newcommand{\OO} {\mathcal {O}}
\newcommand{\ve}{\vec{e}}
\newcommand{\s}{\sigma}
\newcommand{\la}{\lambda}
\newcommand{\LA}{\Lambda}
\newcommand{\HP}{\hat{P}}
\newcommand{\HL}{\hat{L}}
\newcommand{\GG}{\mathcal{G}}
\newcommand{\ls}{\lambda_\s}
\newcommand{\RR}{\mathcal{R}}
\newcommand{\bo}{\otimes}
\newcommand{\TRK}{\mathrm{Tr}\, \mathbf{K}}
\newcommand{\TRKK}{\mathrm{Tr} (\mathbf{K}^2)}

\def\bra#1{\mathinner{\langle{#1}|}}
\def\ket#1{\mathinner{|{#1}\rangle}}
\def\braket#1{\mathinner{\langle{#1}\rangle}}
\def\Bra#1{\left<#1\right|}
\def\Ket#1{\left|#1\right>}
{\catcode`\|=\active
  \gdef\Braket#1{\left<\mathcode`\|"8000\let|\BraVert {#1}\right>}}
\def\BraVert{\egroup\,\mid@vertical\,\bgroup}


%
 \thispagestyle{empty}   
%
 \noindent
 \begin{minipage}{3cm}%
   \includegraphics*[width=2cm]{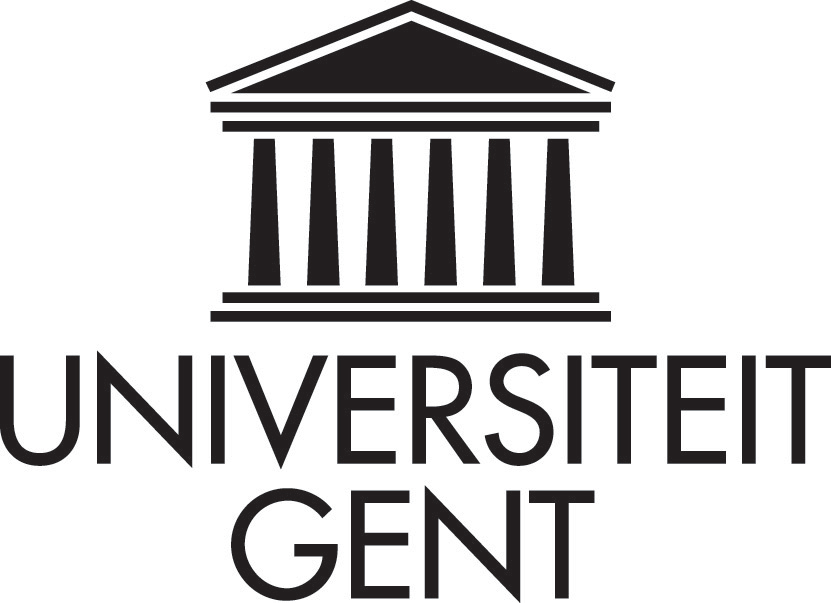}
 \end{minipage}\hfill
 \begin{minipage}{8cm}
 \raggedleft
 \textsf{Universiteit Gent\\
 Faculteit Wetenschappen\\
 Vakgroep Fysica en Sterrenkunde}
 \end{minipage}
%
\bigskip
   \begin{flushleft}
     \Large \textsf{Dynamics of wave fronts and filaments\\ in anisotropic cardiac tissue}\\
     \vspace{0.1in}\large{\textsf{Dynamica van golffronten en filamenten\\ in anisotroop hartweefsel}}
   \end{flushleft}
 \hrule
 \bigskip
   \LARGE\noindent \textsf{Hans J.F.M. Dierckx} \hfill
 \bigskip
%
\bigskip
\begin{figure}[htb]
  \centering%
  \includegraphics*[width=1.0\textwidth]{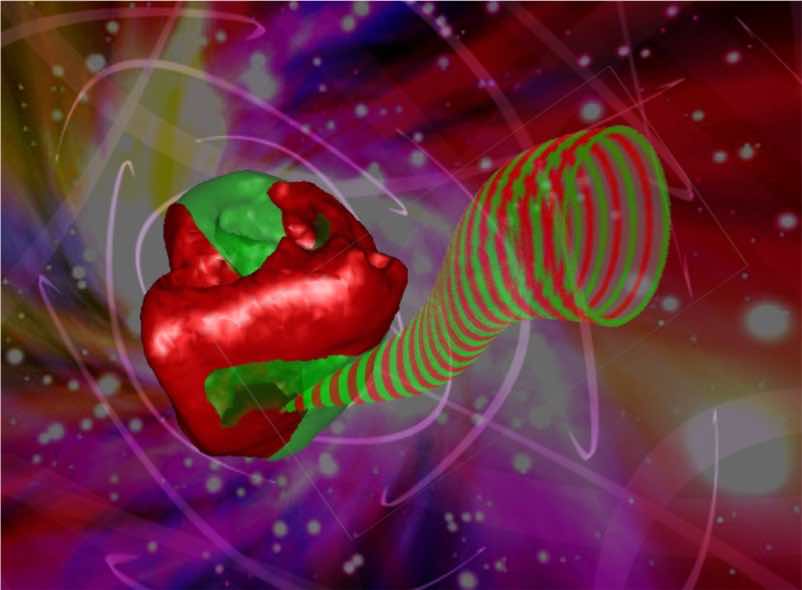}%
\end{figure}%
\normalsize
%
 \normalsize
%
 \vfill
 \begin{minipage}{2.0cm}%
     \includegraphics*[width=3.5cm]{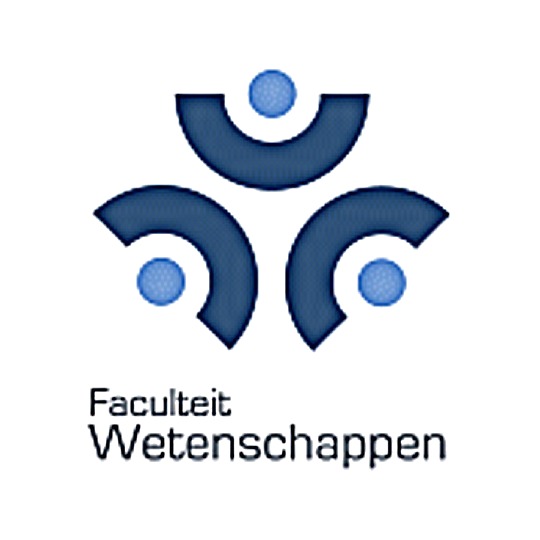}
 \end{minipage}\hfill
 \begin{minipage}{9cm}
 \raggedleft
 \textsf{Proefschrift tot het bekomen van de graad van \\
 Doctor in de Wetenschappen: \\
Fysica \\
 Academiejaar 2009-2010}
 \end{minipage}


\clearpage{\pagestyle{empty}\cleardoublepage}
\thispagestyle{empty}

\normalsize

\noindent
\begin{minipage}{3cm}%
  \includegraphics*[width=2cm]{ugentlogo_zw.jpg}
\end{minipage}\hfill
\begin{minipage}{8cm}
\raggedleft
\textsf{Universiteit Gent\\
Faculteit Wetenschappen\\
Vakgroep Fysica en Sterrenkunde}
\end{minipage}
\\[2cm]

\noindent \begin{tabular}{ @{} l l}
Promotoren: & Prof.\ Dr.\ Henri Verschelde \\
 & Dr.\ Olivier Bernus \\
\end{tabular}
\\[2cm]

\noindent Universiteit Gent \\
\noindent Faculteit Wetenschappen
\\[0.3cm]
\noindent Vakgroep Fysica en Sterrenkunde \\
\noindent Krijgslaan 281, S9 (WE05),
\noindent B-9000 Gent, Belgi\"e
\\[0.3cm]
\noindent Tel.: +32 9 264 47 98 \\
\noindent Fax.: +32 9 264 49 89
\\[4cm]
\noindent Dit werk kwam tot stand met de steun van het \\Fonds Wetenschappelijk Onderzoek-Vlaanderen.

\vfill

\begin{minipage}{2.0cm}%
    \includegraphics*[width=3.5cm]{logofac_3.jpg}
\end{minipage}\hfill
\begin{minipage}{9cm}
\raggedleft
\textsf{Proefschrift tot het behalen van de graad van \\
Doctor in de Wetenschappen: \\
Fysica \\
Academiejaar 2009-2010}
\end{minipage}
\clearpage{\pagestyle{empty}\cleardoublepage}


\hyphenation{gelegenheids-uitstapje}
\frontmatter
\chapter{Dankwoord}
\vspace{0.35in}

\selectlanguage{dutch}
Uiteraard grijp ik deze gelegenheid graag aan om allereerst mijn beide promotoren hartelijk te bedanken.

Henri, zonder jouw onbevangen en enthousiaste onderzoeksvoorstel zou dit werk nooit tot stand kunnen gekomen zijn. Wat begon als een gelegenheidsuitstapje in de bevreemdende wereld van de biofysica werd door jouw inzichten al gauw een reis in vogelvlucht langsheen vele uithoeken van het natuurkundig universum. Bedankt voor de spannende trip!

Olivier, jouw steun van dichtbij en veraf werd niet minder geapprecieerd. Ik waardeer je enorm om mijn gids en coach te zijn tegen de biologische achtergrond van het onderzoek. Bedankt ook voor je gastvrijheid en vertrouwen, en om je uitgebreide netwerk in de gemeenschap van de hartwetenschappers te hebben gedeeld.\\

Het was misschien wat minder te merken tijdens de eindsprint op weg naar dit thesisboek, maar ik heb het gezelschap van mijn naaste collega's ten zeerste op prijs gesteld. Vandaar een dikke merci aan Karel, David, David, Jutho, Nele, Nele en Dirk voor de gepaste onderonsjes. Onder de rubriek vertrokken maar niet vergeten: dank aan Wouter voor de compagnie en een goeie tip over tensoren en aan Jos, om dagdagelijks voor wat leven in de brouwerij te zorgen en bovendien het voortouw te nemen in heel wat sportieve en ontspannende activiteiten. Voorts hou ik eraan om Inge, Anny en Gerbrand te bedanken voor de hulp met administratie en IT. Er is ook nog iemand die er op enkele maanden tijd is in geslaagd m'n bureau aan te vullen met (bijvoorbeeld) een enorme yuccaplant, een lade met suikerrijk proviand om een tweetal weken mee door te komen, een greep uit de elektronica van de 21e eeuw en een aanstekelijke grijnslach. Ben, ik heb met veel plezier mijn stek bij het raam aan jou afgestaan.\\

Gezien het interdisciplinair karakter van mijn onderzoek moest ik ook elders ten rade gaan. Dank aan Steven, Steven en Els van de Gentse MRI groep voor waardevolle discussies en contacten. \selectlanguage{english}I am also grateful to Alan and Steve for being my closest collaborators on medical imaging and together bringing the dQBI idea to practice. Herewith, sincere thanks to Arun Holden and Olivier for making the test runs possible, and to Annick and Erin for their warm hospitality.\\

Throughout my PhD work, I have been lucky to meet some great experts in their scientific discipline. Here I would like to express my gratitude for their willingness to share their opinion, talk over ideas and spend time interacting with freshmen in the field. In particular, I acknowledge Drs. Arkady Pertsov, Marcel Wellner, Vadim Biktashev, Irina Biktasheva, Arun Holden, Sasha Panfilov, Richard Clayton, Darryl Holm, Yves Dedeene, Jacques-Donald Tournier and Valerij Kiselev for helpful discussions. Additional acknowledgements go to Flavio Fenton and Elizabeth Cherry for sharing some DT images and their numerical experience on filament behavior. Oleg Mornev, thank you for your visit to Ghent, during which you have taught me not only on mathematical methods but also on the Russian language and culture. Also, Bruce Searles will understand if I thank him here for having shown to me what our research is really all about.

Furthermore, I have enjoyed sitting together with fellow young researchers on different occasions. Dear Daniel, {\O}yvind, Jos\'e, Lucia, Andy, Pan, Charles, Becky, Bogdan, Diana, Alan, Steve: we should go out for a drink once more.\\

\selectlanguage{dutch}
Natuurlijk zijn er ook nog de vrienden en familie die mij hebben gesteund gedurende de voorbije jaren en tevens voor de gepaste afleiding hebben gezorgd. Dank aan Olaf, Fons, Sanne, Michiel, Sander, Els, Jonas, Stijn, Deborah, Steven en Frederik voor de ontspannende momenten tussendoor. Ik heb er samen met Marlies eveneens van genoten om op regelmatige basis de dansvloer onveilig te maken in het aangename gezelschap van Leander, Annelies, Dennis, Sophie, Frederik en Isolde; merci! Sta me verder ook toe om Ilse, Wouter, Annelien, Lieven en Chris te bedanken voor gezellige weekendactiviteiten en lekker eten.\\

Wie ons goed kent, weet dat mijn ouders een speciale rol innemen voor mij en omgekeerd. Ik ben dan ook trots en dankbaar voor hun voortdurende inzet om hun zoon alle kansen te geven.

Tot slot wil ik graag Marlies bedanken, die mij steeds met enorm veel liefde heeft omringd en gesteund.

\begin{flushright}{\textit{Gent, mei 2010 \\
Hans Dierckx}}
\end{flushright}
\selectlanguage{english}

%
%
%
%
%
%

\clearpage{\pagestyle{empty}\cleardoublepage}


\renewcommand{\contentsname}{Table of Contents} 

\setcounter{tocdepth}{2}
\tableofcontents
\clearpage{\pagestyle{empty}\cleardoublepage}

\listoffigures
\clearpage{\pagestyle{empty}\cleardoublepage}



\chapter*{List of Acronyms}




\textbf{\newline\newline\Large D} \newline
\begin{acronym_expdlist}
\acro{DT}{Diffusion Tensor}
\acro{DTI}{Diffusion Tensor Imaging}
\acro{dQBI}{Dual Q-Ball Imaging}
\acro{DW}{Diffusion Weighted}
\end{acronym_expdlist}

\textbf{\newline\newline\Large E} \newline
\begin{acronym_expdlist}
\acro{EOM}{Equation Of Motion}
\end{acronym_expdlist}

\textbf{\newline\newline\Large F} \newline
\begin{acronym_expdlist}
\acro{FRT}{Funk-Radon Transform}
\end{acronym_expdlist}

\textbf{\newline\newline\Large G} \newline
\begin{acronym_expdlist}
\acro{GM}{Goldstone Mode}
\end{acronym_expdlist}

\textbf{\newline\newline\Large I} \newline
\begin{acronym_expdlist}
\acro{IVS}{Interventricular Septum}
\acro{}{}
\end{acronym_expdlist}

\textbf{\newline\newline\Large L} \newline
\begin{acronym_expdlist}
\acro{LV}{Left Ventricle}
\acro{LVFW}{Left Ventricular Free Wall}
\end{acronym_expdlist}

\textbf{\newline\newline\Large M} \newline
\begin{acronym_expdlist}
\acro{MRI}{Magnetic Resonance Imaging}
\end{acronym_expdlist}

\textbf{\newline\newline\Large O} \newline
\begin{acronym_expdlist}
\acro{ODF}{Orientation Distribution Function}
\end{acronym_expdlist}

\textbf{\newline\newline\Large P} \newline
\begin{acronym_expdlist}
\acro{PDF}{Probability Density Function}
\acro{PM}{Papillary muscle}
\end{acronym_expdlist}

\textbf{\newline\newline\Large Q} \newline
\begin{acronym_expdlist}
\acro{QBI}{Q-Ball Imaging}
\end{acronym_expdlist}

\textbf{\newline\newline\Large R} \newline
\begin{acronym_expdlist}
\acro{RD}{Reaction-Diffusion}
\acro{RDE}{Reaction-Diffusion Equation}
\acro{R.F.}{Radio Frequent}
\acro{RF}{Response Function}
\acro{RTI}{Ricci Tensor Imaging}
\acro{RV}{Right Ventricle}
\acro{RVFW}{Right Ventricular Free Wall}
\end{acronym_expdlist}



\renewcommand{\bibname}{References}     

\selectlanguage{dutch}
\renewcommand{\thesection}{\arabic{section}}    

\renewcommand{\bibname}{Referenties}
\renewcommand\evenpagerightmark{{\scshape\small Nederlandstalige Samenvatting}}
\renewcommand\oddpageleftmark{{\scshape\small Summary in Dutch}}

\chapter[Nederlandstalige samenvatting]%
{Nederlandstalige samenvatting \\\makebox[2.82in]{--Summary in Dutch--}}

\hyphenation{acti-va-tie-golf acti-va-tie-cy-clus voor-ge-steld sys-te-ma-ti-sche an-iso-tro-pie vezel-achtige diffusie-tensor klief-vlak-ken equi-valentie-beginsel over-blij-ven-de uit-ge-voerd actie-poten-tiaal be-we-ring opper-vlakte-span-ning ge-li-ne-a-ri-seer-de ver-ge-lij-king
e-vo-lu-tie-ver-ge-lij-kingen be-we-gings-ver-ge-lij-kingen rich-tings-coef-fi-cient opper-vlakte-span-nings-coef-fi-cient om-wen-te-lings-fre-quen-tie ro-tor-fi-la-ment fi-bril-la-tie be-tref-fen-de fi-la-ment twee-dimensio-na-le}
\def\hyph{-\penalty0\hskip0pt\relax}

Het voorgestelde werk situeert zich in de wiskundige biofysica. Er werd met name bestudeerd hoe de structuur van de hartspier de voortplanting van elektrische depolarisatiegolven be\"invloedt. Aangezien precies deze elektrische prikkels de individuele hartcellen aanzetten tot mechanische samentrekking, geeft afwijkende elektrische activiteit aanleiding tot hartritmestoornissen die de pompfunctie gedeeltelijk of zelfs geheel kunnen uitschakelen.\\

Vanuit wiskundig-fysisch oogpunt kan het hart  worden beschouwd als een exciteerbaar medium. Immers, waar de transmembraanpotentiaal een drempelwaarde overschrijdt, worden de cellen ge\"exciteerd tot een cyclus die bekend staat als een actiepotentiaal. Tijdens de actiepotentiaal depolariseren de cellen en trekken ze samen langsheen hun lange as, waarna ze gedurende korte tijd hun rusttoestand herstellen; een ganse actiepotentiaal duurt ongeveer 300\,ms. In het hart wordt de actiepotentiaal bovendien doorgegeven tussen de cellen door gespecialiseerde verbindingssites (`gap junctions'), wat gemodelleerd kan worden door diffusie van de elektrische transmembraanpotentiaal. Hierdoor ontstaat een heuse golf van elektrische activiteit die als doel heeft de hartspier tot een optimaal geco\"ordineerde contractie te bewegen. Op deze manier ontstaat onder normaal hartritme telkens \'e\'en hartslag.

Op macroscopische schaal kan exciteerbaar biologisch weefsel worden voorgesteld met behulp van een reactie-diffusievergelijking. Hierin treedt de transmembraanpotentiaal op als \'e\'en van de toestandsvariabelen; het is trouwens de enige variabele die diffusie ondergaat op de beschouwde tijdschaal. De wiskundige beschrijving van potentiaaldiffusie wordt bemoeilijkt door de uitgesproken vezelachtige structuur van het hartweefsel. Daarenboven wordt de hartspier door kliefvlakken doorsneden. Men was daarom genoodzaakt in het wiskundige model een anisotrope elektrische diffusietensor op te nemen die in rekening brengt dat depolarisatiegolven ongeveer driemaal sneller reizen langs de spiervezelrichting dan in de dwarse richtingen. Op basis van talloze numerieke simulaties en gesterkt door experimentele metingen werd vervolgens besloten dat inzicht in de rol van anisotropie essentieel is om te begrijpen hoe bepaalde types hartritmestoornissen ontstaan; bij aanvang van dit onderzoeksproject bestond er echter geen systematische methode die analytisch om kon gaan met de generieke anisotropie van de hartspier.\\

In dit werk werd besloten de anisotropie te behandelen vanuit een lokaal equivalentiebeginsel. Voor een blokje met vaste vezelrichting is het namelijk mogelijk om de isotropie te herstellen door de gekozen lengte-eenheid aan te passen in elke hoofdrichting. Deze keuze komt neer op het gebruiken van effectieve reistijden om afstanden weer te geven, wat een vertrouwde techniek is in bijvoorbeeld de optica of bij kaartlezen. Het volstaat om deze operationele definitie van afstand consistent door te voeren in elk klein stukje hartweefsel om de anisotropie in de vergelijkingen volledig weg te werken, ware het niet dat deze procedure de ruimte een intrinsieke kromming oplegt. Hierdoor gedragen patronen van elektrische activiteit in het hart zich alsof ze doorheen een niet-Euclidische ruimte bewegen, waarvan de metrische tensor evenredig is met de inverse van de elektrische diffusietensor.

Gelukkigerwijs kennen fysici nog een ander natuurverschijnsel, namelijk de zwaartekracht, dat tot zijn eenvoudigste vorm gebracht werd door te werken in een intrinsiek gekromde ruimte. De voorgaande beschrijving is natuurlijk Einsteins algemene relativiteitstheorie. Deze stelt onder andere dat licht onafwendbaar rechtdoor straalt, zij het in een ruimtetijd die gekromd wordt door de aanwezigheid van zware massa's. Dit toch wel onverwachte parallellisme houdt in dat een heleboel wiskundige technieken en fysische interpretaties kunnen worden geleend uit hun kosmologische context voor toepassing in de wiskundige biofysica. De evolutievergelijkingen voor excitatiepatronen die bekomen worden in dit werk gelden dan ook meteen voor algemene anisotropie van de hartspier, net omdat op covariante wijze wordt omgegaan met het niet-Euclidische karakter van de ruimte.\\

Als eerste toepassing van het nieuwe formalisme wordt in dit proefschrift het verband bestudeerd tussen de voortplantingssnelheid van actiepotentiaalgolven en de lokale kromming van hun golffront. In de welgekende gelineariseerde vergelijking -- doorgaans de eikonale relatie genoemd -- treedt nu een algemene rich\-tingsco\"effici\"ent op, terwijl de klassieke afleiding steevast als waarde \'e\'en opleverde. Deze co\"effici\"ent heeft de fysische betekenis van een oppervlaktespanning voor het golffront, die uiteraard positief dient te zijn om een eenparige verspreiding van excitatie en contractie in het hart te bewerkstelligen. Deze bewering wordt overigens gestaafd door een variationeel principe af te leiden, waarvan de actie termen bevat die het differentieel ingenomen volume en de totale oppervlakte van het golffront weergeven.
Verder worden in dit werk voor het eerst ook de tweede orde krommingscorrecties voor de beweging van golffronten uitgerekend. Voor isotrope media wijken deze af van de gangbare extrapolatie in de literatuur; in anisotrope media manifesteert zich een nieuw effect dat gelijkaardig is aan gravitationele lenswerking in de kosmologie.

De studie van golffronten wordt afgerond door te kijken hoe snelle periodieke elektrische stimulatie weegt op de voortplanting van actiepotentialen; de berekening levert een algemene uitdrukking op voor de dispersieve oppervlaktespannings\-co\"effici\"ent. Het resultaat kan gebruikt worden om een dimensieloze kritische verhouding af te schatten, die de gevoeligheid van de hartspier beschrijft voor de ontwikkeling van turbulente hartritmes.
\\

Als tweede aanwending van de geometrische interpretatie voor structurele anisotropie worden spiraalgolven beschouwd. Net als golffronten zijn spiraalgolven oplossingen van de reactie-diffusievergelijking in twee dimensies. Spiraalvormige activiteit kan ontstaan wanneer een golffront in twee stukken breekt: het eindpunt van elk van de golffronten vormt dan een fasesingulariteit waarrond het overblijvende golffront zich opkrult. Op deze wijze leidt elk afgebroken golffront tot een spiraalvormig patroon van geactiveerd weefsel dat rond de tip van de spiraalcurve wentelt. Zulke spiraalgolven, die oorspronkelijk bestudeerd werden in de context van oscillerende chemische reacties, vertonen een merkwaardige dynamische stabiliteit. Eens spiraalgolven van elektrische activiteit gevormd worden in het hart verhogen ze de hartslag, aangezien hun omwentelingsfrequentie hoger is dan het natuurlijke hartritme. Deze toestand staat bekend als tachycardie; bovendien verlaagt de effici\"entie van de contractie bij tachycardie.

In werkelijkheid kent de ventrikelwand een eindige dikte waardoor geen spiraalgolven, maar rolgolven optreden. Deze kunnen worden beschouwd als een continue opeenstapeling van spiraalgolven. De tippen van de spiraalgolven vormen dan samen de rotatieas van de rolgolf. Deze draaias, die niet noodzakelijk recht hoeft te zijn, wordt het `filament' van de rolgolf genoemd. Filamenten zijn van groot belang voor de studie van hartritmestoornissen aangezien kon worden aangetoond dat voornamelijk perturbaties die plaatsvinden dichtbij het filament de tijdsevolutie van een ritmestoornis bepalen. Zo kan instabiliteit van \'e\'en rotorfilament resulteren in de voortdurende creatie van nieuwe filamenten. Dit proces stort het hart in een chaotische toestand van elektrische activering, gekend als fibrillatie. Wanneer fibrillatie plaatsgrijpt in de hartventrikels leidt dit tot de dood binnen enkele minuten. Ofschoon zowel tachycardie als fibrillatie reeds uitvoerig gekarakteriseerd werden aan de hand van rotorfilamenten, blijven inzichten betreffende de invloed van anisotropie van de hartspier op de stabiliteit van hartritmes erg beperkt. Men beschikte immers enkel over de laagste orde bewegingsvergelijkingen voor rotorfilamenten in een medium zonder anisotropie.

Op gelijkaardige wijze als voor golffronten worden in dit proefschrift de effectieve bewegingsvergelijkingen afgeleid voor rotorfilamenten in een medium met arbitraire anisotropie. Hieruit blijkt dat de beweging van een filament niet alleen bepaald wordt door zijn kromming en winding ten opzichte van het omringende medium: de intrinsieke kromming van de ruimte ten gevolge van structurele anisotropie werkt namelijk in op het filament onder de vorm van extra getijdenkrachten. De verkregen bewegingsvergelijkingen stellen in laagste orde dat filamenten enkel stationair kunnen zijn als ze langs een geodeet van de gekromde ruimte liggen. Dit resultaat bewijst de geldigheid van het `minimaalprincipe voor rotorfilamenten' dat in 2002 door Wellner, Pertsov en medewerkers werd geponeerd. Tevens voorspellen de hogere orde bewegingsvergelijkingen afwijkingen van het minimaalprincipe door optredende getijdenkrachten.

Uit lineaire stabiliteitsanalyse van de bewegingsvergelijkingen voor filamenten zijn diverse mechanismen af te leiden die rotorfilamenten instabiel kunnen maken en bijgevolg aanleiding geven tot hartfibrillatie. Immers, de effectieve spanning in het filament kan niet alleen negatieve waarden aannemen ten gevolge van excessieve kromming van het filament of winding van de rolgolf omheen het filament, maar eveneens wegens de getijdenkrachten die voortkomen uit de functionele anisotropie van het hartweefsel. De kritische waarden voor deze processen kunnen worden uitgedrukt aan de hand van de dynamische co\"effici\"enten in de bewegingsvergelijking, die op hun beurt berekend worden op basis van de tweedimensionale spiraalgolfoplossing.

In een vereenvoudigd anatomisch model neemt men vaak aan dat de draaiing van spiervezels doorheen de ventrikelwand aan een constant tempo verloopt. Voor deze speciale configuratie wordt de intrinsieke kromming van de ruimte expliciet uitgerekend, zodat het relatieve verschil in de voortplantingssnelheid van excitatiegolven in de verschillende richtingen eindelijk kwantitatief kan worden gekoppeld aan het optreden van hartritmestoornissen.
\\

Een laatste luik van dit proefschrift behelst de medische beeldvorming van structurele anisotropie in de hartspier met behulp van magnetische resonantie beeldvorming (MRI). In het bijzonder wordt een diffusie-MRI techniek vooropgesteld en uitgevoerd om kruisende kliefvlakken in het hart voor het eerst op niet-invasieve wijze in kaart te brengen. Deze methode werd `dual q-ball imaging' (dQBI) genoemd. dQBI is reeds een eerste maal gevalideerd door vergelijking met een standaard histologische techniek die naderhand werd uitgevoerd op hetzelfde hart.\\

Tot slot wordt er in deze scriptie beschreven hoe de intrinsiek gekromde ruimte bepaald kan worden voor een individueel hart op basis van MRI. De betreffende techniek werd Ricci Tensor Imaging (RTI) gedoopt. In combinatie met de wetten voor fronten- en filamentendynamica kan RTI het in de toekomst mogelijk maken om door te lichten of iemand gevaar loopt op hartritmestoornissen of voortijdige hartstilstand ten gevolge van een afwijkende hartstructuur.

\clearpage


\clearpage{\pagestyle{empty}\cleardoublepage}

\renewcommand*{\thesection}{\thechapter.\arabic{section}}       

\selectlanguage{english}
\graphicspath{{chapt_dutch/}{chapt2/}{chapt3/}{chapt4/}}
\renewcommand{\thesection}{\arabic{section}}    

\renewcommand{\bibname}{References}
\renewcommand\evenpagerightmark{{\scshape\small English summary}}
\renewcommand\oddpageleftmark{{\scshape\small English summary}}

\chapter[English summary]%
{English summary}

\hyphenation{re-presented}

The present work resides in the field of mathematical biophysics. Specifically, it analyzes how cardiac structure affects the propagation of waves of electrical depolarization through the heart. Because these waves trigger the individual heart cells to contract mechanically, aberrant electrical activity can lead to heart rhythm disorders that may weaken or even fully suppress the organ's pumping function.\\

In mathematical modeling terms, the heart is considered to be an excitable medium. When the transmembrane potential of a cell is elevated above a given threshold voltage, an activation cycle known as an action potential is initiated. During the action potential, which typically lasts about 300\,ms in a human ventricular cell, the myocardial cell depolarizes and contracts along its long axis. In cardiac tissue, action potentials are conducted between neighboring cells through intercellular gap junctions, a phenomenon that may be modeled by diffusion of the electric transmembrane potential. During normal heart rhythm, the propagating wave of electrical excitation triggers properly timed and coordinated contraction of the heart muscle, which results in a single heartbeat.

Macroscopic wave patterns in excitable biological tissue are commonly represented mathematically using a reaction-diffusion equation, in which the transmembrane potential is represented by one of the state-variables. Notably, only the transmembrane potential undergoes diffusion at the time scale considered. The mathematical description of this electrical diffusion process is made more complex by the pronounced fibrous structure of the heart muscle and the presence of cleavage planes in the tissue, both of which have significant effects on the velocity and direction of wave propagation in cardiac tissue. For this reason, an anisotropic electrical diffusion tensor must be incorporated into the model to represent the fact that the waves of electrical depolarization propagate about three times faster along the myofiber direction than at right angles to it. Both experimental and numerical evidence suggest that insights on the role of functional anisotropy of the tissue may be essential to understand how certain types of arrhythmias are initiated. At the start of this research project, however, no systematic scheme to handle generic tissue anisotropy in an analytical way had been developed.\\

In this work, anisotropy is studied on the basis of a local equivalence principle, because isotropy may be restored in a small tissue block with fixed fiber direction by selecting appropriate length units in each principal direction. This choice is tantamount to measuring distance based on arrival times rather than physical length, which is commonplace in a number of other fields, such as optics and navigation. The consistent application of this operational measure of distance in every patch of myocardial tissue is sufficient to regain local isotropy in the equations, although the procedure imposes an intrinsic geometric curvature on the space considered. In other words, the dynamical patterns of electrical activity in the heart behave as if they reside in a non-Euclidean space, determined by a metric tensor that is proportional to the inverted electrical diffusion tensor.

Fortunately, there exists another natural phenomenon that can be elegantly described within an intrinsically curved space: gravity. In Einstein's general relativity theory, both light rays and test bodies locally travel along straight lines, using a description of spacetime that is curved because of nearby heavy masses. This unexpected parallelism encourages the use of mathematical techniques and physical insights from cosmology for an unconventional biophysical application. Because the evolution equations for the patterns of electrical excitation are derived in this work within a covariant formalism, they are obeyed for any type of tissue anisotropy. \\

As a first application in this dissertation, the novel curved-space formalism is used to investigate the relationship between the propagation speed of action potential waves and the local geometric curvature of the wave fronts. In the linearized equation, which is commonly termed the eikonal relation, the coefficient of linearity derived here is more general than the outcome of the classical proof and may be assigned the meaning of a physical surface tension. It is furthermore shown that the equation of motion for the wave front can be derived from a variational principle; in the physical action terms appear that denote the increase in occupied volume and the total surface area of the wave front.

This work additionally contains the original corrections for wave front motion that are of second order in curvature. For isotropic media, the corrections deviate from those usually found in literature. Remarkably, the structural anisotropy of cardiac tissue induces a net effect on wave front motion that is reminiscent of gravitational lensing in cosmology.

This research next addresses how repeated electrical stimulation affects the propagation of action potentials. Our findings deliver a general expression for the dispersive surface tension coefficient. The outcome may be used to estimate a dimensionless critical ratio that is used to assess vulnerability of the heart to the onset of turbulent cardiac rhythms.\\

The second application of the geometric interpretation of structural anisotropy concerns spiral waves. Together with the wave fronts already described, spiral waves are solutions to the reaction-diffusion equation in two spatial dimensions. Spiral shaped activity may follow from break-up of a wave front: in such a case, each of the endpoints of the broken fronts acts as a phase singularity, around which the remainder of the wave front winds. As a result, each broken wave front gives rise to a spiral shaped pattern of activated tissue that rotates around the spiral's tip. These spiral waves were originally encountered and investigated in the context of oscillating chemical reactions and can exhibit remarkable dynamical stability. Because the rotation frequency of the spiral waves is higher than the heart's natural rhythm, a spiral wave of electrical activity that forms in the heart leads an to an accelerated beating rate. Such state is known as tachycardia and decreases the heart's pumping efficiency.

Due to the finite width of the ventricular wall, three-dimensional scroll waves, rather than two-dimensional spirals, are encountered in the heart. Scroll waves may be envisioned as a continuous stack of spiral waves, with the collection of spiral wave tips forming the rotation axis of the scroll wave. This rotation axis, which may be curved and twisted, is called the `filament' of the scroll wave. Scroll wave filaments are key to the study of cardiac arrhythmias, since it has been demonstrated that mainly perturbations that take place close to the filament affect the temporal evolution of aberrant heart rhythms. In particular, the dynamical instability of a single filament may result in the continuous creation of new filaments, which results in a chaotic state of electrical activation known as cardiac fibrillation. Although both tachycardia and fibrillation have been characterized extensively by means of scroll wave filaments, specific insights on the influence of structural anisotropy on the stability of heart rhythms have been limited, in part because the equations of motion for filaments had only been derived for isotropic media, for a small regime of spiral wave trajectories and in lowest order in curvature and twist.

In a similar manner to the treatment of wave fronts, this dissertation presents a derivation of the effective equations of motion for filaments that are third order in curvature and twist; moreover they hold for media with arbitrary anisotropy. These equations reveal that filament motion depends on both curvature and twist of the filament with respect to the surrounding medium. In addition, the intrinsic curvature of space due to structural anisotropy acts on the filament as a tidal force. The covariant equations of motion state in lowest order that any stationary filament in a medium with generic anisotropy must lie along a geodesic of the curved space, which proves the `minimal principle for rotor filaments' that was put forward by Weller, Pertsov and co-workers in 2002. Furthermore, the higher order equations of motion predict deviations from the minimal principle arising from tidal forces.

From linear stability analysis of the filament equations of motion, several pathways to filament instability may be deduced, each of which is likely to enable the onset of cardiac fibrillation. It is shown that not only excessive extrinsic curvature or twist of the filament, but also the tidal forces that stem from myocardial anisotropy  may cause negative filament tension. The instability thresholds for these processes are expressed using the dynamical coefficients in the equation of motion and can be explicitly calculated from the two-dimensional spiral wave solution.

In a simplified anatomical model, the fiber rotation rate across the ventricular wall is often approximated as constant. For this particular configuration, the intrinsic curvature of space is characterized, such that the relative difference in the conduction velocity of excitation waves in different directions may finally be linked to the development of cardiac arrhythmias in a quantitative way.\\

A separate theme touched upon in this dissertation is the analysis of structural anisotropy in the heart muscle using magnetic resonance imaging (MRI). More precisely, a diffusion-MRI technique is developed and tested to resolve coexisting cleavage planes of different orientation in the heart. The novel  method, termed `dual q-ball imaging' (dQBI), for the first time enables non-invasive discernment of cleavage plane crossings in macroscopic tissue volumes. A preliminary validation was conducted by comparing the outcome from dQBI against standard histology performed on the same heart.\\

Ultimately, this thesis presents an integrative approach to determine the intrinsically curved space for individual hearts using a method called Ricci tensor imaging (RTI). Combined with the laws for wave front and filament dynamics, RTI facilitates the assessment of patient-based risk for cardiac arrhythmias or premature cardiac death as a consequence of aberrant cardiac microstructure.

\clearpage


\clearpage{\pagestyle{empty}\cleardoublepage}

\renewcommand*{\thesection}{\thechapter.\arabic{section}}       

\mainmatter     
\selectlanguage{english}
\renewcommand*{\thesection}{\thechapter.\arabic{section}}

\newcommand\fdtsvrightmarktmp{{\scshape\small Chapter }}
\renewcommand\evenpagerightmark{{\scshape\small\chaptername\ \thechapter}}
\renewcommand\oddpageleftmark{{\scshape\small\leftmark}}

\renewcommand\evenpagerightmark{{\scshape\small Introduction}}
\renewcommand\oddpageleftmark{{\scshape\small Introduction}}

\hyphenation{bio-log-i-cal }

\chapter[Introduction]{Introduction}
\label{chapt:intro}

\section{Introduction}

About every second of our lives, an electrical pulse in our hearts incites the organ to fulfill its pumping function and thereby keeps us alive. The importance of coherent delivery of such activation pulse is particularly felt when heart rhythm disorders arise. For, the failure to timely deliver the electrical trigger in every contractile heart cell obviously leads to unsynchronized contraction of individual heart cells. Conversely, experimental recordings of electrical activity in diseased hearts have shown aberrant activation patterns on their surface.\\

This work does not delve into the intricate electrophysiological processes that underlie heartbeats, nor attempts to solve the partial differential equations that govern the subsequent contraction. Rather, we concentrate on a physical description for the propagation of electrical potentials at the macroscopic level, where distinct wave fronts can be seen to travel through the myocardial wall. The macroscopic properties of cardiac tissue can be efficiently captured by a set of reaction-diffusion (RD) equations \cite{Keener:1998book}. The first state variable in such model is usually chosen to be the local electric potential difference across cell membranes, i.e. the transmembrane potential. The vector of state variables is typically further supplied with intra- and extracellular ion concentrations and
regulatory factors for ion exchange across the cell membrane. The `diffusion' part of the RD system is embodied by adding a diffusion term for the electrical potentials that sustains the spatial spread of the electrical activation.

Reaction-diffusion kinetics are not uniquely reserved to cardiac excitation, as a similar diffusion term appears in the context of oscillating chemical reactions \cite{Winfree:1984b} and pattern formation on animal fur \cite{Murray:1993}. Other waves of different nature that have been studied using reaction-diffusion models include flame front propagation \cite{Zeldovich:1938}, catalytic oxidation on metal surfaces \cite{Jacubith:1990}, the aggregation of social amoebae \cite{Siegert:1992} and the spreading of agricultural techniques \cite{Fort:1999} as well as epidemics \cite{Smith:2001}. In biophysics, RD systems are thought to play a role in the secretion of insulin, in maintaining our biological clocks and with the rhythmic contraction of the uterus when giving birth. Because our analytical theory that is presented below does not specify a particular reaction-diffusion model, the obtained results are expected to be applicable to the other mentioned dynamical systems under minor modifications.\\

Characteristic to the process of cardiac excitation is that conduction takes place in a three-dimensional volume, which allows for much richer dynamics than possible on a two-dimensional surface or in a one-dimensional cable. The second complication lies in the fibrous structure of the heart muscle: as the heart cells are elongated and mainly coupled to each other through their end caps, electrical signals spread about three times as fast in the direction of cell alignment than in the transverse directions \cite{Caldwell:2009}. This effect is known as tissue anisotropy, and developing an analytical description of wave phenomena in the heart that deals with any reasonable arrangement of myofibers formed the main goal of this study.

At this point, a beautiful tie with `mainstream' theoretical physics comes in: tissue anisotropy can be locally removed from the description by appropriate rotation and rescaling of the coordinate axes. Such rescaling leaves us with a background space which is intrinsically curved, and the situation is -at least in the eyes of a theoretical physicist- highly reminiscent to Einstein's general theory of relativity \cite{Einstein:1916}. Consequently, this key remark enables to draw from the mathematical tools designed for working in curved spacetimes. One should therefore not be surprised to encounter covariant differentiation, relativistic spacecraft and Riemann curvature tensors further up in this work, originally applied here in a biophysical context. 

The spacetime analogy is even pushed further, as the time evolution of phase singularity lines (called `filaments' \cite{Clayton:2005}) can be correlated to the cosmic strings that have been hypothesized to exist on galactic scales \cite{Vilenkin:1985}. Importantly, filaments act as the organization centers for the time evolution of cardiac arrhythmias and therefore their study is thought useful for future treatment and prevention of arrhythmias. The bold parallelism between cosmic strings and filaments in the heart paid off immediately: borrowing an expansion from cosmic string literature, our group was able to prove the minimal principle for rotor filaments that had been postulated by Wellner, Pertsov and co-workers in 2002 \cite{Wellner:2002}. Higher order terms in the gradient expansion series provide a more general analytical framework for various facts on filament behavior that have been observed in numerical simulations and experimental context \cite{Berenfeld:1999, Berenfeld:2001, Fenton:1998}. The extended laws of motion that govern filament dynamics in the presence of tissue anisotropy are for the first time derived and discussed in this thesis.\\

A side track of the research performed, which eventually grew into a relatively independent research topic, is the usage of magnetic resonance imaging (MRI) to assess the geometry and microstructure of individual hearts. Rooted in personal experience during Master Thesis research at the Faculty of Engineering in Ghent, an established methodology to probe complex fiber orientation in specific brain regions was applied to obtain detailed images of the cardiac myofiber field. This existing technique that probes fiber orientation is known as q-ball imaging (QBI) \cite{Tuch:2004}, and was for the first time exploited to image complex myocardial fiber structure context within our research project.

Specific to cardiac microstructure is the occurrence of laminar clefts in the tissue \cite{Gilbert:EJCTS}, which are believed to play an important role in the mechanical contraction of the heart \cite{Costa:1997}. Based on physical principles, we adjusted the QBI procedure so that it became sensitive to the orientation of laminar structure instead of fiber organization. The novel technique, which we have termed dual q-ball imaging (dQBI) is the first one that attempts to non-invasively distinguish cleavage planes of different orientation that could coexist at a given point in the tissue. Interestingly, our method indicates that multiple laminar structure could be more prevalent in the cardiac wall than anticipated by former MRI studies; the results are in qualitative agreement with the outcome of histological sectioning. At present, histological sectioning combined with confocal microscopy is considered the gold standard for mapping cardiac microstructure \cite{Sands:2005}. Unfortunately, sectioning methods are inherently destructive and therefore possess limited clinical potential.\\

The two areas of active research, i.e. structural imaging and the asymptotic theory for wave patterns, can be integrated with each other into a methodology for assessing structural susceptibility to electrical instability in individual hearts. The foundations for such method are put down in the last chapter; the basic idea is to predict the forces that could act on occurring rotor filaments using information on local fiber and laminar orientation that has been gathered with MRI.\\

One could argue whether this thesis contributes to medical science in general. Indeed, in this research project no physiological experiments were performed and the numerical simulations executed were limited compared with the contemporary state-of-art computing in the field. Nevertheless, we are convinced to practice useful science, viewed from the following standpoint. Numerous pieces of the puzzle have been collected in many scientific communities, through \textit{in vitro} or numerical experiments, via clinical practice and experience, by putting up models and theories that go with them. Being theoretical physicists, we are hitherto no front-line combatants in clinical advances; as true linkmen we have rather rejoiced putting together the pieces of the puzzle in different patterns, hoping that the general picture someday will show itself. Even before that day, we continue to believe that our interdisciplinary efforts can contribute to better understanding of dangerous arrhythmias. And that could be a first step towards enhanced therapy and wellbeing for the general public.

\section{Outline and scope}

The tight link with cardiological application is yet presented in the first chapter of the main text, which briefly introduces the types of cardiac arrhythmias which this work may be relevant to. Thereafter, a bottom-up approach is followed to reconstruct how the life-keeping (sometimes life-threatening) waves of electrical activity emerge from underlying physiological processes. The myocardial tissue structure at the mesoscopic scale deserves our attention as the emergent tissue anisotropy strongly impacts on the behavior of the patterns of electrical excitation at the tissue and organ level. In the cited derivation of the reaction-diffusion equation that governs cardiac excitation patterns, a distinction is made between the so-called monodomain and bidomain formalisms. The results in this work are derived here only for the monodomain formulation.

Like good quality comic strips in Belgium that tend to divide an interesting story over two albums, the third and fourth chapter make out a diptych on diffusion-weighted MRI of cardiac microstructure. First, physical principles that underly diffusion imaging are summarized, after which it is shown how these give rise to the popular technique of diffusion tensor MRI, which is also called diffusion tensor imaging (DTI) \cite{Basser:1994}. Nowadays, DTI has become the workhorse for the assessment of fibrous and laminar structure in soft biological tissues. Reconstructions of both fibrous and laminar structure using this technique are presented for a rat heart; the data set was collected at the University of Leeds in cooperation with Dr. Olivier Bernus, Dr. Stephen Gilbert and Dr. Alan Benson.

After a discussion on the limited scope of DTI when dealing with microstructure that is more complicated than orthotropic, the fourth chapter steps into the world of high-angular resolution diffusion imaging. In this field, competing techniques aim to resolve more complex microstructure than possible with DTI alone. Fortunately, the contemporary scientific focus for diffusion MRI lies in neurological applications, leaving us the opportunity to adopt the relatively time-efficient method of q-ball imaging \cite{Tuch:2003} for viewing cardiac myofibers with increased angular resolution. Next, we show how to adjust q-ball method to sense laminae instead of fibers; it turns out that one is led to interpret the raw diffusion signal as representing the angular distribution of laminar structure. Despite its conceptual simplicity, the novel technique, which we baptize `dual QBI', is shown to resolve crossing myocardial sheet populations without prior dehydration of the hearts.

After the digression on medical imaging, we return in the fifth chapter to our initial goal, namely to develop a model-independent framework for wave propagation that can deal with generic tissue anisotropy. Importantly, we introduce the notion of an operationally defined distance. Whereas the concept of isochrones is deeply rooted in experimental practice, the step to commit physics in this convention has only sparsely been made. Note that this is the point where the announced parallelism with Einstein's theory of gravity sets in.

As an immediate corollary of the curved space formalism, we revise in chapter six the velocity-curvature relation for wave fronts. Our framework is flexible to such a degree that both curvature of the medium and anisotropy within it can be treated on the same foot. Moreover, as in Mikhailov \cite{Mikhailov:1994}, we obtain a general coefficient of linearity in the velocity-curvature relation, to which the meaning of surface tension can be associated. We originally derive the second order curvature terms in the velocity-curvature relation, including the effect of generic tissue anisotropy. Finally, we restate the propagation of wave fronts as a variational problem, and show how high frequency pacing affects the surface tension of propagating fronts, similar to \cite{Wellner:1997, Pertsov:1997}.

Chapters seven and eight are reserved for filament dynamics in isotropic and anisotropic media, respectively. After defining a coordinate frame that is locally adapted to the filament, higher order correction terms in filament curvature and twist are established for the isotropic case. In the extended equation of motion (EOM) for filaments, twist is seen to couple to translational motion and hence the so-called sproing instability \cite{Henze:1990} can be explained without resorting to a phenomenological model \cite{Echebarria:2006}. The dynamical coefficients that arise in the translational and rotational EOM are obtained as integrals over the Goldstone modes and response functions of the corresponding two-dimensional spiral solution.

Next, we redo our calculations for arbitrary anisotropy types in an proper system of reference, known as Fermi coordinates in a general relativity context \cite{Manasse:1963}. In lowest order, we thus prove the minimal principle that was coined by Wellner and Pertsov, stating that scroll wave filament equilibrate on geodesic curves if one defines the metric as the inverse of the electric diffusion tensor. The next-to-leading order terms that we obtain reveal that local fiber rotation rate explicitly acts on filaments, since different components of the Ricci curvature tensor emerge in the EOM. Our generalized EOM for filaments is valid in the circular core regime, and incorporates various phenomena observed in forward filament simulations described in literature.\\

To conclude, we synthesize the preceding by combining structural imaging with the functional dynamics from chapters six to eight. For, from detailed diffusion MRI measurements, local fiber and laminar orientation can be inferred, which determines the local axes of orthotropy and therefore wave speed. Subsequently, the Riemann curvature tensor can be calculated from the second order spatial derivatives of the electrical diffusion tensor and fed to the filament EOM.\\

The very last chapter presents an overall conclusion to the research conducted. Interestingly, we have been able to establish quite general results, which show dependency on underlying electrophysiology only through the dynamical coefficients in the equations of motion for wave fronts and filaments. Therefore, we are inclined to say that, in spite of the variety of phenomena observed, wave front and filament dynamics in anisotropic cardiac tissue are remarkably universal.

\clearpage

\clearpage{\pagestyle{empty}\cleardoublepage}

\graphicspath{{fig/}}

\renewcommand\evenpagerightmark{{\scshape\small Chapter 2}}
\renewcommand\oddpageleftmark{{\scshape\small Cardiac function}}

\hyphenation{Na-gu-mo atrio-ventricular}

\chapter[Basic cardiac function]{Basic cardiac function}
\label{chapt:struct}


Essential to understanding the cardiac physiology is the observation that all seems to serve the higher goal of reliable, rhythmic contraction as to ensure the organ's pumping function. Coordination of mechanical activity occurs through electrical pulses, called action potentials, which can travel fast between neighboring cells to effectuate nearly synchronous contraction.

We take a bottom-up approach here to see how the physiology of myocardial tissue provides a substrate for the propagating waves of electrical activity. Subsequently, we review the most important patterns of electrical excitation that have been observed during cardiac arrhythmias. A renewed theoretical description of these activation patterns forms the major research goal of this work.

\section{Gross anatomy and function of the heart}

\begin{figure}[h!t] \centering
  \includegraphics[width=1.0 \textwidth]{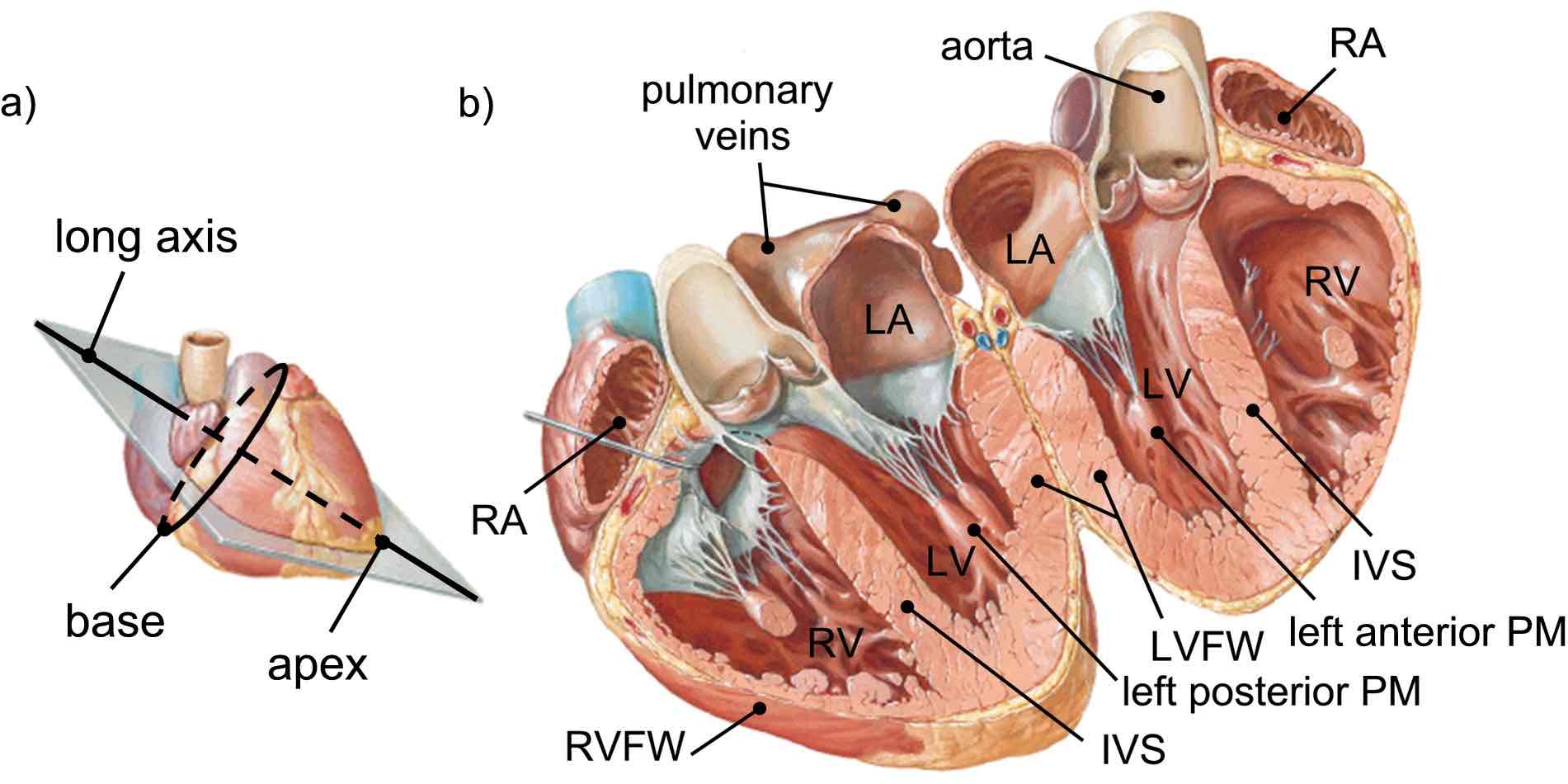}\\
  \caption[Anatomical cross-section of the human heart]{Anatomical cross-section of the human heart, adapted from \cite{Netter:2006}. Panel (a) indicates the plane of section and the natural degree of tilting of the human heart. In panel (b), the four heart chambers are labeled, as well as the papillary muscles (PM) in the LV. Herein, LA and RA denote the left and right atrium, respectively. The lateral sides of both ventricles are bounded by their respective free walls (FW) and the interventricular septum (IVS).} \label{fig:Netter}
\end{figure}

Since William Harvey's discovery in 1628, it is known that the heart muscle pumps the blood around in a closed circulatory system. Only by an uninterrupted unidirectional movement, the blood can fulfill its main functions, of which providing oxygen to the different organs is the most important \cite{Greger:1996}. Mammals and birds have a double-looped circulatory system; therefore the heart pump comprises a left and right halve, each consisting of a ventricle connected to an atrium; see Fig. \ref{fig:Netter}. The intake of blood occurs through the atria, which contract first during a cardiac cycle to fill the ventricles. Thereafter, the ventricles strongly contract to push the blood in its circulatory loop, while valves prevent the blood from flowing back into the atria. One half of the circulatory system is the pulmonary circulation: partially oxygen-depleted blood is taken in the right ventricle (RV) through the right atrium and propelled through the pulmonary artery to the lungs. The oxygenated blood returns to the heart by the pulmonary veins, which end up in the left atrium. This concludes the pulmonary circulation. Next, the blood enters the systemic circulation, which carries it through the left ventricle (LV) and aortic artery, and distributes it to the rest of the body, before it returns to the right atrium. In the heart, both circulation loops are separated by the interventricular septum (IVS) and interatrial septum, which are muscular walls that prevent mixing of oxygenated and oxygen-depleted blood. Adapted to occurring fluid pressures, the left ventricular free wall (LVFW) and interventricular septum are much thicker than the right ventricular free wall (RVFW); the latter is still considerably thicker than the outer atrial walls and interatrial septum.

The smooth outer surface of the heart muscle is known as the epicardium, in contrast to the endocardium which denotes the inner surface. The endocardial surface of the ventricles is manifestly uneven; papillary muscles (PM) are attached to the endocardial wall, and connect to the tricuspid and mitral valves in RV and LV respectively. The apex is the lower tip of the heart, whereas the base denotes the separation zone between atria and ventricles. The heart's long axis runs from apex to base; cross-sections perpendicular the long axis are denoted axial, short-axis or transverse slices. The zone where axial cross-sections reach their largest size is referred to as equatorial.

\section{The cellular basis for electrical activation}

\subsection{Membrane potential and currents}

A living cell cannot escape the fundamental laws of physical equilibrium. Given that the lipid membrane surrounding the cell acts as a barrier with different permeability for the various ions, equilibrium demands a balance between the electrostatic force across the membrane and the diffusive current that results from unequal ion concentrations in the intra- and extracellular spaces. In electrophysiology, transmembrane potential refers to the potential inside the cell, with respect to the extracellular potential, i.e.
\begin{equation}\label{potential}
  V_m  = V_{\rm int} - V_{\rm ext},
\end{equation}
and currents flowing into the cell are considered negative. With these conventions, the Nernst equation can be written for the equilibrium membrane potential $E$ that goes with a single ion species (denoted $I_j$) of charge $q = z e$:
\begin{equation}\label{Nernst}
  E_j = \frac{RT}{z e N_A} \ln \left( \frac{[I_j]_{\rm ext}}{[I_j]_{\rm int}} \right).
\end{equation}
The pre-factor involving the ideal gas constant $R$ and Avogadro's number $N_A$ evaluates to $61.5$ mV at human body temperature $T$ for $z=1$. In reality, the net membrane potential results from an overall equilibrium between all ionic fractions. 

For a typical mammalian heart cell in its resting state, the dominant cation in extracellular space is sodium ($[\rm{Na}^{+}]_{\rm ext} = 145\,$mM), with chlorine being the principal anion ($[\rm{Cl}^{-}]_{\rm ext} = 123\,$mM). Inside the resting cell, potassium has the highest concentration ($[\rm{K}^{+}]_{\rm int} = 140\,$mM), against $4\,$mM in the extracellular space \cite{Boyett:CBH}. Another important ion is calcium, which drives the contractile outbursts of the cell despite its relatively small concentrations: $[\rm{Ca}^{2+}]_{\rm int} = \unit{0.1}{\micro M}$ and $[\rm{Ca}^{2+}]_{\rm ext} = 1.5\,$mM.

The final resting potential of the heart cells balances around $E_m \approx - 80\,{\rm mV}$ and therefore a cell in relaxed state is said to be negatively polarized. Hence, a sudden increase of membrane permeability would cause a transient passive inward current, which strives to depolarize the cell. Such dramatic increase of membrane permeability (for sodium ions) is precisely the mechanism that is in the first stage of the action potential responsible for the outbursts of electric activity in the heart, which is the further topic of the present study.

\subsection{The action potential}

\begin{figure}[h!t] \centering
  \includegraphics[width=0.5\textwidth]{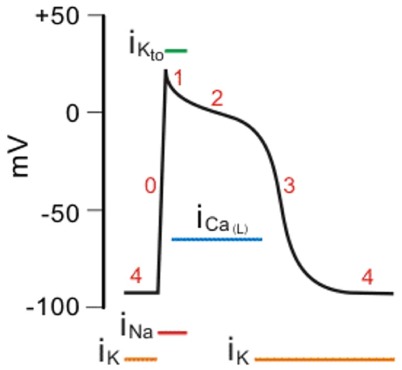}\\
  \caption[Transmembrane voltage during an action potential]{Typical action potential cycle in ventricular myocyte, with indications of the most important currents that sustain the action potential in each phase (adapted from \cite{Klabunde:2004}).}\label{fig:AP}
\end{figure}

The transport of ions through the membrane is not a random process. Membrane channel proteins carry active or passive ion currents, depending on their nature and state.
In cardiac myocytes, the ion channel characteristics are tuned in such a way that the cells loop through a depolarization cycle, which is the action potential that we have used above to discuss the patterns of electrical activity in the heart. A typical action potential of an excitable cell consists of four phases that follow the upstroke \cite{Jalife:BCEC}, as depicted in Fig. \ref{fig:AP}:
\begin{enumerate} \setcounter{enumi}{-1}
  \item \textbf{Action potential upstroke.}
        Once the cell reaches a threshold level of about $-65\,$mV (e.g. by electrode stimulation or excitation by a neighboring cell), the membrane channels for sodium suddenly open up. The abundance of extracellular Na${}^+$ ions rushes into the cell, while the K${}^+$ channels, which previously caused a constant outward current, become shut. This rapid depolarization process invokes a surge in the transmembrane potential that lasts few milliseconds and which is referred to as the action potential upstroke.
  \item \textbf{Rapid repolarization.}
        When $V_m$ reaches the level of $+20\,$mV, the upstroke is terminated by closure of the sodium channels. At the same time, a transient outward current arises, which is mainly due to increased potassium conductance. The initially rapid repolarization gives the action potential its characteristic spike that marks the end of the depolarization phase.
  \item \textbf{Action potential plateau.}
        During upstroke, additional slowly activation currents are also initialized, only becoming noticeable after the transient outward current has died out. In a simplified view, the dominant currents in this phase are an inward calcium current and the delayed rectifying potassium outward current. The balance of these currents yields a nearly constant membrane potential, delivering a plateau phase that lasts $200$ to $300\,$ms.
  \item \textbf{Final repolarization.}
        Inactivation of the calcium channels causes the cell to conclude its repolarization, marking the end of the plateau phase. The still active potassium currents bring the transmembrane potential back to its resting value.
  \item \textbf{Diastolic potential.}
        In myocytes that do not belong to natural pacemaker tissue, the inward rectifying potassium current remains the dominant conductance at rest, and sets the resting membrane potential. This equilibrium is maintained until a new stimulus invokes the next action potential.
\end{enumerate}

The shape and duration of the action potential as well as current densities vary between specialized conducting tissue types in the heart, and moreover with species and age.\\

With each action potential, a time interval is identified during which no new action potential can be triggered. This \textit{absolute refractory period} lasts from the upstroke until about the final repolarization phase. Then follows the \textit{relative refractory period}, during which an additional action potential can be elicited only through a stimulating current that is larger than the one needed to excite a fully recovered cell. Obviously, the duration of the refractory period is determined by the time course of the ion channel conductances.

An immediate consequence of the presence of a refractory period is that colliding waves in such medium annihilate, for it is impossible for one of the waves to travel through the refractory wave tail of the other. The annihilation property is an important distinction with solitons, which are `protected' by conservation laws and therefore exit mutual collisions unaltered.

\section{Models of cardiac excitation}

\subsection{Development and role of models for cardiac excitation}
The common ancestor to modern ionic models of bioelectrical signaling is the Noble-prize winning work by Hodgkin and Huxley, who investigated action potential generation in the squid giant axon \cite{Hodgkin:1952}. Their pioneering model involved activation and inhibition variables for the sodium and potassium currents, and was able to represent action potential formation. The first model to cover cardiac action potentials was established in 1962 by D.~Noble and co-workers \cite{Noble:1962}; since then cell models have grown more and more sophisticated, nowadays involving sometimes over 200 variables. At present, different detailed ionic models have been developed that aim to faithfully represent physiological reality in various cardiac cell types and animal species (see e.g. the reviews \cite{Clayton:2001} and \cite{Clayton:2010}).

Unfortunately, the greater detail in models that strive to faithfully approach physiological reality has not always enhanced predictive power, since a multitude of parameters needs be tuned to match experimentally obtained curves. Additionally, as the information gained on some specific cell processes is rather limited, fitted parameters are occasionally extrapolated towards different temperature or physiological background parameters, between tissue types, or even across various animal species \cite{Niederer:2009}. Also, recent studies have shown considerable robustness of action potential formation against channel modifications, which hints that apparently, we are merely starting to catch sight of nature's many built-in buffers \cite{DNoble:2009}. Whereas the detailed ionic models are undisputedly contributing to the development of anti-arrhythmic drugs and are needed to capture $\rm{Ca}^{2+}$ dynamics during mechanical contraction, they are unlikely to always provide an accurate description of the occurring processes.

One should keep in mind that most detailed cardiac models are being conceived to mimic reality for a given physiological state of the tissue and moreover depend on animal species represented. Also, it cannot be excluded that more intricate feedback loops act in reality, which could make model predictions deviate from experimental behavior. These arguments inspired the saying among cardiac modelers that \textit{``All models are wrong, but some of them may be useful.'} Recent reviews on cardiac modeling include \cite{Cherry:2007} and \cite{Clayton:2010}.   

\subsection{Mathematical formulation of cardiac excitability}
In mathematical terms, a coupled system of first-order differential equations in time may capture how a single cell reacts as state variables change:
\begin{equation}\label{RDE_point_system}
  \dd_t \uu = \mathbf{F}(\uu).
\end{equation}
Usually, the transmembrane voltage of the cell is used as the first state variable, i.e. $u_1 \equiv V_m$. Obviously, the cell's resting state lies at a stable equilibrium point of the system \eqref{RDE_point_system}.

In opposition to the growing complexity within detailed ionic models, model reduction strategies have been used to speed up numerical computations, while preserving the crucial feedback mechanisms (see for an example \cite{Bernus:2002a, tenTusscher:2004b}). Also, the reduced stiffness of the partial differential equations allows one to choose a larger space unit, which also accelerates numerical computations (see e.g. \cite{Bernus:2002b}). Other modelers take even a more drastic viewpoint, since simple low-dimensional models with a modest number variables \cite{Fitzhugh:1961, Aliev:1996, Fenton:2002, BuenoOrovio:2008} and relatively simple reaction kinetics seem to exhibit similar complexity in the emerging patterns as the more detailed models do \footnote{Anticipating our results from Chapters 5-9, we would pretend that any excitable medium which is modeled as a reaction-diffusion system functionally behaves quite independently of the details of the underlying processes as soon as traveling wave solutions are supported.}. In the simpler models of cardiac excitability, the first variable usually stands for the rapidly varying transmembrane potential, while few supplementary variables account for the slower recovery processes. The small number of parameters in the low-dimensional models can still be adjusted to represent measured quantities such as excitability, action potential duration, refractory properties, plateau shape and height.

\begin{figure}[h!b] \centering
  \includegraphics[width=0.8\textwidth]{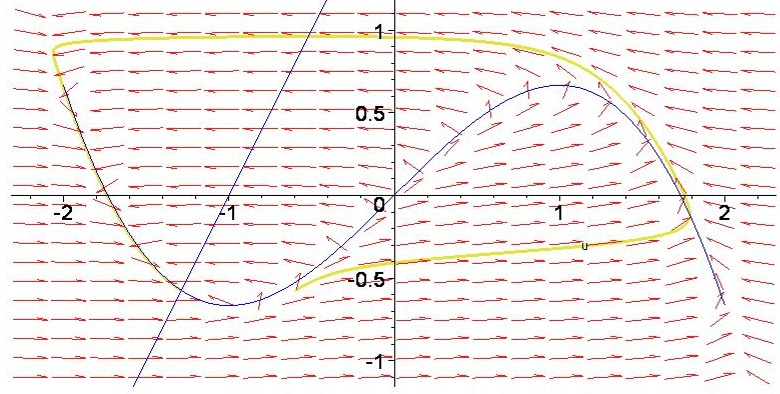}\\
  \caption[Excitation cycle for the FitzHugh-Nagumo system]{Phase portrait (u,v) and excitation cycle for the FitzHugh-Nagumo system with parameters $\beta=0.1$, $\gamma=0.5$ and $\eps=0.2$. Nullclines are drawn in blue.}\label{fig:FHN}
\end{figure}

A frequently used simplification of excitable dynamics is the FitzHugh-Nagumo model \cite{Fitzhugh:1961, Nagumo:1962}. The system is two-dimensional:
\begin{equation}\label{RDE:2dim}
  \begin{cases}
  \dd_t u &= f(u,v),\\
  \dd_t v &= g(u,v),
  \end{cases}
\end{equation}
with reaction functions $f(u,v)$ and $g(u,v)$ that read, in the notation of \cite{Panfilov:CBH},
\bsub \label{RDE:FHN}\begin{eqnarray}
  f(u,v) &=& \frac{1}{\eps} \left(u-\frac{u^3}{3} -v \right), \\
  g(u,v) &=& \eps \left( u + \beta - \gamma v \right),
\end{eqnarray} \esub
where typically $0 < |\beta| < \sqrt{3}$, $0 < \gamma < 1$ and $\eps \ll 1$. The advantage of two-dimensional systems is that their full phase portrait can be drawn, for the FitzHugh-Nagumo system (see Fig. \ref{fig:FHN}). The unique intersection of the nullclines betrays the position of stable equilibrium point, which corresponds to the resting state; a small increase in $u$ causes the system to go through an excitation cycle before returning to the rest state. From Eqs. \eqref{RDE:FHN}, the ratio of timescales for activation ($u$) and recovery ($v$) processes is seen to equal $\eps^2$. Other two-dimensional models of cardiac excitation with continuously differentiable kinetics include the so-called Barkley model \cite{Barkley:1990} and Aliev-Panfilov model \cite{Aliev:1996}. Both mentioned models have a phase portrait which is qualitatively similar to the FitzHugh-Nagumo model.\\

In the context of this work, a distinction should be made between those models that possess continuously differentiable reaction functions and those who have not. Historically, piecewise linear reaction functions have been used to approximate continuous models to enable analytical solutions for e.g. action potential duration \cite{Krinsky:1972}. More recently, low-dimensional models have been constructed in which time constants and gating variables change abruptly when the transmembrane voltage exceeds a particular threshold \cite{Fenton:1998, BuenoOrovio:2008}. The resulting models are suited for numerical simulation, as they evaluate quickly and can be semi-empirically adapted to represent different excitable cell types, animal species and pathological situations. In the light of our analytical theories, however, the class of models with non-continuously differentiable reaction kinetics is less appealing, as the perturbation operator which results from linearization of the reaction functions is ill-posed.

\subsection{Propagation of excitation along the cell membrane \label{sec:bidomA}}

The electrical properties of elongated biological cells are commonly captured by a one-dimensional cable equation. Here, electrical conductances in the intra- ($g_{\rm int}$) and extracellular space ($g_{\rm ext}$) and voltage-dependent membrane currents $i_m$ enter the equations. 

To derive the equation for transmission of cardiac excitation \cite{Holden:CBH}, one observes that the currents $i_{\rm int}$, $i_{\rm ext}$ that run tangential to both sides of the cell membrane (say in the x-direction) are given by
\begin{align}
 i_{\rm ext} &= - g_{\rm ext} \dd_x V_{\rm ext},  &  i_{\rm int} &= - g_{\rm int} \dd_x V_{\rm int}.
\end{align}
Conservation of current additionally imposes that these currents relate to the transmembrane current $i_m$ through $\dd_x i_{\rm ext} = i_m = - \dd_x i_{\rm int}$, with $i_m = c_m \dd_t V_m - I_m$ the transmembrane current due to capacitive effects and non-linear gating processes ($I_m$). Next, one defines the electrical diffusion coefficients for intra- and extracellular space as
\begin{align}
        D_{\rm int} =&  \frac{g_{\rm int}}{c_m} ,&   D_{\rm ext} = &\frac{g_{\rm ext}}{c_m}.
\end{align}
With $V_m = V_{\rm int} - V_{\rm ext}$ now follows, with $I_m/C_m  =F_m$:
\bsub \label{bid1} \begin{eqnarray}
        \dd_t V_m &=&  \dd_x \left( D_{\rm int} \dd_x V_{\rm int}\right) + F_m(V_m, u_k),  \label{bid1a}\\
        - \dd_t V_m &=&  \dd_x \left( D_{\rm ext} \dd_x V_{\rm ext}\right) - F_m(V_m, u_k), \label{bid1b}\\
        \dd_t u_j &=& F_j (V_m, u_k). \label{bid1c}
\end{eqnarray} \esub
Here, the $u_k$, $k =3,4, ..., N_v$ denote the state variables apart from $V_{\rm ext}$ and $V_{\rm int}$ in the particular model used. In modeling literature and practice, it is common to combine Eqs. \eqref{bid1a}, \eqref{bid1b} to a differential equation of the elliptic type
\begin{equation}\label{bidom_ell}
  \dd_x \left( D_{\rm int} \dd_x V_{\rm int} + D_{\rm ext} \dd_x V_{\rm ext}  \right) = 0.
\end{equation}
In forward numerical simulations, iteration of the forward problem is interleaved with solving condition \eqref{bidom_ell} to obtain extracellular potentials. This treatment is known in cardiac modeling as the bidomain description. The bidomain approach proves particularly useful when implementing boundary conditions that only act on the extracellular electric potential, such as electrode currents.


A simplification of the bidomain equations is possible when not considering extracellular current sources or boundary effects. When working in one spatial dimension or with isotropic diffusion coefficients (see below), one can take appropriate linear combinations of \eqref{bid1a}, \eqref{bid1b} to eliminate one of the two electrical potentials as a variable. After introducing an average diffusion coefficient
\begin{equation}\label{Dav}
  \frac{1}{D} = \frac{1}{D_{\rm int}} + \frac{1}{D_{\rm ext}}
\end{equation}
and the diffusion projector $\mathbf{P} = diag(1,0,\ldots,0)$, one comes to the monodomain formulation for one-dimensional cardiac excitation:
\begin{equation}\label{monodom}
 \dd_t \uu = \dd_x \left( D \mathbf{P} \dd_x \uu \right) + \mathbf{F}(\uu).
\end{equation}
with $\uu = (V_m, u_3, ... u_{N_V})$.

In both monodomain and bidomain versions, the spatial coordinate enters the equation through a diffusion term; therefore, both qualify as reaction-diffusion (RD) systems. In physical terms, one can state that the transmission of electrical activity along myocytes in the heart is mediated by passive conduction of the intra- and extracellular spaces. More precisely, a local transmembrane current alters affects the electric potential around it; if the effect is strong enough to bring transmembrane voltage across nearby inward channels above the excitability threshold, these will open up as well and thereby support a traveling wave across the cell.\\

The shape and velocity $c$ of the traveling action potential across a cell can be obtained from substituting $x-ct \rightarrow \xi$ in Eq. \eqref{monodom}:
\begin{equation}\label{trav_wave}
 \dd_\xi \left(D \mathbf{P} \dd_\xi \uu \right) + c \dd_\xi \uu + \mathbf{F}(\uu) =0.
\end{equation}
Hereafter traveling wave solutions are found as solution to the ordinary differential equation \eqref{trav_wave}, with $\dd_\xi \uu (x \rightarrow \pm\infty) = 0$. For the systems relevant to the present context, we assume that the reaction term has been chosen such that the phase portrait permits only a single value $c$ with a unique associated solution $\uu(\xi)$. Hence, we consider a single traveling wave with particular shape and unique velocity.

From Eq. \eqref{trav_wave}, it can moreover be seen that, if the scalar diffusion coefficients get multiplied with a constant factor $a$, the velocity grows by a factor $a^2$, which is written equivalently
\begin{equation}\label{csqD}
  c \propto \sqrt{D}.
\end{equation}

More generally, $D$ acts as the only space constant in the RD model, and therefore determines the typical spatial scale at which excitation patterns develop. The relation of this length scale to anatomical dimensions has profound influence on the stability of heart rhythms, as argued in \cite{Winfree:1998}.

\subsection{Transmission of excitation across cells}

Mammalian myocytes have a more or less cylindrical shape, measuring $50$ to $\unit{150}{\micro\meter}$ in length and $10$ to $\unit{20}{\micro\meter}$ in diameter \cite{Clayton:2010}. Their shape thus justifies the one-dimensional propagation model for the spreading of electrical activation along a single cell discussed above. The myocytes are also longitudinally connected to each other through intercalated disks that ensure reliable mechanical attachment. Importantly, the intercalated disks include gap junction channels that enable intercellular signaling and action potential propagation \cite{Hoyt:1989}. The diffusion of ions and water across gap junctions make heart tissue a functional syncytium, i.e. a network of closely interacting cells.\\

On average, smaller numbers of gap junctions are found in the lateral cell membrane, which allow for communication with neighboring cells in the lateral direction. In places where the myocytes are closely spaced to each other in the lateral direction, a propagating action potential can also be transmitted through intercellular space by its net effect on the on the extracellular potential, which could invoke an action potential in the adjacent cell as well. Nevertheless, as gap junctions are the main pathway for action potential mediation between cells, the spreading of electric activation takes place about three times faster along the long axis of the cells than in transverse directions \cite{Panfilov:CBH}. How to deal with the emergent anisotropy of the tissue with respect to electrical signaling is the most important issue addressed in this work.\\

\begin{figure}[h!t] \centering
  \includegraphics[width=0.6 \textwidth]{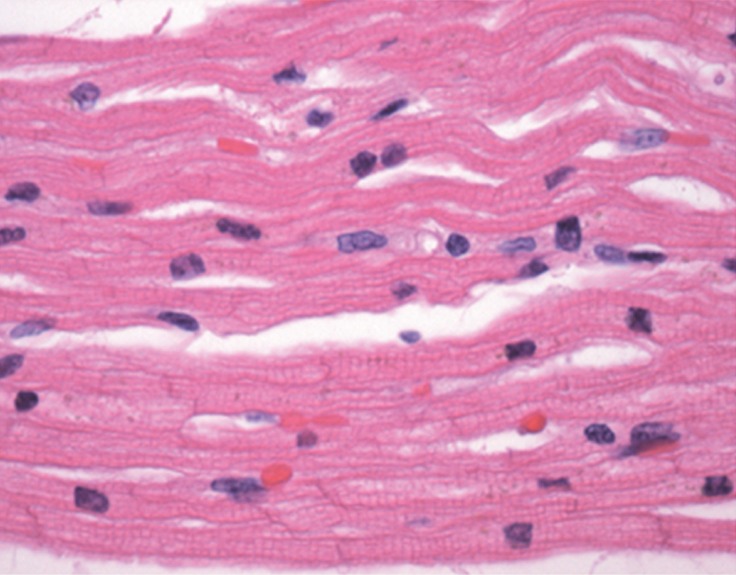}\\
  \caption[Myofibers in ventricular myocardium]{Fibrous structure in healthy ventricular myocardium, as presented in \cite{Kumar:2010}. The strong alignment and longitudinal coupling of the myocytes induces anisotropic conduction velocity for action potentials.}\label{fig:myofibs}
\end{figure}

For the modeling of macroscopic patches of excitable tissue, it is instructive to integrate out the properties of individual cells, i.e. employ a continuum description. Although investigations have shown that action potential propagation across gap junctions is indeed a discrete process \cite{Kleber:2004}, depolarization waves are seen to propagate smoothly at larger scales \cite{Durrer:1970}.\\ 

Promoting the one-dimensional cable equation \eqref{monodom} to a full three-dimension\-al reaction diffusion system is now straightforward: the reaction term is taken as the spatial average of local cell properties and the electric diffusion properties are made dependent of the direction of propagation. In lowest order in an angular expansion, the scalar diffusion coefficient $D$ for the electric potential may be replaced by a symmetric diffusion tensor $\mathbf{D}$ of rank 2. When taking into account local orientation of myocytes, the positive-definite $\mathbf{D}$ obtains the eigenvalues $(D_L, D_T, D_T)$ and therefore locally represents an uniaxial medium. If desired, the formalism can take distinct eigenvalues $(D_L, D_{T1}, D_{T2})$, which confers to tissue with orthotropic structure. Due to the presence of cleavage planes in ventricular tissue, it was recently established that ventricular myocardium behaves as an orthotropic rather than a uniaxial medium \cite{Hooks:2007, Caldwell:2009}.

In the monodomain, we may now write
\begin{equation}\label{RDE_mono}
\dd_t \uu =  \dd_i \left( D^{ij} \dd_j \mathbf{P} \uu \right) + \mathbf{F}(\uu),
\end{equation}
with $P_{mn} = \delta_{m1} \delta_{n1}$. From here on, we adopt the Einstein summation convention, i.e. whenever the same index appears as a superscript and subscript, summation over this index is implicitly understood.

In this work, we will present analytical elaborations on equations of the type \eqref{RDE_mono}, which cover the propagation of excitation sequences in the heart in the monodomain approximation. \\


The bidomain case is more involved, since the intra- and extracellular spaces were experimentally determined to possess unequal diffusion properties \cite{Keener:CBH8}. In the approximation that both compartments share the main principal axes $\vec{v}_a$ (i.e. they are aligned with local cell direction), one obtains
\begin{align}
D^{ij}_{\rm ext} &= \sum_{a=1}^3 D^a_{\rm ext}  v_a^i v_a^j, &  D^{ij}_{\rm int} &= \sum_{a=1}^3 D^a_{\rm int}  v_a^i v_a^j.
\end{align}
Only in the special case where $\mathbf{D}_{\rm ext}$ and $\mathbf{D}_{\rm int}$ are considered with equal anisotropy ratios ($\mathbf{D}_{\rm ext} = \alpha \mathbf{D}_{\rm int}$ for some $\alpha >0$) , the diffusion term can be expressed as in the monodomain case \eqref{RDE_mono}. \\

The laws of motion for activation patterns for which we shall provide an original analytical derivation Chapters \ref{chapt:fronts} to \ref{chapt:filaniso} are obeyed only for monodomain cardiac models, but can easily be extended towards the bidomain case with equal anisotropy ratios. Bidomain models with unequal diffusivity ratios currently fall outside the scope of our theoretical approach.


\section{Heart rhythms}

Having summarized the mechanisms that underly the generation and propagation of action potentials, an overview is presented here of common excitation patterns that are encountered during normal and abnormal cardiac activity.\\

In snapshot views of active myocardium, as in Fig. \ref{fig:reentry} and following, the medium may be divided in an active region and a quiescent region, which are separated be a boundary zone. The boundary zone comprises the wave front and wave tail, depending on whether the cells are depolarizing or repolarizing, respectively.



\subsection{Sinus rhythm: regular heart beat}

During normal heart rhythm, action potentials are generated in the sinoatrial node, a specialized group of cells in the right atrium. These cells spontaneously generate rhythmic activity. 
Excitation spreads from the sinoatrial node all over the atria, which activate in $8$ to $10\,$ms. The activation reaches the atrioventricular node, which is found in the myocardial wall between the right atrium and right ventricle. As the atria and ventricles are electrically isolated from each other by a strip of non-conducting connective tissue, the atrioventricular node is the only pathway for electrical signalling to reach the ventricles. As such, the atrioventricular node accounts for the necessary activation delay of about $20\,$ms between atrial and ventricular activation, due to reduced conduction velocity in nodal tissue ($0.05-0.10\,$m/s). Also, the atrioventricular node acts beneficially as a filter, to prevent abnormal activity taking place in the atria from spreading into the ventricles.

Activation of the relatively large ventricular muscle occurs first through the rapidly conducting Purkinje network, which spread the activation sequence over large parts of the ventricular endocardium. The subsequent activation of the bulk myocardium may in its simplest form be conceived as the propagating plane wave from Fig. \ref{fig:reentry}a.\\

For normal heart rhythm, the study of two and three-dimensional propagation of action potentials is particularly useful to bulk conduction in atrial and ventricular muscle. Moreover, we will see that, during various heart rhythm disorders, the specialized conduction system can be overruled due to refractoriness of the tissue: when sources of unequal frequencies are present the fastest source will dominate, even if arising from an abnormal (ectopic) source.

\subsection{Re-entry}

\begin{figure}[h!t] \centering
  \includegraphics[width=1.0 \textwidth]{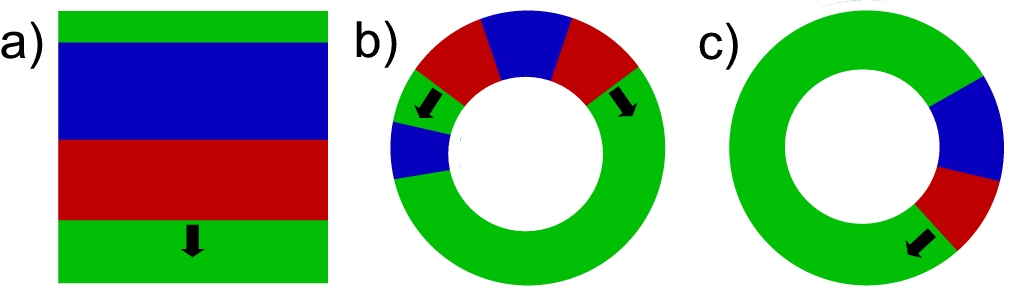}\\
  \caption[Normal versus re-entrant cardiac activation]{The simplest activation pattern in excitable tissue is a plane wave as shown in panel (a), propagating in the direction of the arrow. Red regions denote active medium; blue indicates refractory (recovering) tissue, whereas green coloring represents tissue in its resting state. In a ring of excitable tissue (b), one of both pulses induced by a point stimulus may be eliminated by a patch of recovering tissue, which results in persisting reentrant activity (c).}\label{fig:reentry}
\end{figure}

Of the most important types of cardiac arrhythmias are the so-called re-entrant arrhythmias. The simplest example of re-entry is encountered in ring of excitable tissue, as depicted in Fig. \ref{fig:reentry}b-c.

In response to a point stimulus, two propagating waves are produced, which move in both directions away from the stimulation site. If one of these traveling waves gets blocked by e.g. incomplete recovery of the tissue after the passing of another wave, only one of the traveling waves will survive. A a result, a single wave of excitation will travel around the ring of tissue forever. If the temporal period of the re-entrant cycle is smaller than that of the natural pacemaker, the abnormal activity overcomes normal heat rhythm, leading to an increased heartbeat.

An example of such re-entrant arrhythmia is found in patients with the Wolff-Parkinson-White syndrome. Their hearts exhibit an additional conducting pathway between the atria and ventricles, leading to a severely increased heartbeat due to re-entrant activity. Fortunately, the condition can be remedied by surgical removal of the tissue patch that is responsible for the electrical loophole.


\subsection{Spiral waves}
A more involved re-entrant phenomenon takes place when an activation front in a two-dimensional medium ends on a smooth circular obstacle: the wave front edge traces out the circumference of the obstacle, while the rest of the wave front lags behind in its circular movement. This leads to an overall spiral-like shape for the simultaneously activated group of cells, which is known as a `spiral wave attached to a boundary'. It was noted that, at a great distance of the spiral center, the loops of the spiral are hard to distinguish from a circular wave train; therefore spiral waves can be considered point sources of activation in large enough media. In a cardiological context, a re-entrant wave attached to an inexcitable obstacle (e.g. scar tissue or the onset of a vein or artery) is referred to as anatomical re-entry.

Interestingly, the rotating spiral waves have been observed experimentally and numerically without being attached to an obstacle in the medium. For, in a wide parameter regime, a broken activation front tends to curl around the wave break, with decreased local normal velocity. The remainder of the front keeps on propagating at almost the plane wave speed, and eventually winds up around the wave break, thus creating a spiral wave pattern. The described event is known as functional re-entry, and believed to lie at the base of various heart rhythm disorders, as a spiral wave could form whenever the activation wave's front gets broken. Spiral waves have been observed experimentally in the heart using potentiometric dyes \cite{Davidenko:1993, Pertsov:1993}.

\begin{figure}[h!t] \centering
  \includegraphics[width=0.9\textwidth]{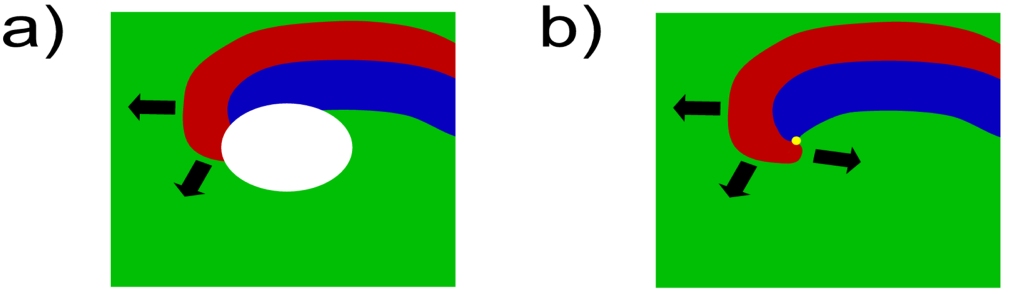}
  \caption[Anatomical vs. functional re-entry]{Anatomical (a) versus functional (b) re-entry. The latter is our first example of a spiral wave, which has a phase singularity (yellow) at its tip. }\label{fig:anat_func_reentry}
\end{figure}

\subsection{Phase singularities}
We will henceforth focuss on functional re-entry. In that case the wave front is composed of cells that activate, whereas the wave back or tail comprises cells that are recovering their resting state. At the unique point where tail and front meet, the normal velocity vanishes and we denote this point as the spiral tip. The thus defined spiral tip has no clearly defined phase in the activation cycle, and for that reason it has been called the phase-change point, or phase singularity point \cite{Gulko:1972, Zykov:1987, Clayton:2005}.

The tip trajectory of a numerically simulated spiral wave in an isotropic homogeneous medium can take various shapes, depending on the used models and parameter regime. In the simplest case, the tip describes a small circle. Other possibilities include a circular (hypo)cycloidal movement or a nearly linear tip trajectory. This phenomenon is denoted `meander' \cite{Winfree:1973}. Our present theoretical study of spiral waves, however, does not account for meandering tip trajectories.

\subsection{Scroll waves and filaments}

So far, we have neglected the third spatial dimension in our discussion of spiral waves. Although the two-dimensional approximation may reasonably represent the thin atrial walls, the thicker ventricular walls form essentially a full three-dimensional medium \cite{Efimov:1999}.

Spiral waves can trivially be generalized to three dimensions by stacking them on top of each other to fill the third dimension. The emerging structures were named `scroll waves'. The phase singularities of the constituent spiral waves aggregate into a line, which is called a (scroll wave) filament. The filament does not need to be a straight line; moreover, filaments are seen to evolve in time \cite{Keener:1988, Clayton:2005}. One substantial constraint is that filaments can only end on the medium boundaries \cite{Pertsov:2000}. Under no-flux boundary conditions for the state variables, filaments are orthogonal to the medium edges at their endpoints. Filaments may also form closed loops, in which case the associated excitation pattern is called a scroll ring.

Rotating spiral waves, scroll waves and their filaments have been seen in a variety of excitable and oscillatory media such as the Belousov-Zhabotinsky chemical reaction \cite{Pertsov:1990b}, self organization of slime molds \cite{Steinbock:1993b}, and vortex solutions of the complex Ginzburg-Landau equation \cite{Abrikosov:1957, Gabbay:1997}. The terminology is thus not restricted to the propagation of bioelectric excitation.

%

\subsection{Fibrillation}

The term fibrillation refers to spontaneous, asynchronous contractions of the cardiac muscle fibers \cite{Gray:1998}, which are believed to originate from turbulent electrical activity. Fibrillation taking place in the atria is a common pathology with elder people. Fortunately, the state is not immediately life-threatening, as the atrioventricular node shields the ventricles from irregular activation.

The situation is different with ventricular fibrillation, for unsynchronized ventricular activation impairs efficient pumping of blood to the body. Consequently, ventricular fibrillation is lethal within few minutes. Not only known heart patients risk developing ventricular fibrillation, as sudden fibrillation events are likely to underly cases of sudden cardiac death of often young people without a history of cardiac disease.
At present, the only known treatment for a fibrillating heart is electrical defibrillation, i.e. administering high-voltage electrical shocks in order to eradicate the chaotic activation pattern.

Due to its life-threatening character, many clinical, experimental and numerical studies have addressed wave dynamics during fibrillation. However, the precise mechanisms which produce and sustain fibrillation are still under discussion \cite{Ideker:2007}. From clinical, experimental and modeling efforts, fibrillation has been linked to the presence of multiple phase singularities \cite{Gray:1998, Clayton:2005}. A recent study indicates that in clinically recorded ventricular fibrillation 9.0 $\pm$ 2.6 sources of electrical activation (i.e. rotor filaments) were identified \cite{tenTusscher:2009}. This number lies about fivefold lower than the number of filaments encountered in fibrillating dog and pig hearts.\\

An often-quoted possible pathway towards fibrillation is filament multiplication. This process takes place whenever a part of a filament hits the medium boundary, which augments the number of filaments in the tissue by one. Or, similarly, a filament may pinch off a scroll ring after self-intersection. Those mechanisms are commonly referred to as filament break-up; a review can be found in \cite{Fenton:2002}. The extended EOM for a single filament that we will derive in the course of chapters 7 and 8 are particularly relevant to this issue, as we quantitatively determine the tissue properties under which an initially stable filament destabilizes. Insights in further temporal evolution necessitate an analytical theory for filament-filament and filament-boundary interactions, which has not been developed yet at the time of this writing.

Recent experimental works \cite{Nash:2006} have concluded that it is unlikely that a single mechanism would account for the various types of fibrillation that have been observed, which could explain why scientific views on fibrillation are still troubled.

\clearpage


\clearpage{\pagestyle{empty}\cleardoublepage}

\graphicspath{{fig/}}

\renewcommand\evenpagerightmark{{\scshape\small Chapter 3}}
\renewcommand\oddpageleftmark{{\scshape\small Cardiac structure and imaging}}

\hyphenation{myo-fiber myo-fi-bers dif-fu-sion diffu-sion-attenuated}

\chapter[Cardiac structure and imaging]{Cardiac structure and imaging}
\label{chapt:imag}

The following part of the text serves to summarize and scrutinize current knowledge on the microstructure of the heart muscle. A popular technique for the non-destructive mapping of fibrous and laminar structure in the heart is \textit{diffusion tensor imaging} (DTI). However, the DTI method is seen to yield highly variable outcome for laminar structure. For that reason, we review the principles of diffusion MRI and DTI in particular. Next, strengths and weaknesses of the DTI formalism are discussed in the light of a simple model on restricted diffusion of water in the tissue. Also, we suggest adaptations to the current methodology in reporting transmural fiber and sheet orientations.

The development and application of a MRI technique that could overcome inherent limitations to DTI will be the subject of the subsequent chapter.

\section{Fiber and sheet structure in the heart muscle}



\subsection{Myocytes define myofibers}

In our treatment of action potential propagation, we have yet touched upon the end-to-end coupling of myocytes. Since the elongated myocytes develop active mechanical stresses along their long axis, such end-to-end coupling enables large macroscopic deformations of the tissue. The lined-up myocytes are furthermore surrounded by a network of highly deformable collagen, which forms a structural reinforcing matrix that is known as the \textit{endomysium}.
Collagen structures which surround groups of myocytes on the other hand, are denoted \textit{perimysium}.\\

We shall make use of the term (myo)fiber only to annotate the local orientation of individual myocytes, without reference to other structure, as in \cite{Gilbert:EJCTS}. We thus employ the concept of myofibers in the sense that `light rays' are used in physics, being a mathematical idealization of reality and possessing no physical thickness. On a higher level of abstraction, our notion of myofibers corresponds to a tangent vector field $\vec{e}_f(\vec{r})$ defined at each point of the myocardium.

\subsection{Spatial organization of myofibers}

\begin{figure}[h!t] \centering
  \includegraphics[width= 0.9 \textwidth]{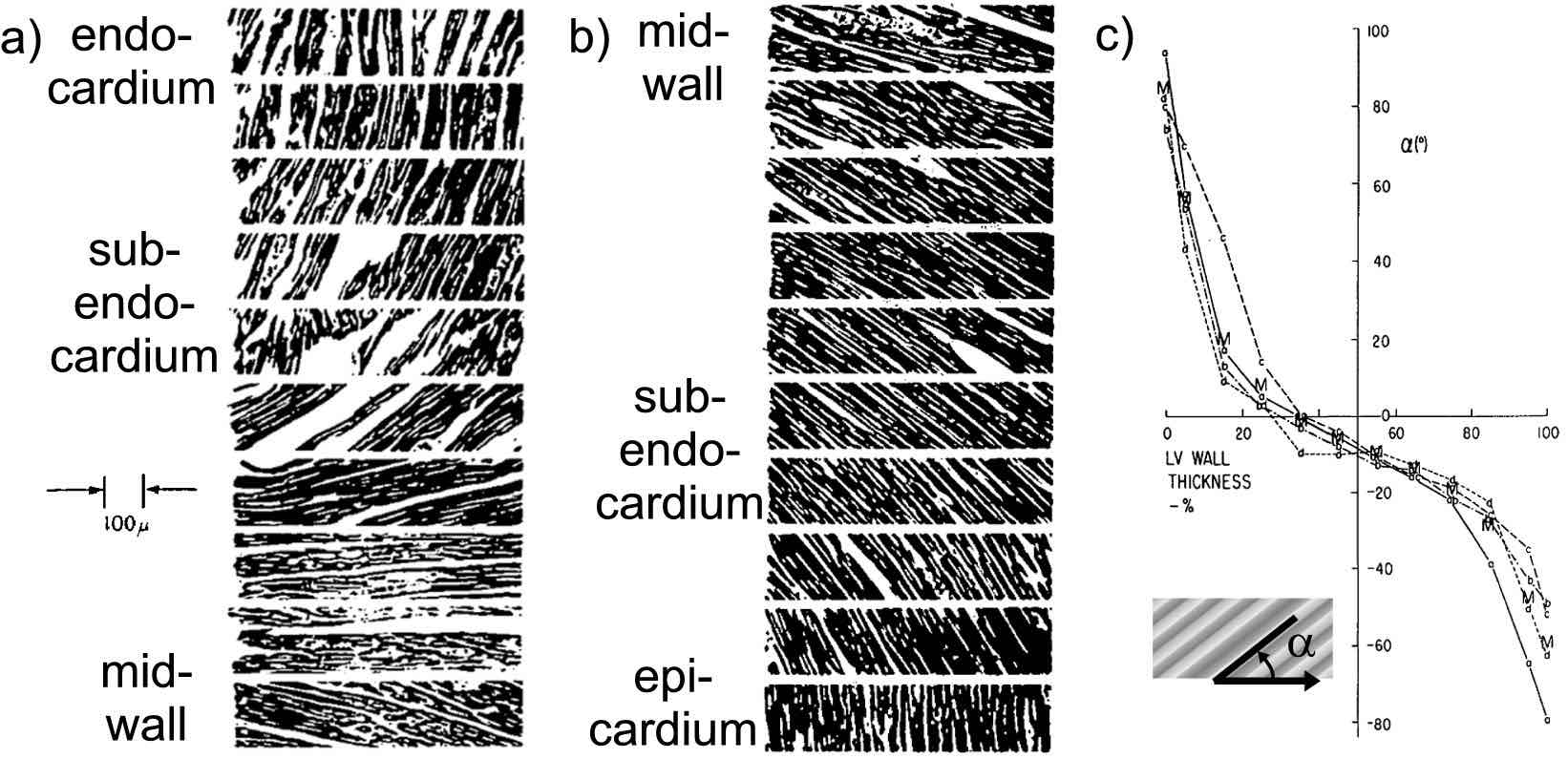}\\
  \caption[Spatial organization of myofibers]{Spatial organization of myofibers adapted from the seminal work of Streeter \etal \cite{Streeter:1969}. Photomicrographs (a-b) were taken of paraffine-fixed slices through the canine left ventricle. The slices were made \unit{5}{\micro\meter} thick and parallel to the epicardial wall at different wall depths. Transmural fiber rotation between consecutive slices is manifest and was quantified using the fiber helix angle $\alpha$, as shown in panel (c). The horizontal axis in (c) runs from endo- to epicardium.
  }\label{fig:Streeter}
\end{figure}

There are various ways to assess myofiber orientation in the heart muscle. The oldest one is anatomical dissection, which can nevertheless be used to make quantitative statements \cite{Streeter:1969}, as quoted here in Fig. \ref{fig:Streeter}. Another possibility is to slice the heart after fixation, and study myofiber direction in each cross-section. Hereby the three-dimensional orientation can be inferred. In small hearts or heart sections that have been made optically transparent \cite{Smith:2008}, three-dimensional fiber structure may be assessed using confocal microscopy. Alternatively, diffusion-MRI based techniques have been developed and applied to map local myofiber orientation \cite{Hsu:1998, Scollan:1998, Sosnovik:2009, Dierckx:2009}. At present, it has even become possible to investigate three-dimensional organization of perimysial collagen in fixed tissue on itself, using an confocal microscopy technique developed in \cite{Sands:2005}.\\

All techniques mentioned above consent on the overall organization of myofibers. In general, myofibers run parallel to the epicardial surface, but the angles they enclose with the axial plane exhibit strong transmural variations. It is commonly said that in most of the ventricular myocardium, the orientation of myofibers changes counterclockwise when going from epi- to endocardium, ranging over about 120$^\circ$. (In physics terms, one could say there is a `left-hand rule' for transmural fiber rotation.) The region where myofibers run within the axial plane is found near the middle of the ventricular wall, and this observation has been used to define either the mid-wall depth \cite{Geerts:2002} or the local axial plane.

Transmural fiber rotation through the bulk of the ventricular mass is sometimes described as simply linear, despite obvious non-linearity that can sometimes be seen in the measured orientations. Another reference to myofiber rotation is `sigmoidal', although we think some experimental evidence could as well support inverse sigmoidal, or other variations thereof \cite{Hsu:1998, Holmes:2000}.
Near the endocardial surface, strong deviations can be expected at the height of the papillary muscles, which run almost perpendicular to the axial plane. When experimental techniques are used that are sensitive to the presence of major blood vessels (e.g. diffusion MRI), we recommend caution when interpreting orientation measurements in regions where the vessels are known to occur, especially in the sub-epicardium.

Some research questions on the myocardial fiber orientation are whether fiber rotation is everywhere continuous, how the fusion sites between RV, LV and septum are structured, and whether a unified picture is also attainable for atrial myofiber organization.

\subsection{Myofibers gather into myocardial sheets}

\begin{figure}[h!t] \centering
  \includegraphics[width=1.0 \textwidth]{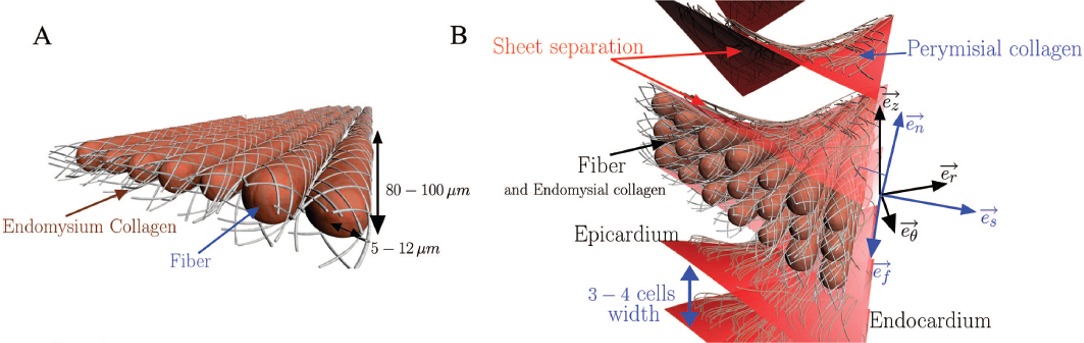}\\
  \caption[Schematic representation of fiber and laminar structure]{Schematic representation of aligned myofibers (A) and myocardial sheet organization (B) adopted from \cite{Rohmer:2007}. The local material axes $(\vec{e}_f, \vec{e}_s, \vec{e}_n)$ can for each point be related to a global cylindrical coordinate system for the heart with basis vectors $(\vec{e}_r, \vec{e}_\phi, \vec{e}_z)$. Such global reference is commonly chosen with the Z axis taken along the heart's long axis, so that $\vec{e}_r, \vec{e}_\phi$ span the local transverse plane. }\label{fig:Rohmer}
\end{figure}

Histological studies have observed that in parts of the ventricular muscle mass, the myocytes are grouped in layers of 3-4 cells thick, which are usually referred to as myocardial sheets (see Fig. \ref{fig:Rohmer}). The myocardial sheets are separated by cleavage planes, which are thought responsible for important mechanical properties of the cardiac muscle. Most notably, the emergent shear stresses in highly contractile tissue with varying material axes are significantly reduced when sliding along cleavage planes is allowed\cite{Costa:1999}. Shear displacements also enable better reduction of the endocardial blood volume during ejection, such that more blood is squeezed out of the ventricles. At the surface of the sheets of tissue, supplementary perimysial collagen structures are present that reinforce the sheets as a whole \cite{Pope:2008}. \\

In what follows, we use the terms `sheet' and `tissue layer' to point to myocardial tissue itself, and reserve `cleavage plane', or `laminar cleft' to indicate the gap that separates consecutive sheets of tissue. The term (myo)lamina is used at the abstract level to denote local orientation of the myocardial sheets, similar to the meaning of `myofiber' in our text. Reporting laminar orientation is mathematically equivalent to communicating the normal vector $\vec{e}_n$ to the local myolamina. The vector field $\vec{e}_n$ is not necessary tangential (i.e. need not possess a set of integral curves) and is not defined whenever neither or multiple laminar orientations are present in a given region of the myocardium.

Note that the plane parallel to the myocardial sheets is spanned by $\vec{e}_f, \vec{e}_s$, with $\vec{e}_s = \vec{e}_n \times \vec{e}_f$. The triad $(\vec{e}_f, \vec{e}_s, \vec{e}_n)$ then defines the local material axes of the orthotropic tissue.

\subsection{Spatial organization of myocardial sheets}

In mid-wall portions of large mammals, ventricular sheet organization can be witnessed with the naked eye in fresh-cut hearts. Recent microscopy studies also describe the presence of cleavage planes in most of the ventricular mass, although no sheet structure was identified in the subepicardial region of rat hearts in \cite{Pope:2008} based on the assessment of perimysial collagen structures. Detailed recording of myocardial laminar organization has been carried out with histological sectioning. This standard technique, however, dehydrates the tissue during fixation and therefore might introduce additional cracks in the tissue which could wrongfully be classified as anatomical cleavage planes. The ultra-milling technique adopted by Sands \etal \cite{Sands:2005} may be considered the current gold standard for assessing laminar orientation, as the protocol allows three-dimensional sectioning of extended volumes with high spatial resolution. Moreover, the collagen structures can be resolved, which enables to discriminate between fixation artefacts and anatomical reality. Present limitations of the ultra-milling technique are its destructive nature which preempts all clinical potential, the limited volume of tissue samples and the requirement to freeze the tissue prior to imaging.

Promising alternatives to sectioning are found in medical imaging techniques. Cleavage planes of finite width in the micrometer regime fall within the scope of micro-CT and high-resolved anatomical MRI. Microstructural laminar orientation has been reported in literature using diffusion tensor imaging as well. Unfortunately, the outcome of these studies is highly variable between individual hearts; see e.g. \cite{Gilbert:EJCTS} for a review and references therein. After an extensive DTI study on canine hearts \cite{Helm:2005}, the authors of that paper concluded that there is a bimodal distribution of cleavage planes between individual dog hearts.

A possible source of much current controversy was uncovered in \cite{Sands:2005}. With an automated confocal microscopy method, an extended mid-wall zone was identified with two distinct local laminar orientations in the lateral free wall of rat LV.  When viewed in a radial-longitudinal plane, the co-existing groups of cleavage planes are found to intersect the axial plane at angles of roughly $\pm 45^\circ$. Where two cleavage planes intersect, X-shaped clefts are observed.

It is important to realize that, as we will explain shortly, the DTI technique is not equipped to distinguish cleavage plane intersections for it can only associate a unique normal vector to the myolaminae. This is our incentive to first recapitulate DTI in the remainder of this chapter and promote an approach that reaches beyond this limitation in the next chapter.

\begin{figure}[h!t] \centering
  \includegraphics[width=1.0 \textwidth]{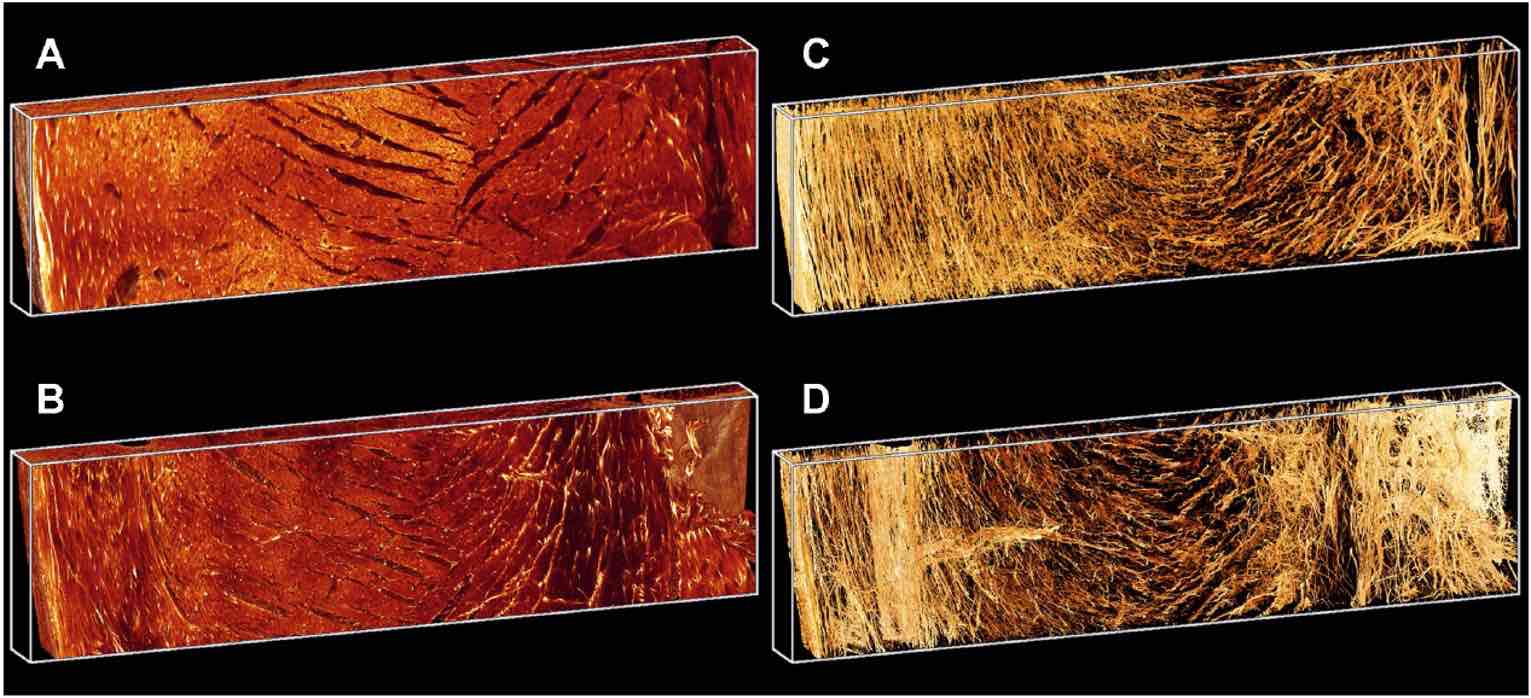}\\
  \caption[Microscopy view on myocardial tissue structure]{Confocal microscopy reconstruction of myocardial laminar (A,C) and collagen (B,D) structure in a transmural wedge of the left ventricle of two rats, as presented in \cite{Pope:2008}. The rendered region measures $4 \times 1 \times 0.25\,$mm and has \unit{1}{\micro\meter} (A,B) and \unit{1.22}{\micro\meter}  (C,D) isotropic resolution. The epicardial surface is situated on the left. The presence of cleavage planes in the dehydrated tissue is apparent, as well as the collagen structures that bound the sheets of tissue. In the mid-wall region, a region is found where to families of non-aligned cleavage planes meet.
  }\label{fig:sheets_Sands}
\end{figure}

\section{Physical principles of diffusion MRI}

\subsection{Magnetic Resonance Imaging in a nutshell}

Up to now, the technique of magnetic resonance imaging (MRI) is one of the few fields where quantum physics has invoked a widespread technology that would not have been possible otherwise. In a similar way, Global Positioning System devices are often quoted as the place where everyday life is touched most by Einstein's theory of gravity.

In short, medical MRI machines manipulate the spin of hydrogen nuclei in the samples they investigate by using magnetic fields and radiofrequent waves. Since the largest fraction of hydrogen atoms in biological tissue is found in water molecules, MRI methods effectively probe the way in which water molecules are embedded in the tissue.

Within the bore of a clinical MRI scanner, a homogenous static magnetic field $\vec{B}_0$ with typical strength $1.5 - 3\,$T is generated and maintained by superconductive coils -  quantum physics once more! For our own scans, we have made use of a non-clinical MRI scanner that operates at a higher field strength ($9.4$\,T) with a smaller field of view. However, the working principles are identical.

When the external $B_0$-field is applied, proton spins precess at the Larmor frequency $\nu_L= \bar{\gamma} B_0$, with $\bar{\gamma} = \gamma/(2\pi) = 42.58\,$MHz/T the gyromagnetic frequency for protons. Furthermore, the strong $\vec{B}_0$ field will align most of the spins parallel to the magnetic field. In the resulting ground state, the sum of all spins in a given volume add up to yield a net magnetization $\vec{M}_0$. The art of MRI science then seeks to interact with the collection of oriented spins by selectively exciting them with pulses of radiofrequent (R.F.) waves. In MRI literature, these R.F. electromagnetic waves are catalogued according to their net effect on a collection of spins, e.g. a $90^\circ$ R.F. pulse makes the proton spin rotate over ninety degrees around an axis of choice.

\subsection{The spin-echo experiment \label{sec:SE}}

\begin{figure}[h!t] \centering
  \includegraphics[width=1.0\textwidth]{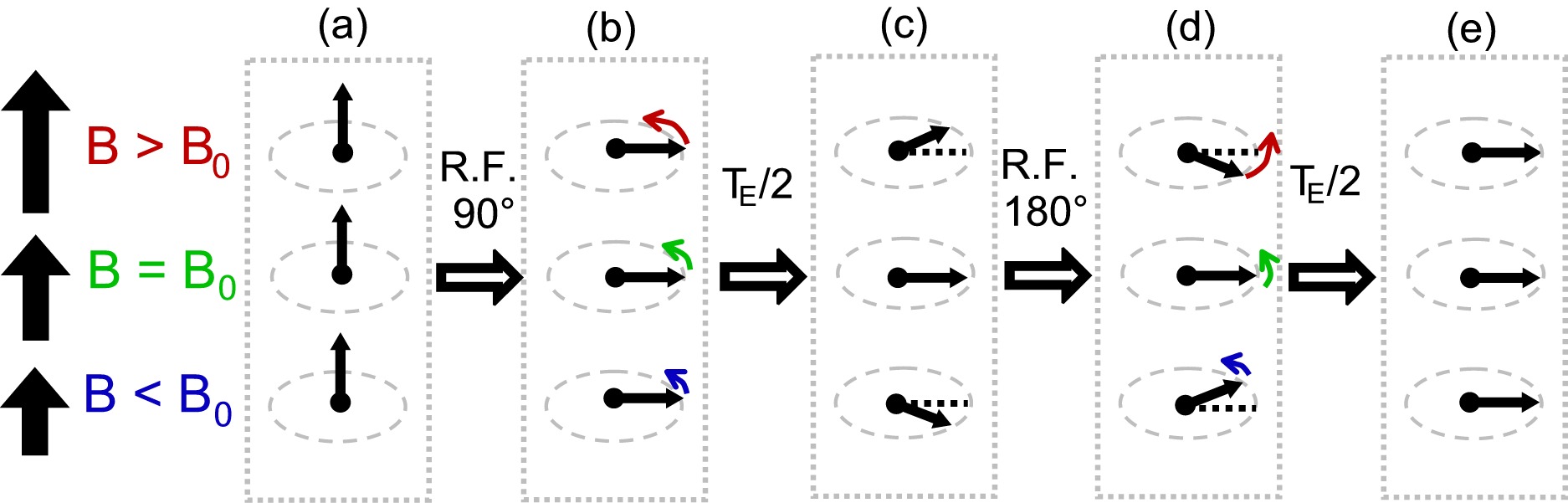}\\
  \caption[Echo formation in a spin-echo MRI experiment]{Echo formation in a spin-echo MRI experiment. Temporal evolution is depicted for three representative spins, which experience a static magnetic field which is larger than (top), equal to (middle) or smaller than (bottom) the $\vec{B}_0$ field.} \label{fig:spin-echo}
\end{figure}

We will now sketch a typical spin-echo experiment, which is also schematically depicted in Fig. \ref{fig:spin-echo}. First, suppose that an ensemble of hydrogen nucleus spins is aligned with $\vec{B}_0 = B_0 \vec{e}_z$ to give an initial magnetization $\vec{M}_0 = M_0 \vec{e}_z$. This situation corresponds to time frame (a) in Fig. \ref{fig:spin-echo}. Playing a $90^\circ$ R.F. pulse brings all spins within the transverse (X-Y) plane (b), in which they will precess at the Larmor frequency. At this time, the collection of coherently rotating spins produce a non-vanishing net magnetization. Following the working principle of an electrical generator, an inductive current will be invoked in a nearby receiving coil. However, this signal is not used in a spin-echo experiment. If no action is taken, the spins gradually recover their ground state; this exponential process occurs with a time constant $T_1$.

Spatial inhomogeneities or gradients in the magnetic field as well as spin-spin interaction cause the initially perfectly coherent collection of spins to dephase (c) with tissue-dependent time constant $T_2$. In the spin-echo sequence, a so-called refocusing $180^\circ$ pulse is played at a time $T_E/2$, such that the spins are rotated over $180^\circ$ around the Y axis (d). The net effect of field inhomogeneities is now reversed such that after a time $T_E$, the spins are approximately in phase again (e), so that a net current is induced in the receiver coil. This signal is dependent of (or, `weighed with') the local relaxation times in the tissue, which brings relative contrast between soft tissues in MR images. The time interval $T_E$ is appropriately called the echo time of the sequence.

An echo is formed at all time instances where the total accumulated phase is equal for all spins, regardless of their position. This requires that the time-integrated magnetic field that a particular spin experiences vanishes for $t = T_E$. In practice, local magnetic gradients $\vec{g}(\vec{r},t)$ are used for for image formation so that the criterion for echo formation demands that
\begin{equation}\label{echo_formation}
  \phi(\vec{r}, t) = \gamma \int_0^t \vec{g}(\vec{r},t') \cdot \vec{r}(t') \mathrm{d} t'
\end{equation}
should vanishe at $t=T_E$. The $180^\circ$ refocussing pulse has the equivalent effect of changing the sign of $\vec{g}$ upon application.

The information acquired during a spin-echo experiment lies in the comparison of $M$ to $M_0$, which is a measure to the spin relaxation processes that have taken place during the echo time $T_E$. Because the spin relaxation times $T_1, T_2$ vary with unequal tissue types, a MRI measurement can discriminate between soft tissues. Essentially, the resulting contrast in an MRI sequence depends on the proton density of the tissue $\rho(\vec{r})$, the spin-lattice relaxation (time constant $T_1$) and spin-spin relaxation (time constant $T_2$). Unlike e.g. X-ray tomography, the contrast can be tuned at wish as to enhance contrast between particular tissue types. Hence, MRI images can be hard to interpret if not accompanied by a description of imaging parameters. The high tunability of MRI sequences and the non-invasive character of the imaging have made MRI the leading technology for the \textit{in vivo} medical imaging of soft tissues.

\subsection{Image formation and resolution}

We will not digress here on how spin packets can be selectively excited and read out to actually produce spatially resolved images. Details on the Fourier encoding steps can be found in any decent MRI handbook or online course (e.g. \cite{Haacke, Hornak:site}). It turns out that spatial resolution of MRI images is limited by available readout bandwidth (hence the use of high $\vec{B}_0$ fields) and the volume of the smallest image point (`voxel'), which determines signal strength. At this time, the typical resolution is ($1-5\,$mm)$^3$ for clinical systems and ($0.1-0.5\,$mm)$^3$for high-field ($9-11\,$T) spectrometers. The most advanced anatomical MRI technique currently reach spatial resolution up to $0.05\,$mm. So, how can one resolve cardiac microstructure at length scales of a few microns?

The answer is to engage the protons themselves in the job: in an aqueous solution, water molecules undergo Brownian motion (here comes Einstein again); as their random movement is hindered by microstructural elements (e.g. cell membranes and collagen structures), a MRI signal which is made sensitive to the diffusive displacement of protons will indirectly pick up microstructural information! Magnetic resonance techniques that take advantage of this principle are designated diffusion- weighted MRI (DW-MRI) or diffusion-MRI.

\subsection{Diffusion weighting and path integrals \label{sec:pathint}}

The principle of diffusion-weighted MRI can be exposed in various ways. Here, we follow course material \cite{Kiselev:course}, due the mathematical rigor and physical nature in the derivation of the fundamental laws in diffusion-MRI.

\begin{figure}[h!b] \centering
  \includegraphics[width=0.8\textwidth]{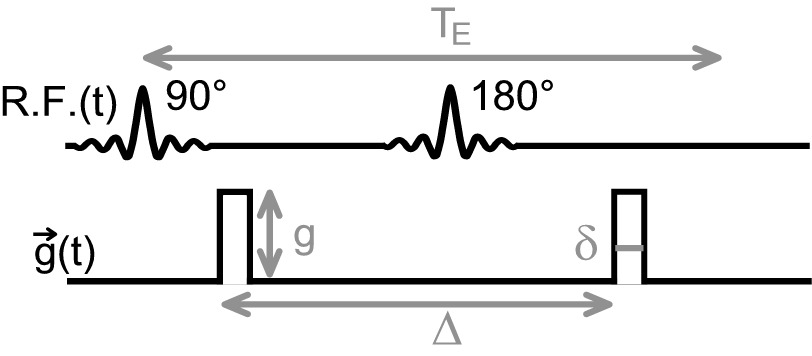}\\
  \caption[Diffusion-weighting gradients in diffusion-MRI]{Time course of R.F. pulses and diffusion encoding gradients in a diffusion-weighting spin-echo sequence.}\label{fig:dwg}
\end{figure}

The basic idea in DW-MRI is to apply a supplementary magnetic field gradient $\vec{g}(t)$ along a particular direction, during the time interval where the MRI echo is formed \cite{Bammer:2003}. The timing of the diffusion-weighting gradients with the radiofrequent pulses is depicted in Fig. \ref{fig:dwg}. Obviously, a first requisite imposed by Eq. \eqref{echo_formation} is that the time-integrated diffusion-encoding gradients before and after the refocusing R.F. pulses should be equal.

The Brownian motion of the water molecules now introduces explicit time-dependence in the position of each given spin $\vec{r}(t)$. Using partial integration and the time-integrated diffusion gradient strength
\begin{equation}\label{intg}
  \vec{G}(t) =    \int_0^t \vec{g}(t') \mathrm{d} t',
\end{equation}
the net phase in the transverse plane at time $t=T_E$ may be written
\begin{equation}\label{phasevel}
  \phi(T_E) = \gamma  \int_0^{T_E} \vec{g}(t) \cdot \vec{r}(t) \mathrm{d}t =  -  \gamma  \int_0^{T_E} \vec{G}(t) \cdot \vec{v}(t) \mathrm{d}t.
\end{equation}
In complex number notation, the magnetization of each spin in the XY-plane is proportional to $e^{i \phi(T_E)}$. After refocussing, this quantity determines the echo strength and therefore the amplitude of the MRI readout signal. With a non-vanishing diffusion encoding gradient, the acquired phase of each proton obviously depends on the random-walk trajectory that it followed during the echo time (typically few tens of milliseconds). The effect of diffusion processes may be filtered out by comparing the formed echo with the echo strength $S_0$ in the absence of diffusion encoding gradients:
\begin{equation}\label{path_integral}
  \frac{S[\vec{g}(t)]}{S_0} = \sum_k e^{i \phi_k(T_E)} = \int_C [\mathrm{d} C] \exp \left( - i  \gamma  \int_0^{T_E} \vec{G}(t) \cdot \vec{v}(t) \mathrm{d} t \right) .
\end{equation}
The sum over all possible spins should be taken over all possible trajectories $C$ that the water molecules can follow in the considered piece of tissue. As a result, the physically observable signal strength $S$ is found as the path integral of a complex action! This is not a mere coincidence, as the use of path integrals to describe diffusive processes precedes their successful application in quantum theory \cite{Wiener:1923, Feynman:1948}. In MRI literature, it is common to conceal the explicit path integral \eqref{path_integral} by writing simple brackets $\langle \cdot \rangle$ instead, which denote averaging over all possible paths, with start points in the voxel considered.

There are several ways to handle the general path integral relation \eqref{path_integral}. One general option is to apply a cumulant expansion, which states that an averaged exponential function may be written as a sum of its statistically independent moments (i.e. cumulants) \cite{Kiselev:course}:
\begin{equation}\label{cumulant_exponential}
        \ln  \langle e^{i k x} \rangle = \sum_{n=1}^\infty \frac{(ik^n)}{n!} \langle x^n \rangle_c.
\end{equation}
Applied to the complex action \eqref{path_integral}, one has in the absence of bulk motion that $\langle \vec{v} \rangle =0$. The cumulant approach generates an expansion in terms of correlation functions:
\begin{equation}\label{SS0}
  \ln\left(\frac{S}{S_0}\right) = - \frac{ \gamma ^2}{2} \int_0^{T_E}  \int_0^{T_E} G_i (t_1) G_j(t_2) \  \langle v^i(t_1) v^j(t_2) \rangle_c \ \mathrm{d} t_1 \mathrm{d} t_2 + \ldots
\end{equation}
where the second cumulant
\begin{equation} \label{cum2}
\langle v^i(t_1) v^j(t_2) \rangle_c = \langle v^i(t_1) v^j(t_2) \rangle - \langle v^i(t_1)\rangle \langle v^j(t_2) \rangle
 \end{equation}
matches the second raw momentum in the absence of bulk motion. Furthermore, the two-point correlator $\langle v^i(t_1) v^j(t_2) \rangle$ only contributes if $t_1 = t_2$, from which a tensor may be defined that has the dimension of a diffusion coefficient:
\begin{equation}\label{def_DT1}
  \langle v^i(t_1) v^j(t_2) \rangle = 2 D^{ij} \delta(t_2-t_1).
\end{equation}
Here a diffusion tensor emerges that captures the anisotropic diffusion of water molecules. Although we use the same notation for both, this proton diffusion tensor should not be confounded with the electrical diffusion tensor that is used in the description of cardiac excitation waves. Nevertheless, both tensors may be expected to align with the local material axes of the tissue, which enables to draw from the proton diffusion tensor to estimate the local orientation of the electrical diffusion tensor. This important observation explains why this dissertation has chapters 3,4 and 9 dedicated to diffusion MRI in the first place. \\

Besides using \eqref{def_DT1}, Eq. \eqref{SS0} may be further rearranged. All reference to the diffusive gradient programming is commonly wrapped into the so-called b-matrix:
\begin{equation}\label{defbmat}
  b_{ij} = \gamma ^2 \int_0^{T_E} G_i (t) G_j(t) \mathrm{d} t.
\end{equation}
In case the gradients vary only in amplitude, not direction, i.e. $\vec{g}(t) = g(t) \vec{u}$, the scalar b-value fully captures the intensity of diffusion weighting:
\begin{equation}\label{def_bval}
  b_{ij} = b u_i u_j, \qquad \text{with}\quad  b = \gamma^2 \int_0^{T_E} G^2 (t) \mathrm{d} t.
\end{equation}
Re-expressing Eq. \eqref{SS0} in terms of the b-value and the proton diffusion tensor brings us to a fundamental relation in diffusion MRI:
\begin{equation}\label{SS0BD}
 \ln \left( \frac{S(\vec{u})}{S_0} \right) = - b \sum_{i,j=1}^3 D^{ij} u_i u_j  + \OO(b^2).
\end{equation}
The dominant angular dependence of the diffusion-attenuated signal is given by the first term of Eq. \eqref{SS0BD}. Writing $E= S/S_0$ for the diffusion-induced attenuation of the MRI signal finally delivers
\begin{equation}\label{atten_DWMRI}
E(\vec{u}) = S/S_0 \exp\left( - b\  \vec{u} \cdot \mathbf{D} \cdot \vec{u} \right).
\end{equation}
Strictly speaking, relation \eqref{atten_DWMRI} only holds exactly for anisotropic Gaussian diffusion. Such process is completely characterized by giving its propagator $ p(\vec{r}_0, \vec{r}_1, \tau)$ for diffusion from position $\vec{r}_0$ to $\vec{r}_1$ in a time interval $\tau$:
\begin{equation}\label{gaussdiff_aniso}
  p(\vec{r}_0, \vec{r}_1, \tau) = \frac{1}{(4\pi \tau)^{3/2} |\mathbf{D}|^{1/2}} \exp\left( - \frac{(\vec{r}-\vec{r}_0) \cdot \mathbf{D}^{-1} \cdot (\vec{r}-\vec{r}_0) }{4\tau} \right).
\end{equation}
Nevertheless, the proton diffusion tensor $\mathbf{D}$ is commonly estimated from \eqref{atten_DWMRI}, which justifies the terminology `apparent diffusion tensor' in such case.

\section{Diffusion tensor MRI}

\subsection{Effective vs. apparent diffusion coefficients}
Equation \eqref{atten_DWMRI} forms the basis of the most common diffusion MRI technique, i.e. diffusion tensor imaging (abbreviated DTI or DT-MRI). The method was coined by Basser \etal to probe microstructure of anisotropic tissues\cite{Basser:1994, Basser:1996}. The underlying idea is that the mean distance traveled by water molecules in a given direction is decreased due to the presence of obstacles which hinder the water molecules in their Brownian movement.

Let us consider a piece of tissue with proton density $\rho(\vec{r})$. In a diffusion process with propagator $p(\vec{r}_0 , \vec{r}_1, \tau)$, the average squared displacement of a water molecule in the direction $\vec{u}$ averaged over a voxel $V$ is then given by
\begin{equation}\label{rms}
  \langle r^2 \rangle(\vec{u},\tau) = \int_V \mathrm{d}^3 r_0 \int_V \mathrm{d}^3 r_1 \rho(\vec{r}_0) p(\vec{r}_0, \vec{r}_1, \tau) \left[ (\vec{r}_1- \vec{r}_0) \cdot \vec{u}\right]^2.
\end{equation}
Herewith, one can define the direction and time dependent effective diffusion coefficient
\begin{equation}\label{def_EDC}
  D_{\rm eff}(\vec{u}, \tau) = \frac{ \langle r^2 \rangle (\vec{u},\tau) }{2 \tau}.
\end{equation}
In lowest order, the weighting direction affects the effective diffusion coefficient in quadratic order:
\begin{equation} \label{def_DT2}
  D_{\rm eff}(\vec{u}, \tau) = \vec{u} \cdot \mathbf{D}_{\rm eff}(\tau) \cdot \vec{u} + \OO(\vec{u}^4).
\end{equation}
One can check that definition \eqref{def_DT2} corresponds with Eqs. \eqref{def_DT1} and \eqref{SS0BD}  up to the order given. It it important to note that the \textit{effective} diffusion tensor is a physical quantity that depends only on the diffusion process (i.e. tissue microstructure and diffusion time $\tau$). In contrast, the \textit{apparent} diffusion coefficient and \textit{apparent} diffusion tensor are slightly different quantities, for they are obtained from estimating diffusivity using Eq. \eqref{atten_DWMRI}. Therefore, they depend on imaging parameters (e.g. b-value) as well:
\begin{equation}\label{def_ADC}
  D_{\rm app}(\vec{u}, \tau, b) = - \frac{1}{b} \ln \left(\frac{S(\vec{u}, \tau, b)}{S_0} \right) \approx \vec{u} \cdot \mathbf{D}_{\rm app}(\tau, b) \cdot \vec{u}.
\end{equation}
Only $D_{\rm app}$ and $\mathbf{D}_{\rm app}$ are accessible in diffusion weighted (DW) experiments; for that reason we will drop the subscripts with the proton diffusion tensor henceforth.

Definition \eqref{def_ADC} of the DT at once implies how to obtain it in practice: repeated measurements of $S(\vec{u},b)$ in $N_g$ independent directions $\vec{u}$, together with $N_0 \geq 1$ acquisitions of $S_0$ allow to estimate the components of $\mathbf{D}_{\rm app}$ by linear regression. We will indicate the total number of images taken by $N_b = N_g+N_0$. In DTI, the minimal number of images required amounts to $N_b = N_g+N_0 = 6 +1$, as six independent tensor components need to be quantified, although taking more diffusion-encoding gradient directions $\vec{u}_i$ increases robustness of the tensor fitting. Evidently, the acquisition time scales proportional to $N_b$.

\subsection{Diagonalization of the diffusion tensor}

From its definition \eqref{def_ADC}, the proton diffusion tensor is real and symmetric. Hence it can be brought to diagonal form by a rotation: $\mathbf{D} = \mathbf{R}^T \boldsymbol{\Lambda} \mathbf{R}$. As the effective diffusion coefficient \eqref{def_DT2} is always positive, $\mathbf{D}$ is positive-definite, and therefore has positive eigenvalues. These eigenvalues are known as the principal diffusivities, for they equal the diffusion coefficients measured along the principal axes of diffusion given by the eigenvectors of $\mathbf{D}$. Commonly, the principal diffusivities are sorted before labeling, i.e. $D_1 \geq D_2 \geq D_3$, with associated eigenvectors $\vec{e}_1, \vec{e}_2,\vec{e}_3$.

To characterize the nature of measured diffusivity, several rotationally invariant measures have been developed. We quote two of them: the mean diffusivity and fractional anisotropy \cite{Westin:2002}:
\bsub \begin{eqnarray}
 D_{\rm av} &=&\frac{1}{3} (D_1 + D_2 + D_3) = \frac{1}{3} \mathrm{Tr}\, \mathbf{D},\\
{\rm FA} &=&  \sqrt{\frac{3}{2} \, . \, \frac{ (D_1- D_{\rm av})^2 + (D_2- D_{\rm av})^2 + (D_3- D_{\rm av})^2   }{D_1^2+D_2^2 + D_3^2}}.
\end{eqnarray} \esub

\subsection{Mapping fibrous and laminar orientation with DTI \label{sec:DTIfib}}

\begin{figure}[h!b] \centering
  \includegraphics[width=0.9 \textwidth]{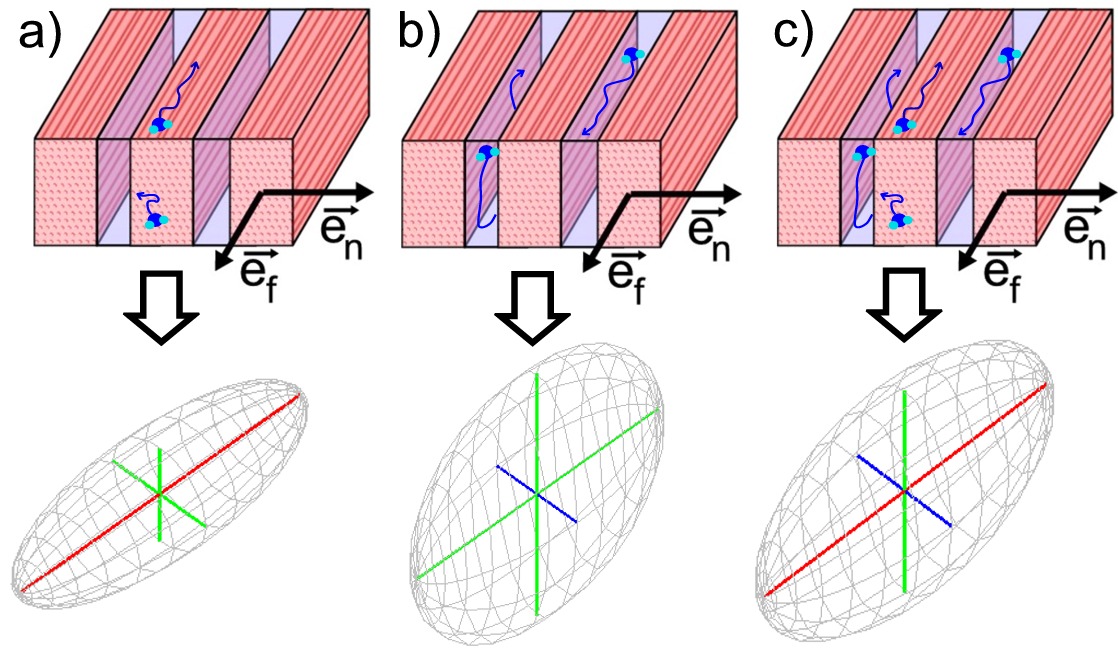}
  \caption[Restricted diffusion and diffusion tensor imaging]{Restricted diffusion and DTI. Considering only diffusion of water within the fiber compartment (a) or contained in the cleavage plane (b) yields in both cases a hypothetical axially symmetric diffusion tensor (lower row). The diffusion tensor reconstructed while taking into account both compartments (c) has three distinct eigenvalues.} \label{fig:diff_in_fib_sh}
\end{figure}

Consider now a cubic voxel filled with anisotropic tissue consisting of well-aligned fibers, e.g. a piece of muscle or a bundle of axons in brain white matter. Denote the voxel-averaged fiber direction by $\vec{e}_f$. Assuming qualitatively that diffusion of water takes place nearly isotropically in the bulk of the intra-and extracellular spaces and that exchange between these compartments is negligible at the time scale considered, diffusion is maximal along the fiber direction, as sketched in Fig.~\ref{fig:diff_in_fib_sh}a. Hence follows the first paradigm of DTI:
\bsub \label{e_DTI}
\begin{equation}\label{ef_DTI}
  \vec{e}_f \approx \vec{e}_1.
\end{equation}
Moreover, one expects that $D_1 \gg D_2 \approx D_3$ in this situation. Relation \eqref{ef_DTI} has been checked extensively through diffusion measurements on real and synthetic samples. Noteworthy is the experimental validation for cardiac myofiber structure against histology in \cite{Scollan:1998, Holmes:2000}.

In this project, we are concerned about whether a similar approach is feasible to probe the local orientation of cleavage planes in the ventricular wall. Thereto, imagine a voxel volume in which only plane-parallel impermeable barriers are present. The orientation of these planar barriers is unambiguously described by their common normal vector $\vec{e}_n$. In this set-up, diffusion is most restricted in the direction of $\vec{e}_n$ (see Fig. \ref{fig:diff_in_fib_sh}), which implies that
\begin{equation}\label{en_DTI}
  \vec{e}_n \approx \vec{e}_3.
\end{equation}
The fact that collections of myofibers define myocardial sheets ensures that $\vec{e}_f \perp \vec{e}_n$, which is consistent with orthogonality of $\vec{e}_1$ and $\vec{e}_3$. With \eqref{ef_DTI}, \eqref{en_DTI} all material axes are fixed, since the direction within the myocardial sheet that forms a right angle with the fiber direction is given by
\begin{equation}\label{es_DTI}
  \vec{e}_s = \vec{e}_n \times \vec{e}_f \ \ \approx \ \ \vec{e}_2 = \vec{e}_3 \times \vec{e}_1.
\end{equation}
\esub
In reality, a voxel containing myocardial cleavage planes always has a considerable fiber compartment volume as well, such that the overall signal results from the superposition of the signals created in both compartments. The measured proton diffusion tensor (Fig. \ref{fig:diff_in_fib_sh}c) is therefore expected to exhibit three distinct eigenvalues $D_1 > D_2 > D_3$, similar to the electrical diffusion tensor orthotropic tissue. From the superposition of uniaxially symmetric diffusion profiles, one may infer that correspondences \eqref{ef_DTI}-\eqref{en_DTI} simultaneously hold in layered myocardial tissue.\\

The estimates \eqref{e_DTI} have been validated in myocardium for regions where a single laminar orientation could be identified \cite{Tseng:2003}. Nevertheless, other studies report a highly variable outcome of sheet orientation as obtained with DTI \cite{Benson:2003, Helm:2005, Gilbert:EJCTS}, which has prevented DTI from taking up a dominant role in the non-invasive imaging of laminar cardiac microstructure.

\subsection{DTI in practice}

It is no coincidence that DTI has grown the most popular present-day technique for the medical imaging of soft anisotropic tissue. Currently, DTI has become the workhorse for the non-invasive tracking of neuronal fibers in the brain, to aid diagnosis and support surgical planning. Probing the anatomical microstructure of patient hearts with DW-MRI has not yet made it to clinical practice, due to the unceasing motion of the heart. In the last few years, scientific progress is being made in DTI of beating hearts in human patients \cite{Reese:1995, Gamper:2007}.\\

For samples with limited movement (e.g. brain, skeletal muscle, \textit{ex vivo} hearts), the DTI method is easily called from built-in scanner software. Besides to the excitation pulse sequence, slice selection and voxel size, the important free parameters $b$-value and diffusion time $\tau$ can be adjusted. Due to relaxation losses of transverse magnetization (see section \ref{sec:SE}), the upper bound for the available echo time given by the minimum of $T_1$ and $T_2$. In fixed canine myocardium, these were in \cite{Helm:2005} measured to be $T_1 = 410\,$ms, $T_2 = 65\,$ms, such that $T_E \leq 60\,$ms poses a practical upper limit to the echo time at typically desired spatial resolution.

Diffusion MRI methods commonly operate in the regime where $bD\approx1$ as a trade-off between angular contrast and signal attenuation, which are both governed by \eqref{atten_DWMRI}. Knowing that at  $25^\circ$C the free diffusion coefficient of water amounts to $\unit{2.2}{\micro}$m$^2/$ms, the effective restricted diffusion coefficients will lie somewhat below this value. Hence, the b-value is most often chosen in the range $400\,$s/mm$^2$ - $3000\,$s/mm$^2$ for \textit{in vivo} applications. Experimentally obtained diffusion coefficients in fixed canine myocardium were measured $D_1 \approx \unit{1.0}{\micro}$m$^2/$ms, $D_2\approx \unit{0.55}{\micro}$m$^2/$ms, $D_3 \approx \unit{0.45}{\micro}$m$^2/$ms for the imaging parameters in \cite{Helm:2005}, while using $b= 1500\,$s/mm$^2$.\\

Additionally, the desired number of gradient encoding directions is selected. Obviously, these gradient directions should be linearly independent, and preferably uniformly distributed in angular space. As DTI requires at least 6 and typically 10 to 20 gradient directions, the sampling scheme is often based on the symmetries of a cube. Due to even symmetry of the diffusion propagator, the sense in which a diffusion encoding gradient is applied does not affect the diffusion-induced signal attenuation. Therefore, the diffusion encoding gradient directions are often chosen from a unit hemisphere.

The acquisition time $T_{\rm acq}$ for a DTI method that scans slice per slice (i.e. a 2D acquisition) depends linearly on the number of slices $N_{\rm sl}$, the time to scan a single slice in one diffusion direction $T_{\rm sl}$, and the total number of images $N_b = N_g + N_0$:
\bsub \begin{eqnarray}\label{Tacq_DWI}
 T_{\rm acq} &=& N_b N_{\rm sl} T_{\rm sl} \\
 T_{\rm sl} &=& T_{\rm prep} + T_E + T_{\rm read}
\end{eqnarray} \esub
The scan time of a single slice is divided over preparation and readout time, which depend on the choice of pulse sequence; for short-lived diffusion gradients, the diffusion time $\tau$ equals the echo time $T_E$.

From \eqref{Tacq_DWI}, the main burden of diffusion MRI strategies can yet be noticed: compared to non-diffusion weighted (`anatomical') MRI, it takes a factor $N_b$ longer to collect a diffusion-weighted image. Nonetheless, the power of diffusion MRI lies in the hierarchy of scales, for microstructure can be assessed at the typical length scale of the diffusion length $L_D = \sqrt{D_0 \tau}$, in voxels of size $a$. When supposing that the time to probe one voxel ($T_v$) is identical for diffusion MRI and a hypothetical anatomical method, the time needed to scan a volume $V$ at spatial resolution $L_D$ with the anatomical method would be $T_{\rm anat} = (V/L_D^3) T_v$. On the other hand, diffusion-MRI can resolve the same structure using larger voxel size $a$, from which follows that $T_{\rm DW-MRI} = (V/a^3)T_v N_b$. The time gain of diffusion MRI with voxel size $a$ compared to an anatomical method which would have to work at a spatial resolution $L_D$ therefore scales as
\begin{equation}\label{scaling_DWMRI}
  \frac{T_{\rm anat.}}{T_{\rm DW-MRI}} \propto \frac{1}{N_b} \left( \frac{a}{L_D}\right)^3.
\end{equation}
For example, with a diffusion length of \unit{10}{\micro\meter} and voxel size 0.5\,mm and 12+1 gradient directions, the time gain ratio \eqref{scaling_DWMRI} amounts to $1.0\,10^{4}$.

Using a Bruker 9.4\,T spectroscope equipped with imaging gradients at the University of Leeds, we have acquired DTI images of entire rat hearts with 12+3 gradient directions in 2h45m, for an image matrix size of  64x50x90.

\subsection{DW-MRI signal from multiple laminar orientation \label{sec:DTIlam}}

The diffusion attenuated MR signal that arises in a voxel which contains multiple fibrous or laminar orientations $\vec{e}_{f,j}$ and $\vec{e}_{n,j}$ may be approximated by the linear superposition of all diffusion compartments involved. Here, one neglects the fraction of water molecules that enters another diffusion compartment during the echo formation time.

In layered myocardium, three compartments are seen to contribute to the diffu\-sion-attenuated MRI signal, i.e. an roughly uniaxial contribution originating within the sheet of myofibers, a nearly isotropic component that stems from the bulk of the water-filled cleavage plane and a term that describes restricted diffusion close to the sheet-void boundary. All compartments are assumed to contribute with weights $p_i$, being the product of the proton density $\rho_i$ and partial volume $V_{i}$ of the compartment; the weights $p_i$ are moreover normalized to add up to one. The diffusion-induced signal attenuation caused by multiple fiber and/or laminar orientations may then be represented as
\begin{equation}\label{simple_sheet_model}
    E(\vec{u})= \sum_{j=1}^{N_{\rm fib} } p_{\rm fib,j}  E_{\rm fib,j}(\vec{u}, \vec{e}_{f,j}) + \sum_{j=1}^{N_{\rm sh} } p_{\rm sh,j}  E_{\rm sh,j}(\vec{u}, \vec{e}_{n,j}) + p_{\rm iso} E_{iso}.
\end{equation}
All listed signal attenuation factors $E$ decrease with longer diffusion time; moreover the partial volume fraction of the compartments fundamentally varies with the diffusion time, as $p_{\rm sh} \propto L_D$.

Importantly, it was remarked in \cite{Frohlich:2006} that extracting physical values for partial volumes and diffusion coefficients based on \eqref{simple_sheet_model} is not possible. For, such fitting in a given direction would require regression of $2 N_{\rm fib} + 2 N_{\rm sh} + 1$ variables, whereas current technology only allows to probe the cumulant expansion \eqref{SS0BD} up to second order, from which only two independent coefficients may be gained.


\subsection{Feasibility of DW-MRI for identifying local sheet orientation}

In the view of the simple analysis \eqref{simple_sheet_model}, the relations \eqref{e_DTI} might indeed be obeyed and could offer a basis for a useful DW-MRI strategy, if following conditions are satisfied:
\begin{enumerate}
\item \textbf{Prevalence.} There should be cleavage planes present in the tissue of the given length scale (gap thickness of $1-\unit{50}{\micro\meter}$) and with sufficient partial volume contribution.
\item \textbf{Parallel barriers.} The surfaces of sheets of myofibers need to be relatively flat at the scale of microns and sufficiently impermeable to limit exchange between diffusion compartments.
\item \textbf{Sufficient signal to noise ratio (SNR).} The oriented myocardial laminae have to be sufficiently abundant to bring signal strength above DW-MRI noise level.
\item \textbf{Signal strength above background processes.}
    The signal generated by the cleavage planes should evenly exceed other well-aligned sources of in-plane diffusion (e.g. outer edges of the sample considered and collagen or capillary networks within the myocardial sheets).
\item \textbf{Alignment of laminae within a voxel.} The cleavage planes should be parallel throughout a voxel to avoid partial volume effects.
\end{enumerate}
The first two criteria are tied to inherent anatomical reality; the latter pair can be conceded to by tuning imaging parameters. Condition (5) can be weakened by selecting a small enough voxel size without dropping below detection threshold. When more than one laminar direction exists at the same anatomical site, however, this concern cannot be reconciled with the diffusion tensor formalism.

\subsection{Limitations of DTI}

From the quoted derivation in \ref{sec:pathint}, it is clear that DTI covers the leading-order term in MRI signal formation, in both angular expansion and the diffusion weighting strength $b$. It should therefore not surprise us that several limitations are inherent to the method \cite{Hagmann:2006}.

To start with, the series \eqref{SS0BD} is known to be exact for Gaussian anisotropic diffusion. This situation is found reasonably fulfilled in anisotropic or porous tissues in the limit of long diffusion times, where the cloud of diffusion protons effectively takes the shape of a the Gaussian anisotropic distribution. For smaller diffusion times, however, the medium heterogeneities start being noticeable, and Eq. \eqref{SS0BD} becomes an approximation.

The higher order corrections to \eqref{SS0BD} in the diffusion strength do not necessarily threaten DTI at moderate or high b-values. Indeed, it suffices to preserve the inverse monotonic correlation between signal strength and directional diffusivity of the medium to establish the principal axes in the tissue. However, when not working with low b-values, it is not possible anymore to quantitatively determine diffusion coefficients, due to the effects of the $b^2$ terms. The qualitative nature of diffusion measurements currently presents an important restriction to all DW-MRI techniques.

The easiest DTI restriction to relieve is the constraint that all angular dependence is caught in a rank-two tensor. With sampling in a large number of gradient directions ($N_g = 60-200$), the signal attenuation due to proton diffusion can be fully sampled on the unit sphere, and higher angular resolution is likely to be achieved. We will pursue this path in the following chapter.

\section[Methodology for reporting orientations of myofibers and -laminae]{Methodology for reporting orientations \\of myofibers and -laminae}

Since our reporting of fibrous and laminar structure in some aspects differs from other works, we indicate suggested modifications here and intend to justify these.

\subsection{Imaging planes and reference systems for the heart}

\begin{figure}[H] \centering
  \includegraphics[width=0.75\textwidth]{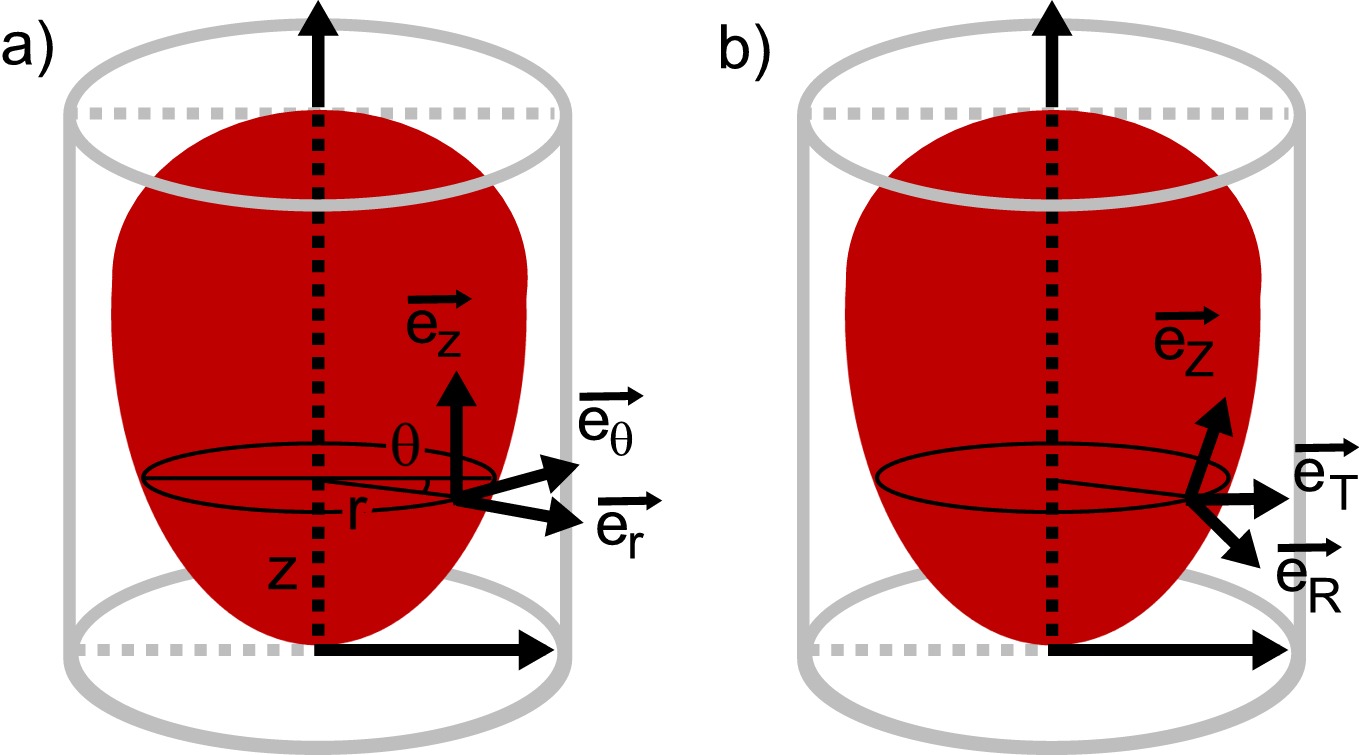}\\
  \caption[Cylindrical coordinates for the heart]{Coordinate systems to denote local directions in the heart:
  cylindrical coordinates (a) and a wall-bound tangential reference frame (b).}\label{fig:im_coords}
\end{figure}

Due to the loose outer cylindrical symmetry of the heart, one often works in a cylindrical coordinate system with the origin in the apex and the Z-axis along the heart's long axis (see Fig. \ref{fig:im_coords}a). An orthonormal set of of unit vectors $\vec{e}_r, \vec{e}_\theta, \vec{e}_z$ can then be defined all over the heart; these reference vectors were yet depicted in Fig. \ref{fig:Rohmer}.
An even better choice in the light of inter-individual comparison would be to redefine $\vec{e}_r \rightarrow \vec{e}_R$ everywhere perpendicular to the epicardial surface, for such wall-bound system would make the referencing less susceptible to deformations of the imaged heart. For simplicity, we have used the cylindrical reference system.

\subsection{Definition of fiber angles}
\begin{figure}[h!t] \centering
  \includegraphics[width=1.0\textwidth]{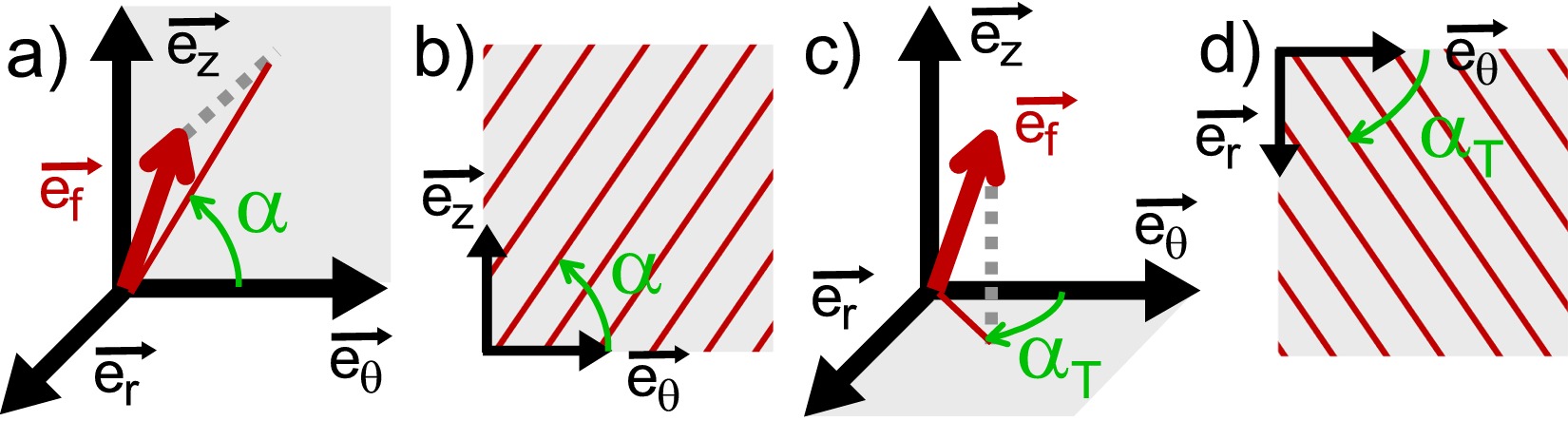}\\
  \caption[Definition of fiber angles]{Definition of fiber angles for reporting local myofiber orientation. The fiber helix angle (a) indicates fiber direction in the circumferential-longitudinal plane (b), whereas the fiber transverse angle (c) refers to apparent fiber direction in the transverse plane (d).}\label{fig:def_fibang}
\end{figure}

Local myofiber orientation is clear as soon as a tangent vector $\pm \vec{e}_f$ to the myofiber direction is given. Inherently, fiber direction is determined by two degrees of freedom, in the cardiac community usually chosen to be the fiber helix angle $\alpha_H$ (or $\alpha$) and fiber transverse angle $\alpha_T$. These `fiber angles' have been defined as \cite{Costa:1997, Geerts:2002}.
\begin{align} \label{fiber_angles}
  \tan(\alpha_H) &= \frac{\vec{e}_f \cdot \vec{e}_z}{ \vec{e}_f \cdot \vec{e}_\theta}, &
  \tan(\alpha_T) &= \frac{\vec{e}_f \cdot \vec{e}_r}{ \vec{e}_f \cdot \vec{e}_\theta}.
\end{align}
Note that the definition of $\alpha_T$ is particularly sensitive to the goodness-of-fit of the cylindrical coordinate frame if such referencing is used. In anatomical cross-sections, the fiber angles \eqref{fiber_angles} may be directly interpreted as the apparent angles between the myofibers and a reference direction; see also Fig. \ref{fig:def_fibang}.

\subsection{Redefinition of sheet angles}

The situation for reporting on myocardial sheet orientation in literature is less transparent. Although the orientation of a plane in three-dimensional Euclidean space is fully characterized by its normal vector and hence only exhibits two degrees of freedom, a variety of sheet angles has been employed in cardiac literature \cite{Costa:1997, Gilbert:EJCTS}. Like fiber angles, sheet angles are mostly defined via projections of a characteristic unit vector onto the cylindrical coordinate planes. The large number of sheet angles arises since some authors project the in-plane vector $\vec{e}_s$, while others use $\vec{e}_n$; this yet leads to six possible projections. 

We here argue that it is more convenient to define sheet angles with respect to the normal vector $\vec{e}_n$ than with the vector $\vec{e}_s$ that lies within the myocardial sheet plane. The basic argument is that angular measures based on $\vec{e}_s$ alone do not allow to infer myocardial sheet orientation, as additional knowledge on the fiber direction is required in such case.

As an example, consider the measurement of cleavage plane orientation in a given cross-section, e.g. as obtained in histological sectioning. In the plane of intersection, which we may freely call the XY plane, laminar orientation is captured by a vector within the XY plane that is tangential to the myocardial sheet direction. In other words, one looks for a specific linear combination $p\vec{e}_f + q\vec{e}_s$ that makes the component transverse to the slice (Z) disappear. Hence follows $p = a e_s^z$, $q=- a e_f^z$ for real $a$, from which we find
\begin{equation}\label{intersection_histology_1}
  (v_x, v_y, v_z) = a (e_f^x e_s^z - e_s^x e_f^z , e_f^y e_s^z - e_s^y e_f^z, 0),
\end{equation}
which clearly depends on both $\vec{e}_f$ and $\vec{e}_s$. As an alternative, we derive the same direction from the sheet normal vector $\vec{e}_n$ by stating that $\vec{n} \perp \vec{r}$ for all $\vec{r}$ in the sheet plane. This implies $\vec{n}\cdot \vec{v} = 0$, or
\begin{equation}\label{intersection_histology_2}
  (v_x, v_y, v_z) = a (n_y , -n_x, 0),
\end{equation}
 which is the same solution as \eqref{intersection_histology_1} since $\vec{e}_n = \vec{e}_f \times \vec{e}_s$. The advantage of formulation \eqref{intersection_histology_2} is that now a single vector fully fixes the cleavage plane direction as seen on a histological slice. Moreover, this approach hold for any projection plane: given a cross-section of the heart, the intersected cleavage planes are found in such plane by projecting $\vec{e}_n$ onto that plane, and rotating over $90^\circ$.

 In cardiac literature, the visible intersection of a cleavage plane with the transverse plane and longitudinal-radial plane is commonly assessed with the $\beta'$ and $\beta''$ sheet angles, respectively \cite{Costa:1997, Gilbert:EJCTS}:
\begin{align} \label{beta_acc_angles}
  \tan(\beta') &= \frac{\vec{e}_s \cdot \vec{e}_z}{ \vec{e}_s \cdot \vec{e}_r}, &
  \tan(\beta'') &= \frac{\vec{e}_s \cdot \vec{e}_\theta}{ \vec{e}_s \cdot \vec{e}_r}.
\end{align}
For the arguments given above, the sheet angles $\beta'$ and $\beta''$ must not be interpreted as the apparent intersection angle of the cleavage plane in the relevant image plane. This complication makes validation against histology a non-trivial task.

\begin{figure}[h!t] \centering
  \includegraphics[width=1.0\textwidth]{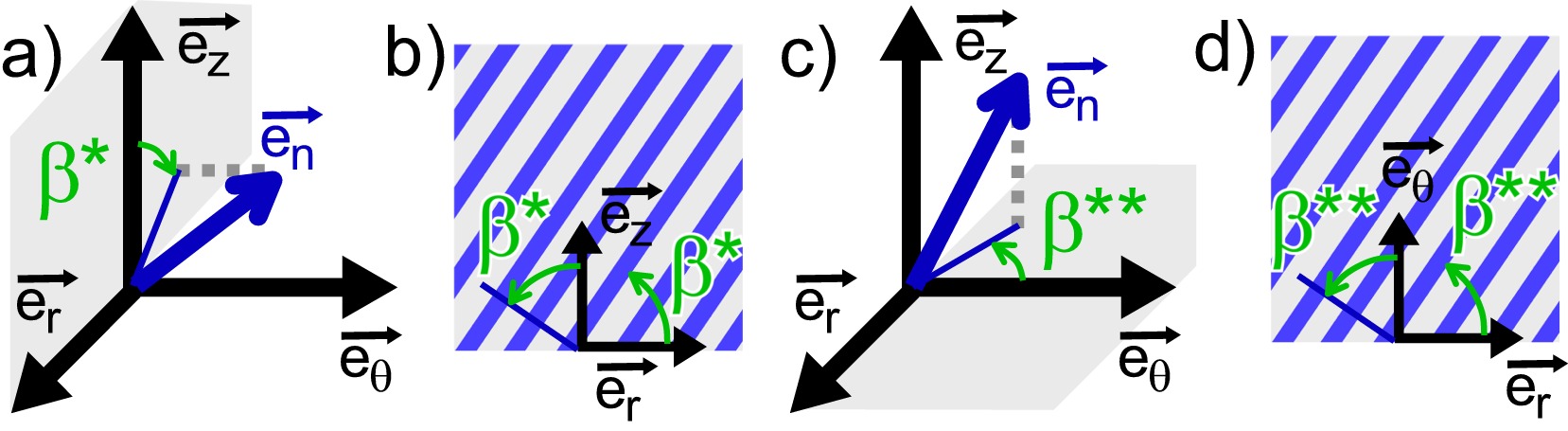}\\
  \caption[Redefinition of sheet angles]{Definition of sheet angles for reporting local myofiber orientation. The sheet angle $\beta^*$(a) indicates the apparent orientation of laminar myocardial structure within the longitudinal-radial plane (b), while the sheet angle $\beta^{**}$ (c) refers to visible cleavage plane orientation in the transverse plane (d).}\label{fig:def_shang}
  \vspace{0.5cm}
\end{figure}

The appropriate sheet intersection angles with respect to the transverse and longitudinal-radial plane may be most simply defined using the normal vector to the laminae. In other words, we propose
\begin{align} \label{beta_star_angles}
  \tan(\beta^*) &= - \frac{\vec{e}_n \cdot \vec{e}_r}{ \vec{e}_n \cdot \vec{e}_z}, &
  \tan(\beta^{**}) &= - \frac{\vec{e}_n \cdot \vec{e}_r}{ \vec{e}_n \cdot \vec{e}_\theta},
\end{align}
to replace the $\beta'$ and $\beta''$ sheet angles based on $\vec{e}_2$ (and therefore probed using the second DT eigenvector).
One can check that our $\beta^*$ coincides with the angle $\phi$ defined in \cite{Helm:2005}; the angle $\beta$ that was used in \cite{Rohmer:2007} equals $\beta^*-90^\circ$. A visual representation is given in Fig. \ref{fig:def_shang}.

If needed, the laminar structure in a tangential plane, i.e. in a slice parallel to the epicardial surface, may be analogously captured by a `sheet helix angle':
\begin{equation} \label{beta_H angle}
  \tan(\beta_H) = - \frac{\vec{e}_n \cdot \vec{e}_\theta}{ \vec{e}_n \cdot \vec{e}_z}.
\end{equation}
Since the sheet angle $\beta_H$ only differs from the fiber helix angle in so far as the myofibers do not lie in a tangential plane, this quantitative measure is of less importance than the sheet angles $\beta^*$ and $\beta^{**}$ defined above.

\subsection{Pitfalls in representing fiber and sheet angles}

When presenting graphs or color plots of fiber and sheet angles, we emphasize that it is essential to keep in mind that the $sense$ of the unit vectors $\vec{e}_f$, $\vec{e}_s$, $\vec{e}_n$ is unphysical, as only directions (tangent lines) need to be specified. Therefore, fiber and sheet angles must be defined irrespective of the sense of these vectors, which is currently satisfied in literature. In addition, we are convinced that the unphysical nature of the sense of the mentioned vector fields also implies the following:
\begin{enumerate}
 \item \textbf{No drawing of arrows in vector maps of $\vec{e}_f$, $\vec{e}_s$ and $\vec{e}_n$.}  Rather, one may draw both senses on vector maps. When performing interpolation of these vector fields, some might need to be flipped to point in the same direction; this procedure can at best only be performed locally.
  \item \textbf{Fiber and sheet angles are periodic functions over $180^\circ$.} Independently of the chosen interval to which the angle values have been mapped (e.g. $[-90^\circ,+90^\circ]$ or $[0,180^\circ]$), the cyclic identification needs be respected, especially when taking moving averages of angles close to the (virtual) borders of the angular interval. In the particular case of transmural profiles, either the angular range must be widened when angular values exceed a border value, or the periodicity should be made clear by continuing the profile curve after jumping over $\pm 180^\circ$ to the other side of the box.
 \item \textbf{Use cyclic colormaps when encoding fiber or sheet angles.} Nowadays, it is common practice to use a blue-to-red colormap that ranges from $-90^\circ$ to $+90^\circ$ to indicate fiber and sheet angles (see e.g.\cite{Gilbert:EJCTS}). As a consequence, a striking red/blue interface appears in these images, even though no sharp gradients are present in the tissue structure. We opt to remedy this artefact by adopting a cyclic hsv-colormap (hue-saturation-value), where blue is found at $-60^\circ$ and red at $+60^\circ$. A continuous color change over magenta tints then accounts for a smooth transition at the value $\pm 90^\circ$.
 \item \textbf{Fiber and sheet angles that use projection on a plane cannot be consistently defined for all orientations.} In the worst case scenario, the vector $\vec{e}_f$ or $\vec{e}_n$ could be orthogonal to the plane in which the fiber or sheet angles have been defined. In such case, the definition of the angle contains the indeterminate fraction(0/0) and therefore may deliver any value of the fiber or sheet angle. If occurring, this situation is easily recognized by considering the fiber or sheet angles that have another plane of projection. This indetermination owing to the user's definition must be distinguished from fundamental uncertainty in sheet and fiber angle that results from the anatomical reality.
\end{enumerate}

\section{DTI results for fiber and sheet orientation}

\subsection{Materials and methods for DTI acquisition}

DTI imaging of two rat hearts was performed at the University of Leeds on a 9.4\,T Bruker spectrometer (Ettlingen, Germany) equipped with imaging gradients. The hearts were prepared by Dr. Stephen Gilbert, and the sample was fixed in its container and scanner bore by Dr. Alan Benson. With both scans, the hearts' long axes were approximately aligned with the $\vec{B}_0$ field.\\

We decided to use a 3D phase encoding spin-echo sequence with $T_E = 15\,$ms and repetition time $T_R = 550\,$ms. A pair of short-living diffusion-weighing gradients of duration $\delta = 2\,$ms was applied with an intergradient interval of $\Delta = 7\,$ms to yield a nominal b-value of $3000\,$s/mm$^2$. Diffusion-encoded information was used from $N_g = 16$ gradient directions and the geometric average of $N_0 =3$ non-DW images. Voxel size was \unit{300}{\micro\meter} isotropic, with final image matrix size $64\times 50\times 90$. Data taking for use in the DTI reconstruction took in total 3 hours 28 minutes.\\

Afterwards, diffusion tensors were reconstructed using custom-written Matlab software (The Mathworks Inc., Natick, MA). A threshold on the $b_0$ image was first used to select only those voxels that contained myocardial tissue. Next, the diffusion-weighted measurements in $N_g = 16$ directions $\vec{u}_k, k=1, ..., N_g$ were organized in a set of $N_g$ linear equations, following Eq. \eqref{SS0BD}: $\ln(S_k/S_0) = D^{ij} b u_i^k u_j^k$. A least squares fitting process is then straightforward to estimate the six independent tensor components $D^{ij}$ in each voxel. For all voxels, the resulting diffusion tensor was diagonalized and the eigenvalues were sorted. Finally, the local material axes were inferred based on the correspondence \eqref{e_DTI}.\\

We present slices here from a DT-imaged rat heart, with the fiber and sheet angles color-coded using a cyclic colormap. Transmural profiles are presented for a sector through the LVFW of $20^\circ$ wide and three voxels thick, containing $N=147$ voxels.

\subsection{Fiber orientation with DTI \label{sec:DTIrat}}

\begin{figure}[h!t] \centering
\mbox{
  \raisebox{2.75cm}{a)} \includegraphics[width=0.4 \textwidth]{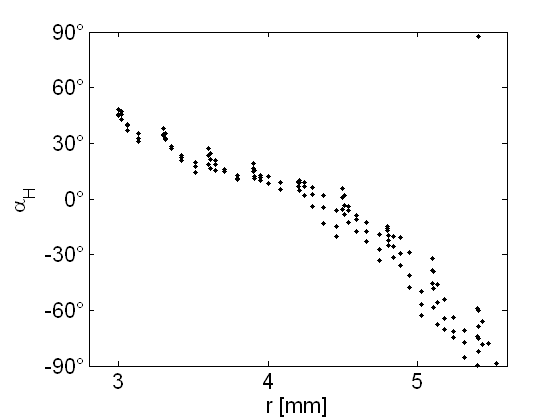}
   \raisebox{2.75cm}{b)}\includegraphics[width=0.4 \textwidth]{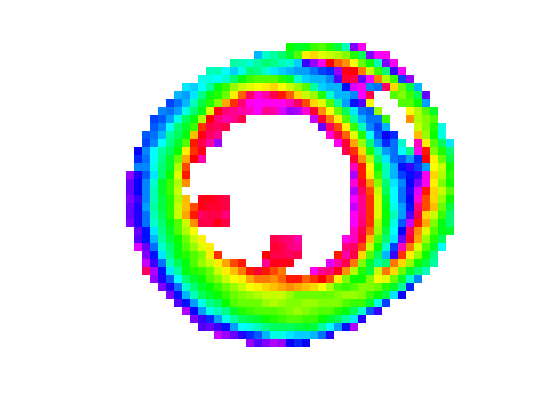}
   \raisebox{0.75cm}{       \includegraphics[width=0.1 \textwidth]{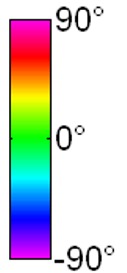}}
   }
  \caption[DTI result: fiber helix angle]{Fiber helix angle $\alpha_H$ reconstructed from DTI. a) Transmural course of $\alpha_H$ through the LVFW, with distance measured from the slice centroid. b) Axial slice with cyclic colormap representing $\alpha_H$.}\label{fig:DTI_results_AH}
\end{figure}

\begin{figure}[h!t] \centering
\mbox{
  \raisebox{2.75cm}{a)} \includegraphics[width=0.4 \textwidth]{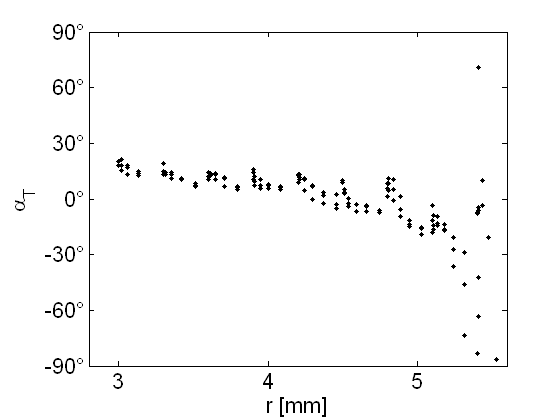}
   \raisebox{2.75cm}{b)}\includegraphics[width=0.4 \textwidth]{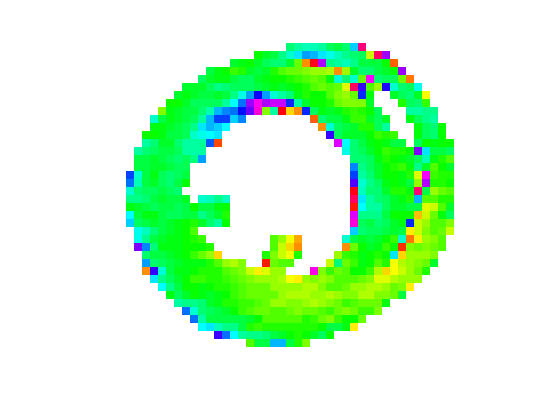}
   \raisebox{0.75cm}{       \includegraphics[width=0.1 \textwidth]{colorbar9090.jpg}}
   }
 \caption[DTI result: fiber transverse angle]{Fiber transverse angle $\alpha_T$ reconstructed from DTI. a) Transmural course of $\alpha_T$ through the LVFW, with distance measured from the slice centroid. b) Axial slice with cyclic colormap representing $\alpha_T$.}\label{fig:DTI_results_AT}
\end{figure}

The fiber orientation in an axial cross-section of the heart is fully captured by listing the fiber helix angle $\alpha_H$ and fiber transverse angle $\alpha_T$, as presented in Figs. \ref{fig:DTI_results_AH} and \ref{fig:DTI_results_AT}. Note that throughout the ventricles, a gradual fiber counterclockwise fiber rotation is witnessed when going from epi- to endocardium. In the papillary muscles, the myofibers run nearly parallel to the long axis, as may be seen from $\alpha_H$ taking values close to $\pm 90^\circ$. Due to the use of a cyclic colormap, artificial discontinuities that are due to color-coding have been avoided.\\

The fiber transverse angle slowly decreases from $\alpha_T = 20^\circ$ to $\alpha_T = -20^\circ$ when moving outward in the LVFW. Near the epicardial surface ($r>5.5\,$mm), $\alpha_T$ is ill-posed, since the projection of $\vec{e}_f$ onto the transverse plane degenerates to a point, given that $\alpha_H \approx \pm 90\circ$ in that region.

\subsection{Laminar orientation with DTI}

In our results from DTI, we present both transmural sheet angle profiles for sheet angles defined using $\vec{e}_s$ (i.e. $\beta'$, $\beta''$) and sheet angles $\beta^*$, $\beta^{**}$ that we have based on $\vec{e}_n$ in Eqs. \eqref{beta_star_angles}.\\

\begin{figure}[h!t] \centering
\mbox{
  \raisebox{2.75cm}{a)} \includegraphics[width=0.4 \textwidth]{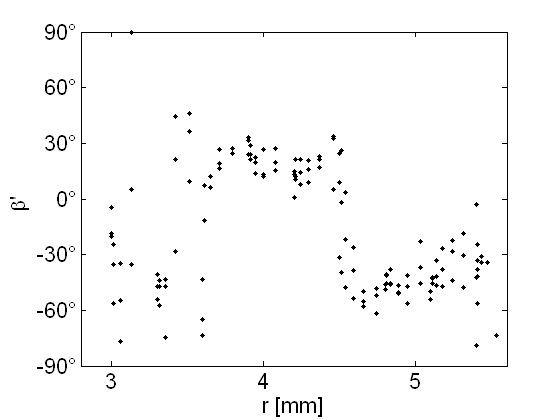}
   \raisebox{2.75cm}{b)}\includegraphics[width=0.4 \textwidth]{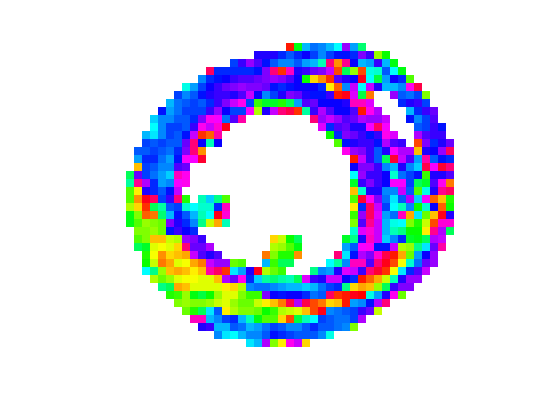}
   \raisebox{0.75cm}{       \includegraphics[width=0.1 \textwidth]{colorbar9090.jpg}}
   }
   \mbox{
  \raisebox{2.75cm}{c)} \includegraphics[width=0.4 \textwidth]{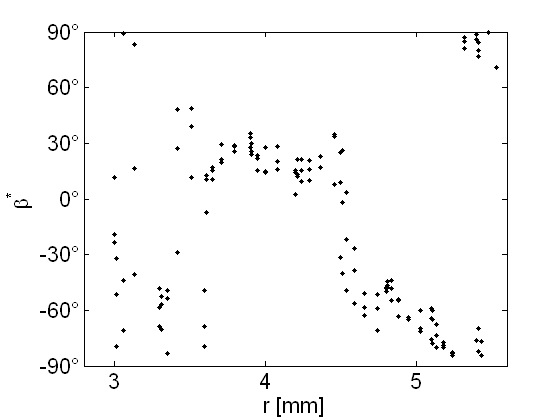}
   \raisebox{2.75cm}{d)}\includegraphics[width=0.4 \textwidth]{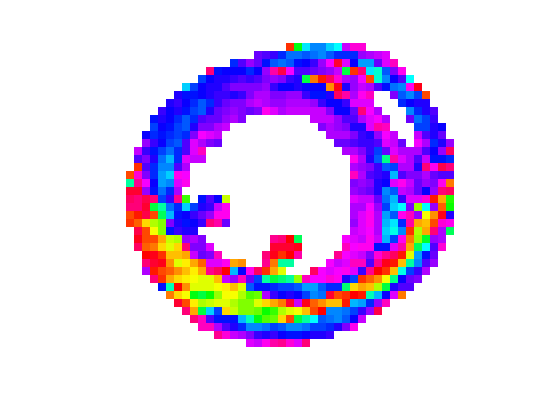}
   \raisebox{0.75cm}{       \includegraphics[width=0.1 \textwidth]{colorbar9090.jpg}}
   }
  \caption[DTI result: sheet angles $\beta'$ and $\beta^*$]{Orientation of cleavage planes in the radial-longitudinal direction viewed with DTI using the $\beta'$ (a-b) and $\beta^*$ (c-d) sheet angles. Panels (a),(c) denote the transmural course through the LVFW, with distance measured from the slice centroid. Panels (b),(d) show the axial slice with a cyclic colormap. }\label{fig:DTI_results_BA}
\end{figure}

We first we consider the inclination of cleavage planes in the radial-longitudinal plane; DTI outcome is depicted in Fig. \eqref{fig:DTI_results_BA}. Both angles $\beta'$ and $\beta^*$ indicate an inclination of $-60^\circ$ to $-80^\circ$ in the anterior and septal wall portions of the LV. The situation in the lateral and posterior LVFW is different, as a discontinuity is encountered in the central part of the myocardial wall on the transmural profiles. No abrupt change in orientation is associated in the epicardial part of the $\beta^*$ profile (outer right in panel \ref{fig:DTI_results_BA}c), as all angles defined are cyclic with period $180^\circ$.

\begin{figure}[H] \centering
\mbox{
  \raisebox{2.75cm}{a)} \includegraphics[width=0.4 \textwidth]{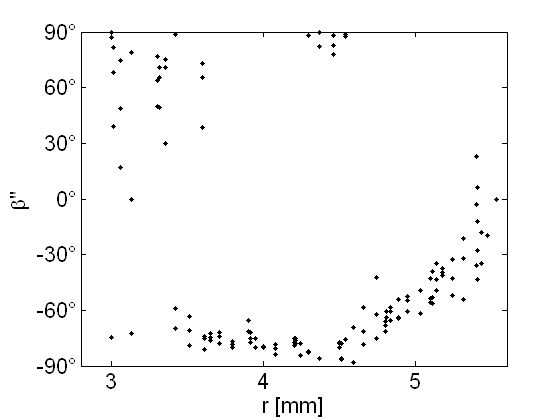}
   \raisebox{2.75cm}{b)}\includegraphics[width=0.4 \textwidth]{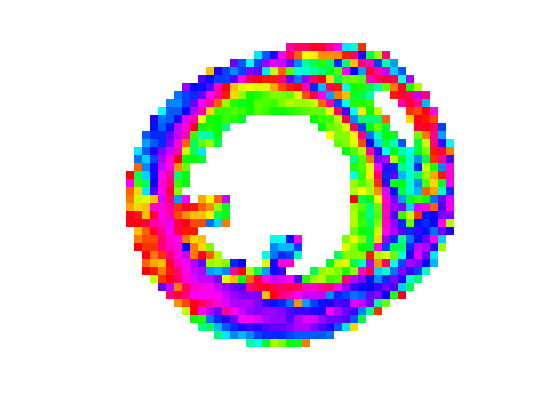}
   \raisebox{0.75cm}{       \includegraphics[width=0.1 \textwidth]{colorbar9090.jpg}}
   }
   \mbox{
  \raisebox{2.75cm}{a)} \includegraphics[width=0.4 \textwidth]{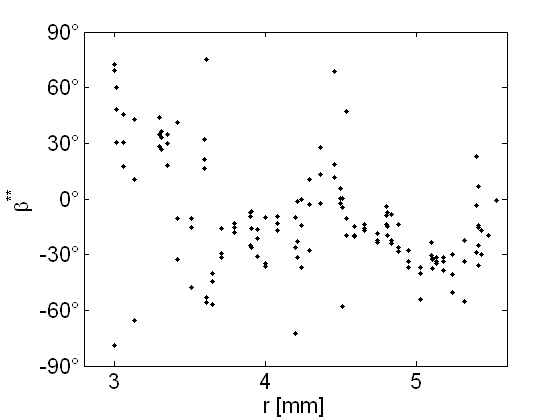}
   \raisebox{2.75cm}{b)}\includegraphics[width=0.4 \textwidth]{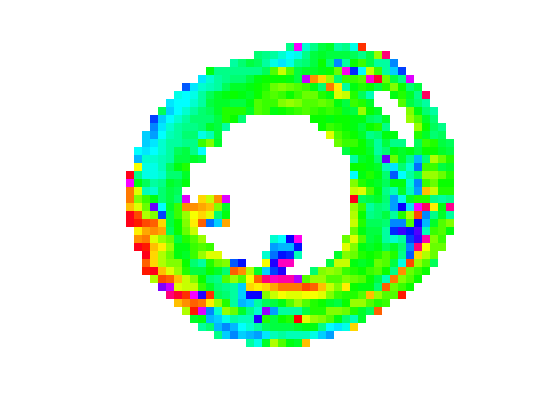}
   \raisebox{0.75cm}{       \includegraphics[width=0.1 \textwidth]{colorbar9090.jpg}}
   }
  \caption[DTI result: sheet angles $\beta''$ and $\beta^{**}$]{Orientation of cleavage planes in the radial-circumferential direction viewed with DTI using the $\beta''$ (a-b) and $\beta^{**}$ (c-d) sheet angles. Panels (a),(c) denote the transmural course through the LVFW, with distance measured from the slice centroid. Panels (b),(d) show the axial slice with a cyclic colormap. }\label{fig:DTI_results_BAA}
\end{figure}

Secondly, we render the DTI results for laminar orientation in the transverse plane using $\beta''$ and $\beta^{**}$ angles in Fig. \eqref{fig:DTI_results_BAA}. Note that the numerical outcome significantly differs between the $\beta''$ and $\beta^{**}$ angles. As we argued before, in such case only the $\beta^{**}$ sheet angle may be associated with the visible texture of a transverse cross-section. A subendocardial discontinuity was found in both the $\beta''$ and $\beta^{**}$ transmural profiles.

\begin{figure}[h!t] \centering
\mbox{
  \raisebox{2.75cm}{a)} \includegraphics[width=0.4 \textwidth]{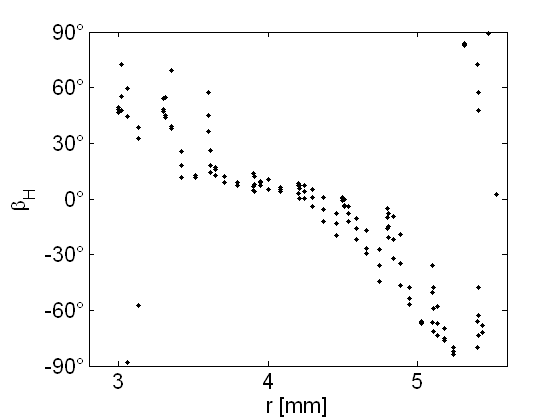}
   \raisebox{2.75cm}{b)}\includegraphics[width=0.4 \textwidth]{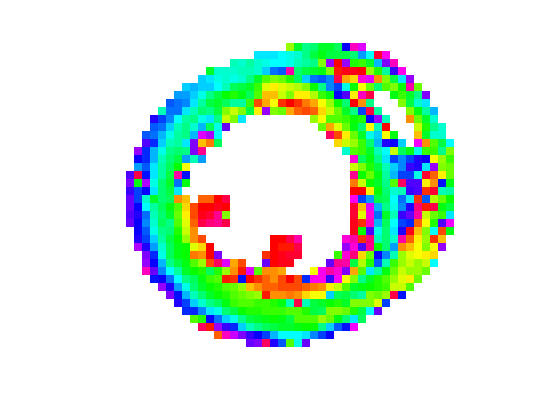}
   \raisebox{0.75cm}{       \includegraphics[width=0.1 \textwidth]{colorbar9090.jpg}}
   }
  \caption[DTI result: sheet angle $\beta_H$]{Orientation of cleavage planes in the radial-circumferential direction viewed with DTI using the $\beta_H$ sheet angle as defined in Eq. \eqref{beta_H angle}. a) Transmural course through the LVFW, with distance measured from the slice centroid. b) Axial slice with cyclic colormap representing $\beta_H$.}\label{fig:DTI_results_BH}
\end{figure}

Lastly, the $\beta_H$ helix angle displayed in \ref{fig:DTI_results_BH} captures the orientation of myocardial sheets in the tangential plane. Due to the approximate alignment of fibers within this plane, the values for $\beta_H$ closely resemble those the fiber helix angle $\alpha_H$ from Fig. \ref{fig:DTI_results_AH}, as we had anticipated.

%
%
%
%
%
%


\clearpage{\pagestyle{empty}\cleardoublepage}


\renewcommand\evenpagerightmark{{\scshape\small Chapter 4}}
\renewcommand\oddpageleftmark{{\scshape\small Dual q-ball imaging}}

\hyphenation{inter-gradient geo-met-ri-cal}

\chapter[Dual QBI]{Dual q-ball imaging}
\label{chapt:QBI}

In this chapter, we outline how one may surpass the limited angular resolution that is inherent to DTI. To start with, we apply the existing q-ball method to probe myocardial structure and resolve complex fiber structure in specific anatomical regions of the heart. Thereafter, we formulate a variation to the q-ball method so that it manages to resolve crossing cleavage planes instead of myofibers. The combination of both techniques is called dual q-ball imaging (dQBI) and exemplified here on entire rat ventricles. We provide preliminary validation against histology as well, which shows a reasonable quantitative match.

\section[Review of conventional q-ball imaging for fiber structure]{Review of conventional q-ball imaging\\ for fiber structure}

To introduce the high angular resolution diffusion techniques, it is instructive to first expose the formalism of q-space, after which the q-ball imaging method for the imaging of fibers can be readily unfolded.

\subsection{Fourier formulation of diffusion MRI}
The general expression \eqref{path_integral} for signal formation in diffusion MRI simplifies significantly if a pair of equal short-living gradients is used of constant gradient strength $g$ and duration $\delta$. (See e.g. \cite{Mitra:1995} and Fig. \ref{fig:dwg}). If the duration $\delta$ of the pulses in the diffusion encoding gradient pair is much shorter than their time lag $\Delta$, it is said that the narrow pulse criterion is satisfied:
\begin{equation}\label{npc}
  \delta \ll \Delta.
\end{equation}
In such case, each of the diffusion encoding gradients may be represented by a Dirac delta distribution:
\begin{equation}\label{narrow_pulse_grads}
  \vec{g}(t) = g \delta\  \left[ \delta\left(t- t_0\right)  -  \delta\left(t- t_1\right) \right] \vec{u}.
\end{equation}
The path integral \eqref{path_integral} now breaks down into the propagator equation
\begin{eqnarray}
\langle e^{- i\phi} \rangle &=& \int d^3 r_0 \int d^3 r_1 \rho(\vec{r}_0) p(\vec{r}_0, \vec{r}_1, t_1-t_0) e^{i \gamma \delta \vec{g} \cdot (\vec{r}_1- \vec{r}_0)}.
\end{eqnarray}
With a change of integration variable $\vec{r}_1 \rightarrow \vec{r} = \vec{r}_1- \vec{r}_0$ and denoting the intergradient interval $t_1-t_0$ as $\Delta$, one finds that
\begin{equation}\label{DW_fourier}
E = S/S_0 = \langle e^{- i\phi} \rangle = \int d^3 r P(\vec{r}, \Delta) e^{i \vec{q} \cdot \vec{r}} = \FF[P(\vec{r}, \Delta)](\vec{q}).
\end{equation}
Here, the propagator $P(\vec{r}, \Delta)$ represents the voxel-averaged probability that a proton within the voxel is found displaced over  $\vec{r}$ after a time $\Delta$. This propagator is commonly named either the net spin displacement function, voxel-averaged spin propagator, or probability density function (PDF):
\begin{eqnarray} \label{defPDF}
P(\vec{r}_1-\vec{r}_0, t_1-t_0) &=& \int d^3 r_0 \rho(\vec{r}_0) p(\vec{r}_0, \vec{r}_1, t_1-t_0).
\end{eqnarray}
Hence, with a pair of short-lived diffusion encoding gradients ($\delta \ll \Delta$), the diffusion signal attenuation \eqref{DW_fourier} gains a particular form: it is nothing else than the three-dimensional Fourier transform of the PDF!

Fourier-conjugated to the net displacement $\vec{r} = \vec{r}_1- \vec{r}_0$ is the variable $\vec{q}$ \footnote{Some authors use $\vec{q} =(2\pi)^{-1} \gamma \delta \vec{g}$, e.g. \cite{Tuch:2004}. The factor $2\pi$ is then manifest in the definition of the Fourier transform.}:
\begin{equation}\label{def_qvar}
  \vec{q} = 2 \pi \bar{\gamma} \delta \vec{g} = \gamma \delta \vec{g}.
\end{equation}
Otherwise stated, the magnitude of $\vec{q}$ is proportional to the diffusion gradient amplitude $g$ and its direction corresponds to the direction $\vec{u}$ along which the diffusion encoding is applied. The Larmor frequency for protons $\gamma$ acts as a proportionality constant that makes the dimensions consistent. Equation \eqref{def_qvar} enables to refer to $\vec{q}$ as the diffusion wave vector\footnote{The q-space formalism must not be confounded with k-space in MRI applications: the imaging gradients $\vec{k}$ are Fourier-conjugate to the absolute position $\vec{x}$, while the diffusion encoding gradients $\vec{q}$ form a Fourier pair with the displacement (relative position) of the protons $\vec{r}$.}. Note that in the case of short gradient pulses, the b-value \eqref{def_bval} may be taken
\begin{equation}\label{bval_shortgrads}
    b \approx q^2 \Delta.
\end{equation}

From Eq. \eqref{DW_fourier}, it was deduced that the full PDF can hypothetically be recovered by sampling a large fraction of q-space and performing an inverse Fourier transform \cite{Basser:2002, Tuch:2003, Sosnovik:2009}. This method is known as diffusion spectrum imaging. However, resolving spatial details requires sampling at large q-vectors and is therefore technically demanding; also, dense sampling of a considerable part of q-space is relatively time consuming. Moreover, the narrow pulse criterion is only seldom satisfied with acquisitions on clinical scanners.

In comparison, one could say that DTI samples q-space in only $N_g$ points, at a fixed distance from the origin. There exists a method that lies in between q-space imaging and DTI through economical sampling in q-space: q-ball imaging.

\subsection{Theory of q-ball imaging}

In 2003, an approach to relieve the sampling burden of diffusion spectrum imaging was proposed by Tuch \cite{Tuch:2003}. To this purpose, not the entire PDF was aimed to be reconstructed, but solely its angular content. Tuch defined the fiber orientation distribution function (ODF) as the normalized radial projection of the PDF:
\begin{equation}\label{def_fiberODF}
    \psi(\vec{u}) = \frac{2}{Z} \int\limits_0^\infty P(r\vec{u}) \mathrm{d}r =  \frac{1}{Z} \int\limits_{-\infty}^\infty P(r\vec{u}) \mathrm{d}r,
\end{equation}
with Z a normalization constant such that the integral of $\psi$ over the unit sphere sums to one. The second expression in \eqref{def_fiberODF} follows from the first when assuming even symmetry of the propagator $P(\vec{r})$. It was furthermore shown that the ODF $\psi(\vec{u})$ could be reasonably estimated from a series of acquisitions with the same gradient strength $q_s$ but with different gradient directions; for that reason the new method was termed \textit{q-ball imaging} (QBI). Reconstruction of the fiber ODF involves the Funk-Radon transform (FRT). The FRT, which is also called the Funk transform, or spherical radon transform \cite{Helgason:1999} is a linear map from the unit sphere to itself for which the function value in a point $\vec{u}$ is given by the sum over the corresponding equator:
\begin{equation}\label{FRTdef}
    \mathcal{G}[f (\vec{w})] = \int_{S_2} f(\vec{v}) \delta(\vec{v} \cdot \vec{w}) d \Omega.
\end{equation}
With the FRT, one can write following estimator for the ODF \cite{Tuch:2003}.
\begin{equation}\label{hatpsiE}
    \hat{\psi}(\vec{u}) = \frac{1}{Z} \mathcal{G}[E(q_s \vec{u})].
\end{equation}
In our treatment, we will consistently put a hat on the ODF symbol when intending an estimated distribution function. The estimator \eqref{hatpsiE} is in fact the true ODF convoluted with the zeroth order Bessel function of the first kind $J_0$. The analytical proof given in \cite{Tuch:2003, Tuch:2004} will be extended in this work for the imaging of laminar structure in section \ref{sec:proof_dQBI}. In cylindrical coordinates $(r, \theta, z)$ with the diffusion gradient axis along $Z$, one obtains:
\begin{equation}\label{odfbessel}
     \mathcal{G}[E(q_s \vec{u})] = 2 \pi q_s \int d \theta \int r dr \int dz P(r,\theta,z) J_0(q_s r).
\end{equation}
It was moreover argued that the approximation $\hat{\psi}(\vec{u}) \approx \psi(\vec{u})$ holds as far as the function $J_0(q_s r)$ is concentrated near the origin. Larger values of $q_s$ lead therefore to better estimates of the PDF, but again at the cost of lower signal strength, since $b \propto q^2$ due to Eq. \eqref{bval_shortgrads}.

\subsection{Meaning and visualization of the fiber ODF}

Recalling section \ref{sec:DTIfib}, the restricted diffusion in tissue with a single predominant fiber orientation has a PDF with largest extent in the fiber direction $\vec{e}_f$ due to maximal diffusivity in that direction. Through Eq. \eqref{def_fiberODF}, the fiber ODF $\psi(\vec{u})$ and its estimator $\hat{\psi}(\vec{u})$ are argued to be monomodal distributions (with antipodal symmetry) that reach their maximum in the direction of fiber alignment.

\begin{figure}[H] \centering
  \mbox{
  \raisebox{3cm}{a)} \includegraphics[width=0.15\textwidth]{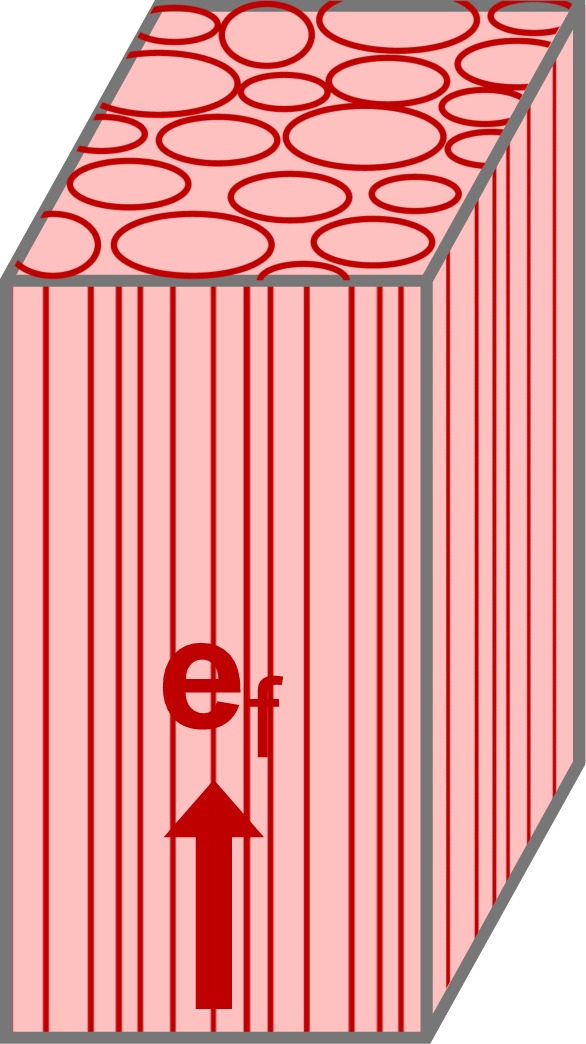}
  \raisebox{3cm}{b)} \includegraphics[width=0.08\textwidth]{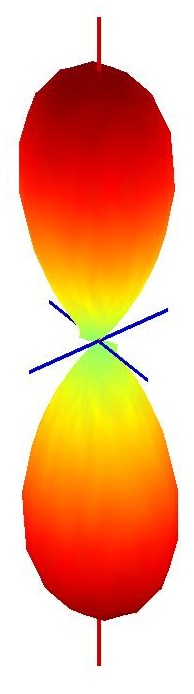}
  \raisebox{3cm}{c)} \includegraphics[width=0.25\textwidth]{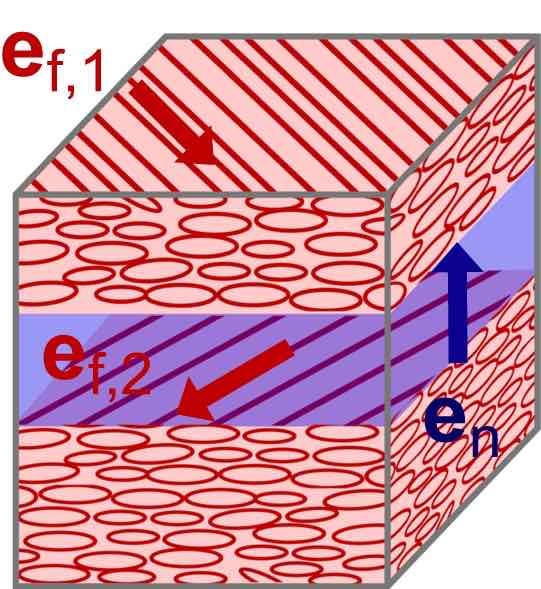}
  \raisebox{3cm}{d)} \includegraphics[width=0.35\textwidth]{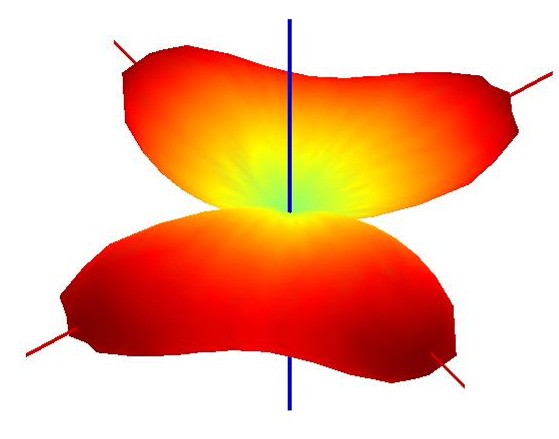}
  }
  \caption[Visualization of the fiber ODF]{Representation of the fiber ODF as a spherical function $r=\psi(\vec{u})$. The local maxima of the fiber ODF point in the underlying fiber direction. Therefore, voxel (a) with a single fiber direction (which is also detectable with DTI) generates a monomodal fiber ODF(b), whereas a voxel (c) containing myofiber of multiple orientation yields a fiber ODF with local maxima that point in the respective fiber directions (d).}\label{fig:vis_ODF}
\end{figure}

By linearity of the QBI method (which is embodied by the the Funk-Radon transform), the ODF associated to a voxel with more than one fiber orientation is simply the partial volume-weighted sum of the ODFs, in the limit where water exchange between compartments is negligible. Therefrom, it can be expected that closely spaced groups of fibers with different orientation generate a cross-shaped fiber ODF, as depicted in \ref{fig:vis_ODF}b. Also, when few distinct fiber orientations are present, their directions $\vec{e}_{f,i} $ may be inferred from the local maxima of the fiber ODF.

In summary, QBI replaces the DTI relation \eqref{ef_DTI}, i.e. $\vec{e}_f \approx \vec{e}_1$ by
\begin{equation}\label{QBI_fibs}
\psi(\vec{u}) \ \text{reaches a local maximum in}\ \vec{e}_{f,i}.
\end{equation}

\subsection{Implementation of QBI}

\begin{figure}[H] \centering
\mbox {
  \raisebox{2.8cm}{a)}
  \includegraphics[width=0.3\textwidth]{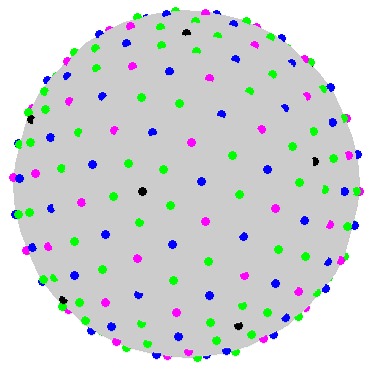} \hspace{1cm}
    \raisebox{2.8cm}{b)}
    \includegraphics[width=0.3\textwidth]{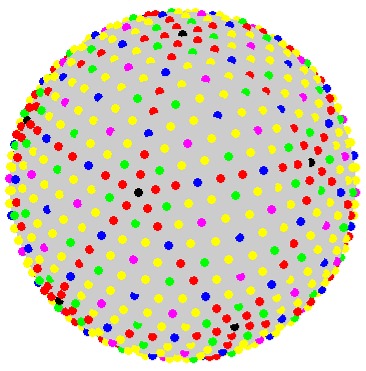}
}
  \caption[The q-ball: discrete sampling directions on a sphere]{Sampling directions on the unit sphere for $2N = 2.\,136$ (a) and $2.N = 2.\, 406$ (b). The vertices of the initial regular icosahedron are depicted by the black dots; colors indicate equivalent positions due to symmetry.} \label{fig:qballdirs}
\end{figure}

The data acquisition for QBI hardly differs from taking a DTI scan. That is, only the set of gradient encoding directions $N_g$ needs be increased with an order of magnitude. To construct a set of directions nearly uniformly spaced on the unit sphere, we have started from a regular icosahedron. When $n$ points are added to each edge of the triangular faces and all intersection points within the face are drawn, $3/5 + 3n/2 + n(n-1)/2$ points lie in each face. Given that there are 20 faces in a icosahedron, we obtain, after dividing by two to exclude antipodal directions, a gradient direction set of size
\begin{equation}\label{ngrads}
  a_n = 6 + 5 n(n+2).
\end{equation}
From this series, we have used $a_3=81$ and $a_8=406$ in our implementation, which is illustrated Fig. \eqref{fig:qballdirs}. An additional gradient direction set was made by adding the barycenters of the set with $n=2$ to the gradient directions, which increased the number of vertices with $10(n+1)^2$ to $136$. Subsequent tessellation of this set delivers the previously mentioned set with $a_8 = 406$ directions. Note that the average span between adjacent gradient directions can be approximated by $2 / \sqrt{ 2 N_g}$, which yields $14^\circ$ for $N_g=136$ and $8^\circ$ for $N_g=406$.

The tensor fitting process in DTI is in QBI replaced by taking the Funk-Radon transform, for which various implementations have been suggested in literature. Some authors \cite{Descoteaux:2007, Hess:2006} favor the use of spherical harmonics $Y_{\ell m}(\vec{u})$, since the FRT becomes diagonal in this basis:
\begin{equation}\label{diag_FRT}
  \GG[Y_{\ell m}(\vec{u})] = 2 \pi P_\ell(0) Y_{\ell m}(\vec{u}),
\end{equation}
with $P_\ell$ a Legendre polynomial. Additionally, the use of spherical harmonics exhibits implicit low-pass filtering and even parity of the ODF may be readily imposed. However, the order at which the spherical harmonics expansion of the estimated ODFs should be cut off is still not clear.

For that reason, we have preferred to stay close to Tuch's numerical realization of the FRT \cite{Tuch:2004}, which makes use of spherical radial basis functions on the unit sphere. For each voxel, the q-ball data set  $S(\vec{u}_i), i=1,..., N_g$ was via the discrete implementation of the FRT cast onto a denser set of $N_q$ uniformly distributed directions. In particular, our measurements with $N_g=136$ were mapped by the FRT onto $N_q = 406$ directions; both sets of directions were displayed in Fig. \ref{fig:qballdirs}. A similar interpolation is found evenly in DTI and the spherical harmonics implementation of QBI and serves to increase the effective angular resolution.

\subsection{Advantages and drawbacks of QBI}

Without doubt, the main advantage of QBI compared to DTI is the ability to describe diffusion patterns of complex angular structure. Even more, QBI does not assume a diffusion model and respects the linear superposition of diffusion compartments. From a practical side, QBI is easily implemented on a MRI scanner (adjust the gradient table) and image reconstruction is straightforward. Also, the resulting fiber ODFs are easy-to-understand, as their three-dimensional representation $r= \psi(\vec{u})$ `points' in the direction of highest probability for underlying fiber orientation.

Weaknesses of QBI can be summarized as being still a time-consuming acquisition process, having still a limited angular resolution and not yet offering quantitative diffusion data. Also, given that QBI cannot determine the position of structures within a voxel, no distinction can be made between closely spaced myofiber structures of different orientation and a single meshwork of fibers.

\section[Application of QBI for probing myocardial fiber structure]{Application of QBI for probing\\ myocardial fiber structure \label{sec:QBIapplied}}

We present to our knowledge the only application of QBI to assess fibrous structure in the heart with increased angular resolution. Since the myofiber field is well-defined in typical ventricular anatomy, only few voxels are expected to yield more that one fiber direction when using decent spatial resolution. Nonetheless, we have observed more complex fiber structure in specific anatomical regions such as the LV-RV-IVS fusion site and the onset of the papillary muscles to the endocardium.

Very recently, Sosnovik \etal have used diffusion spectrum imaging to investigate non-trivial fibrous structure near infarction zones \cite{Sosnovik:2009}.

\subsection{Conventional QBI on synthetic myofiber data}

We have first tested whether QBI is likely to resolve crossing myofibers, given the reduced anisotropy ratio in the proton diffusion tensor in myocardial tissue compared to brain white matter, where QBI is mostly being used. To this aim we have assumed simple anisotropic Gaussian diffusion compartments with principal diffusivities
\begin{align} \label{Dfib_synth}
 \{D^{\rm fib}_i \} &= (1.0, 0.5, 0.5)\unit{}{\micro\meter}^2\rm{/ms},&  \{D^{\rm sh}_i \} &= (2.2, 2.2, 0.8) \unit{}{\micro\meter}^2\rm{/ms},
\end{align}
from which a hypothetical MRI signal was synthesized as a sum of negative exponentials. The partial volume attributed to the myofiber and cleavage plane compartments was 50\% each. Reconstructing the DT from this hypothetical signal delivered as eigenvalues $(1.22,\  0.78,\,  0.65)  \unit{}{\micro\meter}^2\rm{/ms}$, which closely matches experimental values $(1.14\pm0.16,\  0.79\pm0.16,\  0.67\pm0.15) \unit{}{\micro\meter}^2\rm{/ms}$ that we have obtained in \textit{ex vivo} rat ventricle.

\begin{figure}[H] \centering
\mbox{
  \raisebox{2.75cm}{a)} \includegraphics[width=0.45 \textwidth]{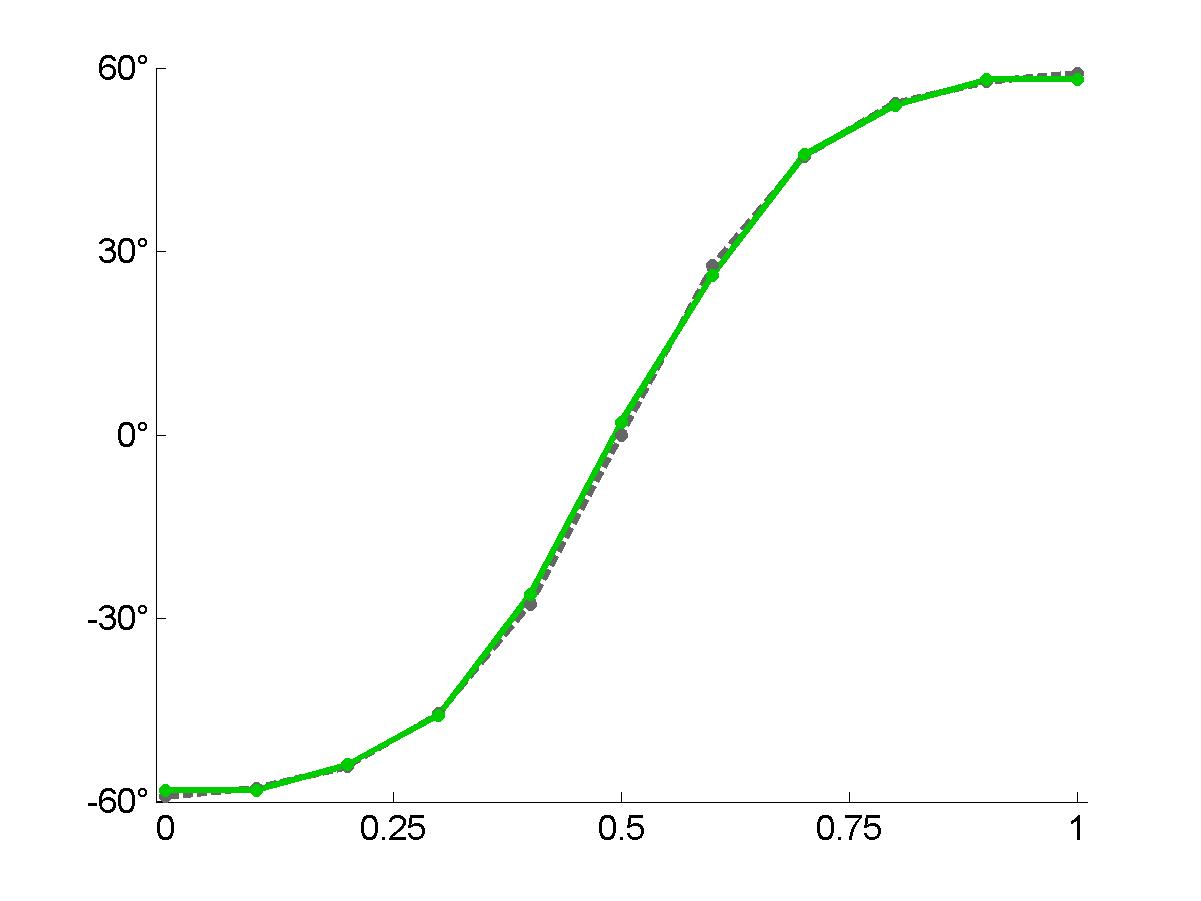}
   \raisebox{2.75cm}{b)}\includegraphics[width=0.45 \textwidth]{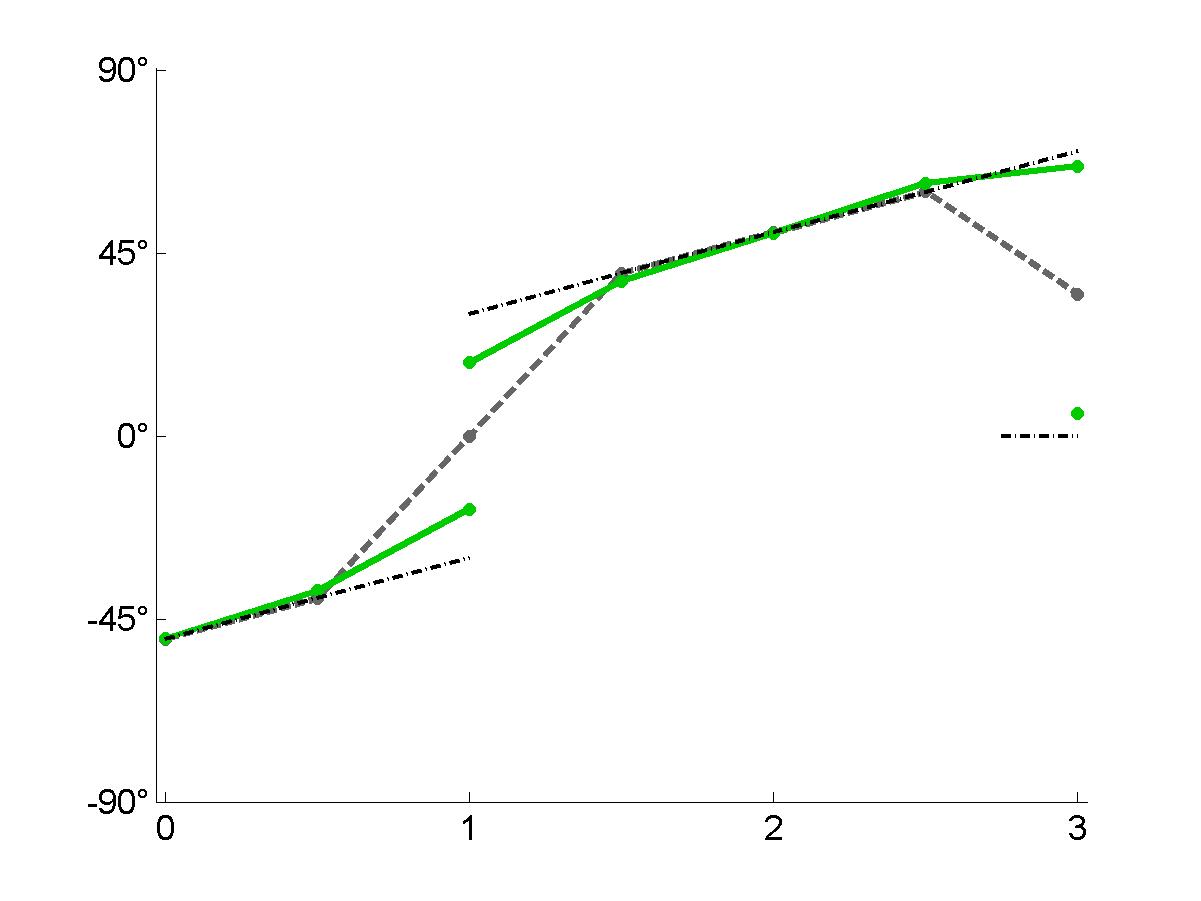}
   }\\
\raisebox{2.cm}{c)}   \includegraphics[width=0.95 \textwidth]{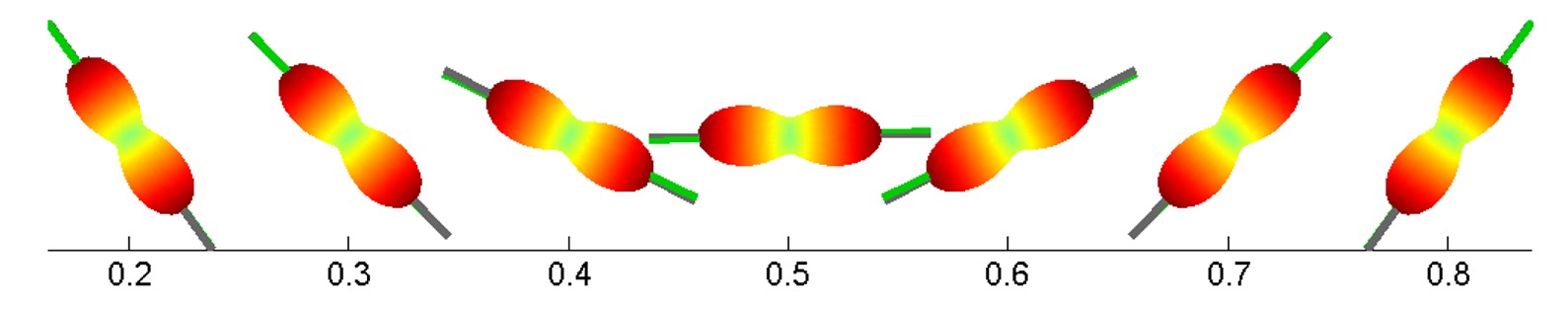}\\
\raisebox{2.cm}{d)}   \includegraphics[width=0.95 \textwidth]{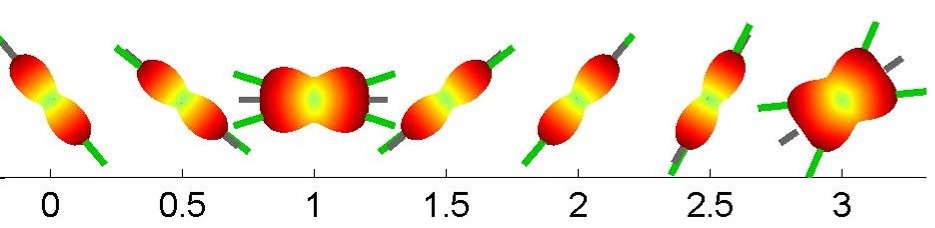}
  \caption[QBI result for synthetic myofiber data]{QBI result for synthetic myofiber data, for simple sigmoidal myofiber rotation (a,c) and more complex organization (b,d). In the transmural profiles (a-b) for fiber helix angle $\alpha_H$, the horizontal coordinate may be interpreted as epicardial depth in mm; the data points show ground truth (black), DTI outcome (gray) and QBI outcome (green). The fiber ODFs (c-d) have been rotated over $90^\circ$ towards the viewing direction. }\label{fig:QBIfibturn}
\end{figure}

With the signal generated by the values \eqref{Dfib_synth}, we simulated transmural fiber rotation and measured the hypothetical fiber helix angle $\alpha_H$. This was not only done for simple sigmoidal rotation over $120^\circ$, but also for a profile that includes abrupt fiber rotation in a mid-wall region, and distinct fiber orientation on the endocardial surface. Both geometries are represented in panels (a,c) and (b,d) of Fig. \ref{fig:QBIfibturn}. In the transmural profiles, the black dash-dot line indicates ground truth, green denotes QBI outcome and the DTI result is given in gray. It can be seen that, with complex fiber organization, QBI is expected to significantly better perform than DTI.

\subsection{Fiber structure in the LVFW}

We here present the QBI counterpart of the axial slice represented in section \ref{sec:DTIrat}. Scanning parameters were identical to those listed in \ref{sec:DTIrat}. In fact, the DTI reconstruction ($N_g=16$) was conducted from a subset of the $N_g = 136$ uniformly distributed gradient directions that we now use for QBI. The total acquisition time for QBI rises linearly with $N_b$; it amounted to 25h24m for a single heart. \\

\begin{figure}[h!t] \centering
\mbox{
  \raisebox{2.75cm}{a)} \includegraphics[width=0.4 \textwidth]{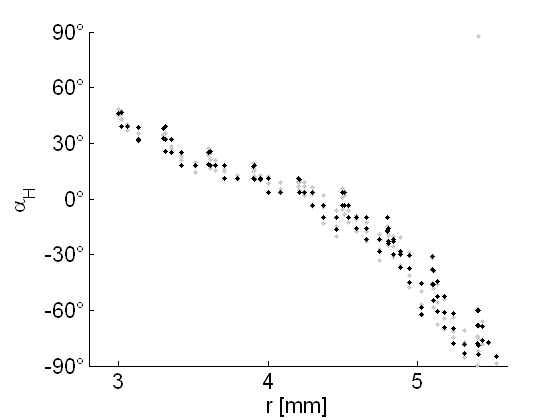}
   \raisebox{2.75cm}{b)}\includegraphics[width=0.4 \textwidth]{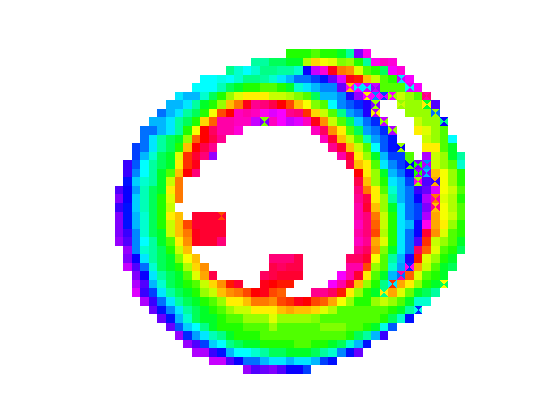}
   \raisebox{0.75cm}{       \includegraphics[width=0.1 \textwidth]{colorbar9090.jpg}}
   }
  \caption[QBI result: fiber helix angle]{Fiber helix angle $\alpha_H$ reconstructed from QBI. a) Transmural course of $\alpha_H$ through the LVFW, with distance measured from the slice centroid.
  b) Axial slice with cyclic colormap representing $\alpha_H$.}\label{fig:QBI_results_AH}
\end{figure}

\begin{figure}[h!t] \centering
\mbox{
  \raisebox{2.75cm}{a)} \includegraphics[width=0.4 \textwidth]{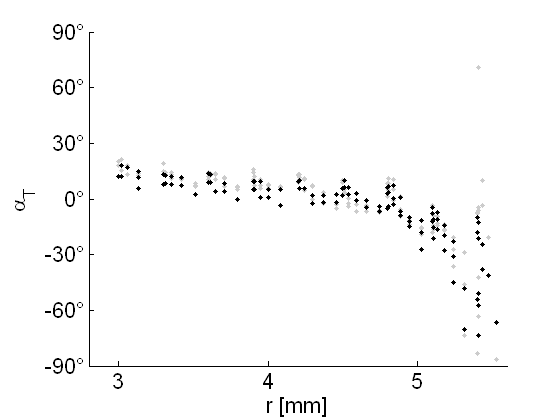}
   \raisebox{2.75cm}{b)}\includegraphics[width=0.4 \textwidth]{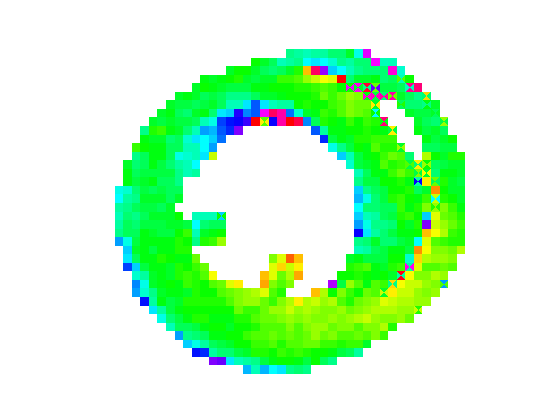}
   \raisebox{0.75cm}{       \includegraphics[width=0.1 \textwidth]{colorbar9090.jpg}}
   }
 \caption[QBI result: fiber transverse angle]{Fiber transverse angle $\alpha_T$ reconstructed from QBI. a) Transmural course of $\alpha_T$ through the LVFW, with distance measured from the slice centroid. b) Axial slice with cyclic colormap representing $\alpha_T$.}\label{fig:QBI_results_AT}
\end{figure}

With the fiber directions extracted in each voxel with QBI, one may evenly compute the fiber angles and represent them in a slice view. The result is shown in Figs. \ref{fig:QBI_results_AH}-\ref{fig:QBI_results_AT}. Voxels that contain more than one fiber direction are colored accordingly in the color-encoded slice. The selected sector in the LVFW did not hold multimodal ODFs; therefore, the transmural profiles obtained by QBI (black dots) are nearly identical to the DTI result, which is rendered by in gray.

\subsection{Detailed study of ventricular fiber structure}

\begin{figure}[h!t] \centering
  \includegraphics[width=1.0\textwidth]{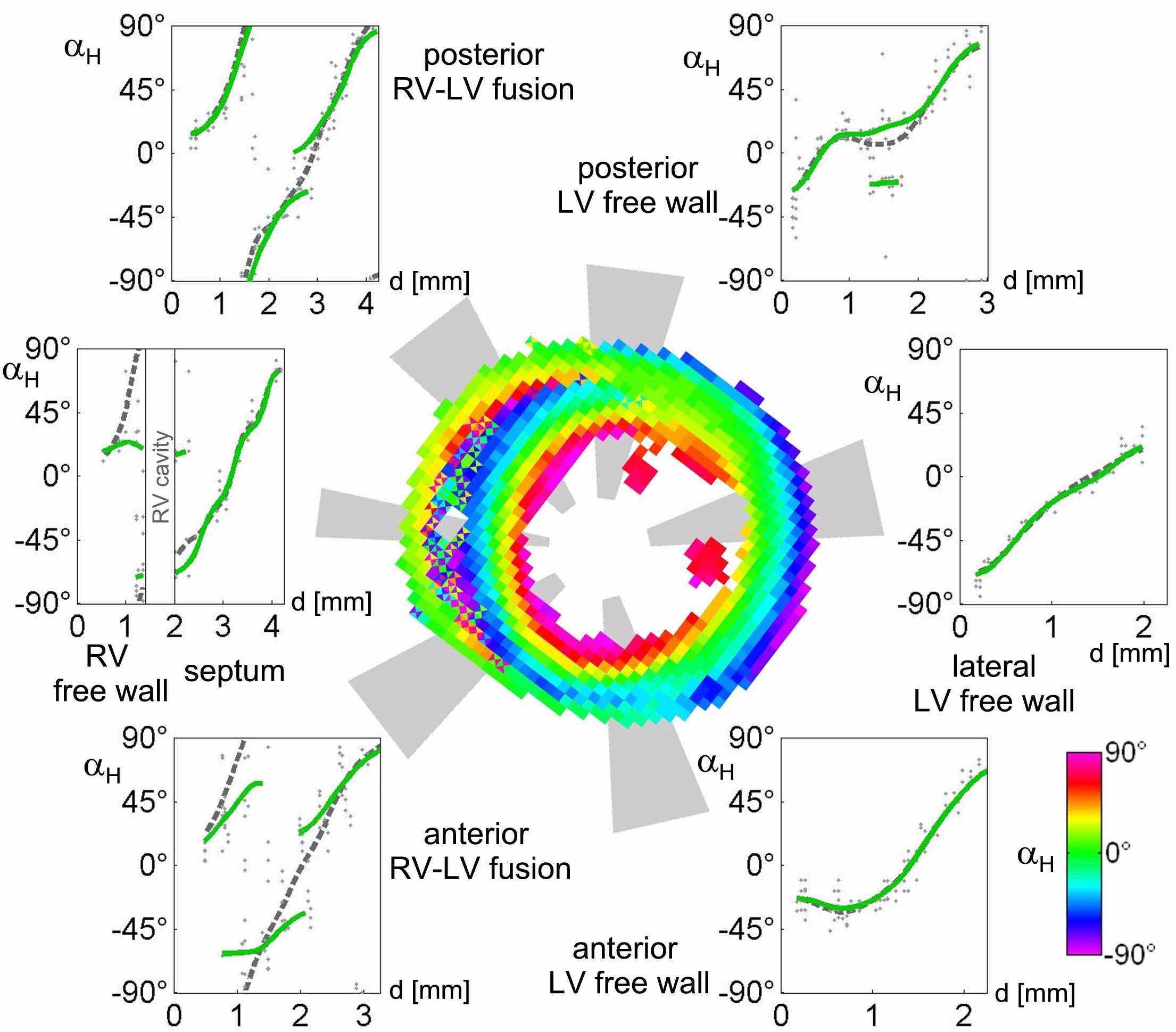}\\
  \caption[Fiber QBI results for six ventricular sectors]{Fiber QBI results in six ventricular sectors in terms of the fiber helix angle $\alpha_H$. In the transmural profiles, distance $d$ represents distance from the epicardium. Green curves denote the QBI outcome, which may be compared to the DTI result given in gray dotted lines.
  }\label{fig:dQBI_fibs}
\end{figure}

We have also performed a more detailed study with QBI of the fiber direction in six distinct sectors of the same heart \cite{Dierckx:2010a}. Figure \ref{fig:dQBI_fibs} displays an overview of our findings. Note that the viewing angle is different from before in this composite figure; moreover the transmural position is given here by epicardial depth $d$ in millimeters.

The chosen sectors for analysis were the anterior, lateral and posterior portions of the LVFW, a sector through the IVS and RV, and the anterior and posterior fusion sites where LV, RV and IVS merge. Distinct fiber populations identified by fiber QBI were semi-automatically segmented, after which a moving-average could be taken to generate averaged transmural profiles (green) that take into account the distinct myofiber orientations if present. DTI outcome is presented by the gray dotted lines for comparison.

Strikingly, only the anterior, lateral LV wall and IVS were found to exhibit a gradual fiber rotation. In the posterior LVFW, the transmural profile shows an additional fiber direction in mid-wall, which can be traced back to the posterior LV-RV fusion. In the DTI profile, this anatomical feature causes an local decrease of $\alpha_H$; with QBI, the transmural course remains monotonous. In both fusion sites, QBI enables to see how the discontinuity in the transmural fiber course sets in, as an abrupt change of fiber rotation develops in mid-wall; this feature evolves in the circumferential direction to become the RV cavity.\\

In the following section we will expose how one may likewise assess laminar structure with increased angular resolution.

\section{Dual q-ball imaging for laminar structure}

Based on geometric arguments, we demonstrate how concepts from conventional QBI may be translated to QBI for laminar structure. Obviously, a first correlation is that $\vec{e}_f$ needs be replaced by $\vec{e}_n$ in our description. To proceed, we need the notion of a laminar ODF.

\subsection{Introduction of the laminar ODF}

\begin{figure}[h!b] \centering
  \mbox{
  \raisebox{3.05cm}{a)}\includegraphics[width=0.35\textwidth]{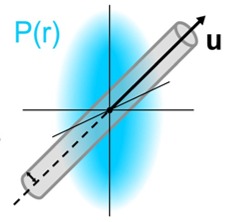}\hspace{1cm}
  \raisebox{3.05cm}{b)}\includegraphics[width=0.4\textwidth]{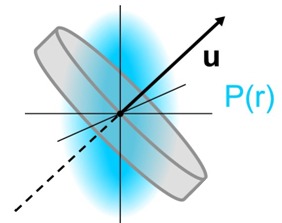}
  }\\
  \caption[Duality in the definitions of the fiber and laminar ODF]{Duality in the definitions of the fiber (a) and laminar ODF (b). 
  }\label{fig:FODF_LODF_proj}
\end{figure}

For isotropic diffusion that is restricted between parallel impermeable barriers with spacing $a$, the PDF will take the shape of a thin disc after a long enough diffusion time $tau$ or diffusion length $L_D$
\begin{equation}\label{long_diff_time}
  L_D = \sqrt{D \tau} \gg a.
\end{equation}
This thin disc is oriented parallel to the occurring barriers, i.e. orthogonal to the normal vector to the plane barriers. On that account, we define the laminar orientation distribution function $\xi(\vec{u})$ as a counterpart to the fiber ODF \eqref{def_fiberODF}:
\begin{eqnarray}
  \xi(\vec{u}) &=& \frac{1}{Z'} \int_{\vec{u}_\perp} P(\vec{r}) \mathrm{d} S \label{def_laminarODF}
\end{eqnarray}
The difference in the integration subspace between the fiber and laminar ODFs is illustrated in Fig. \ref{fig:FODF_LODF_proj}.

When a voxel contains multiple, non-aligned compartments in which diffusion takes place between closely spaced parallel barriers \eqref{long_diff_time}, the resulting laminar ODF arises as the linear superposition of the laminar ODFs of the compartments, in the limit where exchange between compartments is negligible at the considered time scale $\tau$. As a consequence, the laminar ODF that is produced by water contained in between two non-aligned myocardial cleavage planes is likely to yield a laminar orientation distribution $\xi(\vec{u})$ which exhibits local maxima in the directions normal to the cleavage planes. Therefore the DTI relation \eqref{en_DTI}, i.e. $\vec{e}_n \approx \vec{e}_3$ is superseded by
\begin{equation}\label{dQBI_sheets}
 \xi(\vec{u})\ \text{reaches a local maximum in}\  \vec{e}_{n,i}.
\end{equation}
in dual QBI of laminar structure.

At this stage already, the tight mathematical correspondence to the fiber ODF $\psi(\vec{u})$ becomes clear: both approaches correlate through a projective duality of the vector space $\mathbb{R}^3$, in which each line is mapped onto the plane orthogonal to it and \text{vice versa}. This mathematical duality allows us to easily deploy a framework suited for imaging myocardial sheets, which we therefrom term `dual QBI', abbreviated dQBI.\\

\begin{figure}[h!b] \centering
\mbox{
  \raisebox{3.05cm}{a)} \includegraphics[width=0.3 \textwidth]{geom_FODF_cr.jpg}\
   \raisebox{3.05cm}{b)}\includegraphics[width=0.42 \textwidth]{FODF_f.jpg}\
   \raisebox{3.05cm}{c)}\includegraphics[width=0.15 \textwidth]{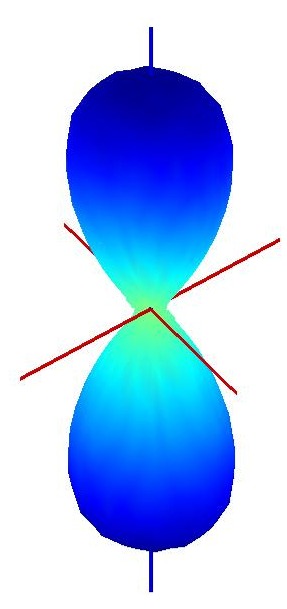}
   }
\mbox{
  \raisebox{3.05cm}{d)} \includegraphics[width=0.30 \textwidth]{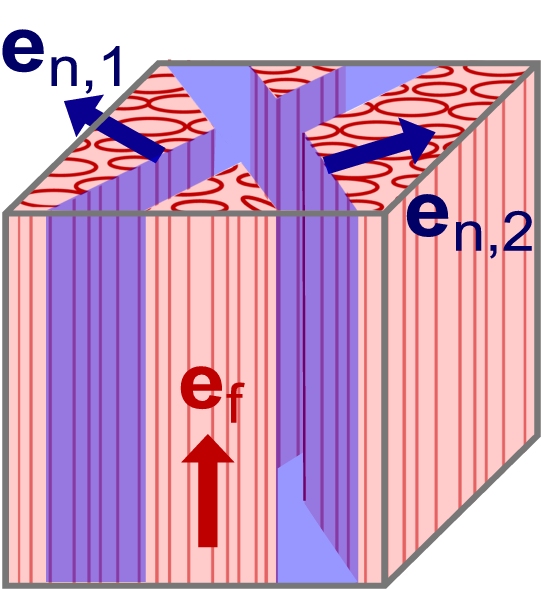}\
   \raisebox{3.05cm}{e)}\includegraphics[width=0.1 \textwidth]{FODF_s.jpg}\
   \raisebox{3.05cm}{f)} \includegraphics[width=0.5 \textwidth]{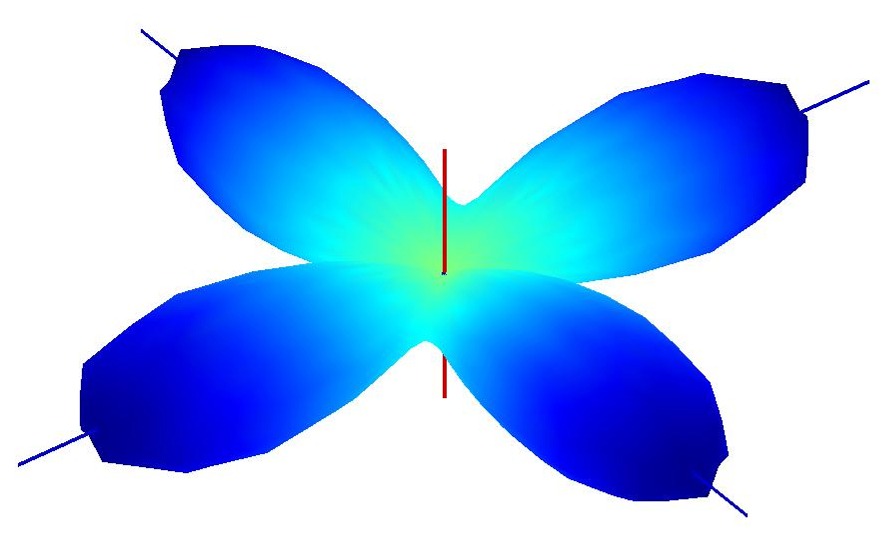}
   }
 \caption[Dual QBI result: fiber vs. laminar ODF]{Dual QBI result: fiber ODF (b,e) vs. laminar ODF (c,f), for a voxel containing either a double population of fibers separated by a cleavage plane (a), or coexisting cleavage planes of different orientation (d). }\label{fig:FODFvsLODF}
\end{figure}

To emphasize that dual QBI complements conventional QBI for fibers, we have presented both fiber and laminar ODFs for the geometries of either crossing myofibers or coexisting cleavage planes; see Fig. \ref{fig:FODFvsLODF}.


\subsection{An estimator to the laminar ODF}

For practical applications, an analogue of the estimator \eqref{hatpsiE} is required. Substitution of the PDF in definition \eqref{def_laminarODF} while using the Fourier relation \eqref{DW_fourier} yields immediately
\begin{eqnarray}\label{xicalc}
 \xi(x,y,z) &=& \frac{1}{Z'} \int\limits_{-\infty}^\infty \mathrm{d}x\  \int\limits_{-\infty}^\infty \mathrm{d}y\  \int_{\mathbb{R}^3} \mathrm{d}^3 q e^{i\left(q_xx+q_yy+q_zz \right)} E(q_x,q_y,q_z) \nonumber\\
   &=& \int\limits_{-\infty}^\infty \mathrm{d}q_z E(0,0,q_z) e^{i q_z z}.
\end{eqnarray}
Hence we observe that the laminar ODF in a given direction $\vec{u}$ is proportional to the one-dimensional Fourier transform of the signal $E(q\vec{u})$. Since the Fourier transform is easily inverted, one promptly finds
\begin{eqnarray}
  E(\qq) &=& Z' \mathcal{F}^{-1}[\xi(\rr)](\qq). \label{LODF_fourier}
\end{eqnarray}
Due to antipodal symmetry, only the real (i.e. non-imaginary) component of the right-hand side in \eqref{LODF_fourier} is retained. In this manner, we have derived that the diffusion attenuated signal itself can be considered as the laminar ODF multiplied with a cosine kernel:
\begin{equation}\label{lodfcos}
    E(0,0, q_z) = \int \xi(0,0, u_z) \cos(q_z z) \mathrm{d}z.
\end{equation}
We thus have established the dual version of relation \eqref{odfbessel}, where a Bessel function was found as modulation kernel. From \eqref{lodfcos} we obtain the amazingly simple estimator:
\begin{equation}\label{xihat_def}
    \hat{\xi}(\vec{u}) = \frac{1}{Z'} E(\vec{q}),
\end{equation}
for which we claim that $\hat{\xi}(\vec{u}) \approx \xi(\vec{u})$.

%
%

\subsection{Unified framework for both QBI and dual QBI \label{sec:proof_dQBI}}

We now provide a general formalism which includes the laminar and fiber ODF theory as special cases, as a generalization of the proof given in \cite{Tuch:2003}. Our derivation illustrates the true dual nature of the newly proposed method. The formalism relies on two mathematical theorems on Fourier transformation:
\begin{enumerate}
  \item \textbf{Projection slice theorem.} \\
   Consider a m-dimensional linear sub-manifold $\alpha$ of n-dimensional Euclidean space. (Hence, for $m=2$, $\alpha$ is a plane through the origin, and it is a line through the origin for $m=1$.) The projection-slice theorem then states that the m-dimensional Fourier transform $\FF_n$ of the projection of an function $f(\rr)$ onto the linear submanifold $\alpha$ of dimension $m$ is equal to an m-dimensional slice of the n-dimensional Fourier transform $\FF_n$ of that function consisting of an m-dimensional linear submanifold through the origin in the Fourier space which is parallel to the projection submanifold\footnote{In this context, the projection of a scalar function is understood as a line integral of this function, in contrast to the projection of a vector onto a reference triad, e.g. while defining fiber and sheet angles.}.

  If we use $\Pi_\alpha$ for the projection on $\alpha$, and $\Sigma_\alpha$ for taking the slice parallel to $\alpha$, we may write in the language of operators:
  \begin{equation} \label{PStheorem}
     \FF_m \circ \Pi_\alpha  = \Sigma_\alpha \circ \FF_n.
  \end{equation}
  Note that this theorem goes evenly for the inverse Fourier transform. The projection slice theorem lies at the basis of many imaging techniques.

  \item \textbf{Convolution theorem.} (See e.g.\cite{Haacke}.)\\
  The Fourier transform of the product of two functions is the convolution of the Fourier transforms of each function. In one dimension, one has
  \begin{equation}
    \FF[f(x).g(x)](q) = \FF[f(x)](q) \ast \FF[g(x)](q),
  \end{equation}
  with the convolution product $\ast$ defined as
  \begin{equation}
   f(x) \ast g(x) = \int_{-\infty}^\infty f(x-y) g(y) dy.
  \end{equation}
 \end{enumerate}

\noindent To get started with the common formalism for QBI and dQBI, we reformulate the fiber and laminar ODFs as projections of the net spin displacement function $P(\vec{r})$, given a direction with unit vector $\vec{u}$:
\begin{eqnarray}
  Z \psi(\vec{u}) &=& \int\limits_{\vec{u}_\|} P(\vec{r}) d\ell  = \left. \left[\Pi_{\vec{u}_\perp}  P(\vec{r}) \right] \right|_{\vec{r}=0}, \\
  Z' \xi(\vec{u}) &=& \int\limits_{\vec{u}_\perp} P(\vec{r}) d S  =  \left. \left[\Pi_{\vec{u}_\|} P(\vec{r}) \right] \right|_{\vec{r}=0}.
\end{eqnarray}
Using the Fourier slice theorem for $n=3$, and $\alpha$ equal to $\vec{u}_\perp$ or $\vec{u}_\|$, we have by virtue of the projection-slice theorem \eqref{PStheorem}:
\begin{equation}
\FF^{-1} [ \Pi_\alpha P(\vec{r}) ] = \Sigma_\alpha \FF^{-1}[P(\vec{r})] = \Sigma_\alpha E(\vec{q})  \label{deriv1}.
\end{equation}
We have to take into account, however, that with QBI no MRI signal is recorded for $\| \vec{q} \| \neq q_s$. This truncation may be performed by multiplying both sides of \eqref{deriv1} with a delta distribution $\delta^m (\| \vec{q} \| - q_s)$:
\begin{equation}
\FF^{-1} [ \Pi_\alpha P(\vec{r}) ] \delta^m (\| \vec{q} \| - q_s) = \left( \Sigma_\alpha E(\vec{q})\right) \delta^m (\| \vec{q} \| - q_s). \label{deriv2}
\end{equation}
When a Fourier transform is taken on both sides, the product on the left-hand side becomes a convolution with $M_m(\vec{r}) = \FF[\delta^m (\| \vec{q} \| - q_s)]$:
\begin{equation} \Pi_\alpha P(\vec{r}) \ast M_m(\vec{r}) = \FF\left[\left( \Sigma_\alpha E(\vec{q})\right) \delta^m (\| \vec{q} \| - q_s). \right]\label{deriv3}\end{equation}
Evaluating both sides in the origin $\vec{r}=0$ turns the remaining convolution and Fourier integrals into simple integrals:
\begin{equation}
\int\limits_{\alpha} \Pi_\alpha P(\vec{r}) M_m(\vec{r})d^m S  = \int\limits_{\alpha} \left( \Sigma_\alpha E(\vec{q})\right) \delta^m (\| \vec{q} \| - q_s) d^m S. \label{deriv4}
\end{equation}
In summary, the modulated version of the distribution function $\Pi_\alpha P(\vec{r})$ can be obtained by integrating the QBI signal over the subspace $\alpha$.

From \eqref{deriv4}, one may recover both QBI and dual QBI. For $\alpha = \vec{u}_\perp$, $m$ equals $2$ and we find
\begin{equation} M_2 (\vec{r}) = \int d^2q e^{-2\pi i \vec{q} \cdot \vec{r}} \delta^2 (\| \vec{q} \| - q_s) = 2\pi q_s J_0(2\pi q_s r) \label{Mr_odf}
\end{equation}
and the integral in (\ref{deriv4}) comes down to the familiar spherical Radon transform \eqref{FRTdef}. We now see that the transformation which was put forward by Tuch is in fact a natural choice, given the limited range of the acquired data.

Alternatively, we could have chosen $\alpha = \vec{u}_\|$ (or $m=1$) to obtain
\begin{equation}
M_1(\vec{r}) = \int dq e^{-2\pi iqz} \delta (|q| - q_s) = e^{-2\pi i q_s z} + e^{2\pi i q_s z}  = 2 \cos(2\pi q_s z). \label{Mr_sndf}
\end{equation}
The corresponding integral transform for the MRI signal simplifies to $2 E(q_s \vec{u})$.

As promised, we have now retrieved all results cited previously as results of the same expression \eqref{deriv4}. 
%

\subsection{Analytical expression for the laminar ODF}

\begin{figure}[h!b] \centering
  \mbox{
  \raisebox{1.99cm}{a)} \includegraphics[width=5cm]{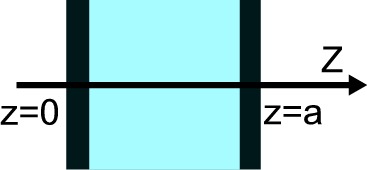}
  \raisebox{1.99cm}{b)} \includegraphics[width=5cm]{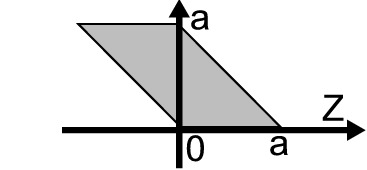}
  }
  \caption[Diffusion between parallel barriers]{a) Diffusion between impermeable infinity parallel barriers of spacing $a$. b) Integration domain for obtaining Eq. \eqref{free_diffusion3}.}\label{fig:parbar}
\end{figure}

Let us check whether the proposed formula \eqref{xihat_def} exhibits the desired dependence on gradient direction and imaging parameters. As a simple model that nevertheless possesses an analytical solution, we choose free diffusion of water between two infinite plane barriers with spacing $a$, as indicated in panel (a) of Fig. \eqref{fig:parbar}. The geometry can be thought to represent a single myocardial cleavage plane filled with water; we therefore later assume the length parameter $a$ to lie within the range 1-\unit{50}{\micro\meter}.

\begin{figure}[h!t] \centering
  \includegraphics[width=0.8 \textwidth ]{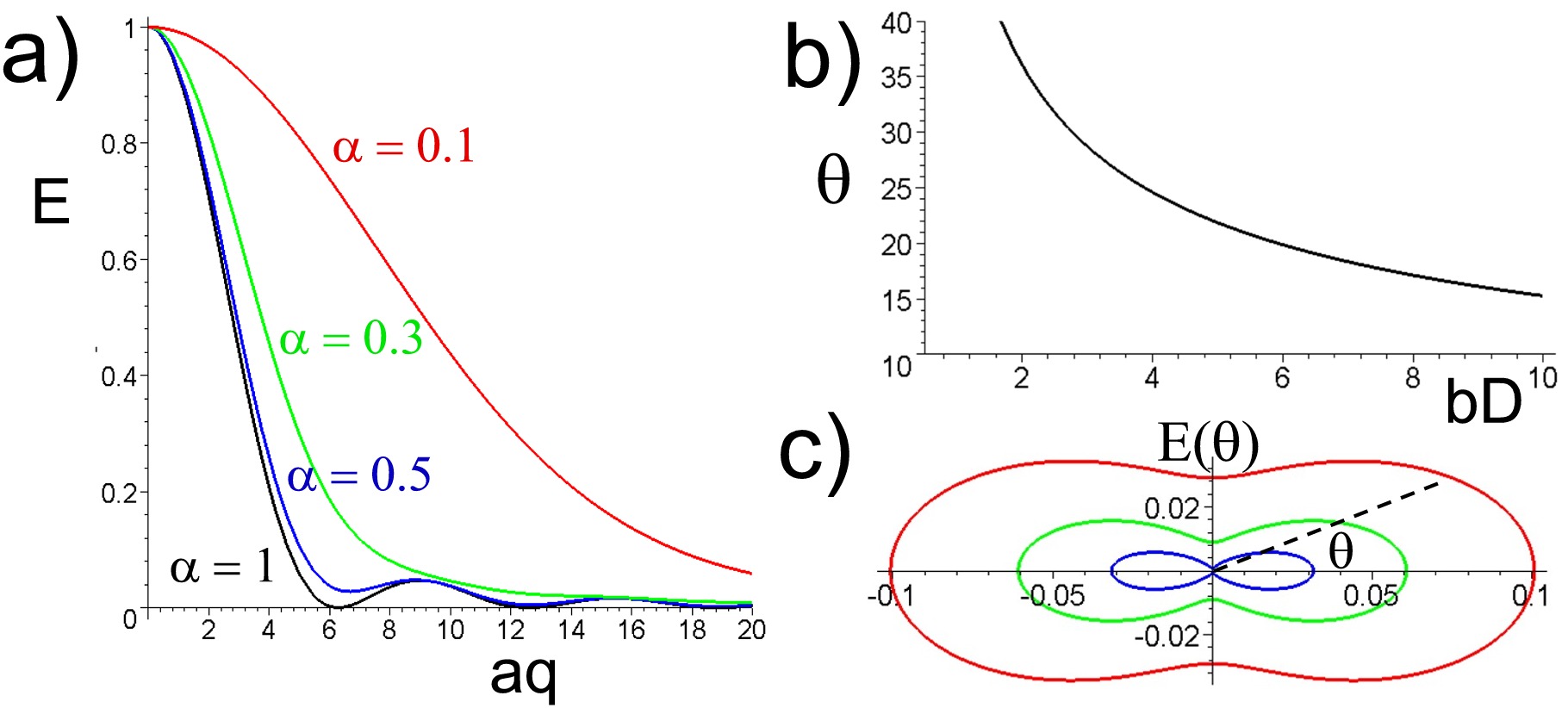}
  \caption[Analytical estimates to the laminar ODF]{Analytical estimates to the laminar ODF $\xi(\vec{u})$. a) Signal strength along the normal direction of the cleavage plane as a function of $qa$. b) Decrease in width of the laminar ODF with increasing $bD_0$ (with $\theta$ in degrees). c) Angular dependence of $\xi(\vec{u})$, for b-values $1500$, $2000$ and $3000\,$s/mm$^2$ given in red, green and blue. Curves (c) were drawn for $a=\unit{25}{\micro\meter}$, $\Delta = 10\,$ms, $\delta = 2\,$ms.
  }\label{fig:diffrac1}
\end{figure}

We take on Cartesian coordinates with the Z-axis normal to the barriers, such that the liquid-filled void corresponds to $0 < z < a$, as in \cite{Deswiet:1994}. A uniform proton density per unit surface implies $\rho(\vec{r}_0, t_0) = \chi(z_0) / a $, with $\chi(z) = H(z) H(a-z)$ the characteristic function for the interval given in terms of the Heaviside step function $H(z)$. To calculate the PDF, one needs additionally the Green function for the diffusion equation. Because diffusion is statistically independent along orthogonal directions, the Green function can be written as a product $P^L P^T$ of the free diffusion propagator $P^T$ in x and y directions
\begin{equation}\label{free_diffusion1}
  P^T(x_0,y_0; x, y, \tau) = \frac{1}{4\pi D_0 \tau} \exp\left( -\frac{(x-x_0)^2+(y-y_0)^2}{4D_0 \tau}\right)
\end{equation}
and the diffusion propagator $P^L$ along z. The latter is obtained in an eigenfunction expansion with Neumann boundary conditions:
\begin{equation}\label{free_diffusion2}
P^L(z_0; z, \tau) = \chi(z_0) \chi(z) \left(\frac{1}{a} + \frac{2}{a} \sum \limits_{n=1}^\infty e^{-n^2 \pi^2 D_0 \tau / a^2} \cos \frac{n\pi z_0 }{a} \cos \frac{n\pi z}{a} \right).
\end{equation}
A coordinate change $\vec{r}-\vec{r_0} \rightarrow \vec{R}$ allows to calculate the probability displacement or PDF \eqref{defPDF}. The integration interval for $z_0$ is found to run over $[-Z,a]$ for $Z<0$ and $[0, a-Z]$ for $Z>0$, as depicted in Fig. \ref{fig:parbar}b. Since the resulting PDF is is symmetric around $Z=0$, we may write (with $L_D = \sqrt{D_0 \tau}$):
\begin{multline} \label{free_diffusion3}
P(\vec{R}, \tau) =  \frac{\chi( |Z|) }{ 4 a^2 \pi L_D^2} e^{ -\frac{ X^2+ Y^2 }{4L_D^2}}   \left[ (a-|Z|) + \sum \limits_{n=1}^\infty  e^{-n^2 \pi^2 L_D^2 / a^2} \, .  \right. \\
\left. \left((a-|Z|) \cos \frac{n \pi |Z|}{a} - \frac{a}{n \pi} \sin \frac{n \pi |Z|}{a} \right)  \right].
\end{multline}
We now calculate the signal strength for a diffusion-encoding gradient vector $\vec{q}$ that encloses an angle $\theta$ with the positive Z-axis. Without losing generality, we may consider $\vec{q}$ in the XZ-plane: $\vec{q}\cdot \vec{R} = qX\sin\theta + qZ \cos \theta$. In the narrow pulse regime, the signal strength is obtained after Fourier transformation \eqref{DW_fourier}:
\begin{eqnarray} \label{Sq_pb}
E(\vec{q}, \tau) &=&  e^{- q^2 L_D^2 \sin^2 \theta} \left[ \frac{\sin^2 \left(\frac{a q \cos \theta }{2}\right) }{  \left(\frac{a q \cos \theta }{2}\right)^2 }  + \sum \limits_{n=1}^\infty e^{-n^2 \pi^2 L_D^2 / a^2} \, .  \right. \qquad\\
&& \left.  \frac{4 a^2 q^2 \cos^2 \theta }{\left( n^2 \pi^2 - a^2 q^2 \cos^2 \theta\right)^2} \left(1- (-1)^n \cos \left( a q \cos \theta \right) \right)  \right]. \nn
\end{eqnarray}
For $\theta=0^\circ$, this result is quoted in \cite{Deswiet:1994}. Note that the dimensionless parameters that describe the diffusion-weighed signal are
\begin{align} \label{diffrac_dimpar}
b D_0 &= q^2 D_0 \tau = q^2 L_D^2,&  \alpha &= L_D / a, & a q &=\sqrt{b D_0} /  \alpha.
\end{align}
In our interpretation $E(q\vec{u}) = \hat{\xi}(\vec{u})$, several deductions can be made from Eq. \eqref{Sq_pb}:

\begin{enumerate}
\item \textbf{Limits of short and long diffusion times}\\
In the limit where the diffusion length $L_D$ is much larger than the cleavage plane separation $a$ (i.e. large $\alpha$), the signal intensity normal to the barriers (i.e. $\theta=0$) is given by the single slit diffraction pattern
\begin{equation}\label{diffrac_longtau}
  \hat{\xi}(\vec{e}_z) = \frac{1}{Z} E(q \vec{e}_z) = \frac{1}{Z} \left( \frac{\sin(aq/2)}{aq/2} \right)^2.
\end{equation}
Hence, diffraction lobes arise for $aq > \pi$. For short diffusion times, however, diffusion is nearly Gaussian, and a monotonous $E(q)$ curve may be expected for $\theta=0^\circ$. As we operate in the regime $L_D = \unit{5}{\micro\meter}$, $a = 5 - \unit{50}{\micro\meter}$, the latter case is more likely to be fulfilled. Note that, for a cylindrical packing of rods, a combined numerical and experimental study was performed in \cite{BarShir:2008}, which demonstrated that diffraction effects indeed set in when $\alpha \gg 1$ (with $a$ denoting the cylinder radius in their case). Figure \ref{fig:diffrac1}a displays the theoretical MRI signal $E(q)$ in the direction normal to the bounding barriers for different values of $\alpha$.
\item \textbf{Angular contrast}\\
In order to interpret the measured signal as an estimator to the local cleavage plane orientation, $E(\vec{q})$ must be strongly attenuated in the direction along the sheet. From Eq. \eqref{Sq_pb}, we note that the pre-factor $e^{- q^2 L_D^2 \sin^2 \theta}$ effectuates this attenuation and therefore determines angular contrast. With \eqref{diffrac_dimpar}, we infer that the angular contrast is ensured as soon as a high enough $b$-value is taken. For example, with $b = 3000\,$s/mm$^2$, the reduction amounts to $1.10^{-3}$, such that the signal is heavily suppressed in the directions tangential to the cleavage plane, as desired. The full width at half maximum of the signal due to this pre-factor is obtained from $\sin \theta = \sqrt{\frac{\ln 2}{bD_0}}$; this curve is also presented in Fig. \ref{fig:diffrac1}b. As a result, the optimal b-value arises as a compromise between angular constrast and signal strength; see Fig. \ref{fig:diffrac1}c.

The overall pre-factor in Eq. \eqref{Sq_pb} furthermore disarms potential diffractive maxima that could arise for $\theta \neq 0$. For, diffractive maxima only  manifest for $aq > 2\pi$ and $L_D > a$, which would also imply that $ bD_0  = q^2 L_D^2 > (2\pi)^2$. In practice, b-values as high as $18. 10^3\,$s/mm$^2$ are not used for imaging purposes. In summary, we have reasoned that the laminar ODF estimator \eqref{xihat_def} associated to restricted diffusion between parallel barriers always exhibits a single maximum in the direction normal to the boundaries.

\item \textbf{Sensitivity to different length scales}\\
From the signal attenuation given in Fig. \ref{fig:diffrac1}a, it can be seen that the largest contribution to the signal is generated for $ aq < 2\pi$. For that reason, the estimated laminar ODF $\hat{\xi}(\vec{u})$ will be most sensitive to the cleavage planes which have widths
\begin{equation}\label{crit_width}
  a < \frac{2 \pi }{q} = 2 \pi \frac{L_D}{\sqrt{bD_0}} = 2\pi \sqrt{\frac{\tau}{b}}.
\end{equation}
In our scans, we had $\tau = 7.0\,$ms and $b=3000\,$s/mm$^2$; therefore most sensitivity was established for $a< \unit{10}{\micro\meter}$.
\end{enumerate}

\section[Application of dual QBI for mapping myocardial laminar structure]{Application of dual QBI for mapping of\\ myocardial laminar structure\label{sec:dQBIapplied}}


\subsection{Dual QBI on synthetic data of myolaminae}

To first test whether crossing cleavage planes may be resolved by the proposed method, we have used the model \eqref{Dfib_synth} with Gaussian anisotropic diffusion to investigate coexisting cleavage planes as well. We let the normalized transmural coordinate coincide with the partial volume of one of the cleavage plane populations; the resulting transmural profile in terms of the $\beta^*$ sheet angle is depicted in Fig. \ref{fig:dQBI_sheets}. The intersection angle was set to $80^\circ$.

\begin{figure}[h!t] \centering
  \raisebox{6cm}{a)} \includegraphics[width=0.8 \textwidth]{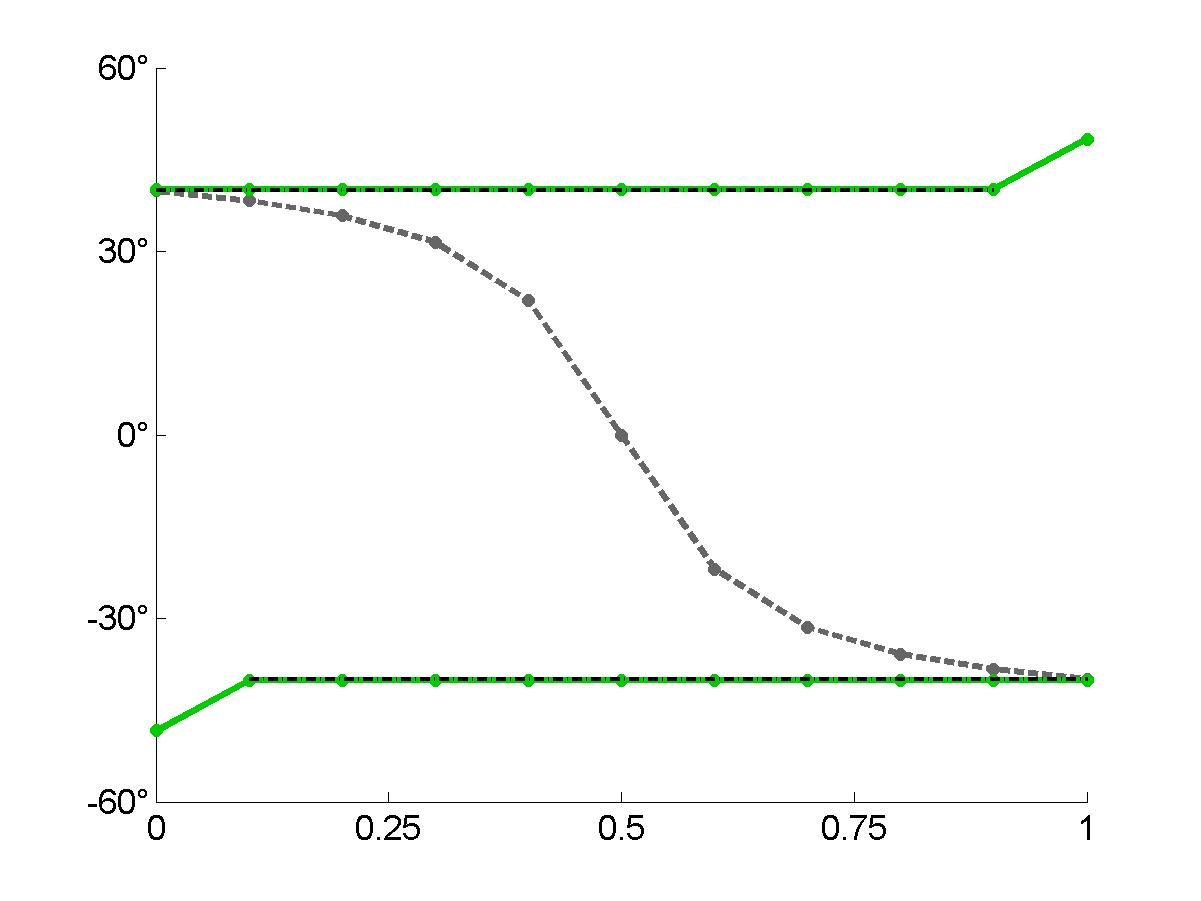}\\
   \raisebox{2cm}{b)}\includegraphics[width=0.9 \textwidth]{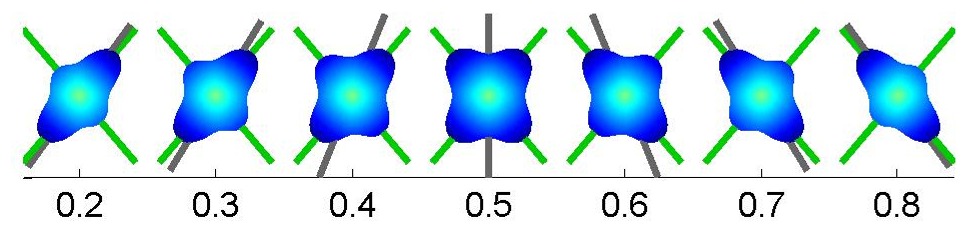}
 \caption[Dual QBI result for synthetic data of myolaminae]{QBI result for synthetic data of myolaminae, with gradual overlap between the populations. In the transmural profiles (a) for sheet angle $\beta^*$, the horizontal coordinate may be interpreted as normalized transmural position; the data points show ground truth (black), DTI outcome (gray) and QBI outcome (green). Corresponding laminar ODFs are shown in (b).}\label{fig:dQBIsheets}
\end{figure}

From this example we infer that the dQBI formalism holds the power to discriminate a continuous rotation of myolaminae from a gradual overlap of distinct cleavage plane populations. Since histological evidence has pointed out that the merging of cleavage plane populations is present in the ventricular wall, dual QBI may prove particularly useful to investigate the complex laminar structure within the ventricular wall.

\subsection{Laminar structure in the LVFW}

\begin{figure}[h!t] \centering
\mbox{
  \raisebox{2.75cm}{a)} \includegraphics[width=0.4 \textwidth]{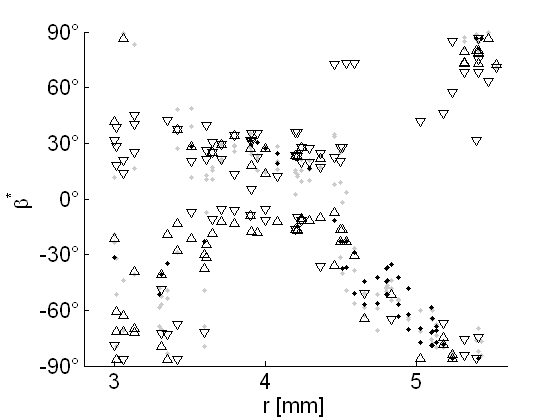}
   \raisebox{2.75cm}{b)}\includegraphics[width=0.4 \textwidth]{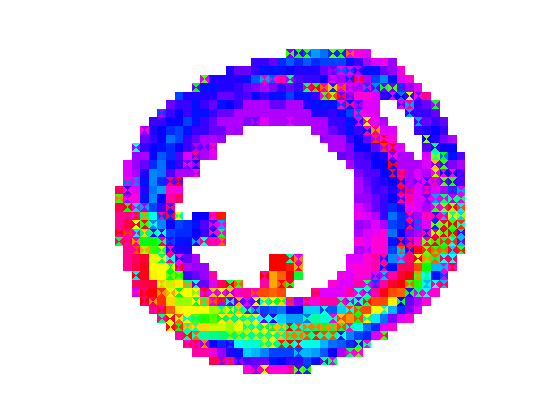}
   \raisebox{0.6cm}{       \includegraphics[width=0.1 \textwidth]{colorbar9090.jpg}}
   }
  \caption[Dual QBI result: sheet angle $\beta^*$]{Sheet angle $\beta^*$ reconstructed with dual QBI. a) Transmural course of $\beta^*$ through the LVFW, with distance measured from the slice centroid. Black dots indicate the dQBI result in voxels where one laminar direction was found. With multiple laminar orientation, the largest maximum of the laminar ODF was rendered with $\bigtriangleup$, and the second largest with $\bigtriangledown$. For comparison, the unique direction predicted by DTI is depicted by the gray dots.  b) Axial slice with cyclic colormap representing $\beta^*$.}\label{fig:dQBI_results_BS}
\end{figure}

\begin{figure}[h!t] \centering
\mbox{
  \raisebox{2.75cm}{a)} \includegraphics[width=0.4 \textwidth]{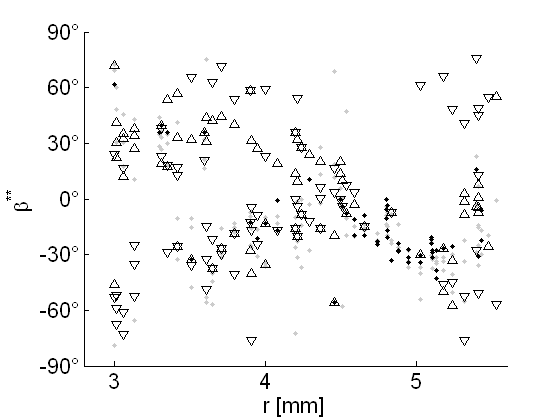}
   \raisebox{2.75cm}{b)}\includegraphics[width=0.4 \textwidth]{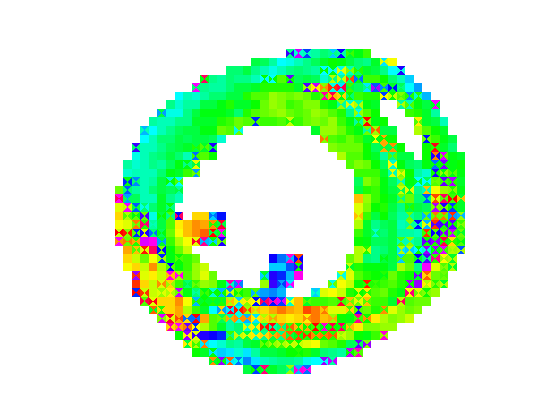}
   \raisebox{0.6cm}{       \includegraphics[width=0.1 \textwidth]{colorbar9090.jpg}}
   }
 \caption[Dual QBI result: sheet angle $\beta^{**}$]{Sheet angle $\beta^{**}$ reconstructed with dual QBI. a) Transmural course of $\beta^{**}$ through the LVFW, with distance measured from the slice centroid. Same symbols were used as in the previous figure. b) Axial slice with cyclic colormap representing $\beta^{**}$.}\label{fig:dQBI_results_BSS}
\end{figure}

The high-angular resolution analysis conducted on a lateral sector of the ventricular wall in a rat heart may now be extended to include laminar microstructure. Importantly, both QBI and dual QBI start from the same data set; for that reason no additional scanning was required. The laminar structure in terms of the $\beta^*$ and $\beta^{**}$ angles is presented in Figs. \ref{fig:dQBI_results_BS}-\ref{fig:dQBI_results_BSS}. Voxels that contain more than one laminar direction have been doubly colored in the axial slice view.

In a large part of the LVFW preparation, manifold laminar structure was identified by dual QBI. In the transmural profiles, voxels in which a single normal direction was found are depicted by black dots. In case of multiple laminar structure, upper ($\bigtriangleup$) and lower ($\bigtriangledown$) triangles indicate orientation of the first and second local maximum of the calculated laminar ODFs. For reference, the DTI outcome is also displayed in the scatter plots by gray dots.

Inspecting the entire transverse slices depicted, one notices that manifold laminar structure seems to be prevalent in the lateral and posterior wall sections of the rat ventricle. Zooming to the LVFW, two cleavage plane populations are witnessed in the inner half of the myocardial wall, with unequal inclination in both the transverse and longitudinal-radial planes. Here we see the source of discontinuity in the DTI-derived sheet angle profile \ref{fig:DTI_results_BA}: the onset of the second population was left unnoticed by DTI means; only when this population gains dominance in partial volume terms, the recorded angle suddenly adapts to the novel alternative orientation.

\subsection{Detailed study of ventricular myocardial structure}

 The orientation of myolaminae was also investigated in the same representative sectors of the ventricle as in \ref{fig:dQBI_fibs}  \cite{Dierckx:2010a}. Here, the transmural inclination of cleavage planes is quantified using the $\beta$ sheet angle from \cite{Rohmer:2007}. As noted before, this sheet angle has $\beta= \beta^*-90^\circ$.\\
 
\begin{figure}[h!t] \centering
  \includegraphics[width=1.0\textwidth]{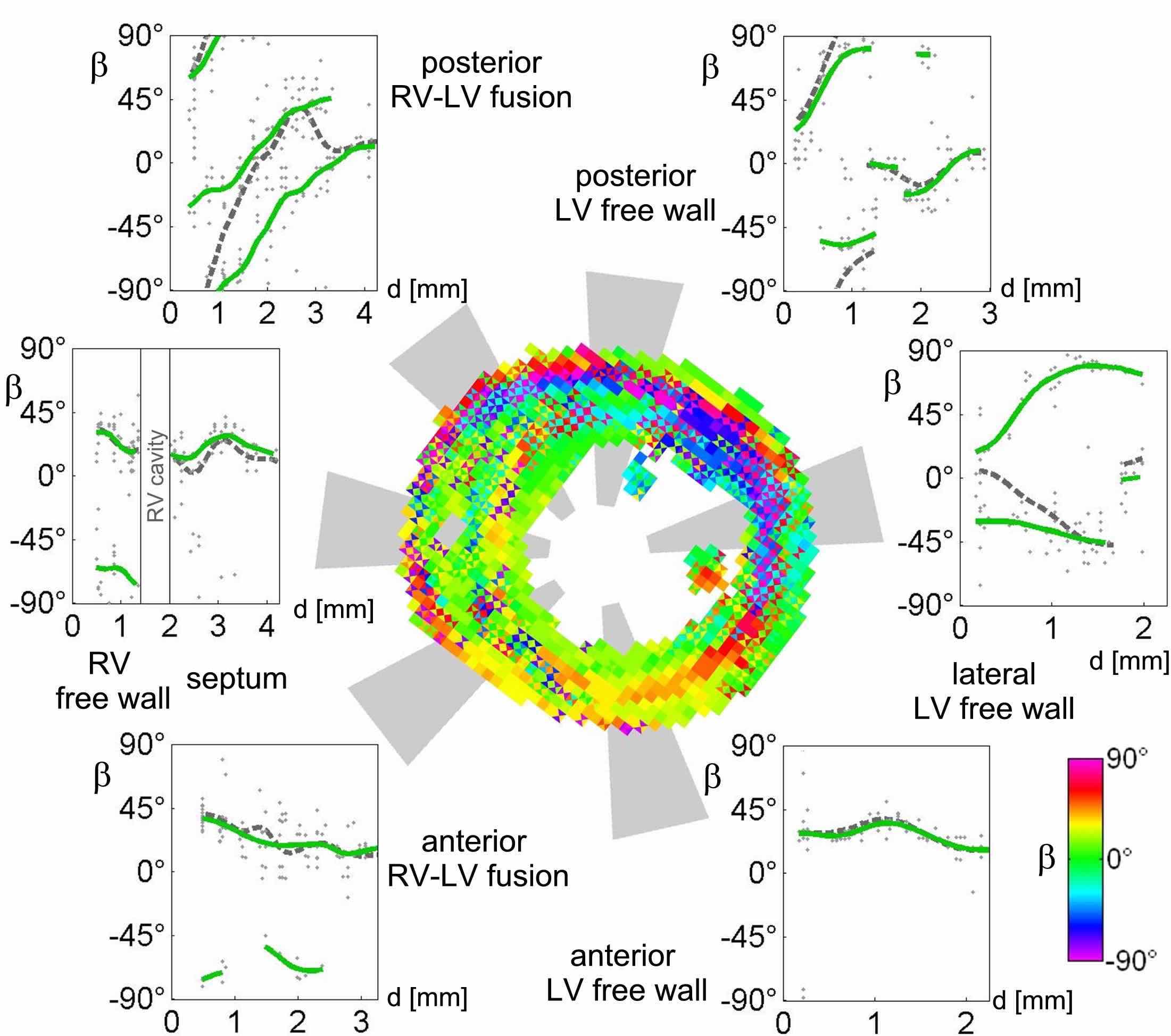}\\
  \caption[Dual QBI result for six ventricular sectors]{Dual QBI result for six ventricular sectors in terms of the sheet angle $\beta$. In the transmural profiles, distance $d$ represents distance from the epicardium. Green curves denote the QBI outcome, which may be compared to the DTI result given in gray dotted lines.
  }\label{fig:dQBI_sheets}
\end{figure}

The quantitative results confirm the trend that we had yet identified in Figs. \ref{fig:dQBI_results_BS} and \ref{fig:dQBI_results_BSS}: only in the anterior section of the ventricles, an unambiguous laminar orientation is encountered. When moving from the lateral LVFW section towards the posterior LV-RV fusion site, a double laminar orientation is manifest over nearly the entire wall thickness.

Note furthermore that the mid-wall zone in the posterior LV wall that has $\beta \approx 0^\circ$ is almost certainly an artefact of the imaging method. For, from the fiber analysis presented in Fig. \ref{fig:QBI_results_AH}, we know that two oblique fiber directions are present in this region. As the cleavage plane signal is easily suppressed by the fiber signal (due to higher attenuation according to $e^{-bD}$), the direction of smallest diffusion will lie around $\vec{e}_r$, leading to the apparent sheet angle $\beta \approx 0^\circ$. In other words, the dQBI method is at its current scope only able to resolve crossing cleavage planes if the fiber direction is well defined. Fortunately, crossing cleavage planes are for geometric reasons only consistent with a single fiber direction, which should run parallel to both cleavage planes.

\section{Dual QBI on multiple shells}

\subsection{Fiber ODF redefined as a true probability measure \label{sec:ODFV1}}

In 2009, two groups independently proposed an amendment for fiber imaging with QBI: Tristan-Vega \etal  \cite{TristanVega:2009, TristanVega:2010} and Aganj \etal \cite{Aganj:2009, Aganj:2009b}. In these papers, it was recalled that the (fiber) ODF \eqref{def_fiberODF} as defined by Tuch is not a probability distribution in the strictest sense. Indeed, the marginal probability that the net spin displacement takes place in a small cone $\mathcal{K}$ that spans a solid angle $d\Omega$ in the direction $\vec{u}$ is given by
\begin{equation}\label{marg_displ}
  p_m(\vec{u}) d \Omega = \int_{\vec{r} \in \mathcal{K}} P( \vec{r} ) r^2 \sin \theta d^3 r
\end{equation}
in not further specified spherical coordinates $(r, \phi, \theta)$. Obviously, definition \eqref{marg_displ} leads to $\int p_m(\vec{u}) d\Omega =1$.

\begin{figure}[h!b] \centering
  \includegraphics[width=0.8\textwidth]{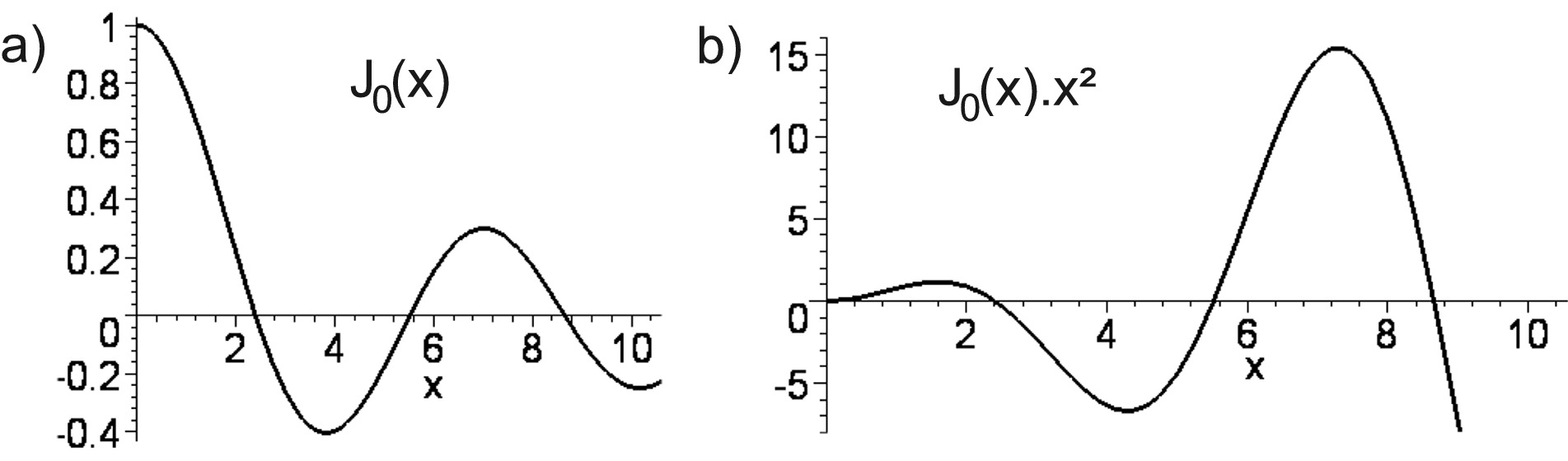}\\
  \caption[Bessel function convolution kernels]{Bessel convolution kernels for fiber QBI. a) Correct convolution kernel $J_0(x)$. b) Improper convolution kernel $J_0(x).x^2$.
  }\label{fig:bessel2x}
\end{figure}

Inspired by Eq. \eqref{marg_displ}, we now introduce an alternative definition fiber ODF $\psi^V$ that includes the spherical volume element:
\begin{equation}\label{def_FODFV}
  \psi^V(\vec{u}) = \int_{-\infty}^\infty P(s \vec{u}) s^2 ds = \int_{\vec{r} \in \vec{u}_{||}} P(\vec{r}) \left( \vec{r} \cdot \vec{u} \right)^2 d r.
\end{equation}
Importantly, our formulation \eqref{def_FODFV} differs from \cite{Aganj:2009, TristanVega:2009}, where it is stated
\begin{equation}\label{def_OPDF}
  \Phi(\vec{u}) = \int_{\vec{r} \in \vec{u}_{||}} P(\vec{r}) \left( \vec{r} \right)^2 d r.
\end{equation}
The slight difference with \eqref{def_FODFV} is not relevant for the theoretical concept of the modified ODF, but will come into play when calculation its estimator. Obtaining the estimator $\hat{\Phi}$ is straightforward after noticing that multiplication with a coordinate $x^j$ translates as a derivative action $ i \dd_{q_j}$ in the Fourier domain. Hence, both Tristan-Vega \etal and Aganj \etal were led to
\begin{equation}\label{phihat}
  \hat{\Phi}(\vec{u}) = \frac{1}{Z} \GG[ - \nabla^2_q E(q_s \vec{u})]
\end{equation}
with $\nabla^2_q$ the Laplacian operator in q-space. Remarkably, in neither publication the statement $\hat{\Phi}(\vec{u}) \approx \Phi(\vec{u})$ was further explored. Following the appendix of \cite{Tuch:2004}, we obtain in cylindrical coordinates $(\rho, \phi, z)$ and $(\kappa, \varphi, q_z)$, with the direction of interest along the z and $q_z$ axes:
\begin{eqnarray}
  \GG[ - \nabla^2_q E(q_s \vec{u})] &=& \GG\left[ \int_{\mathbb{R}^3} d^3 r P(\vec{r}) (\rho^2 + z^2)  e^{i \vec{q} \cdot \vec{r} } \right] \nn \\
  &=& \int_{\mathbb{R}^3} d^3 r P(\vec{r}) (\rho^2 + z^2) \int_0^{2\pi} d\varphi \, q_s e^{i q_s \rho \cos(\phi-\varphi)} \nn\\
  &=& 2\pi q_s \int_{\mathbb{R}^3} d^3 r P(\vec{r}) (\rho^2 + z^2) J_0(q_s \rho). \label{deriv_Eq}
\end{eqnarray}
The last expression tells us that the estimator \eqref{phihat} consists of two contributions:
\begin{equation}\label{phihat2}
   Z \hat{\Phi}(\vec{u}) =  2\pi q_s \int_{\mathbb{R}^3} d^3 r P(\vec{r}) z^2 J_0(q_s \rho) +  2\pi q_s \int_{\mathbb{R}^3} d^3 r P(\vec{r}) \rho^2 J_0(q_s \rho).
\end{equation}
The first term is the desired marginal probability \eqref{def_OPDF} modulated with a Bessel kernel as in standard QBI; the second term entails modulation of the conventional fiber ODF with the function $\rho^2 J_0(q_s \rho)$, depicted in Fig. \ref{fig:bessel2x}b. This function has a prominent negative lobe and grows unbounded for large $\rho$. In the published results \cite{TristanVega:2009} this term leads to negative probabilities. The presence of lobes of negative probability was disregarded in the paper, as the authors took the absolute value of the probabilities to handle the inconsistency.

We now argue that using the correction expression of the volume-weighed fiber ODF, i.e. Eq. \eqref{def_FODFV} offers a correct way to incorporate the spherical volume element. As the weighing factor should only apply to the direction $\vec{u}$, the estimator for the volume-weighed fiber ODF reads
 \begin{equation}\label{psivhat}
  \hat{\psi}^V(\vec{u}) = \frac{1}{Z} \GG[ - \nabla^2_{\vec{u}} E(q_s \vec{u})].
\end{equation}
With this stipulation, we re-execute derivation \eqref{deriv_Eq} to obtain
\begin{equation} \label{psivhat_approx}
 Z \hat{\psi}^V(\vec{u}) = 2\pi q_s \int_{\mathbb{R}^3} d^3 r P(\vec{r}) z^2 J_0(q_s \rho)
\end{equation}
without the spurious $\rho^2 J_0(q_s \rho)$ term.

\subsection{Two-shell estimates for fiber QBI \label{sec:ODFV2} }

Expression \eqref{psivhat} is rather awkward in numerical calculations. An alternative expression may be obtained based upon the commutation properties of spherical Radon transform $\GG$ and the Laplacian in spherical coordinates. In spherical q-space coordinates, the Laplacian may be split in terms with only radial or angular derivatives:
\begin{equation}\label{lapl_sphere}
  \nabla^2_q = \dd^2_q + \nabla^2_T = \dd^2_q + \frac{2}{q} \dd_q + \frac{1}{q^2} \nabla^2_{LB}
\end{equation}
with $\nabla^2_{LB}$ the Laplace-Beltrami operator on the sphere. The part of the Laplacian containing only tangential derivatives to the sphere has been denoted $\nabla^2_T$. Consider now the operator $ \GG \nabla^2_{\vec{u}}$ from \eqref{psivhat}, with the direction of interest along the Z-axis, and $\varphi$ the azimuthal angle in the XY plane. In such case, one has in the equatorial plane:
\begin{align}
 \nabla^2_{\vec{u}} &= \dd^2_{q_z}, & \GG \dd_\varphi &= 0, & \nabla^2_T = \dd^2_{q_z} + \frac{1}{q} \dd_q + \frac{1}{q^2} \dd^2_\varphi.
\end{align}
These expressions lead to:
\begin{eqnarray}\label{GGnab}
     \GG \nabla^2_{\vec{u}} &=& \GG \left( \nabla^2_T - \frac{1}{q} \dd_q \right)  = \GG \left(\frac{1}{q^2} \nabla^2_{LB} + \frac{1}{q} \dd_q\right) \nn \\
      &=&  \left(\frac{1}{q^2} \nabla^2_{LB} + \frac{1}{q} \dd_q \right) \GG.
\end{eqnarray}
Here, we have used that $\GG$ commutes with $\nabla^2_{LB}$ because both operators are diagonal in a spherical harmonics basis; see Eq. \eqref{diag_FRT}. The interchange of $\dd_q$ and $\GG$ is evenly allowed, since the FRT operates on each q-shell separately. The commutation property \eqref{GGnab} enables us to restate \eqref{psivhat} as
\bsub \begin{eqnarray}\label{psivhat2}
  \hat{\psi}^V(\vec{u}) &=& \frac{1}{Z}  \left( - \frac{1}{q^2_s} \nabla^2_{LB} \GG[E(q_s\vec{u})] - \frac{1}{q_s} \dd_{q_s} \GG[E(q_s\vec{u})] \right) \\
   &=& \frac{1}{Z} \left( - \frac{1}{q^2_s} \nabla^2_{LB} \hat{\psi}(\vec{u}) - \frac{1}{q_s} \dd_{q_s} \hat{\psi}(\vec{u}) \right).
\end{eqnarray} \esub
We conclude that the volume-weighed fiber ODF $\hat{\psi}^V(\vec{u})$ may be estimated by the Laplace-Beltrami sharpened ordinary ODF estimator $\hat{\psi}$, plus a term involving the change of the $\hat{\psi}$ with varying $q_s$. An additional factor $q_s^2$ may be absorbed in the normalization factor:
\begin{eqnarray}\label{psivhat3}
  \hat{\psi}^V(\vec{u}) &=& \frac{1}{\tilde{Z}} \left( - \nabla^2_{LB} \GG[E(q_s \vec{u})] - q_s \dd_{q_s} \GG[E(q_s \vec{u})] \right).
\end{eqnarray}

The Laplace-Beltrami term can be evaluated from the DW-MRI attenuation obtained at a single q-value. However, quantifying the term with the radial derivative either necessitates sampling at unequal q-values (for fixed $\Delta$) or assuming a particular dependence $E(q_s)$. In high angular resolution diffusion imaging, it is popular to hypothesize a mono-exponential signal decay, as a generalization of the DT formalism \cite{Ozarslan:2006, TristanVega:2009, Aganj:2009}:
\begin{equation}\label{monoexp}
  E(q, \vec{u})  \approx e^{- b(q^2) D(\vec{u}) }.
\end{equation}
Given that $b \propto q^2$, one may reason that at low b-values $q \dd_q E(\vec{q}) \approx 2 E(\vec{q}) \ln E(\vec{q})$, which leads to
\begin{eqnarray}\label{psivhat4}
  \hat{\psi}^V(\vec{u}) &\approx& \frac{1}{\tilde{Z}} \left( - \nabla^2_{LB} \GG[E(q_s \vec{u})] - 2 \GG[E(q_s \vec{u}) \ln E(q_s \vec{u}) ] \right).
\end{eqnarray}
Diffusion has been characterized, however, as non mono-exponential in many biological tissues, including myocardium \cite{Hsu:2001}. This fact preempts the use of \eqref{monoexp} at moderate b-values. Recall also that the QBI method and its amendments have been designed to handle complex diffusion processes. For closely spaced myofiber bundles, the condition \eqref{monoexp} is therefore likely to be violated due to the presence of multiple diffusion compartments.


A possible solution to assess \eqref{psivhat3} without taking on a diffusion model is to sample the DW-MRI signal at two distinct q-shells, and explicitly calculate the radial derivative in \eqref{psivhat3}. Unfortunately, this process doubles the acquisition time, and numerical differentiation between closely spaced q-shells is likely to be prone to noise artefacts.

\subsection{Volume-weighed laminar ODFs}

Here, we `dualize' the results from paragraphs \ref{sec:ODFV1} and \ref{sec:ODFV2} to measure cleavage plane orientations instead of fiber directions. The volume-weighed laminar ODF that includes the spherical volume element reads, in Cartesian coordinates with $\vec{u}$ along the Z-axis:
\begin{equation}\label{def_LODFV1}
  \xi^V(\vec{u}) = \int dx dy P(x,y,0) \left(x^2+y^2 \right).
\end{equation}
Without reference to coordinates, this definition may be stated
\begin{equation} \label{def_LODFV2}
  \xi^V(\vec{u}) = \int_{\vec{r} \in \vec{u}_{\perp}} P(\vec{r}) \left( \vec{r} \times \vec{u} \right)^2 d^2 r.
\end{equation}
Based on Fourier transformation properties, we conjecture that $ \xi^V(\vec{u}) \approx \hat{\xi}^V(\vec{u})$ with
\begin{equation} \label{xivhat1}
  \hat{\xi}^V(\vec{u}) = - \frac{1}{Z} \nabla^2_T E(q \vec{u}).
\end{equation}
Explicit calculation in cylindrical coordinates $(\rho, \phi, z)$ indeed confirms that
\begin{eqnarray}
 - \nabla^2_T E(q_s \vec{u}) &=& \int_{\mathbb{R}^3} d^3 r P(\vec{r}) \rho^2 e^{i q_s z}
                                              =  \int_{\mathbb{R}^3} d^3 r P(\vec{r}) \rho^2 \cos q_s z. \label{deriv_nabTEq}
\end{eqnarray}
Here too, the estimator to the volume-weighed ODF \eqref{xivhat1} may be rephrased using terms with either angular or radial derivatives: \begin{equation}\label{xivhat2}
  \hat{\xi}^V(\vec{u}) = \frac{1}{\tilde{Z} }  \left( - \nabla^2_{LB} E(q_s \vec{u}) -2q_s \dd_{q_s} E(q_s \vec{u}) \right).
\end{equation}
The different factor $2$ compared to \eqref{psivhat4} stems from the dimension of the subspace of integration in the definition of the fiber ODF versus the laminar ODF.

\section{Histological validation of dual QBI}

\subsection{Methods}

The rat hearts that were subjected to a q-ball image protocol were in the same position subjected to a high resolution (non-DW) MRI scan with voxel size $\unit{75}{\micro\meter} \times \unit{75}{\micro\meter} \times \unit{150}{\micro\meter}$. After imaging, the rat hearts were subjected to histological sectioning, which was carried out by Dr. Stephen Gilbert from the university of Leeds. One of both hearts was sliced perpendicular to the long axis in order to acquire axial cross-sections. The second heart was sectioned in the coronal direction, intersecting both ventricles. Digital pictures were taken with a pixel size of \unit{10}{\micro\meter}.

Registration of the histological images to the high-resolution MRI image matrix was performed manually by identifying anatomical landmarks in both data sets. The virtual section through the MRI data set was then iteratively improved to match the histology slice as good as possible, in spite of the deformations induced by the fixation and slicing procedures.

Using the diffusion-weighted data set, the normal vectors to the myolaminae derived from DTI and dQBI were interpolated to the location of the histological slice. Using the $\beta^*$ and $\beta^{**}$ sheet angles for axial and longitudinal slices respectively, the apparent texture of the cut surface due to the presence of cleavage planes was predicted based on the MRI data.


\subsection{Results from histological validation}

\begin{figure}[h!b] \centering
  \includegraphics[width=0.75\textwidth]{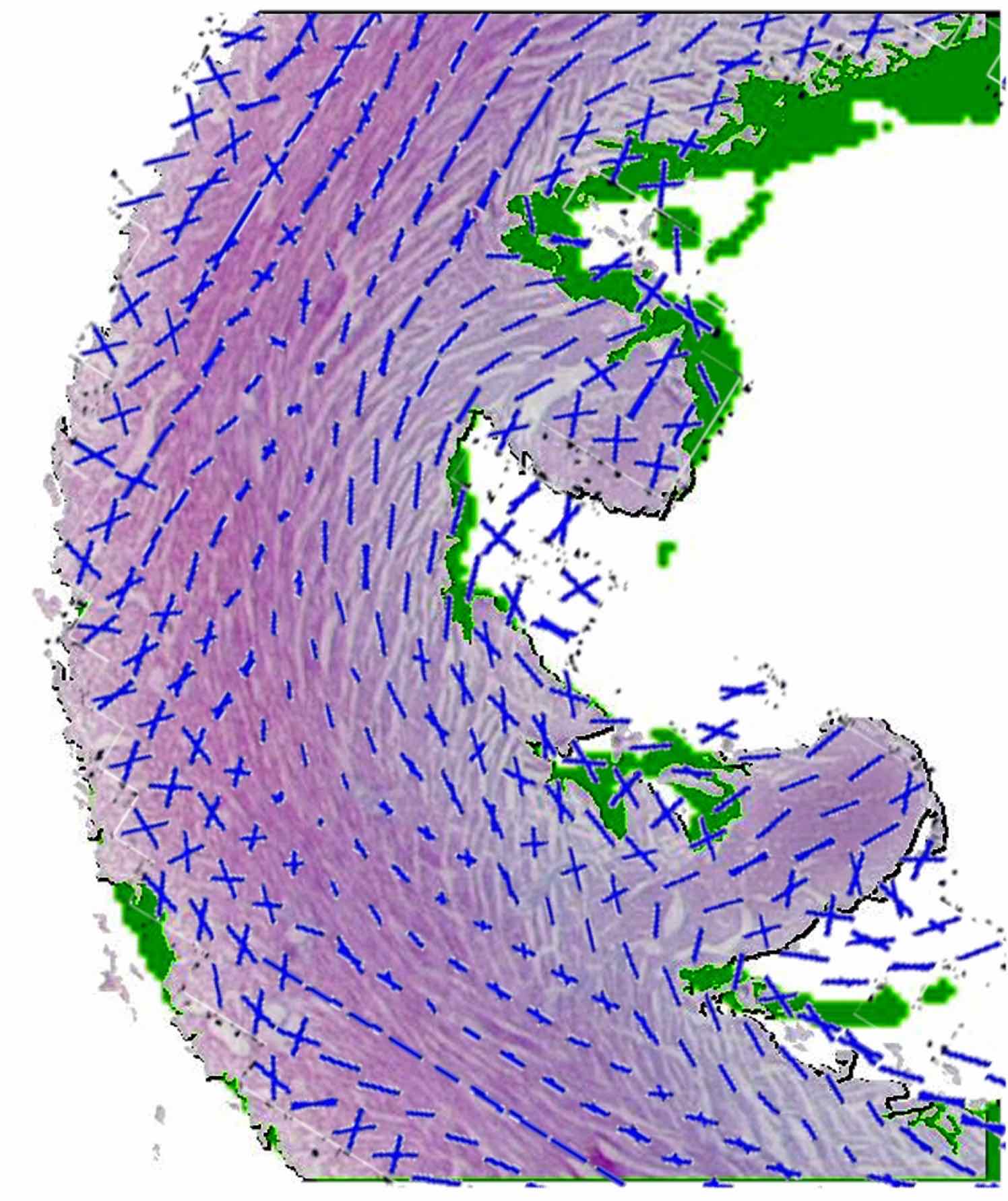}
  \caption[Histological validation of dQBI: axial view]{Histological validation of dQBI: axial view on the left ventricular wall. The reconstructed cleavage plane orientation using the dQBI method given by the blue rods should be compared to the texture of the underlying histological image. The green border zone indicates incomplete registration of the two data sets.
  }\label{fig:hist_axial}
\end{figure}

In the axial histological slice, crossing cleavage planes could be discerned near the endocardial surface. Figure \ref{fig:hist_axial} displays an overlay of the reconstructed directions with the dual QBI method. In regions with one cleavage plane orientation, a clear match is found. Moreover, in the subendocardial zone, the diffusion-weighted image is able to reconstruct the coexisting populations of ill-aligned cleavage planes as well.

\begin{figure}[h!b] \centering
  \raisebox{3.5cm}{a)} \includegraphics[width=0.8\textwidth]{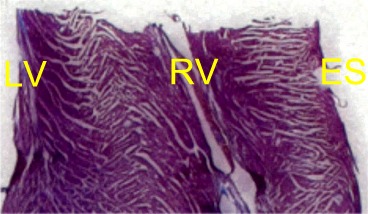}\\
  \raisebox{3.5cm}{b)}  \includegraphics[width=0.8\textwidth]{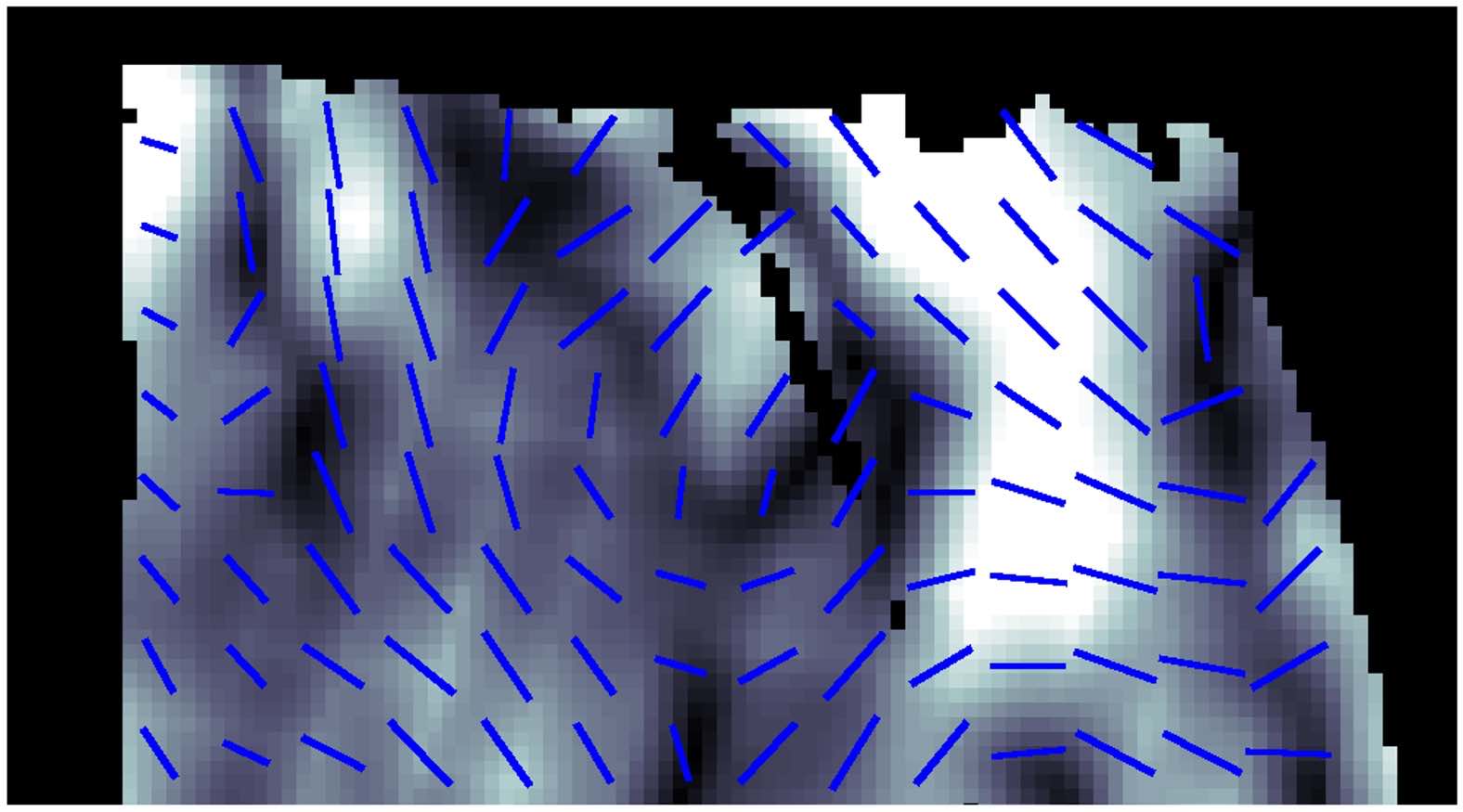}\\
  \raisebox{3.5cm}{c)}  \includegraphics[width=0.8\textwidth]{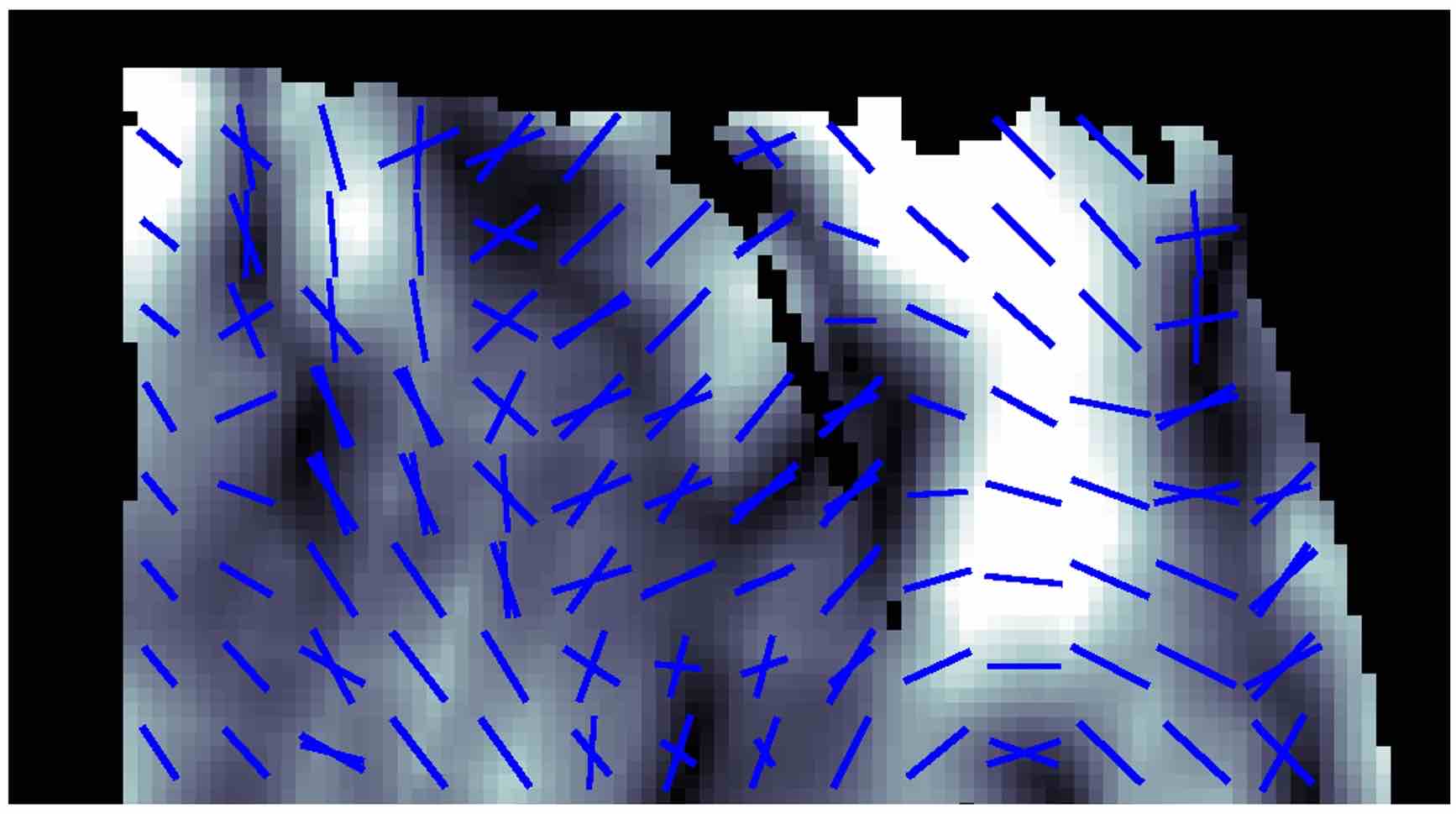}
  \caption[Histological validation of dQBI: longitudinal view]{Histological validation of dQBI: longitudinal view from LV cavity to epicardial surface (ES). The reconstructed cleavage plane orientation using the DTI (b) and dQBI (c) methods should be compared to the texture of the underlying histological image. Panels (b) and (c) are shaded according to the relative difference between second and third DT eigenvalues.
  }\label{fig:hist_long}
\end{figure}

The longitudinal slice could only locally be registered to the MRI data set due to larger deformation in the preparation process. Therefore, Fig. \ref{fig:hist_long} offers a transmural view through the intraventricular septum and right ventricular wall. Here too, dual QBI resolves the crossing cleavage planes in the central part of the septum. However, a fully quantitative comparison requires more robust registration than the one that was used here.

\section{Scope and outlook of dual QBI}

The methods described in the course of this chapter have enabled us to view crossing myocardial cleavage planes in the heart for the first time using a diffusion-weighed MRI technique. We have at the same time witnessed some important limitations to the method.\\

First, cleavage plane crossings can be obscured by the fiber signal in regions where myofibers of different orientation are found (see Fig. \ref{fig:dQBI_sheets}). Second, planes of preferred diffusion have been identified in the subepicardium, seemingly opposing the confocal microscopy results described by Pope \etal in \cite{Pope:2008}. Both issues may be attributed to the fact that dual QBI in its present form does not estimate absolute probabilities of diffusion in a particular direction. For that reason, stronger or weaker diffusion processes may mimic the appearance of cleavage planes, as in the mentioned examples. The underlying reason is that our methods have been designed on geometrical grounds, without attempting to separate distinct compartments in which diffusion takes place. A future enhancement of dual QBI could therefore address unraveling signal contributions from the fiber and sheet compartments, if possible.

Third, the angular resolution of QBI and dual QBI is still limited. In fact, we have been working with diffusive displacement orientation distribution functions, not fiber (or sheet) orientation probabilities. For diffusion in fibrous tissue, it has been noted that the diffusive displacement of water can be regarded as the spherical convolution of the fiber orientation probability with the signal from a hypothetical voxel with perfectly aligned fibers\cite{Tournier:2007}. This observation lies at the base of a numerical deconvolution technique which has been shown to significantly improve angular resolving power in fibrous tissue\cite{Tournier:2007}. It can therefore be expected that the imaging of myolaminae may evenly benefit from a similar reconstruction scheme.

Fourth, one has to admit that we barely know how the water diffusion processes in fixated and fresh myocardium take place. Ameliorated imaging strategies thus require better knowledge of diffusion propagator, after which dependence of imaging parameters may be investigated.

Lastly, it is remains to be seen how dual QBI performs when performed with imaging parameters typical for a clinical scanner. Obviously, this step needs be taken in order to enable studying larger hearts, such as human, with the proposed technique.




%

%


\clearpage{\pagestyle{empty}\cleardoublepage}


\renewcommand\evenpagerightmark{{\scshape\small Chapter 5}}
\renewcommand\oddpageleftmark{{\scshape\small Curved space formalism}}

\hyphenation{ani-so-tro-py Min-kow-ski}

\chapter[Curved space formalism]{A curved space formalism\\ for handling tissue anisotropy}
\label{chapt:activ}

This chapter constitutes an important stepping stone for our treatment of activation wave dynamics in anisotropic media. Inspired by analogies in pathfinding, optics and gravitation, we are led to re-think the notion of distance in an electrophysiological context. The emergent metric tensor, which determines distance between nearby points, is found to be proportional to the inverse electrical diffusion tensor of the medium and confirms the findings of Wellner and Pertsov \cite{Wellner:2002}.

Given the complex fibrous and laminar architecture of the heart muscle, endowing space with the mentioned metric tensor results in a non-Euclidean space. The situation is reminiscent of Einstein's theory of gravity, where mass, not anisotropy, induces local spacetime curvature.  An introduction on the used differential geometry concepts is contained in appendix \ref{app:diff_geom}. In the main text, we discuss how curvature invariants encompass anisotropy effects. For the particular case of intramural fiber rotation, we calculate the metric and curvature tensors for use in upcoming chapters. As a result, we establish that transmural fiber rotation makes the myocardium a hyperbolic space, i.e. with negative intrinsic curvature.


\section{Operational definition of distance}

\subsection{Anisotropy and rescaling \label{sec:local_resc}}

Let us first consider tissue anisotropy in its simplest appearance: a homogeneous block of myofibers that are all aligned along a given coordinate axis, i.e. the electrical diffusion tensor does not depend on position. Due to unequal signal velocities along the block's principal axes, the RDE \eqref{RDE:2dim} contains an anisotropic diffusion term. In this elementary example, isotropy is trivially restored by a coordinate transformation $(x,y,z) \rightarrow (x',y',z')$:
\begin{align} \label{global_resc}
  x &= x' \sqrt{\frac{D^{xx}}{D_0}},&          y &= y' \sqrt{\frac{D^{yy}}{D_0}},&   z &= z' \sqrt{\frac{D^{zz}}{D_0}},
\end{align}
with $D_0$ a constant with the dimension of a diffusion coefficient. The global transformation \eqref{global_resc} reduces the diffusion term in the RDE \eqref{RDE_mono} to $D_0$ times the standard Laplacian operator; see also \cite{Winfree:1998}. For more complicated fiber arrangements, this procedure has been carried out in a local and perturbative way in a variety of works \cite{Morozov:1999, Berenfeld:1999, Berenfeld:2001, Wellner:2002, Setayeshgar:2002, Davydov:2002, Davydov:2004}. We now proceed to show that this strategy can be generalized and formalized within a framework of non-Euclidean geometry.

\subsection{Notion of distance in physics}

What does the concept of distance mean to a physicist? With the advent of general relativity theory, the concept of absolute distance had to be given up, as different observers disagree on measured lengths. Absolute distance was traded in for an operational definition of distance, based on time measurements by an observer. In particular, the distance to an object may be measured by sending a light signal (traveling at the speed of light $c$) to the object and receiving the reflected beam after a time $2 \Delta t$. The distance from the observer to the object is now \textit{defined} as $\Delta x =c \Delta t$.

In various domains of physics, the notion of distance is intimately linked to the recording of time intervals. Even the SI unit of length, the meter, is defined as the distance traveled by light in vacuum within a fixed amount of time. Also, astronomical distances are expressed in lightyears. For light propagation through a medium, the absence of a constant signaling velocity may be handled by introducing the dimensionless refractory index $n$:
\begin{equation}\label{def_n}
  c = \frac{c_0}{n}.
\end{equation}
Furthermore the optical path length $s$ turns out to be proportional to arrival time:
\begin{equation}\label{def_opl}
  ds = n\, dx = \frac{c_0}{c} dx =c_0 dt.
\end{equation}
Phenomena that occur in excitable media with a discontinuity in the diffusion coefficient have already been treated with the concept of a refractory index in \cite{Pechenik:1998, Mornev:2004}.

\subsection{Measuring distances in the heart}

Does an equivalent of the speed of light exist in the heart? We are convinced that such analogue does exist, namely the unique propagation velocity of a planar activation front, which we shall also denote $c$. In general, this plane wave speed depends on position, direction and local electrophysiology parameters. As with the optics example, we choose a reference speed $c_0$, representing plane wave speed in an isotropic medium with scalar electrical diffusion coefficient $D_0$. With spatially homogeneous reaction kinetics, the effective distance between nearby points can be written down with respect to the propagation of action potentials:
\begin{equation}\label{def_el_length0}
  ds^2 = c_0^2 dt^2 = \left( \frac{c_0}{c} \right)^2 dx^2 = dx \frac{D_0}{D^{xx}} dx.
\end{equation}
Extended towards two or three dimensions, the infinitesimal length segment $ds$ is commonly denoted as a quadratic form:
\begin{equation}\label{def_el_length}
  ds^2 =  D_0 d\vec{r} \cdot \mathbf{D}^{-1} \cdot d\vec{r} = D_0 (D^{-1})_{ij} dx^i dx^j.
\end{equation}
The tensor $D_0 \mathbf{D}^{-1}$ is seen to fulfill the role of a metric tensor $\mathbf{g}$ in differential geometry terms. In essence, the metric tensor prescribes how coordinates $dx^i$ relate to the physical distance $ds$ between two nearby points:
\begin{equation}\label{def_metric_diffgeom}
  ds^2 =  g_{ij} dx^i dx^j.
\end{equation}
In Euclidean space, the metric tensor reduces to unity and with $g_{ij}=\delta_{ij}$, which reduces Eq. \eqref{def_metric_diffgeom} to the familiar Pythagorean theorem. The fundamental relation
\begin{equation}\label{def_metric}
  g_{ij} = D_0 (\mathbf{D}^{-1})_{ij}
\end{equation}
is responsible for rephrasing tissue anisotropy in the curved space formalism adopted in this work. The collected equations \eqref{def_metric_diffgeom}-\eqref{def_metric} designate a distance measure in the heart, which we shall henceforth call the \textit{operational measure of distance}.

In what follows, the inverse of $g_{ij}$ is denoted $g^{ij}$:
\begin{equation}\label{gijgjk}
  g_{ij} g^{jk} = \delta_i^{\hs k}.
\end{equation}
Applying this concept to the myocardial context immediately yields, from \eqref{def_metric},
\begin{equation}\label{def_metric_contra}
  g^{jk} = D_0^{-1} D^{jk}.
\end{equation}

\subsection{Operational distance anomaly for curved wavefronts \label{sec:anomaly}}

We emphasize here that the operational definition for distances in the heart only holds locally. For, given two remote points A and B, the travel time for an initially plane wave from A to B may differ from the time required to go from B to A. The culprit is the geometric curvature of the wave front affecting its propagation speed. In an isotropic medium with local diffusion coefficient $D$, the dependence upon wave front curvature $K$ (i.e. convexity or concavity) is mostly quoted as \cite{Zykov:1997}:
\begin{equation}\label{vc_1}
  c = c_0 - \gamma D K.
\end{equation}
In the next chapter, we will elaborate on the linearity coefficient $\gamma$ (which is generally taken positive), and generalize Eq. \eqref{vc_1} for use in anisotropic contexts.

\begin{figure} \centering
\mbox{
\raisebox{2.4cm}{a)}  \includegraphics[width=0.42\textwidth]{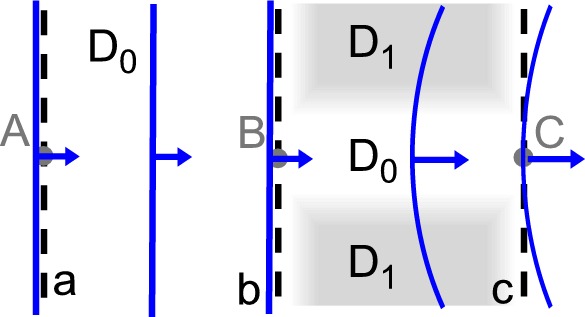} \hspace{0.3cm}
\raisebox{2.4cm}{b)}  \includegraphics[width=0.42\textwidth]{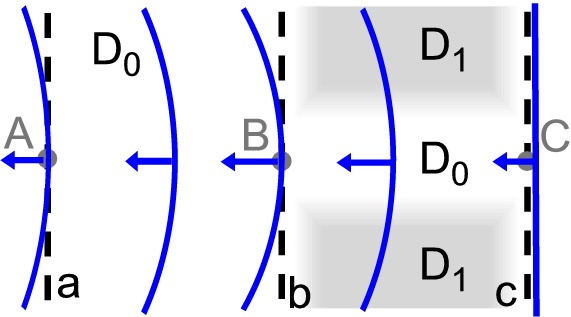}
}
  \caption[Asymmetry in arrival time measurements]{Asymmetry in arrival time measurements for an initially plane wave in an unbounded 2D geometry with $D_1>D_0$. a) Plane wave initiated in A. b) Plane wave started in C.} \label{fig:asymmetry}
\end{figure}

Let us, for the present purpose, consider the infinite two-dimensional medium depicted in Fig. \ref{fig:asymmetry}. When a plane wave is initiated through point A in the plane $a$, it will reach B within a time $T_{A\rightarrow B} = |x_A-x_B|/c_0$ without a change of shape. Due to the increased diffusion coefficients in the zone between B and C, the wave front center will lag behind its lateral parts, thus shaping a concave wave front. This front on average propagates faster than $c_0$ due to relation \eqref{vc_1}, and will have concave shape when reaching the plane trough C. The elapsed time between the activation of points B and C is denoted $T_{B \rightarrow C}$. For propagation in the reverse direction, symmetry implies that $T_{C\rightarrow B} = T_{B\rightarrow C}$. After the passage of the concave wave trough B, it continues its journey from B to A at a velocity larger than the plane wave velocity $c_0$, such that $T_{B\rightarrow A} < T_{A\rightarrow B}$. For this configuration, we have now clearly shown that
\begin{equation}\label{asymm}
  T_{A \rightarrow C} > T_{C\rightarrow A}.
\end{equation}

The result \eqref{asymm} is not in contradiction with \eqref{def_el_length}, but precisely the reason why the definition of operational distance needed be given in terms of differentials. The correct way to study aforementioned example is to record arrival times for plane waves that travel between points that lie infinitesimally close to each other. Such approach comes down to working in the limit of negligible wave front curvature.

\subsection{Hamilton-Jacobi interpretation of operational distance}

To continue, we demonstrate that our operational definition of distance also extremizes arrival time in directions that do not coincide with local axes of anisotropy. In the Hamilton-Jacobi formulation of optics, the distance between two infinitesimally close points is determined by the plane wave that takes \textit{longest} to travel from one point to the other. For, the plane waves that minimize the elapsed time before activation of the second point have both points included in their wave front. Therefore, the lowest possible travel time always equals zero.

Let us first study the isotropic case where, in two dimensions, a family of plane waves with normal vector $(\cos \theta, \sin \theta)$ travels from the origin to a point with Cartesian coordinates $(dx, dy)$ and polar coordinates $(dr, \phi)$. The elapsed time between activation of both points is readily calculated as $dT = (dr/c) \cos (\theta-\phi)$ with $|\theta-\phi| < \pi/2$. The plane wave which forms the upper bound to $dT$ is identified as $\theta = \phi$ and the operational distance therefore equals $ds = c\ dT_{max} = dr$, confirming one's everyday notion of distance.

Next, imagine the same configuration in an anisotropic medium with fibers aligned along the X-axis such that $D^{xx} = D_L$ and $D^{yy} = D_T$. In such a medium, a plane wave that propagates under an angle $\theta$ with the X-axis has velocity $c^2 = c_0^2 (D (\theta)/D_0 )$, while the Cartesian distance remains $dr \cos(\theta-\phi)$. The squared arrival time thus reads
\begin{equation}\label{sqT}
  dT^2 = \frac{dr^2}{c_0^2} \frac{D_0 \cos^2(\theta-\phi)}{ D_L \cos^2 \theta + D_T \sin^2 \theta}.
\end{equation}
The angle $\theta$ which maximizes travel time is found from $\dd_\theta (dT^2) = 0$, which is solved to
\begin{equation}\label{tantheta}
 \tan \theta = \frac{D_L}{D_T} \tan \phi = \frac{D_L}{D_T} \frac{dy}{dx}.
\end{equation}
Plugging this result into equation \eqref{sqT} yields, after some arithmetics:
\begin{equation}\label{sqT_metriek}
 ds^2 = c_0^2 dT^2 = D_0 \left( \frac{dx^2}{D_L} + \frac{dy^2}{D_T} \right),
\end{equation}
matching our operational definition of distance. To illustrate the methodology of working in non-Euclidean space, we perform the rescaling
\begin{align}
 dx' &= dx \sqrt{D_0/D_L}, &  dy' &= dy \sqrt{D_0/D_T},
\end{align}
such that isotropy of propagation speed is restored with $c'=c_0$. After regaining isotropy in the rescaled space, one simply puts
$dT = dr'/c'$ which immediately leads to expression \eqref{sqT_metriek}. Note that, in the rescaled space, the normal vector to the wave front encloses an angle $\psi \neq \theta$ with the fiber direction which has
\begin{equation}
    \tan \psi = (dy'/dx') = \sqrt{D_L/D_T} (dy/dx).
\end{equation}

\begin{figure}[h!b] \centering
  \includegraphics[width=0.9 \textwidth]{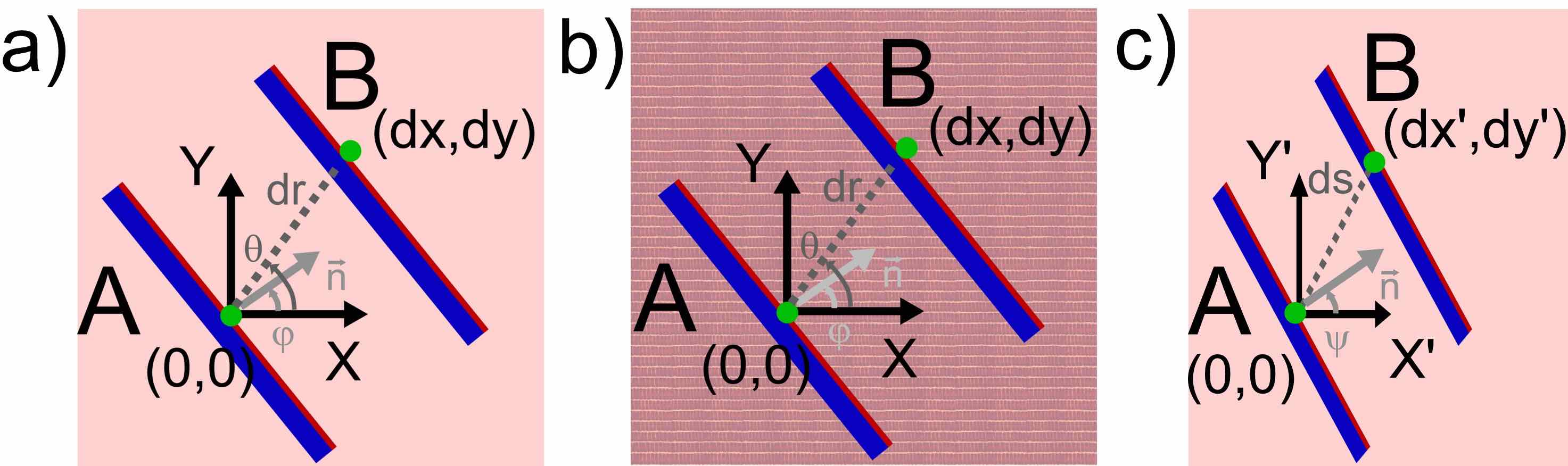}\\
  \caption[Plane waves connection two nearby points]{Plane waves connecting two given points in isotropic (a), anisotropic (b) and rescaled (c) space. }\label{fig: pw_rescaled}
\end{figure}

\subsection{The road map analogy}

\begin{figure}[h!b] \centering
    \mbox{
\raisebox{5cm}{a)}\  \includegraphics[height=5.5cm]{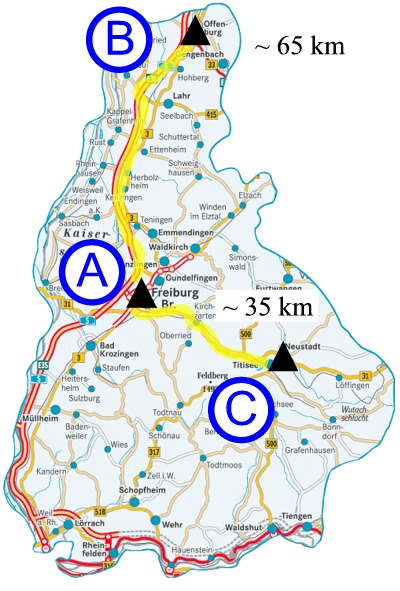} \hspace{1cm}
\raisebox{5cm}{b)}\  \includegraphics[height=5.5cm]{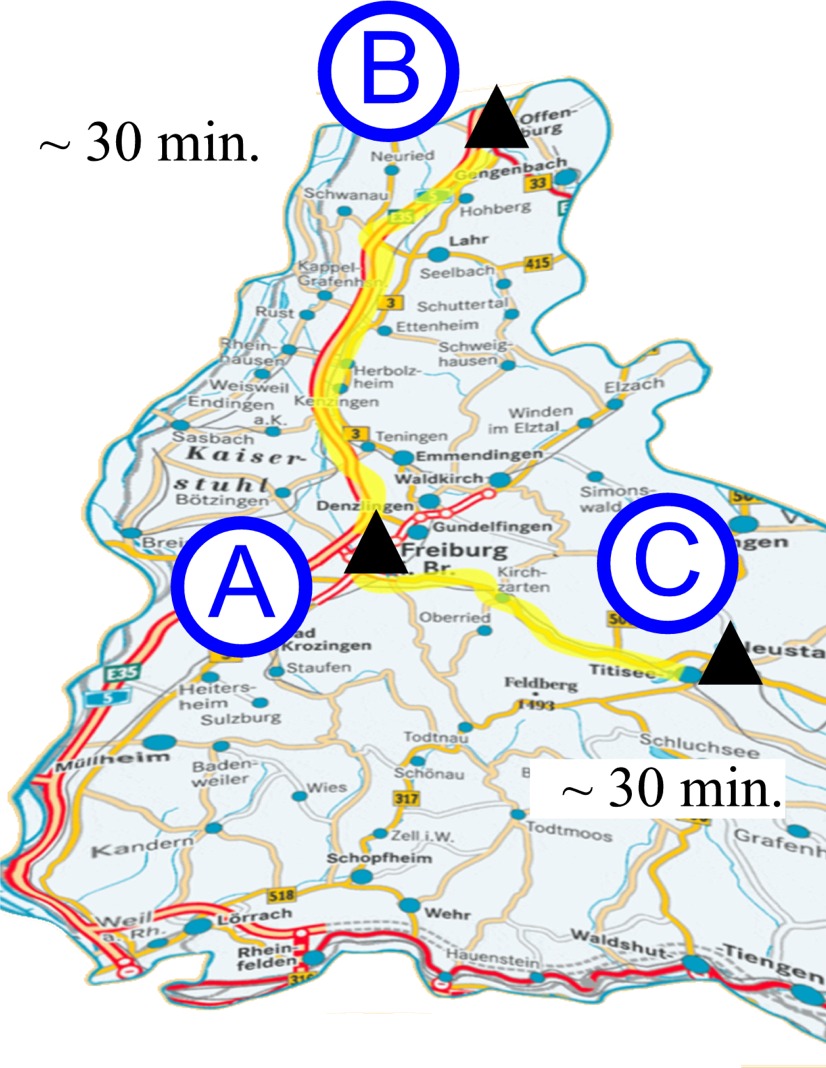}
}
\caption[The road map analogy]{The road map analogy illustrates how regional differences in propagation velocity (a) give rise to the concept of a deformed space (b) when one measures distances in terms of travel time.}\label{fig:roadmap}
\end{figure}

To get an intuitive notion on how anisotropy and inhomogeneity of signaling speed result in a curved space, one may appeal to a road map analogy as is depicted in Fig. \ref{fig:roadmap}. Conventionally, the distance between the pairs of places (A,B) or (A,C) is logged in kilometers, as shown in panel (a). Nevertheless, in practical situations one may wish express the proximity of another city by the time it takes to get there. One could even try to draw a map according to this new definition of distance; such map (of which panel (b) gives an artist's impression) will look severely distorted due to the distinct maximum speeds permitted on different roads.

In general, the resulting map cannot be charted on a flat surface: only by embedding in the third dimension it can be concretized. The property that prevents the map from being drawn on a plane surface is called \textit{intrinsic curvature} and is fully determined by intrinsic characteristics of the road map (i.e. its metric tensor).

On the other hand, for large travel maps the curving of the earth's surface can no longer be neglected. The manner in which the map relates to positions in the surrounding (three-dimensional) space is denoted the \textit{extrinsic curvature}. \\

In a similar way, the anisotropy in the heart muscle with respect to conduction velocity may be accounted for. In the presence of local myofiber rotation, the emerging three-dimensional structure chart can not be realized in Euclidean three-dimensional space; thereby the heart muscle too gains intrinsic curvature. In the same view, extrinsic curvature of a piece of heart tissue originates from the geometric curvature of the myocardial walls with respect to the surrounding Euclidean space.

In the next section, we will briefly outline how curved spaces are commonly treated mathematically using differential geometry.

\section{Differential geometry tools}

\subsection{Local equivalence}
Since the metric tensor defined by \eqref{def_el_length} is symmetric and positive-definite, one can diagonalize this tensor and rescale local coordinates akin to \eqref{global_resc} to make the space locally Euclidean in that point, i.e.
\begin{equation} \label{loc_eq}
g_{ij}(x^m) = \delta_{ij} + \OO(x^2).
 \end{equation}
Exploiting this local equivalence is similar to choosing a local free-falling (Minkowski) frame in the context of general relativity theory. 

For generic tissue anisotropy, the local equivalence \eqref{loc_eq} can only be established in a single point of space at the same time. The closest-to-Euclidean reference frame around a single curve is known as a local Fermi frame \cite{MTW}, and we will use such coordinate system to parameterize the immediate neighborhood of  scroll wave filaments in chapter \ref{chapt:filaniso}. For hypersurfaces such as activation fronts, a nearly Euclidean frame may be achieved using Gauss normal coordinates; see chapter \ref{chapt:fronts} for more details.

\subsection{Geodesics}
The simplest class of curves in Euclidean space are straight lines. In a curved space, one can only enforce \textit{local} straightness via the local equivalence principle. The resulting curves are known as geodesics, and these are the closest analogues to straight lines in Euclidean space (which here stands for an isotropic excitable medium). Geodesics are moreover proven to be the curves of extremal length connecting two given endpoints. That geodesics can be curves of either minimal of maximal length is easily seen by regarding a great circle on a sphere, which is an example of a geodesic curve.

Geodesics have been established to fulfill a special role in anisotropic excitable media. For, in the pioneering work by Wellner \etal \cite{Wellner:2002}, it was argued by both local rescaling of a specific anisotropy type and numerical simulations that the equilibrium configuration of a scroll wave filament with fixed endpoints takes the shape of a geodesic, when looking upon space using a metric tensor given by the correspondence \eqref{def_metric}. This statement was termed the `minimal principle for rotor filaments'; the proof given was only valid for a restricted class of anisotropy and for not too intensely curved filaments.

From earlier studies of scroll wave filaments in isotropic media, it is known that they equilibrate to a straight line. In the light of local equivalence, it should therefore not surprise us that Wellner's minimal principle is stated in terms of geodesics with regard to the metric tensor \eqref{def_metric}. Also, since geodesics trace out paths of extremal effective length, the equilibrated filament at the same time extremizes the travel time for an activation pulse connecting both anchored filament endpoints. This idea was formally proven and discussed using Hamilton-Jacobi theory in \cite{tenTusscher:2004}.


\subsection{Tensors}
An important corollary of working in curved space is that coordinates are degraded to mere labels, as they only represent distance between points via the metric tensor \eqref{def_metric}. As a consequence, there is more freedom in choosing suitable coordinates. In our description of scroll wave filaments, for example, we let the filament coincide with one of the coordinate lines.

Mathematical objects that bear physical meaning irrespective of the coordinate system used are known as \textit{tensors}. When going from a coordinate system $x^\mu$ to $x'^\nu$, a tensor quantity of rank $m+n$ changes as
\begin{equation}\label{def_tensor}
    T\, {}^{\mu'_1 ... \mu'_m}_{\hspace{25pt}  \nu'_1 ... \nu'_n} = \mathop{\sum_{\mu_1, ..., \mu_m}}_{\nu_1, ..., \nu_n}\frac{\dd x'^{\mu'_1}}{\dd x^{\mu_1}} .\, ... \,. \frac{\dd x'^{\mu'_m}}{\dd x'^{\mu_m}}
    . \frac{\dd x^{\nu_1}}{\dd x'^{\nu'_1}}  .\, ... \,. \frac{\dd x^{\nu_n}}{\dd x'^{\nu'_n}} . T\,{}^{\mu_1 ...\mu_m}_{\hspace{25pt} \nu_1 \nu_2 ... \nu_n}. \quad
\end{equation}
In the view of the defining property for tensors \eqref{def_tensor}, the placement of upper and lower indices does matter. Quantities with upper indices, such as coordinate differentials $dx^\mu$ are said to transform in a contravariant way; tensors with lower indices exhibit covariant transformation properties, i.e. they behave like derivatives $\dd_\mu$.

Notably, the chain rule for derivatives guarantees that the result of summation over a mixed contra- and covariant index pair is invariant under coordinate transformations. Hence, only such combination can bear physical meaning that supersedes the notion of coordinates. Since all index pairs of physically meaningful quantities should appear in such combinations, the summation sign is commonly omitted; this rule is known as the Einstein summation convention. Taking the sum of contra-and covariant is called contraction; contracting a tensor decreases its rank by two. A tensor of rank 0, i.e. without indices, is a scalar (scalar function), which can obviously also carry physical information. In physics calculations, it is possible to transform indices of one type to the other by means of the metric tensor in either its covariant or contravariant form (Eqs. \eqref{def_metric}-\eqref{def_metric_contra}):
\begin{align}
 T^i &= g^{ij} T_j, &   T_i &= g_{ij} T^j.
 \end{align}
A first tensorial example was met in the previous chapter: the definition of the proton diffusion tensor $D^{ij}$ transcends local rotations. Given a three-dimensional rotation with rotation matrix $R^i_{\hs j} = \dd x^i / \dd x'^j$, the tensor transformation property \eqref{def_tensor} yields in Euclidean space (where $g_{ij} = \delta_{ij}$): $\mathbf{D'} =  \mathbf{R} \mathbf{D} \mathbf{R}^T$. In this way one retrieves the well-known transformation law for rank-2 tensors under rotation. Contraction of the proton diffusion tensor results in $D^i_{\hs i} = D^{ij} \delta_{ij} = \mathrm{Tr}(\mathbf{D})$. The result is the diffusion tensor trace, which is used in diffusion imaging as a rotationally invariant scalar quantity.

Second, consider the covariant metric tensor $g_{ij}$ in Cartesian coordinates, which is given by \eqref{def_metric} for our present biophysical application. The metric tensor components are readily reexpressed in another coordinate system $x^\mu$ using Eq. \eqref{def_tensor}:
\begin{equation}\label{transf_g}
  g_{\mu \nu} = \frac{\dd x^i}{\dd x^\mu} g_{ij} \frac{\dd x^j}{\dd x^\nu}.
\end{equation}

\subsection{Covariant differentiation}
Because a tensor's representation in components is inherently tied to a local basis, one should take into account the change of basis vectors when taking spatial derivatives of a tensor quantity. In practice, this is done by replacing `ordinary' spatial derivatives (denoted $\delta_\mu \cdot$ or ${\cdot}_{,\mu}$) with `covariant' (i.e. coordinate invariant) spatial derivatives (written $\DD_\mu\cdot$ or ${\cdot}_{;\mu}$):
\bsub \label{def_covderiv} \begin{eqnarray}
 A^\nu_{\hs ; \mu} = \DD_\mu A^{\nu}  &=& \dd_\mu A^\nu + \Gamma^{\nu}_{\mu \la} A^\la,\\
 A_{\nu ; \mu} =  \DD_\mu A_{\nu} &=& \dd_\mu A_\nu - \Gamma^{\la}_{\mu \nu} A_\la.
\end{eqnarray} \esub
Here, the Christoffel symbols $\Gamma$ account for the change in basis vectors; they are uniquely constructed as linear combinations of first order spatial derivatives of the metric tensor, as listed in Eq. \eqref{christ} in appendix \ref{app:diff_geom}.

\subsection{Curvature tensors}
In view of the local equivalence principle \eqref{loc_eq}, curvature of space manifests itself in the spatial derivatives of the metric tensor from the second order onwards. It turns out that a single geometrical (i.e. tensorial) object captures the curvature of space in a given point, the Riemann curvature tensor \cite{MTW}:
\begin{equation}\label{def_riemann}
 R^\mu_{\ \nu \alpha \beta} = \dd_\alpha \Gamma^\mu_{\nu \beta} - \dd_\beta \Gamma^\mu_{\nu \alpha} + \Gamma^\mu_{\alpha\lambda} \Gamma^{\lambda}_{\nu \beta} - \Gamma^\mu_{\beta\lambda} \Gamma^{\lambda}_{\nu \alpha}.
\end{equation}
Due to symmetries (see appendix \ref{app:diff_geom}), this tensor only has six degrees of freedom in three dimensions. Contraction over the first and third index delivers the Ricci curvature tensor:
\begin{equation}\label{def_ricci}
    R_{\mu \nu} = R^\alpha_{\hs \mu \alpha \nu},
\end{equation}
which is symmetric, and therefore also has six components in three-dimensional space. Only in three dimensions (and lower), all Riemann tensor components may thus be expressed as linear combinations of the Ricci tensor components; the formula is quoted in Eq. \eqref{Riemann2Ricci}.

\begin{figure} \centering
  \includegraphics[width=0.8\textwidth]{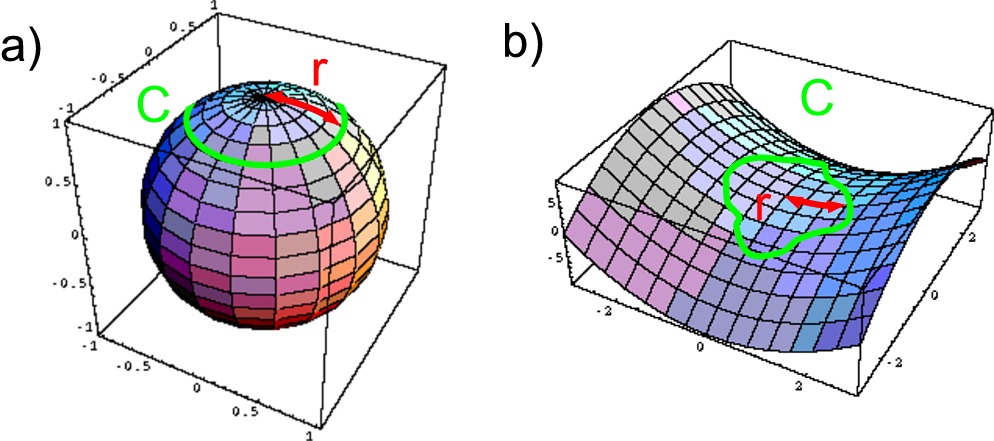}\\
  \caption[Circumference of geodesic circles]{Dependence of the circumference $C$ of small circles on the curvature of the embedding surface. For positively curved space (a), $C < 2\pi r$, and the reverse is true for hyperbolic spaces (b).}\label{fig:circumfer}
\end{figure}

The trace of the Ricci tensor defines the Ricci curvature scalar
\begin{equation}\label{def_RR}
   \RR = R^\mu_{\ \mu} = g^{\nu\mu} R_{\nu\mu}.
\end{equation}
The sign of the Ricci scalar indicates whether the space in that point is locally hyperbolic ($\RR<0$), spherical $(\RR>0)$ or flat ($\RR=0$). In hyperbolic spaces, which are often referred to as saddle-like, initially parallel geodesics tend to diverge, whereas they converge in spherical spaces. The sign of the Ricci scalar may be further exemplified by calculating the circumference $\mathcal{C}$ of the curve with radius $a$ on a curved surface, as sketched in Fig. \ref{fig:circumfer}:
\begin{equation}\label{circumfer}
 \mathcal{C} = 2 \pi a\left[ 1 - \frac{1}{12} a^2 \RR + \OO(\RR^2a^4)\   \right].
\end{equation}
For a sphere of radius $R_0$, one calculates that $\RR = 2 / R_0^2$. Thus, geodesics always converge on the sphere and the circumference of a circle with radius $a$ (drawn on the sphere) is smaller than $2\pi a$ by virtue of Eq. \eqref{circumfer}. In chapter \ref{chapt:filaniso}, we will quantify how much the rotation period of a scroll wave accommodates to the circumferential deficit.

\section{Curvature measures for rotational anisotropy \label{sec:rot_aniso}}

\subsection[A simple model for ventricular myofiber organization]{Rotational anisotropy as a simple model for ventricular myofiber organization}
Assessing the influence of tissue anisotropy on the dynamics of activation waves and rotor filaments in cardiac tissue has mostly been performed in a simplified model of myofiber architecture, which is known as rotational anisotropy \cite{Streeter:1969, Panfilov:1995, Fenton:1998}. In this geometry, the myofiber direction is taken everywhere parallel to the epicardial and endocardial borders, which are represented by infinite parallel planes where no-flux boundary conditions apply. The transmural fiber rotation is assumed with a constant turning rate (pitch) $\mu$ such that the fiber helix angle alpha ranges over about $\Delta \alpha = 120^\circ$, similar to the observed rotation angles in ventricular myocardium.
Moreover, although myofiber rotation in reality occurs in three dimensions at a non-constant rate, a sufficiently small block of tissue can be to some extent approximated with local rotational anisotropy of given pitch and orientation.

\begin{figure}[h!t] \centering
  \includegraphics[width=0.75\textwidth]{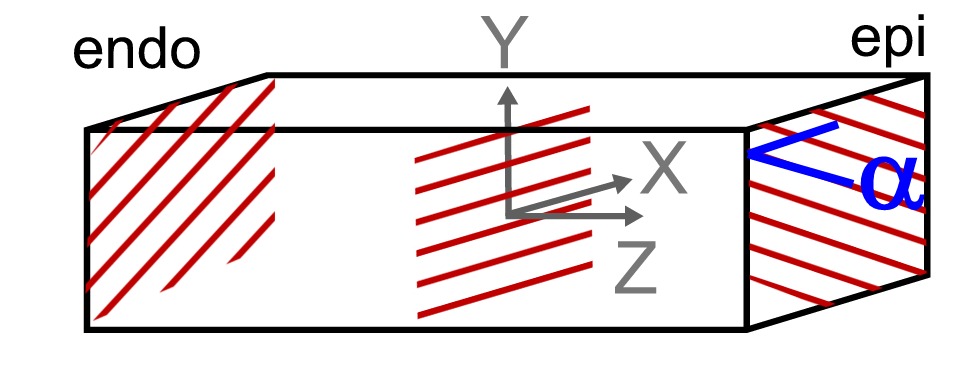}\\
  \caption[Rotational anisotropy configuration]{Geometry for rotational anisotropy. The fiber helix angle has been annotated $\alpha$ here.}\label{fig:RA_geom}
\end{figure}

In numerical simulations of scroll waves in a medium with rotational anisotropy, several remarkable facts on filament behavior have been observed:
\begin{enumerate}
  \item Filaments that are initiated parallel to the epicardial surface (i.e. infinitely long intramural filaments) were observed to drift transmurally towards an epicardial depth in which they are aligned either parallel or perpendicular to local myofiber orientation \cite{Wellner:2000}, depending on the used reaction kinetics.
  \item A straight transmural filament orthogonal to the medium boundaries which is stable at low myofiber rotation rate or anisotropy ratio loses stability when the pitch of fiber rotation is increased \cite{Fenton:1998}.
  \item The destabilizing effects of rotational anisotropy depend on the used electrophysiological model. See e.g. \cite{Rappel:2001} for a numerical comparison of two detailed cardiac models.
  \item Straight intramural filaments of finite length (i.e. in a bounded medium) not only undergo drift but also rotate in their evolution process to align with local myofiber direction \cite{Berenfeld:2001, Wellner:2000}.
  \item In a specific parameter regime, twist develops in localized sections of transmural filaments, and seemingly propagates as a whole along the filament  \cite{Fenton:2002}. The entities with concentrated twist were annotated \textit{twistons}. However, the dynamical origin of these `twistons' remains unclear.
\end{enumerate}
We will offer an analytical description for the first three of these facts in chapter \ref{chapt:filaniso}. Thereto, it is instructive to first calculate the space curvature that fronts and filaments experience in a medium with rotational anisotropy.

\subsection{Explicit curvature measures for rotational anisotropy \label{sec:curv_RA}}

For definiteness, we choose the transmural fiber rotation to take place around the Z-axis of a Cartesian coordinate frame. We additionally lay the X-axis along the local myofiber orientation at $z=0$, conferring to the sketched situation in Fig. \ref{fig:RA_geom}. Furthermore, orthotropic anisotropy is admitted, i.e. the electric diffusion tensor has distinct eigenvalues ($D_L, D_P, D_T$) with $D_L$ the electrical diffusivity in the myofiber direction, and $D_T$ the electrical diffusion coefficient in the transmural direction. The third constant $D_P$ represents electrical diffusivity in the planes parallel to XY in the direction orthogonal to the local myofiber orientation. In what follows, we only assume that
\begin{align}
 D_L &\geq D_P >0,& D_L &\geq D_T >0
\end{align}
and therefore orthotropic media with rotational anisotropy are implicitly considered. We convene that positive myofiber pitch $\mu$ indicates right-handed fiber rotation, such that in anatomical reality one has $\mu<0$. Obviously, one has for the fiber helix angle that $\alpha = \mu z$.

With $g^{ij} = D_0^{-1} D^{ij}$ one obtains
\begin{equation}\label{RA_gcon}
  (g^{ij}) = D_0^{-1} \left(
             \begin{array}{ccc}
               D_L \cos^2 \alpha + D_P \sin^2 \alpha & (D_L-D_P) \cos \alpha \sin \alpha & 0 \\
               (D_L-D_P) \cos \alpha \sin \alpha & D_L \sin^2 \alpha + D_P \cos^2 \alpha & 0 \\
               0 & 0 & D_T \\
             \end{array}
           \right).\qquad
\end{equation}
The covariant metric tensor follows from matrix inversion of \eqref{RA_gcon}:
\begin{equation}\label{RA_gcov}
  (g_{ij}) = D_0 \left(
             \begin{array}{ccc}
               D_L^{-1} \cos^2 \alpha + D_P^{-1} \sin^2 \alpha & (D_L^{-1}-D_P^{-1}) \cos \alpha \sin \alpha & 0 \\
               (D_L^{-1}-D_P^{-1}) \cos \alpha \sin \alpha & D_L^{-1} \sin^2 \alpha + D_P^{-1} \cos^2 \alpha & 0 \\
               0 & 0 & D_T^{-1} \\
             \end{array}
           \right).\qquad
\end{equation}
The metric tensor completely determines the curvature tensors $R_{ijkl}$ and $R_{ij}$. Of the six independent components of the Riemann tensor, we only quote $R_{xyxy}$:
\begin{equation}\label{RA_R1212}
  R_{xyxy} = \frac{\mu^2}{4} \left(D_P^{-1}-D_L^{-1} \right)^2 D_0 D_T.
\end{equation}
The Ricci curvature tensor for rotational anisotropy is computed equal to
\begin{eqnarray}\label{RA_Ricci_cov}
  (R_{ij}) =  - \frac{\mu^2}{2} D_T \frac{(D_L^2-D_P^2)}{D_L D_P}. \hspace{0.6\textwidth}
  \\ \left(
             \begin{array}{ccc}
               D_L^{-1} \cos^2 \alpha - D_P^{-1} \sin^2 \alpha & (D_L^{-1} + D_P^{-1}) \cos \alpha \sin \alpha & 0 \\
               (D_L^{-1} + D_P^{-1}) \cos \alpha \sin \alpha & D_L^{-1} \sin^2 \alpha - D_P^{-1} \cos^2 \alpha & 0 \\
               0 & 0 & D_T^{-1} \\
             \end{array}
           \right).\nn
\end{eqnarray}
By comparison of \eqref{RA_Ricci_cov} and \eqref{RA_gcov}, the Ricci tensor may be seen to exhibit following eigenvalues $\{R_i\}$ and eigenvectors $\{\vec{v}_i\}$:
\begin{equation}\label{RA_eig_Ricci}
     \begin{array}{r@{=}l r@{=}l}
       \RR_1 & - \left( \frac{\mu^2}{2} \frac{(D_L^2-D_P^2)}{D_L D_P}\right) \frac{D_T}{D_L}& \vec{v}_1  & (\cos \alpha, \sin \alpha, 0), \\
       \RR_2 & \hspace{1 em} \left( \frac{\mu^2}{2} \frac{(D_L^2-D_P^2)}{D_L D_P}\right) \frac{D_T}{D_P} &\vec{v}_2 & (-\sin \alpha, \cos \alpha, 0), \\
       \RR_3 & - \left( \frac{\mu^2}{2} \frac{(D_L^2-D_P^2)}{D_L D_P}\right)  & \vec{v}_3 & (0, 0, 1).\\
     \end{array}
\end{equation}
Apparently, the Ricci tensor has two strictly negative eigenvalues in myocardium with constant fiber rotation rate.

When calculating the Ricci curvature scalar, the arbitrarily chosen constant $D_0$ comes in. This issue can be dealt with by introducing the rescaled pitch
\begin{equation}
 \mu' = \frac{\Delta \alpha}{\Delta z'} = \mu \sqrt{\frac{D_T}{D_0}},
\end{equation}
such that the Ricci curvature scalar substantiates to
\begin{equation} \label{RA_RR}
 \RR = - \frac{\mu^2}{2} \frac{(D_L-D_P)^2 D_T}{D_L D_P D_0} = - \frac{{\mu'}^2}{2} \frac{(D_L-D_P)^2}{D_L D_P}
\end{equation}
for a medium with rotational anisotropy of constant pitch $\mu$.

\subsection[Discussion of intrinsic curvature with rotational anisotropy]{Discussion of intrinsic curvature for a medium with rotational anisotropy}

From the foregoing analysis follows that the model configuration of rotational anisotropy delivers a space that has always negative intrinsic scalar curvature. As the myocardial fiber rotation may be reasonably called the dominant effect on intrinsic cardiac geometry, we also expect that initially parallel geodesics will diverge with realistic ventricular anatomy.
%

We can also compare the magnitude of the curvature measures relative to the thickness $L$ of the myocardial wall. For fiber rotation over $120^\circ$ one has $\mu = \frac{2\pi}{3L}$. In terms of the Ricci tensor eigenvalues, one may associate the intrinsic curvature radii $r_i = |\RR_i|^{-1/2}$. For a typical anisotropy ratio of $9:1:1$ for myocardial ventricular tissue, we thus obtain
\begin{align}
 r_1 &= 0.68\, L,&  r_2&=r_3 = 0.23\, L.
\end{align}
Hence, the typical radii of curvature for excitable tissue with the given anisotropy ratio are comparable to the medium size, which indicates that intrinsic curvature of functional myocardium is of significant importance.

\clearpage{\pagestyle{empty}\cleardoublepage}

\graphicspath{{fig/}{fig/fig_fronts/}}

\renewcommand\evenpagerightmark{{\scshape\small Chapter 6}}
\renewcommand\oddpageleftmark{{\scshape\small Geometric theory for wave fronts}}

\renewcommand{\bibname}{References}

\hyphenation{mani-fest-ly where-as}

\chapter[Wave front dynamics]{Geometric theory for wave fronts}
\label{chapt:fronts}


In this chapter we use the new curved space formalism to adapt the equation of motion for wave fronts to a general anisotropic context. Our approach results in an extended velocity-curvature relation for wave fronts in anisotropic media which is consistent with earlier descriptions in literature. The coefficient of linearity in the velocity-curvature relation is additionally shown to differ from one when not all state-variables have equal diffusion coefficients. We also demonstrate for the first time that the equation of motion may be derived from a variational principle of elegant geometrical nature. Preliminary numerical validation of the coefficient of linearity in the velocity-curvature equation for several low dimensional ionic models is presented as well.

By imposing periodic boundary conditions, we enlarge our description to encompass not only excitable but also oscillatory media. This case also applies to high-frequency myocardial pacing, which was treated for two specific models in \cite{Pertsov:1997, Wellner:1997} by finite renormalization theory.

\section{Velocity-curvature relation in literature}

\begin{figure}[h!b] \centering
  \includegraphics[width=0.9\textwidth]{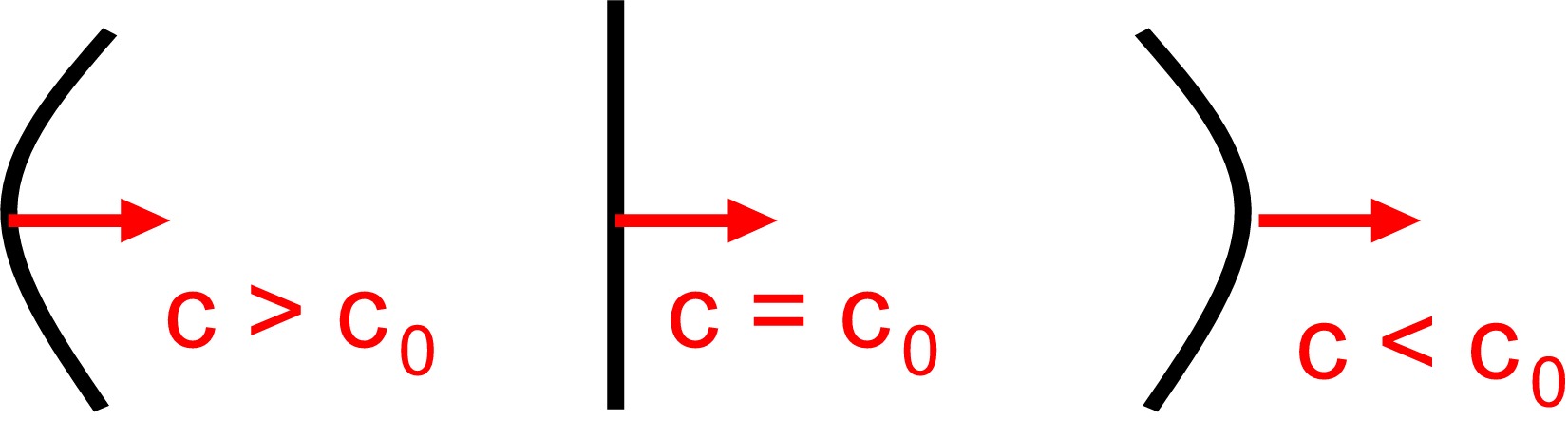}\\
  \caption[Dependence of the plane-wave velocity on wave front curvature]{Dependence of the plane-wave velocity on curvature of the wave front.}\label{fig:vc_simple}
\end{figure}

An important consequence of the strong non-linearity of the RDE for excitable media is that the emanating waves do not obey a superposition principle as in electromagnetism. Specifically, the local propagation velocity of an excitation wave depends on its geometrical curvature, as is illustrated in Fig. \ref{fig:vc_simple}. This particularity has yet been mentioned in our discussion of the non-reciprocity of the operational distance in paragraph \ref{sec:anomaly}.\\

In physiological terms, the curvature effect has been explained based on passive electrical conduction properties: an excitable myocyte in rest will easier activate when more cells around it offer a stimulating electrical current. Therefrom, it might be expected that a concave wave front advances more quickly than a convex one. Attributing a positive curvature $K$ to a convex wave front, one could anticipate a linear relationship between the propagation velocity and curvature of a wave front (with $A>0$):
\begin{equation}\label{vc_0}
  c = c_0 - A K,
\end{equation}
where $c_0$ is the plane wave speed in the (isotropic) medium considered. Note that, from an evolutionary standpoint, Eq. \eqref{vc_0} furnishes the depolarization waves with a stabilizing property: parts of the wave front that have advanced at a quicker pace are slowed down due to the curvature effect, whereas parts of the wave front that lag behind are (literally) stimulated to speed up. This idea may be verified in mathematical terms; see paragraph \ref{sec:stab_fronts} below.

Two further repercussions of Eq. \eqref{vc_0} are relevant to the cardiac context. First, the wave fronts only propagate if their curvature is smaller than a critical curvature $K_{cr}$, which would make $c$ vanish. This situation is relevant with electrical stimulation through electrodes, as their radius should be large enough in order to start an propagating activation wave.
Second, the slower propagation of convex fronts lies at the basis of spiral wave formation: the curvature effect makes the tip stall, whereas the more distant parts of the wave front exhibit a larger propagation speed due to their lower geometrical curvature. From Eq. \eqref{vc_0} alone, the asymptotic shape of a rotating activation pattern was proven to be an Archimedean spiral \cite{Zykov:1997}.\\

Looking closer to Eq. \eqref{vc_0}, the (mostly) positive constant $A$ is seen to bear the dimension of a diffusion coefficient. From dimensional arguments, $A$ may therefore be taken proportional to the electrical diffusion coefficient $D_0$ in the medium. Numerical calculations reported in \cite{Zykov:1997}, showed that a model-dependent `correction factor' (\textit{sic}) needs be included: \begin{equation}\label{vc_2}
  c = c_0 - \gamma D_0 K,
\end{equation}
although the theoretical elaborations in \cite{Keener:1986} delivered $\gamma \approx 1$ based on matched asymptotic expansions.

For specific RD models, explicit prescriptions for $\gamma$ have been established \cite{Panfilov:1995b, Pertsov:1997}. However, only Mikhailov \etal provided in the appendix of \cite{Mikhailov:1994} an explicit result that is valid for generic RD models. In our treatment, we will retrieve their result (see Eq. \eqref{defgamma}) and moreover generalize it for temporally periodic activation waves. Also, we have calculated, for the first time to our knowledge, the second order curvature corrections to the velocity-curvature relation \eqref{vc_2}.\\

In the previous chapter we have already mentioned that rotational anisotropy in the heart impacts on the orientation of the isochrone surfaces after point stimulation on the epicardial surface. Also, the effect of anisotropy can invoke cusp waves in ventricular tissue, which were investigated in \cite{Bernus:2004} by theoretical, numerical and experimental means. The linearized velocity curvature relation for wave fronts \eqref{vc_2} has been reformulated in a kinematical approach for two-dimensional media of spherical \cite{Zykov:1996} and  non-uniformly curved surfaces \cite{Davydov:2000}. With our covariant gradient expansion approach, we not only offer an elegant form for the resulting equations, but also provide additional corrections to these results and quantify the model-dependent constants in the EOM.

\section[Geometric derivation of the velocity-curvature relation]{Geometric derivation of the \\velocity-curvature relation for wave fronts}

\subsection{A collection of one-dimensional solutions \label{sec:fronts}}

We start from the monodomain reaction-diffusion equation \eqref{RDE_mono}:
\begin{equation}\label{RDE_chapt_fronts}
  \dd_t \uu = \dd_i\left(D^{ij}\dd_j \mathbf{P} \uu \right)+ \mathbf{\bPhi}(\uu),
\end{equation}
The generic reaction kinetics are assumed to support a unique plane wave solution $\uu_0(\xi) = \uu_0(x-ct)$ when the medium is homogenous and isotropic with scalar diffusion coefficient $D_0$. Hence $\uu_0$ satisfies
\bsub \begin{eqnarray}\label{planewave}
 & D_0 \mathbf{P} \dd^2_\xi \uu_0 + c_0 \dd_\xi \uu_0 + \mathbf{\bPhi}(\uu_0) =0,& \\
  &\lim \limits_{\xi\rightarrow \pm\infty} \uu_0'(\xi) = 0,& \label{bc_u0fronts}
\end{eqnarray} \esub
where $c_0$ denotes the velocity of the plane wave in a medium with isotropic electric diffusivity $D_0$. Equation \eqref{planewave} is recognized as the familiar cable equation for one-dimensional propagation.

For all times $t$, we define the wave front as the locus where one of the state variables, say the first ($u_1$), reaches a threshold value $u_1^{\rm th}$. After we will have introduced the adjoint Goldstone modes for translation, we shall discuss how specifying a  threshold value affects the resulting EOM. For now, we choose the electrical transmembrane potential $V_m$ as the state variable and $u^1_{th}$ as the arithmetic average of the medium's resting potential and the potential after upstroke. To distinguish the wave front from its tail, we additionally impose that $\dd_t V>0$ where the threshold voltage is reached. The resulting surface $\Sigma_0$ has one dimension less than the surrounding space and may therefore be termed a hypersurface. The hypersurface $\Sigma_0$ is parameterized by two coordinates $\s_1,\s_2$, and can be oriented by making the surface normal $\vec{n}$ point in the same sense as $-\nabla V$. Note that after collision of wavefronts the surface may not possess a well-defined normal vector on the seam where the fronts have met. However, the wave front never ceases to be an orientable surface.\\

In the treatment of wave fronts in this chapter, we do not consider wave breaks. We therefore only allow the wave front hypersurface to terminate on the medium boundaries, where the no flux condition for currents prescribes
\begin{equation}\label{boundary}
    N_i D^{ij} \bP \dd_j \uu = 0
\end{equation}
with $\vec{N}$ an outward normal vector to the medium. As the ventricular myofibers run parallel to the outer surface of the heart, the electrical diffusion tensor gains a diagonal shape relative to the normal vector at the medium boundaries, which reduces condition \eqref{boundary} to the Neumann condition $\dd_N V_m = 0$. However, a rigorous implementation of realistic boundary conditions is only possible in the bidomain formalism. As we presently intend to develop a local theory, we shall not consider boundary effects here.\\

\begin{figure}[h!b] \centering
  \mbox{
  \raisebox{3.25cm}{a)}\includegraphics[height=3.95cm]{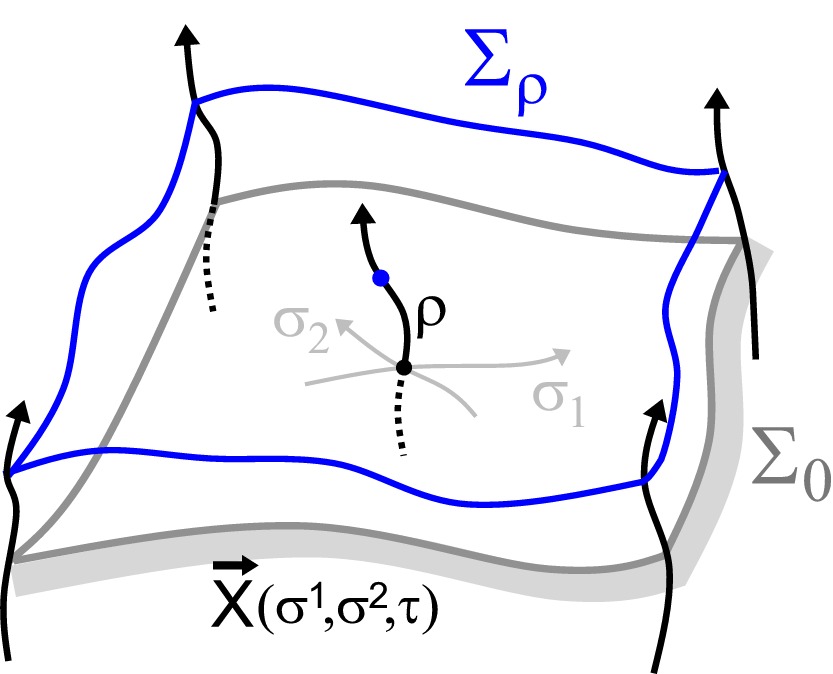} \hspace{0.6cm}
  \raisebox{3.25cm}{b)} \hspace{0.3cm}\includegraphics[height=3.95cm]{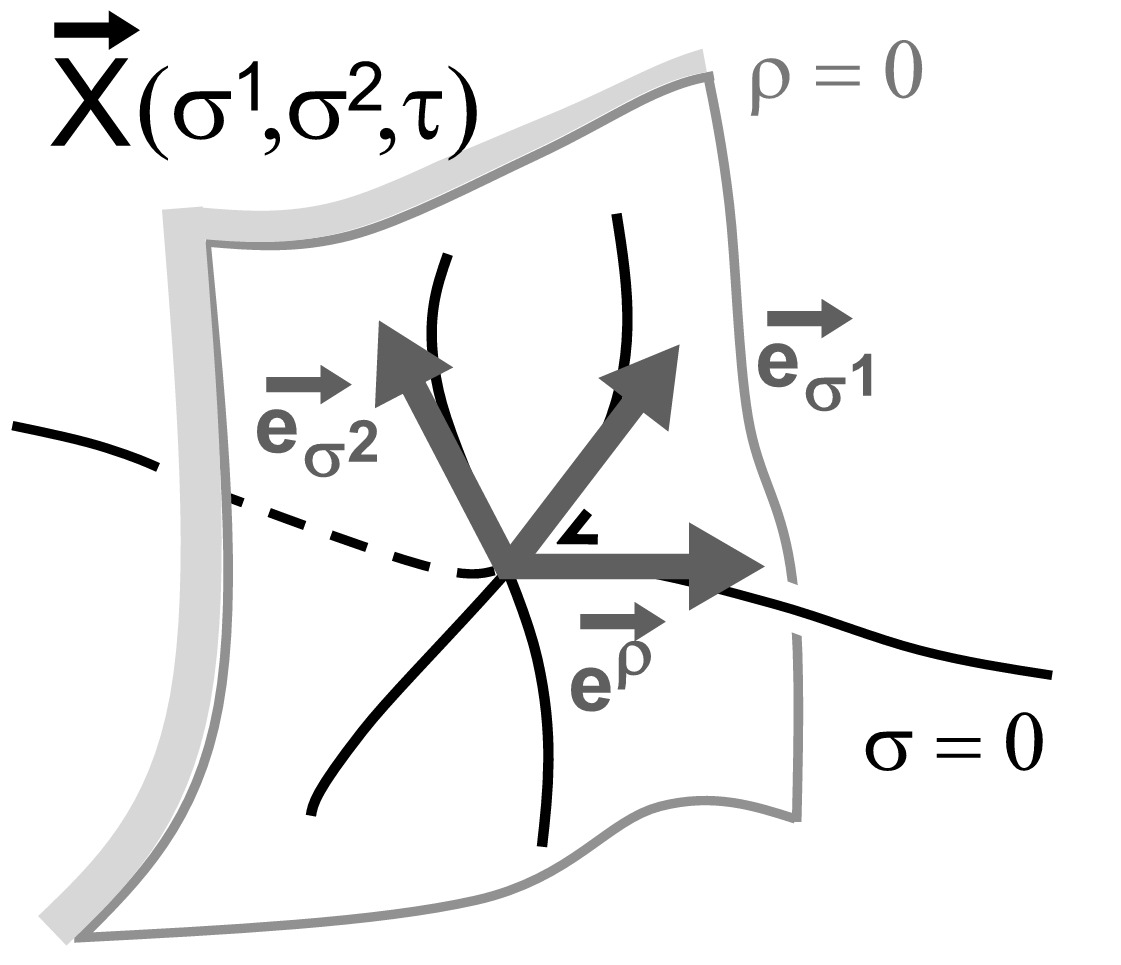}
  }
  \caption[Coordinate frame adapted to the wave front shape. ]{Construction of a coordinate frame adapted to the wave front shape. a) Gaussian normal coordinates generated by geodesics orthogonal to the wave fronts. b) Reference triad in a given point of the wave front.} \label{fig:coord_front}
\end{figure}

The main ansatz in our theory is that a two-dimensional pack of one-dimensional solutions $\uu_0$ from \eqref{planewave} yields a good approximation to the three-dimensional activation pattern. The range in which this procedure may be performed is expected to be constrained by the ratio of the wave front thickness $d$ (see \ref{sec:localfront} for a formal estimate) to the wave front's radius of curvature.
For isotropic media, the wave front curvature measure $K$ from Eq. \eqref{vc_2} that affects its propagation speed in lowest order equals twice the mean curvature, i.e.
\begin{equation}\label{def_K_isotropic}
  K = \frac{1}{R_1} + \frac{1}{R_2} = K_1 + K_2,
\end{equation}
where $R_1, R_2$ are the principal radii of curvature at a given point of the surface. We will shortly generalize this definition to an anisotropic context (see subsection \ref{sec:curv2}), and use the same annotation $K$ for both contexts. Thus, writing $K$ for the scalar parameter that reflects extrinsic wave front curvature, we now introduce the dimensionless formal expansion parameter
\begin{equation}\label{fronts_def_la}
  \la = d K.
\end{equation}
To substantiate the idea of assembling the wave front from a collection of plane wave solutions, a coordinate frame is constructed that resembles local Cartesian coordinates in a neighborhood of $\Sigma_0$, for all time instances $t$. That is, in every point of the wave front hypersurface (with coordinates $\s^1, \s^2$) the unique geodesic is traced that cuts the surface perpendicularly, as outlined in Fig. \ref{fig:coord_front}a. The arc length parameter $\rho$ along each of these geodesics is used as the third spatial coordinate, with $\rho = 0$ for the points on the wave front. Furthermore, we shall write $\tau$ as the time coordinate in the co-moving frame, as opposed to the `laboratory' time $t$. In the context of wave fronts, we will write henceforth indices $A,B,\ldots$ to indicate the coordinates $\s^A = \s^1,\s^2$ and reserve $\mu, \nu, \ldots$ to range over all three spatial coordinates $\s^1,\s^2, \rho$.

In the actual curvilinear context, a natural set of base vectors is found in the triad $\vec{e}_\mu, \vec{e}^\nu$, defined as
\begin{align}
    \vec{e}_\rho &= \dd_\rho \vec{x},  &    \vec{e}_{A} &= \dd_A \vec{x}, &
    \vec{e}^\rho &= \vec{\nabla} \rho,  &    \vec{e}^{A} &= \vec{\nabla} \s^A \label{triad}.
\end{align}
From the chain rule for derivatives, this basis is verified to satisfy the biorthogonality condition
\begin{equation}\label{biorth_fronts}
    \vec{e}_\mu \cdot \vec{e}^{\,\nu} = \delta_\mu^\nu.
\end{equation}

Next, we lay out the known solution $\uu_0(\rho)$ along the geodesic that intersects $\Sigma_0$ orthogonally, drawn in Fig. \ref{fig:coord_front}b. Since the coordinate frame approaches local Euclidean space in the best possible way, the thus constructed activation pattern serves as the lowest order approximation to an exact solution of \eqref{RDE_chapt_fronts}. Note that the procedure given yields an 1-parameter family of branes $\Sigma_\rho$, one of which corresponds to the actual wave front $\Sigma_0$. Recalling that the plane wave solution needs no corrections, our ansatz may be quoted
\begin{equation}\label{fronts_ansatz}
    \uu(\rho, \s^1,\s^2,\tau) = \uu_0(\rho) + \la \tilde{\uu}(\rho, \s^1,\s^2,\tau)+ \OO(\la^2).
\end{equation}
 We will also assume now -- and ascertain later -- that the perturbative correction $\tuu$ varies slowly along the wave front:
\begin{equation}\label{dstuu_fronts}
    \dd_\s \tuu = \dd_\rho \tuu . \OO(\la).
\end{equation}

\subsection{Gaussian normal coordinates for the wave front}

One may have noticed that the conceived coordinate system from the previous paragraph is nothing less but a frame of Gaussian normal coordinates, for each time $\tau$ \cite{MTW}. The Gauss lemma then ensures that all surfaces with $\rho$ constant intersect the lines of constant $\s^1,\s^2$ perpendicularly. Adding the fact that $\rho$ measures arc length along the geodesics, we obtain
\begin{equation}\label{metric:GNC}
    g_{\mu\nu} = \left(
                   \begin{array}{ccc}
                     1 & 0 & 0 \\
                     \begin{array}{c} 0\\ 0 \end{array} & \multicolumn{2}{c} {g_{AB}}
                   \end{array}
                 \right).
\end{equation}
With this choice of coordinates, the components $g_{\mu \rho}$ equal $\delta_{\mu \rho}$ up to all orders in $\rho$ by virtue of the Gauss lemma. The expansion \eqref{metric:GNC} evidently only holds in the neighborhood of the wave front, i.e. where geodesics which leave the wave front orthogonally do not yet intersect.

Given a Gauss normal frame that has been constructed for a time instance $t=0$, we let the frame advance together with the wave front with a local normal velocity $c_M$. (We will later substantiate $c_M$ to equal successive approximations to the local wave front speed $c(\s^1, \s^2, \tau)$). 
Had the space been Euclidean, we could have denoted the position $\vec{X}_f$ of the wave front as $\vec{X}_f = \vec{X}_m + \vec{X}$, with $\vec{X}$ the wave front position in the moving frame. In anisotropic media, however, only the differentiated version may be employed:
\begin{equation}\label{wavefrontspeed}
  \dd_t \vec{X}_f = \dd_t \vec{X}_m + \dd_t \vec{X} = c_M \vec{n} + \dd_t \vec{X}.
\end{equation}
Hence, the Taylor expansion of an arbitrary position near $\Sigma_0$ becomes:
\begin{equation}\label{decposition}
    x^i = X^i(\s_1,\s_2,\tau) + \rho n^i (\s_1,\s_2,\tau)+ \ldots
\end{equation}
Together with the trivial time coordinate change, Eq. \eqref{decposition} makes out a coordinate transformation:
\begin{equation}\label{coord_tf}
  \begin{cases}
    x^i &= x^i(\rho, \s^1, \s^2, \tau),  \\
    t &= \tau.
  \end{cases}
\end{equation}

Using the chain rule, the transformation laws for derivatives that follow from Eqs. \eqref{coord_tf} are easily deduced. Adopting the notation \eqref{triad}, one finds
\bsub \begin{eqnarray}
        \dd_\mu &=& \vec{e}_\mu \cdot \nabla, \\
        \dd_\tau &=& \dd_t + \dd_\tau \vec{x} \cdot \vec{e}^\mu \dd_\mu.
\end{eqnarray} \esub
One can moreover check that with Eqs. \eqref{triad}, the geodesic equation $\DD_\rho \vec{e}_{\rho}=0, \forall \rho$ and the definition $g_{\mu\nu} = \vec{e}_\mu \cdot \vec{e}_\nu$ directly leads to expression \eqref{metric:GNC} for the metric tensor. Moreover, $g_{\rho\nu} = \delta_{\rho\nu}$ even implies that the higher order corrections to Eq. \eqref{decposition} must vanish and $\vec{e}^\rho = \vec{e}_\rho =\vec{n}$.
The `bulk metric' $g_{ij}$ gives rise to an induced metric $h_{AB}$ on each of the surfaces of constant $\s$:
\begin{equation}\label{inducedmetric0}
    h_{AB} = e_A^i g_{ij} e_B^j
\end{equation}
With our choice of coordinates, we have $e_A^\mu = \delta_A^\mu$, so relation (\ref{inducedmetric0}) simplifies to
\begin{equation}\label{inducedmetric}
    h_{AB}(\rho, \s_1, \s_2) = g_{AB}(\rho, \s_1, \s_2)
\end{equation}
regardless of the explicit parametrization $(\sigma_1, \sigma_2)$ inside the wave front $\Sigma_0$.

\subsection{Extrinsic curvature tensor for (1+2)-dimensional space \label{sec:curv2}}

Surfaces which are embedded in Euclidean space, e.g. wave fronts in isotropic media, have an intrinsic curvature that compensates their bending in the surrounding space. One could for example imagine a sphere, which becomes more curved in the imbedding space as its intrinsic curvature grows. In the view of the curved space formalism, a wave front's intrinsic curvature needs not equal its extrinsic curvature anymore in an anisotropic medium, as space itself may be curved due to anisotropy effects.\\

The object that captures the intrinsic curvature of the wave front is the induced metric $h_{AB}$ that was given in \eqref{inducedmetric0}. Solely based on this quantity, one may calculate the intrinsic curvature of the two-dimensional wave front surface using Eq. \eqref{def_riemann}, where indices run over $\s^1, \s^2$. The resulting quantity quantity is denoted $^{(2)}R_{ABCD}$, to emphasize that curvature of space is only measured in the two-dimensional wave front. Likewise, curvature of three-dimensional space can be written $^{(3)}R_{ABCD}$, but the superscript $(3)$ is commonly omitted; see \cite{MTW} and section \ref{sec:curv2} in the appendix. From expression \eqref{R2ABCD} further follows that $^{(2)}R_{ABCD}$ may promptly be rephrased in terms of the wave front's Gaussian curvature $K_G$.

The curvature of the wave front with respect to the surrounding space is typified by the rate of change of the normal vector to the wave front, as one moves on it:
\begin{equation}\label{def_Kext}
  \DD_A \vec{e}_\rho = K_A^{\hs B} \vec{e}_B.
\end{equation}
Using the definition of the Christoffel symbols \eqref{christ} immediately follows that
\begin{equation}\label{def_Kext2}
  K_A^{\hs B} = \Gamma^B_{A\rho}.
\end{equation}
We have adopted a different sign convention from \cite{MTW} to ensure that expanding spherical wave fronts exhibit positive extrinsic curvature, in accordance to the sign in \eqref{vc_1}. The tensor $K_{AB}$ is named the \textit{extrinsic curvature tensor} and proven to be symmetric \cite{MTW}. The promised scalar quantity from paragraph \ref{sec:fronts} that generalizes $K$ (`twice the mean curvature') to a non-Euclidean context, is identified here as the trace of the extrinsic curvature tensor $K_{AB}$. For that reason, we will henceforth also denote
\begin{equation}\label{def_TRK}
  \TRK = K^A_{\hs A} = h^{AB} K_{AB}
\end{equation}
simply as $K$. Our expansion parameter $\la$ given by \eqref{fronts_def_la} should consequently be interpreted with $K \equiv \TRK$.

For the special case of Gaussian normal coordinates, it follows from Eqs. \eqref{metric:GNC} and \eqref{def_Kext2} that
\begin{equation}\label{Kext2h}
  K_{AB} = \frac{1}{2} \dd_\rho h_{AB}.
\end{equation}
Since $\rho$ denotes geodesic arc length perpendicular to the front, Eq. \eqref{def_Kext2} also implies
\begin{equation}\label{Kdeth}
K =  K_A^{\hs A} = \Gamma^A_{A\rho} =  \Gamma^\mu_{\mu \rho} = \frac{1}{\sqrt{h}} \dd_\rho \sqrt{h}.
\end{equation}
The Riemann and Ricci curvature tensors for the full three-dimensional space can be decomposed in the intrinsic wave front curvature $^{(2)} R_{ABCD}$ and extrinsic curvature terms; applicable expressions are quoted from \cite{MTW} in Eqs. \eqref{Riemann_front}.

\subsection{Eigenfunctions to the linear perturbation operator}

When the RDE is restated in the Gaussian normal coordinate frame, an important role is reserved for the linear perturbation operator associated to \eqref{RDE_chapt_fronts}:
\begin{equation}\label{def_HL_fr}
    \HL = D_0 \mathbf{P} \dd^2_\rho + c_0 \dd_\rho + \bPhi'(\uu_0).
\end{equation}
Importantly, the generic reaction term $\bPhi'(\uu_0)$ forbids the operator $\HL$ to be brought to self-adjoint form.

At this point, it is worth noting that the decomposition of $\uu$ according to \eqref{fronts_ansatz} is a priori only fixed up to a term proportional to $\dd_\rho \uu_0$. For, if $\uu_{sol}$ is a solution to RDE \eqref{RDE_chapt_fronts}, translational invariance of this equation allows other solutions $\uu_{sol} + \epsilon^i \dd_i \uu_{sol}$ with $\epsilon^i$ a small but otherwise arbitrary vector. Plugging such shifted solution in the RDE uncovers that the spatial derivative of the one-dimensional solution to the RDE is an eigenfunction of $\HL$ with eigenvalue zero, i.e. a zero mode:
\begin{equation}\label{HL_zero_fr}
    \HL \dd_\rho \uu_0  = \boldsymbol{0},
\end{equation}
Adopting the bracket notation that is customary in quantum mechanics, Eq. \eqref{HL_zero_fr} becomes
\begin{equation}\label{HL2}
    \HL \ket{\dd_\rho \uu_0}  = \ket{ \boldsymbol{0} },
\end{equation}
where we have written $\ket{ \boldsymbol{0} }$ to denote the function that is identically equal to zero for all positions and state variables. We will further denote $ \ket{\dd_\rho \uu_0} $ as $\ket{\bpsi_\rho}$, i.e. the translational zero mode. To remove the indetermination in the definition of $\uu_0$, we make use of the standard inner product on the Hilbert space of square-integrable functions on the real axis:
\begin{equation}\label{def_inner1D}
 \langle \mathbf{f} \mid \mathbf{g} \rangle = \langle \mathbf{f} , \mathbf{g} \rangle = \int \limits_{-\infty}^\infty  \mathbf{f}^H \mathbf{g} \ \mathrm{d}\rho
\end{equation}
where a Hermitian conjugate is included in the definition. Afterwards one solves the adjoint linearized eigenvalue problem
\begin{equation}\label{adjsol}
 \HL^\dagger \bY^\rho = \boldsymbol{0}  \Leftrightarrow \bra{\bY^\rho} \HL  = \bra{ \boldsymbol{0} }
\end{equation}
to obtain an eigenfunction $\bra{\bY^\rho}$ which we normalize such that
\begin{equation}\label{def_Yleft}
  \langle \bY^\rho \mid \bpsi_\rho \rangle = 1.
\end{equation}
Since the traveling wave solution $\uu_0$ is localized in space around the wave front, we assume the integral in the left-hand side of \eqref{def_Yleft} to converge without further knowledge about the extent of the left-eigenfunction $Y^\rho(\rho)$. Now we are able to remove the translational indetermination in \eqref{fronts_ansatz} by imposing the gauge condition
\begin{equation} \label{gauge_fronts}
\langle \bY^\rho \mid \tuu \rangle = 0.
\end{equation}

\subsection{Diffusion term as a covariant Laplacian}

A crucial observation within the curved space formalism advocated in the previous chapter is that the diffusion term in Eq. \eqref{RDE_chapt_fronts} reduces to a covariant Laplacian when the electrical diffusion tensor has constant determinant. For, with $|G| = det(g_{ij})$ and $|D| = det(D^{ij})$, one obtains from Eq. \eqref{covlap} and \eqref{def_metric}:
\begin{eqnarray}\label{diffterm_covlap}
  D_0 \DD_i \DD^i  &=& \frac{D_0}{ \sqrt{|G|} } \dd_i \left( \sqrt{|G|} g^{ij} \dd_j \right)
                                  = \frac{ \sqrt{|D|} }{ \sqrt{D_0} } \dd_i \left( \frac{ D_0^{3/2} }{ \sqrt{|D|} } D^{ij} \dd_j \right) \nn \\
                                &=& \dd_i \left( D^{ij} \dd_j \right) - \frac{1}{2}  \frac{\dd_i |D|}{ |D|} D^{ij} \dd_j.
\end{eqnarray}
In any homogeneous medium, albeit anisotropic, $|D|$ is a constant. We later treat small spatial inhomogeneities in $|D|$ perturbatively in section \ref{sec:metricdriftfronts}; they are found to cause an additional drift of the wave front. Note that in the electrophysiological context of myocardial activation waves, the common class of media that are built from a prototype fiber (and sheet) with spatially varying orientation obey the constant determinant condition.

\subsection{Reformulation of the RDE in Gauss normal coordinates\label{sec:reex_fronts}}
We are now ready to derive the actual equation of motion for the wave front surface in the monodomain description. We start by elaborating each term in the reaction-diffusion equation \eqref{RDE_chapt_fronts}, using the ansatz \eqref{fronts_ansatz}, expansion \eqref{decposition} and the chain rule \eqref{ddt}. For the time-derivative term we obtain
\begin{equation}
  \dd_t \uu = \la \dd_\tau \tuu -\left( \vec e^\rho \cdot \dot{\vec{X}}- c\right) \left(\dd_\rho \uu_0 + \la \dd_\rho \tuu \right)
  -\la \rho \left(\vec{e}^A \cdot \dot{\vec{e}}_\rho(0) \right) \dd_A \tuu+ \OO(\la^2),
\end{equation}
where we have used the fact that $\vec{e}^\rho \cdot \dot{\vec{e}}_\rho =0$ given the biorthogonality condition \eqref{biorth_fronts}. In lowest order we also calculate that $\vec{e}^A(0) \cdot \dot{\vec{e}}_\rho(0) =- h^{AB} \vec{e}_\rho(0) \cdot \dd_B \dot{\vec{X}}$, which relates to local rotation of the wave front. Altogether, this term is $\OO(\la^3)$ and can be omitted in the calculation.

Next, the nonlinear reaction term is expanded as
\begin{eqnarray}
  \bPhi(\uu) &=& \bPhi(\uu_0) + \la \bPhi'(\uu_0) \tuu + \frac{\la^2}{2} \tuu \bPhi''(\uu_0) \tuu + \OO(\la^3).
\end{eqnarray}
Here, the derivatives of the reaction term are taken with respect to the state-variables $u_m, m \in \{1,2, \ldots, N_V\}$; more specifically one has
\begin{equation}
        \left(\tuu \bPhi''(\uu_0) \tuu \right)_k = \sum_{m,n=1}^{N_V} \left[\frac{\dd^2 F_k}{\dd u_m \dd u_n}\right](\uu_0)\  \tilde{u}_m \tilde{u}_n.
\end{equation}

Under the condition that the electrical diffusion tensor has constant determinant, we may reexpress the diffusion term based on \eqref{diffterm_covlap}. Adopting the notation $|g| = det(g_{\mu\nu})$ leads up to second order in $\la$ to
\begin{eqnarray} \label{diffterm_fr2}
 \dd_i\left(D^{ij}\dd_j \mathbf{P} \uu \right) &=& D_0 \DD_\mu \DD^\mu \mathbf{P} \uu
     = D_0 \frac{1}{\sqrt{|g|}} \dd_\mu\left( \sqrt{|g|} g^{\mu\nu} \dd_\nu \mathbf{P} \uu \right)  \\
     &=& D_0 \Gamma^A_{\rho A} \mathbf{P} \dd_\rho (\uu_0 +\la \tuu) + D_0 \mathbf{P} \dd_\rho^2 (\uu_0 +\la \tuu) \nn \\
     && + \la  D_0 \mathbf{P} \Gamma^C_{C A} h^{AB} \dd_B \tuu + D_0 \mathbf{P} \la \dd_A h^{AB} \dd_B \tuu + \OO(\la^3). \nn
 \end{eqnarray}
Retaining terms up to linear order in $\tuu$, the original RDE \eqref{RDE_chapt_fronts} rephrased in the Gaussian normal coordinates system reads, (with ansatz \eqref{fronts_ansatz}):
\begin{eqnarray}\label{all3}
   \la \left(\dd_\tau - \HL \right) \tuu &=& \left( \vec e^\rho\cdot \dot{\vec{X}} \right) \left(\dd_\rho \uu_0 +  \la \dd_\rho \tuu \right)
   +  \Gamma^A_{A\rho} D_0 \mathbf{P} \left(\dd_\rho \uu_0+ \la \dd_\rho \tuu \right)  \nn \\
   && + \frac{\la^2}{2} \tuu \bPhi''(\uu_0) \tuu + \OO(\la^3).
\end{eqnarray}
We have taken into account here that $ \uu_0 (\rho)$ is an exact plane wave solution to the RDE in the isotropic case.\\

From the velocity-curvature relation in the isotropic case, it is furthermore known that the velocity correction $\dot{\vec{X}}$ to the plane wave speed $c$ is proportional to the extrinsic curvature of the wave front. Hence we anticipate that
\begin{equation}\label{dotX_small}
  \dot{\vec{X}}\cdot \vec{e}^\rho = \OO(\la).
\end{equation}
Similarly our work flow in \cite{Verschelde:2007}, we project Eq. \eqref{all3} onto the translational mode $\bra{\bY^\rho}$. The left hand side of the equation vanishes since $\bra{\bY^\rho}$ is a time-independent zero mode of $\HL$. The resulting equation of motion now reads
\begin{multline}\label{eomfr1}
    \vec e^\rho\cdot \dot{\vec{X}} \left(1 + \la \langle \bY^\rho \mid \dd_\rho \tuu \rangle \right) \\ =  - D_0 \bra{\bY^\rho}  \Gamma^A_{A\rho}  \HP \ket{\dd_\rho \uu_0 + \la \dd_\rho \tuu} - \frac{\la^2}{2} \tuu \bPhi''(\uu_0) \tuu + \OO(\la^3).
\end{multline}

\subsection{Lowest order equation of motion \label{sec:eomfr}}
In lowest order in $\la$, one finds from Eq. \eqref{eomfr1}
\begin{equation}  \label{introTRK}
     \Gamma^A_{A \rho} = K_{AB} h^{AB} = \TRK = K,
\end{equation}
with all quantities evaluated on the wave front. Up to first order in curvature effects, Eq. \eqref{eomfr1} then implies
\begin{equation}\label{eomfr_lowest0}
    \vec e^\rho\cdot \dot{\vec{X}} = - D_0 K  \bra{\bY^\rho} \HP \ket{\bpsi_\rho}.
\end{equation}
The integral on the right-hand side evaluates to a dimensionless numerical coefficient, similar to the dynamical coefficients obtained in the study of scroll wave filaments. We shall write this coefficient, which depends on the medium properties, as
\begin{equation}\label{defgamma}
    \gamma = \bra{\bY^\rho} \HP \ket{\bpsi_\rho} = \frac{\int_{-\infty}^\infty (\bY^\rho)^H \mathbf{P} \bpsi_\rho \mathrm{d} \rho }{\int_{-\infty}^\infty (\bY^\rho)^H \bpsi_\rho \mathrm{d} \rho}.
\end{equation}
Our result \eqref{defgamma} is strongly related to the expression obtained in \cite{Mikhailov:1994}, where critical broken wave fronts were considered. In such context, the overlap integrals needed be calculated over the full two-dimensional solution that represents a rigidly translating broken wave front.

The velocity component parallel to the wave front $\vec{e}_A \cdot\dot{\vec{X}}$ corresponds to reparameterization and therefore bears no physical meaning. For that reason we may freely impose the common assumption
\begin{equation}\label{dotXperp}
  \dot{\vec{X}} \cdot \vec{e}_A = 0,
\end{equation}
which we consider equivalent to a transverse (or Coulomb) gauge. Now, add this gauge condition to \eqref{eomfr_lowest0} to obtain
\begin{equation}\label{eikonal4}
    \dot{\vec{X}} = - \gamma D_0 K \, \vec{e}_\rho.
\end{equation}
In an inertial frame of reference, the time derivative should be taken with respect to $t$. Applying the chain rule \eqref{ddt} and gauging away the tangential velocity term yields
\begin{equation}\label{eikonal5}
    \dd_t \vec{X} = c_0\vec{e}_\rho - \gamma D_0 K \vec{e}_\rho.
\end{equation}
Hence, we have derived following velocity-curvature relation for wave fronts which covers also the anisotropic case:
\begin{equation}\label{eomfr_lowest1}
    c = c_0  - \gamma D_0 K = c_0 - \gamma D^{ij} K_{ij}.
\end{equation}
Should the medium happen to be isotropic, one must put $D^{ij} = D_0 \delta^{ij}$. The EOM then confirms the known result
\begin{equation}\label{eomfr_lowest_iso}
    c =  c_0 - \gamma D_0 K,
\end{equation}
in which case $K$ automatically designates the sum of the principal curvatures of the wave front. Interestingly, we have obtained a general coefficient of linearity $\gamma$ as in \cite{Mikhailov:1994}, in contrast to approaches based on matched asymptotic expansions, which tend to end up with $\gamma=1$. There is a remaining ambiguity in Eqs. \eqref{eomfr_lowest1}-\eqref{eomfr_lowest_iso} however: the factors $\gamma,$ $D_0$ are only determined up to a constant factor. Thereby, one can only determine the product  $\gamma D_0$ from experimental or numerical observations on the velocity-curvature effect.

\subsection{Elaborations on the correction term $\tuu$ \label{sec:tuu}}

Returning to Eq. \eqref{all3}, following linear evolution equation for the correction $\tuu$ is found:
\begin{eqnarray}\label{eq_tuu}
  \la \dd_\tau \tuu &=& \la \HL \tuu +  (\vec e^\rho\cdot \dot{\vec{X}} + D_0 K \mathbf{P}) ( \dd_\rho \uu_0 + \la \dd_\rho \tuu) + \frac{\la^2}{2} \tuu \bPhi''(\uu_0) \tuu + \OO(\la^3) \nn \\
  &=& \la \HL \tuu + D_0 K (  \mathbf{P} - \gamma \mathbf{I}) \dd_\rho \uu_0 + \OO(\la^2).
\end{eqnarray}
Let us first recall to the previously silent assumption that the plane wave solution is stable with respect to small perturbations. This condition implies that $\HL$ has only eigenvalues $\zeta_{[j]}$ with negative real part. Note also that the unique right eigenfunction with eigenvalue $0$ cannot be part of $\tuu$, for it was gauged away in \eqref{gauge_fronts} for all times $\tau$. Therefore, the contribution $\HL \tuu$ in Eq. \eqref{eq_tuu} comprises a relaxation term with a dominant time constant that equals $ - 1 / \mathrm{Re}(\zeta_{[j]})$, with $\zeta_{[j]}$ the eigenvalue to $\HL$ (given $\uu_0$) with largest real part.

The explicit form of the source term in Eq. \eqref{eq_tuu} unveils that the lowest order perturbative correction $\tuu$ vanishes either when $K=0$ or when $ \HP= \mathbf{I}$. The first condition is fulfilled for plane wave fronts; in this case the solution $\uu_0$ is indeed known to be exact. Even for curved wave fronts, the lowest order correction $\tuu$ is killed if $\mathbf{P}$ is proportional to the identity matrix:
\begin{equation}\label{Pone}
  \mathbf{P} = \mathbf{I} \Rightarrow \gamma = 1 \Rightarrow  \mathbf{P} = \gamma \mathbf{I}.
\end{equation}
In other words, Eq. \eqref{eq_tuu} leads to the conclusion that the lowest order perturbative correction $\tuu$ arises as a correction for the diffusion mismatch between different state variables.\\

Secondly, Eq. \eqref{eq_tuu} also guarantees that initially small corrections $\tuu$ do not grow faster than linear, with a growth rate bounded by the extrinsic curvature of the wave front:
\begin{equation}\label{growth_front}
  \frac{||\tuu||}{ ||\uu_0||} \leq \frac{D_0 K}{d_f} \tau = \la \frac{D_0}{d_f d} \tau.
\end{equation}
Here, vector norms are taken with respect to the scalar product \eqref{def_inner1D}. Also, $d_f$ denotes the length scale at which the wave's upstroke takes place and $d$ is the characteristic thickness of the wave as employed for defining $\la$ in Eq. \eqref{fronts_def_la}. Hence, we have ascertained that our present analysis holds in a finite time interval around the instance of consideration.\\

Third, the evolution equation Eq. \eqref{eq_tuu} enables to verify assumptions made during the derivation of the EOM. First of all, the source term in \eqref{eq_tuu} is proportional to $K$, from which one finds that $\tuu$ is of order $\la$. This observation justifies our main ansatz \eqref{fronts_ansatz}. Secondly, the correction $\tuu$ only depends on the $\s^1, \s^2$ coordinates through the variation of the wave front curvature, legitimizing Eq. \eqref{dstuu_fronts}.\\

As a fourth elaboration on Eq. \eqref{eq_tuu}, we have sought its exact solution in terms of left and right eigenfunctions to the perturbation operator $\HL$. To that purpose we state
\begin{align} \label{eigmodes_L_fr}
 \HL  \ket{\bpsi_{[i]}} &= \zeta_{[i]}  \ket{\bpsi_{[i]}}, &   \bra{\bY^{[i]}} \HL &= \zeta^*_{[i]}  \bra{\bY^{[i]}},
\end{align}
with additionally $\ket{\bpsi_{[0]}} = \ket{\bpsi_\rho}$, $\bra{\bY^{[0]}} = \bra{\bY^\rho}$ and $\zeta_{[0]} = 0$. Owing to the gauge condition \eqref{gauge_fronts}, $\tuu$ has no zero-mode component. Therefore it is appropriate to construct an invertible operator $\HL_0$ by excluding the zero mode components from the spectral decomposition of $\HL$:
\begin{equation}\label{def_L0}
  \HL_0 = \HL - \ket{\bpsi_\rho} \bra{ \bY^\rho} = \sum_{[i] \neq [0]} \zeta_{[i]} \ket{\bpsi_{[i]}} \bra{\bY^{[i]}}.
\end{equation}
Based on an eigenfunction expansion $\tuu = \sum_{[i] \neq [0]} C_{[i]} \ket{\bpsi_{[i]}}$, the full solution to \eqref{eq_tuu} is determined by
\begin{equation}\label{coeff_vgl}
  \dot{C}_{[i]} = \zeta_{[i]} C_{[i]} + D_0 K(\tau) \bra{\bY^{[i]}} \HP \ket{\bpsi_{[i]}}.
\end{equation}
The solution for all times $\tau$ to this ordinary differential equation is found from convolution with the inhomogeneous term:
\begin{equation}\label{coeff_vgl2}
  C_{[i]} = C_{[i]}(0) + \bra{\bY^{[i]}} \HP \ket{\bpsi_\rho} \int_0^\tau D_0 K(\tau') e^{ \zeta_{[i]} (\tau-\tau')} \mathrm{d} \tau'.
\end{equation}
This result accounts for `memory' effects in the wave front solution, as the current value of state-variables depend on the full history of the wave's evolution. However, the non-local description of time is inconvenient for the purpose of our further theoretical developments.\\

Fortunately, it is possible to retrieve a simpler form for the lowest order solution to \eqref{eq_tuu} which will prove useful in subsequent calculations. Thereto we remark that the time-derivative of $\tuu$ only acts through the rate-of-change of wave front curvature. For, a steady-state solution to \eqref{eq_tuu}, which we will annotate $\uu_1$, is obtained as
\begin{eqnarray}\label{eq_uu1}
   \uu_1 &=& \HL^{-1}_0 \ D_0 K(0) ( \gamma \mathbf{I} - \mathbf{P} ) \dd_\rho \uu_0 \\
             &=& - D_0 K(0) \sum_{i\neq 0} \frac{1}{\zeta_{[i]}} \mathbf{P} \dd_\rho \uu_0
            = - D_0 K(0)\  \HL^{-1}_0  \mathbf{P} \dd_\rho \uu_0 \nn,
\end{eqnarray}
where $K(0)$ denotes the extrinsic curvature evaluated on the wave front surface $\rho = 0$. From this solution, one finds that the time derivative only acts on the perturbative correction through $\dd_\tau K$, which can be shown to be an effect of order $\la^3$. For, from $\dot{\vec{e}}_A = \dd_A \dot{\vec{X}}$ follows that $\dd_\tau h_{AB} = \OO(\la^2)$. Given that $K = \frac{1}{2} h^{AB} \dd_\rho h_{AB}$, the $\dd_\tau K$ can only be of $\OO(\la^3)$.
Consequently, $\uu_1$ as given by \eqref{eq_uu1} may rightfully be considered the solution to \eqref{eq_tuu} in lowest order in curvature effects. In other words, the lowest order modification $\tuu$ to the unperturbed wave front profile $\uu_0$ can be approximated as $\la \tuu = \la \uu_1 + \OO(\la^2)$ in lowest order in curvature effects.\\

With this knowledge we can take one step deeper in our perturbative scheme and propose
\begin{equation}\label{def_tuu1}
  \tuu(\rho, \s^A, \tau) = \uu_1(\rho, K(\s^A, \tau)) + \la \tuu^{(1)} (\rho, \s^A, \tau).
\end{equation}
In other words, our best estimate to the unknown exact solution to the RDE \eqref{RDE_chapt_fronts} has improved from $\uu_0$ to
\begin{equation}\label{def_uu01}
  \uu_0^{(1)}(\rho, K) = \uu_0(\rho) + \uu_1(\rho, K),
\end{equation}
giving way to the modified ansatz
\begin{equation}\label{fronts_ansatz2}
  \uu(\rho, \s^A, \tau) = \uu^{(1)}_0(\rho, K(\s^A, \tau)) + \la^2 \tuu^{(1)}(\rho, \s^A, \tau).
\end{equation}
In calculating higher order corrections to the velocity-curvature equation, one is confronted with two options. The first is to use the original ansatz \eqref{fronts_ansatz} and quantify several of the second order corrections based on overlap integrals of $\bra{\mathbf{Y}^\rho}$ and $\ket{\uu_1}$. Alternatively, one could opt to redo the calculations with the more precise ansatz \eqref{fronts_ansatz2}, which bears implicit curvature-dependence. Both approaches are consistent in the next order of calculation.

\subsection[The limit of high excitability]{Magnitude of terms in the EOM in the limit of high excitability}

To further study the `field correction' $\tuu$, it is instructive to note that some dynamical coefficients considerably simplify in the case of equal diffusion systems. When $\HP \rightarrow \hat{I}$, one has from \eqref{defgamma} that $\gamma \rightarrow 1$, which is a well known fact in literature.

In another limit, i.e. the limit of high excitability, one also has that $\gamma \rightarrow 1$ \cite{Keener:1991b}. Using the parameter $\veps$ to denote the time-scale ratio between fast and slow variables, the limit of high excitability confers to $\veps \rightarrow 1$. We now hypothesize that in a myocardial context
\begin{equation}\label{hyp_eps}
 \langle\ \cdot\ \mathbf{P}\ \cdot\ \rangle = \langle \ \cdot\  \mathbf{I}\ \cdot \  \rangle + \veps \langle\ \cdot \ (\mathbf{P} -\mathbf{I})\ \cdot \ \rangle.
\end{equation}
Otherwise stated, the fast kinetics are expected to dominate the contribution in the matrix elements, such that expectation values for $\mathbf{P} -\mathbf{I}$ are small. The mentioned case of equal diffusion is fulfilled by the limit $\veps = 0$, although the fundamental cause is of distinct nature.

Formally, we will use $\veps$ as a bookkeeping parameter to annotate that the dynamical coefficients in our equations of motion are expected to take values close to zero or one in the limit of high upstroke velocity. Note that the prevalence of $\veps$ mainly resides in the lower order terms, as the higher order matrix elements are unlikely to disappear owing to orthogonality of the zero modes since coordinate functions $\rho^A$ are included in the spatial integration.

Applied to our foregoing analysis of wave fronts, we have that
\begin{align}
 \gamma &= 1+ \OO(\veps),  & \uu_1 = \OO(\veps).
\end{align}
As a result, the wave modification correction $\tuu$ is of order $\la \veps$ in the limit of high excitability, and vanishes in lowest order for equal diffusion systems.

\subsection{Higher order corrections to the EOM for fronts}

In this paragraph we advance the gradient expansion to include quadratic curvature effects. Thereto we retake the EOM up to second order in $\la$, i.e. Eq. \eqref{eomfr1}. Before bringing the quantity $ \Gamma^A_{A\rho} = h^{AB} K_{AB}$ outside the integration brackets, we perform Taylor expansion in the direction perpendicular to the wave front:
\begin{equation}\label{exp_TRK}
  K = K(0) + \rho \left. \dd_\rho K\right(0) + \OO(\rho^2).
\end{equation}
Herein, $\dd_\rho K$ equals $\DD_\rho K$ since $K$ is a scalar curvature invariant. The matrix element accompanying this curvature correction is given by
\begin{equation}\label{def_eta}
  \eta = \bra{\bY^\rho} \HP \rho \ket{\bpsi_\rho}.
\end{equation}
To better separate extrinsic curvature of the wave front and intrinsic curvature of the surrounding space, we infer from Eq. \eqref{Riemann_front3} that
\begin{eqnarray} \label{drKAB}
    \dd_\rho K^A_{\hs B} = - R^{\rho A}_{\hs\hs\rho B} - K^{AC} K_{CB},
\end{eqnarray}
whence
\begin{eqnarray} \label{drTRK}
    \dd_\rho K =  - R^\rho_{\hs \rho} - \TRKK.
\end{eqnarray}

This expression can be related to the scalar curvature of three-dimensional space using Eq. \eqref{Riemann_front3}. For the sake of clarity, we indicate relevant dimensions here:
\begin{eqnarray}\label{Ricci_2+1}
  {}^{(3)} \RR &=& {}^{(3)}R^{A\rho}_{\ \ A \rho} + {}^{(3)}R^{\rho A}_{\ \ \rho A} + {}^{(3)}R^{AB}_{\hs \hs AB} \nn \\
                    &=& 2\  {}^{(3)}R^{\rho}_{\hs \rho} + {}^{(2)}R^{AB}_{\hs \hs AB} + (K_B^{\hs A} K^B_{\hs A} - K^B_{\hs B} K^A_{\hs A}) \nn \\
                    &=& 2\  {}^{(3)}R^{\rho}_{\hs \rho} + 2 K_G + \TRKK - K^2.
\end{eqnarray}
After going to a local reference system where $K_{AB}$ is diagonal, one observes that $K^2 = (K_1+K_2)^2 = \TRKK + 2 K_G$, so that Eq. reduces to $\RR = 2\  {}^{(3)}R^{\rho}_{\hs \rho}$. This makes sense, as in a isotropic medium both quantities consistently disappear.
The quantity \eqref{drTRK} thus becomes
\begin{eqnarray} \label{drTRK2}
    \dd_\rho K = -\frac{\RR}{2} - \TRKK.
\end{eqnarray}

Grace to the elaborations on $\tuu$ in the previous paragraph we may also estimate the correction terms in Eq. \eqref{eomfr1}. Herein, we annotate the first order corrections which arise from taking inner products with $\uu_1$ instead of $\uu_0$ with a superscript `$(1)$':
\bsub \label{def_dyncoeff1} \begin{eqnarray}
    \langle \bY^\rho \mid \dd_\rho \tuu \rangle &=& \veps I^{(1)} D_0 K + \OO(\la^2), \\
    \bra{\bY^\rho} \HP \ket{ \dd_\rho \tuu}   &=& \veps \gamma^{(1)} D_0 K + \OO(\la^2),\\
    \langle \bY^\rho \mid \tuu \bPhi''(\uu_0) \tuu \rangle   &=& \veps^2 \varphi^{(1)} D_0^2 K^2 + \OO(\la^2).
\end{eqnarray}\esub
The dynamical coefficients read explicitly
\bsub \begin{eqnarray}
  I^{(1)}  &=& - \bra{\bY^\rho} \dd_\rho \HL_0^{-1} \HP \ket{\bpsi_\rho}, \\
  \gamma^{(1)}  &=& - \bra{\bY^\rho} \HP \dd_\rho \HL_0^{-1} \HP \ket{\bpsi_\rho}, \\
  \varphi^{(1)} &=& \langle \bY^\rho  \mid \bF''(\uu_0)_{mn}   (\HL_0^{-1} \HP \bpsi_\rho)^m  (\HL_0^{-1} \HP \bpsi_\rho)^n \rangle.
\end{eqnarray} \esub
The partial results mentioned in this paragraph turn Eq. \eqref{eomfr1} into a velocity-curvature relation which is valid up to second order in curvature corrections:
\begin{eqnarray}\label{ho_front1}
&&\hspace{-0.5cm} \vec{e}^\rho \cdot \dot{\vec{X}}(1 + \veps I^{(1)} D_0 K ) = \\ &&- \left[\gamma + \left(\veps \gamma^{(1)} + \veps^2 \frac{\varphi^{(1)}}{2}\right) D_0 K\right] D_0 K - \eta D_0 \left( \TRKK + \frac {\RR}{2} \right) + \OO(\la^3). \nn
\end{eqnarray}
This version of the extended velocity-curvature relation may be cast into
\begin{equation}\label{ho_front1bis}
 \vec{e}^\rho \cdot \dot{\vec{X}} = - \left( \gamma + \veps \bar{\gamma}^{(1)} D_0 K\right) D_0 K - \eta D_0 \left( \TRKK + \frac {\RR}{2} \right) + \OO(\la^3)
\end{equation}
with
\begin{equation}\label{def_bargam(1)}
 \veps \bar{\gamma}^{(1)} = \veps \gamma^{(1)} + \veps^2 \frac{\varphi^{(1)}}{2} + \veps \gamma I^{(1)}.
\end{equation}\\

As announced in paragraph \ref{sec:tuu}, it is also possible to use the more accurate ansatz \eqref{fronts_ansatz2}. Thereto, one solves the one-dimensional reaction-diffusion equation with scalar curvature term:
\begin{equation}\label{RDE_DK}
  D_0 \mathbf{P} (\dd^2_\rho + K \dd_\rho) \uu + c \dd_\rho \uu + \bPhi(\uu) = 0,
\end{equation}
which has the solution $\uu = \uu_0^{(1)}(\rho, K)$ and $c = c^{(1)}(K)$. As a consequence, one will also need to refine $\HL^{(1)} = D_0 \dd^2 \rho + K \dd_\rho + c^{(1)} \dd_\rho + \bPhi'(\uu^{(1)}_0)$.
 With this alternative approach, only the terms which have been generated in the diffusion term expansion prevail, albeit with curvature-dependent coefficients:
\begin{equation}\label{ho_front2}
 \vec{e}^\rho \cdot \dot{\vec{X}} = c^{(1)}(K) - \eta^{(1)}(K) D_0 \left( \TRKK + \frac {\RR}{2} \right) + \OO(\la^3).
\end{equation}
Clearly, Eqs. \eqref{ho_front1} and \eqref{ho_front2} yield equivalent behavior of wave fronts up to the order given. For practical calculations, the strategy of \eqref{ho_front1} seems more attractive, as the zero modes $\bY^\rho$, $\bpsi_\rho$ only need to be computed once and may serve to produce all dynamical coefficients that appear in the given EOM.

\section[Discussion of the modified velocity-curvature relations]{Discussion of the modified velocity-\\curvature relations}

\subsection{Surface tension of fronts}

Regardless of the anisotropy of the medium we have formally proven that the linear approximation to the curvature-velocity relation, Eq. \eqref{eomfr_lowest1}, exhibits a coefficient of linearity $\gamma$ that in general differs from one for systems where not all state-variables diffuse at equal rates ($\mathbf{P} \neq a \mathbf{I},\, a>0$). In \cite{Zykov:1997}, the numerical observation that sometimes $\gamma \neq 1$ was not confirmed theoretically, and the $\gamma$ coefficient was termed a `correction factor'. More recently, Mikhailov and co-workers \cite{Mikhailov:1994} performed essentially the same expansion as ours, though based on a half wave front in an isotropic medium and without constructing the Gauss normal coordinates that are needed to establish higher order corrections in curvature. Our explicit prescription \eqref{defgamma} for $\gamma$ is consistent with their findings. Also, systems with vanishing $c_0$ were investigated in \cite{Panfilov:1995b}. Here, a special choice of reaction kinetics allowed to explicitly compute $\gamma$, which we have checked to be consistent with \eqref{defgamma}.\\

In physicist's terms, the parameter $\gamma$ is responsible for the coupling of the lowest order curvature effects to the motion of the excitation pattern. This statement might remind one to the dynamics of scroll wave filaments: the dominant coupling constant for weakly bent filaments was denoted filament tension, as it governs the growth rate of the filament's total length; see \cite{Biktashev:1994} and our amendments in upcoming chapters (sections \ref{sec:filtension} and \ref{sec:filtension_aniso}). In the following, we intend to relate $\gamma$ to the rate of change of the wave front's total surface area, by which the linearity coefficient immediately gains the physical meaning of a \textit{surface tension}.

In a curvilinear context, the total surface area of the wave front is for arbitrary parameterization given by
\begin{equation}\label{totalsurface}
  S = \int \mathrm{d}^2 \s \sqrt{h}.
\end{equation}
Here $h$ stands for the determinant of the induced metric, which changes under reparameterization of the surface as the square of the Jacobian determinant due to Eq. \eqref{transf_g}. Thereby, including the $\sqrt{h}$ in the integrand \eqref{totalsurface} makes the calculated surface area $S$ invariant under a change of coordinates within the wave front; see also \cite{Zwiebach:2004}. Appealing to \eqref{Kdeth} and applying no-flux boundary conditions \eqref{boundary} where the front meets the medium boundaries, we may therefore assert that
\begin{eqnarray}
    \frac{d S}{d t} &=& \int \left(\dd_\rho \sqrt{h}\right) \left(\dd_t \vec{X} \cdot \vec{e}^\rho \right) \mathrm{d}^2 \s \nn\\
                           &=& c_0 \dd_n S - \gamma D_0 \int K^2 \sqrt{h} \mathrm{d}^2 \s + \OO(\la^3). \label{dSdt}
\end{eqnarray}
The first term is rate of change of the front's surface area due to its unequivocal forward motion, which stems from the traveling wave nature of the one-dimensional solution. The second term has fixed sign and indicates that the surface area tends to decrease for positive $\gamma$. This justifies our interpretation of $\gamma$ as a surface tension.

\begin{figure}[h!t] \centering
  \mbox{
\raisebox{2.5cm}{a)}  \includegraphics[height=3 cm]{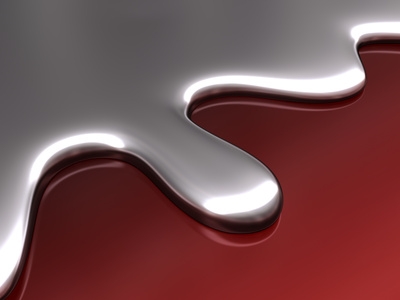}\hspace{0.5cm}
\raisebox{2.5cm}{b)}  \includegraphics[height=3 cm]{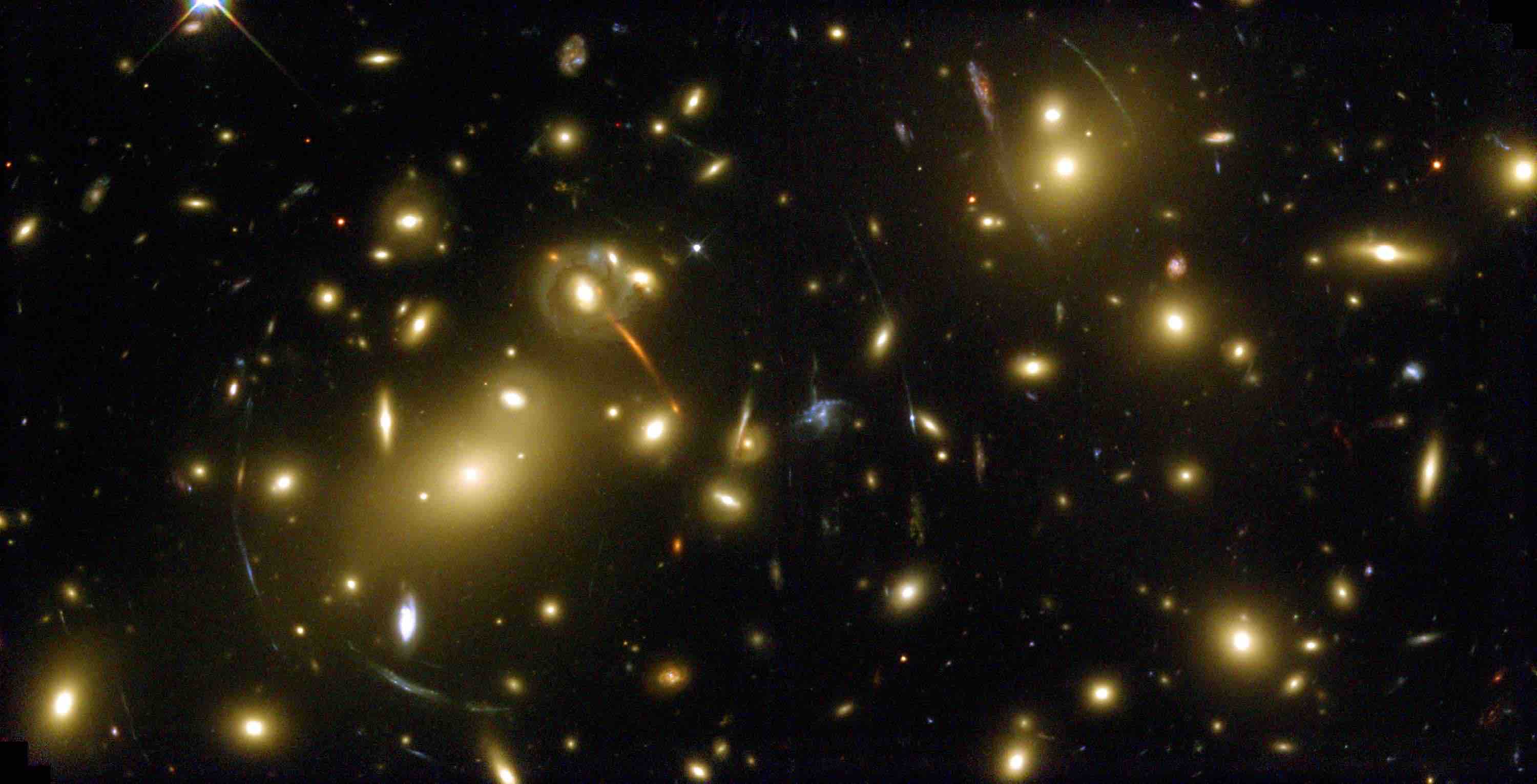}
}
  \caption[Physical systems similar to myocardial activation fronts]{Physical analogues of myocardial activation fronts. a) Surface tension with a `front' of liquid mercury. b) Gravitational lensing effect witnessed by the Hubble space telescope. }\label{fig:analogies}
\end{figure}

In this view, one can loosely compare a cardiac activation wave that reaches the end of a narrow cylindrical pathway with a glass being poured with water at a constant pace. To abolish the effect of gravity in this example, one may evenly pour the liquid onto a flat surface to get a two-dimensional analogue of active myocardium, as illustrated in Fig. \ref{fig:analogies}a.

\subsection{Measurement of the dynamical coefficients from the EOM}

The dimensionless surface tension coefficient $\gamma$ cannot possibly be determined from observing propagating wave fronts in a (real or virtual) active medium since the separation $D^{ij} = D_0 g^{ij}$ may be arbitrarily made. Hence, one always encounters the product $\gamma D_0$ in the equation of motion for wave fronts. For the same reason, estimated quantities from tracking wave front motion will involve this factor. The critical radius of curvature $R_c$, for example, is given by the curvature radius at which $c$ vanishes. For cylindrical $(m=1)$ or spherical waves $(m=2)$, one obtains in lowest order
\begin{equation}\label{Rc_modified}
  R_c = m \frac{\gamma D_0}{c_0},
\end{equation}
thus depending on the dynamical parameter $\gamma D_0$ rather than the space scale constant $D_0$. The same subtlety applies when estimating the space constant $D_0$ from an experimentally obtained epicardial activation pattern: the slope of the velocity-curvature relation is $\gamma D_0 m$ under such circumstance. Taking this slope as the value for $D_0$ in the RD model will not reproduce the same activation pattern if an electrophysiological model is used where $\gamma D_0$ significantly deviates from $D_0$.

An absolute dimensionless experimental evaluation of $\gamma$ from experimental electrophysiology might nevertheless be feasible by estimating $D_0$ based on transmembrane conductivity and specific electrical capacitance of the myocytes, through the definition of the electrical diffusion coefficient in Chapter \ref{chapt:struct}. This procedure, however, is likely to suffer from the mismatch between discrete and continuous descriptions of the myocardial syncytium.\\

From the RDE one observes that $D_0$ is the only constant that determines the spatial scale of the problem. Therefore, different dynamical coefficients that appear in the EOM can be related to each other as all of them contain the same unknown factor $D_0$. Physical scales can be determined from these quotients. In equally diffusive media, for example, the ratio $\eta D_0 / \gamma D_0$ reveals the length scale beyond which the linear approximation to the velocity-curvature relation is valid. Similar ratios appear in our description of rotor filaments, some of which denote instability thresholds for e.g. curvature and twist.

Since the ratios of dynamical coefficients do bear physical meaning, it may be worth trying to identifying them in live myocardium. These electrophysiological determinants might later serve to distinguish between -- and perhaps validate -- competing models of cardiac electrophysiology.  A more modest goal could be to assess only the sign of the dynamical coefficients, which yet suffices to enable or exclude possible pathways to instability, as will be discussed below.

\subsection{Extended velocity-curvature relation in isotropic media}

In isotropic media, the general expression \eqref{ho_front1} becomes
\begin{equation}\label{ho_front_iso}
 c = c_0 - \left( \gamma + \veps \bar{\gamma}^{(1)} D_0 K \right) D_0 K - \eta D_0 \TRKK + \OO(\la^3).
\end{equation}
Importantly, this relation does not match the sometimes quoted intuitive extrapolation
\begin{equation}\label{ho_front_wrong}
 c \stackrel{?}{=} c_0 - \gamma D_0 K + e K^2 + \OO(\la^3),
\end{equation}
due to the presence of the $\TRKK = (K^2 - 2 K_G)$ term in Eq. \eqref{ho_front_iso}! Even more, for systems with high excitability or equal diffusion, $\gamma^{(1)}$ is small or vanishing, such that Eq. \eqref{ho_front_iso} clearly illustrates the inconsistency of \eqref{ho_front_wrong}.

Why do the higher order terms not simply contain powers of $K$? Delving into our derivation and introducing a function
\begin{equation}\label{defdensity}
  E_P(\rho) = [\bY^\rho]^H (\rho) \mathbf{P} \bpsi_{\rho} (\rho),
\end{equation}
we learn from Eq. \eqref{eomfr1} that
\begin{equation}\label{eomfr_intK}
 \dot{\vec{X}} \cdot \vec{e}^\rho = - \int E_P(\rho) \TRK(\rho) \mathrm{d} \rho
\end{equation}
plus correction terms due to the perturbative correction $\tuu$. Apart from the $\tuu$ corrections, the entire expansion series is generated by Taylor expansion of the $\TRK(\rho)$, weighed with $E_P(\rho)$, which we therefore term a \textit{density} function. For that reason, all dynamical coefficients that are not related to modification of the wave profile are given by the raw moments of $E_P(\rho)$, e.g. $\int E_P \mathrm{d} \rho = \gamma$, $\int \rho E_P \mathrm{d} \rho = \eta$ and so on. The associated curvature tensors are $\frac{1}{n!}[\dd_\rho^n \TRK](0)$, which are in isotropic media calculated to equal
\begin{equation}\label{TRK_induction}
 \frac{1}{n!}  \dd^n_\rho \TRK = \frac{(-1)^n}{n} \mathrm{Tr}(\mathbf{K}^n)
\end{equation}
by induction on Eq. \eqref{drKAB}. This explains the occurrence of terms $\mathrm{Tr}(\mathbf{K}^n)$ in the extended equation of motion for wave fronts.

An immediate consequence of the expansion series in $\mathrm{Tr}(\mathbf{K}^n)$, not $\mathrm{Tr}(\mathbf{K})^n$, is that a spherical wave front experiences twice as much curvature effects than a cylindrical wave front, also in higher orders (although still in the limit of negligible $\tuu$ terms). With $m=1$ for a cylinder and $m=2$ for a sphere and $r$ denoting their respective radii, a single formula covers both cases:
\begin{equation}\label{eomfr_cyl_sph}
 c = c_0 - \gamma D_0 \frac{m}{r} - \eta D_0 \frac{m}{r^2} - \bar{\gamma}^{(1)} D_0^2 \frac{m^2}{r^2}.
\end{equation}

\subsection{Linear stability analysis of wave fronts \label{sec:stab_fronts}}

Based on our dynamical equations \eqref{ho_front1bis}, one may undertake linear stability analysis on curved propagating wave fronts.
To start with, let us denote the perturbed wave front position as
\begin{equation}\label{LSA_fronts1}
  \vec{X} = \vec{X}_0 + a x(\s^1, \s^2) \vec{e}_\rho(0),
\end{equation}
for small $a$. Therefrom, the perturbed triad vector are derived; the argument $(0)$ denotes evaluation at the wave front:
\bsub \label{LSA_fronts2}
\begin{eqnarray}
  \vec{e}_A &=& \vec{e}^0_A(0) + a \dd_A x \vec{e}_\rho(0) + \OO(a^2).\\
  \vec{e}^\rho &=& \frac{1}{2} \eps^{AB} \dd_A \vec{X} \times \dd_B \vec{X}  = \vec{e}_\rho^0 - a\dd_A x \vec{e}^A + \OO(a^2).
\end{eqnarray} \esub
The extrinsic curvature tensor becomes
\begin{eqnarray}\label{LSA_fronts3}
  K_{AB} &=& \left(\DD_A \vec{e}_\rho\right) \cdot \vec{e}_B = K_{AB}(0) - a \dd^2_{AB} x + \OO(a^2).
\end{eqnarray}
whereas the Ricci scalar curvature may be expanded as
\begin{equation}\label{Ricci_exp}
  \RR = \RR(0) + ax \dd_\rho \RR(0) + \OO(a^2).
\end{equation}
Fourier decomposition $\s^A \rightarrow p^A = p u^A$ $(A\in\{1,2\})$ for the tangential directions to the wave front brings $\dd_A \rightarrow i p_A$, such that the EOM \eqref{ho_front1bis} implies a linear evolution equation $\dot{x} = \Omega_c x$, of which the real part of the growth rate is
\begin{eqnarray}\label{LSA_fronts4}
\Omega = - p^2 \left( \gamma D_0 + 2 \bar{\gamma}^{(1)} D_0^2 K(0) + 2 \eta D_0 K_{AB}(0) u^A u^B \right) - \frac{\eta}{2} \DD_\rho \RR.
\end{eqnarray}
Remark that the Ricci scalar term shifts the wave front as a whole and therefore in leading order does not affect the dynamical stability of the wave front. Equation \eqref{LSA_fronts4} furthermore predicts that plane waves are linearly stable as soon as $\gamma >0$. For uneven waves, stability demands that the quantity in brackets is positive in all directions. With this, we could define an effective surface tension $\gamma$:
\begin{eqnarray}\label{def_gamma_eff}
 \gamma_{\rm{eff}} = \gamma + 2 \bar{\gamma}^{(1)} D_0 K_0 + 2 \min \limits_{\vec{u}} \left[ \eta K_{AB}(0) u^A u^B \right].
\end{eqnarray}
Depending on the sign of $\eta$, wave fronts are therefore expected to start destabilizing in the directions of either positive or negative extrinsic curvature.

\subsection{Wave fronts in anisotropic media and the lensing effect}

The effect of the curved space formalism is even felt in the zeroth order equation of motion, since we obtain $c =c_0$ instead of writing $c =c_0 \sqrt{\vec{n} \cdot \mathbf{D} \cdot \vec{n} /D_0}$ as commonly (and rightfully) found in other works, e.g. \cite{Winfree:1998, Keener:CBH8}. For, in our case the operational definition of distance automatically takes into account the appropriate rescaling. When applying the newly derived equations of motion, one should evenly keep in mind that distances should be scaled accordingly.

The dominant response of the wave front's extrinsic curvature was in an aniso\-tropic context brought to
\begin{equation}\label{DKaniso}
  DK \rightarrow D_0 \TRK = D_0 g^{ij} K_{ij} = D^{ij} K_{ij}.
\end{equation}
Our systematic treatment confirms earlier work from \cite{Morozov:1999, Davydov:2002}, who explicitly carried through the local rescaling to reach the same conclusion, albeit not in manifestly covariant shape. In fact, attaining \eqref{DKaniso} is unavoidable once the operational definition of distance has been adopted, since the unique curvature scalar with the correct dimension is precisely $\TRK$. By virtue of the covariant formalism, the next-to-leading order term $\TRKK$ immediately makes sense in anisotropic media as well.\\

In addition, non-trivial anisotropy of the medium affects the motion of wave fronts in another way, which we will call the `lensing effect'. The lensing effect is brought about by the intrinsic curvature of electrophysiological space and mediated through the $\RR$ term in the EOM \eqref{ho_front1bis}. The physical origin of the lensing effect traces back to the local equivalence principle: whenever geodesics tend to focus ahead of the wave front, there is no meaningful way to discern their convergence from the local configuration of an inward spherical wave, and the front will move faster. 

In cosmology, a similar phenomenon is encountered in the `gravitational lensing effect', where starlight (that travels along geodesics of space) is deflected near massive bodies, thus creating an optical lens. Because we have found in Chapter \ref{chapt:activ} that electrically active myocardium behaves generally as a negatively curved space due to myofiber rotation, the gravitational lensing effect in myocardium behaves opposite to its cosmological counterpart. 

\subsection{Metric induced drift \label{sec:metricdriftfronts}}

We now investigate the effect of variations of the determinant of the electric diffusion tensor; we had assumed this determinant to be constant in Eq. \eqref{diffterm_fr2} in order to simplify the diffusion term using Eq. \eqref{diffterm_covlap}. Writing $c_{|D|}$ for the wave front velocity under the determinant condition, i.e. the right-hand side of Eq. \eqref{ho_front1bis}, the correction for varying determinant is deduced from \eqref{diffterm_covlap}:
\begin{equation}\label{mdf_1}
  \vec{e}^\rho \cdot \dot{\vec{X}} = c_{|D|} - \frac{D_0}{2} \bra{\bY^\rho} \dd_\mu (\ln |D|) g^{\mu \nu} \ket{\dd_\nu (\uu_0 + \la \tuu)}.
\end{equation}
To treat small spatial variations in perturbation theory, we work in the regime of where spatial deviations from homogeneity take place on the same scale as curvature effects:
\begin{equation}\label{ansatz_metric drift}
 d\ \dd_\mu( \ln |D|) = \OO(\la).
\end{equation}
Herein, $d$ is the same measure of typical wave fronts thickness as in our definition \eqref{fronts_def_la} of the parameter $\la$. With, this \eqref{mdf_1} can be expanded up to second order in $\la$ to yield
\begin{multline}\label{mdf_2}
  \vec{e}^\rho \cdot \dot{\vec{X}} = c_{|D|} - \dd_\rho (\ln |D|) \frac{D_0}{2}  \left( \gamma + \veps \bar{\gamma}^{(1)} D_0 K \right) \\ -  \dd^2_\rho \ln |D| \frac{D_0}{2} \eta + \OO(\la^3).
\end{multline}
without any need to introduce additional dynamical coefficients. Altogether, our most advanced EOM for wave fronts now reads:
\begin{multline}\label{mdf_3}
  c = c_0 - \left( \gamma + \veps \bar{\gamma}^{(1)} D_0 K\right) D_0 K - \eta D_0 \left( \TRKK + \frac {\RR}{2} \right)\\
 - \dd_\rho \ln |D| \frac{D_0}{2}  \left( \gamma + \veps \bar{\gamma}^{(1)} D_0 K \right)  - \dd^2_\rho \ln |D| \frac{D_0}{2} \eta + \OO(\la^3).
\end{multline}

\subsection{Wave front motion on curved surfaces with anisotropy}

Our theory straightforwardly extends towards wave propagation on a two-dimen\-sional surface, which may be moreover curved into the third physical dimension. Real-life examples include oscillation chemical reactions in a thin film on a curved surface and excitation waves in thin tissue slabs, which may be anisotropic themselves. It is not unlikely that the latter case could provide an accurate description to the propagation of excitation in the relatively thin atrial wall.
The particular case of a non-planar surface with isotropic propagation has already been studied by Davydov et al.\cite{Davydov:2002} by analytical, experimental and numerical means in lowest order in curvature.\\

We first characterize the medium, say an uneven piece of atrial tissue. One may define the geometry of the slab by a suitable parametrization $x^i(p^a)$ ($i=1,2,3$ and $a=1,2$) with respect to laboratory Cartesian coordinates which may be written $\left(x(p^1, p^2), y(p^1, p^2), z(p^1, p^2) \right)$. In each point of the slab, one also needs a prescription (or measurement) of the electrical diffusion tensor in a local tangential Cartesian frame $(x', y')$, which is denoted $D^{mn}$. Hereto the metric tensor in local tangential coordinates is directly associated via $g_{mn} = D_0 (D^{-1})_{mn}$. Through the absolute coordinate system $x^i$, one can relate $x'^m$ to $p^A$, to yield
\begin{equation}\label{gAB_curved_surf}
  g_{ab} = \frac{\dd x'^m}{\dd p^a} g_{mn} \frac{\dd x'^n}{\dd p^b}.
\end{equation}
This metric tensor plays the role of the metric tensor with Cartesian indices $g_{ij} = D_0 (D^{-1})_{ij}$ in the beginning of our derivation presented in the previous section, which is also applicable to two spatial dimensions instead of three.

Next, one may as before propose wave front-adapted coordinates $(\rho, \s)$, given an initial configuration of the wave front, whose instantaneous position is now described by a single curve. In the co-moving Gaussian normal coordinates $(\rho, \s, \tau)$, the induced metric $h_{\mu \nu}$ evenly takes a diagonal shape\footnote{In this lower dimensional context, $\mu, \nu$ only take the values 1, 2.}. An important adaptation is that the geometric curvature invariants should be adapted to working in one dimension lower. The geodesic curvature $k$ of the wave front fully captures its extrinsic curvature and therefore replaces the tensor $\mathbf{K}$ in the EOM. Also, the Ricci curvature of a two-dimensional surface boils down to twice the local Gauss curvature $K_G$ of the surface, measured in the operational distance convention. With these modifications, Eq. \eqref{mdf_3} promptly produces the EOM for wave fronts on curved anisotropic excitable surfaces:
\begin{multline}\label{ho_front_surf}
 c = c_0 - \gamma D_0 k - \bar{\gamma}^{(1)} D_0^2 k^2 - \eta D_0 \left( k^2 + K_G \right) \\
        - \frac{D_0}{2} \left[ \left(\gamma + \bar{\gamma}^{(1)} D_0 k \right) \dd_\rho \ln |g_{ab}| + \eta \dd^2_\rho \ln |g_{ab}| \right].
\end{multline}
We remark that the result that Davydov \etal obtained for an isotropic curved medium that
 $c = c_0 - D_0 k,$
which is the simplest case included in Eq. \eqref{ho_front_surf}. Note that they indeed had to write $\gamma=1$ since an equal diffusion system was investigated. Up to the given order, however, the metric-induced drift term was omitted in \cite{Davydov:2002}. 
Additionally, we provide supplementary corrections to the motion of wave fronts in the given configuration that are second order in curvature.

\subsection{Localization property\label{sec:localfront}}

Given our theoretical developments, we may finally address the issue how to localize the sometimes quite extended traveling wave solution. From Eq. \eqref{eomfr_intK} we deduce that the wave front motion is only sensitive to curvature in the region where the function $E_P(\rho)$ given by Eq. \eqref{defdensity} is significantly different from zero. Due to this weighing property, we had yet associated the notion of density upon its definition. Not only curvature effects but all diffusion-related processes are in lowest order expected to couple through the function $E_P$; see e.g. the metric induced drift in Eq. \eqref{mdf_1}. Perturbations that take place irrespective of the diffusion processes, however, are coupled via
\begin{equation}\label{defdensity_I}
  E_I(\rho) = [\bY^\rho]^H (\rho) \bpsi_{\rho} (\rho),
\end{equation}
and the concept of density functions may obviously be generalized to even other operators. Note that only $E_I$ is normalized to unity; nevertheless all density functions can take both positive and negative values.

\begin{figure}[h!b] \centering
  \includegraphics[width=1.0 \textwidth]{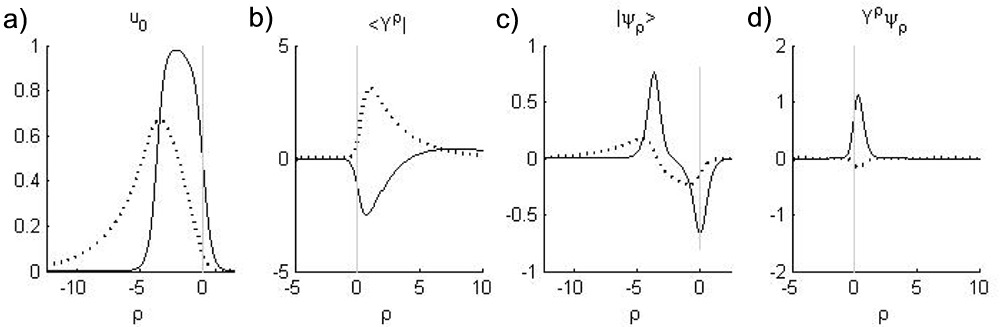}\\
  \caption[Localization of wave fronts]{Numerical inquiry on the localization of wave fronts in the Barkley model \cite{Barkley:1990} with parameters $a=0.75$, $b=0.02$, $\eps = 0.05$, $D_u=1.0$ and $D_v = 0.1$. Spatial profiles are depicted for the both state variables $u$ and $v$ in solid and dashed lines, respectively. From the traveling pulse solution (a), the left (b) and right (c) zero modes were calculated. The density function $E= \bY^\rho \bpsi_\rho$ is displayed in panel (d) for both state variables separately.}\label{fig:loc_fronts}
\end{figure}

We have numerically computed the Goldstone mode $\bpsi_{\rho}$ and its adjoint $\bY^\rho$ for few low dimensional ionic models of cardiac tissue, which have moreover continuously differentiable reaction kinetics; see section \ref{sec:numeric_fronts} for more details. With this, the functions $E_P$, $E_I$ were calculated; they are depicted in Fig. \eqref{fig:loc_fronts}. Since the density functions contain a factor $\dd_\rho \uu_0$, they are expected to predominantly contribute near the wave front and wave back. For isolated pulses, the numerically obtained adjoint modes $\bY^\rho$ appeared to only differ significantly from zero near the wave front, which is consistent with their interpretation as response functions (RFs) with respect to external perturbations, akin to RFs in the context of scroll wave filaments \cite{Biktashev:1995b}.

Given the localized influence of curvature on the behavior of wave fronts, we may finally concretize the `characteristic thickness' of an activation front, which was denoted $d$ in the definition \eqref{fronts_def_la} of our expansion parameter $\la  = K\,d$: we now define $d$ to represent the width of the density function $E_P$:
\begin{equation}\label{def_width_d}
  d^2 = \langle \rho^2 \rangle_c = \int \limits_{-\infty}^{+\infty} E_P(\rho) \rho^2 \mathrm{d} \rho - \left(\  \int \limits_{-\infty}^{+\infty} E_P(\rho) \rho \mathrm{d} \rho \right)^2.
  \end{equation}

One may want to compare this effective width $d$ with the wave's wavelength, which is commonly defined as the distance between wave front and wave back. As the effective width $d$ may be substantially smaller than the wavelength, we have found a (weaker) analogue of the wave-particle duality for excitation patterns, which was discussed in \cite{Biktasheva:2003} for spiral waves.

At this point we also emphasize that choice of an origin on the $\rho$-axis affects the equations of motion in order $\la^2$: a shift of origin over a distance $\rho_0$ leaves the lowest order dynamical coefficient $\gamma$ unaltered, but $\eta$ displaces to
\begin{equation}\label{change_eta}
  \eta(\rho_0) = \int (\rho-\rho_0) E(\rho) \mathrm{d} \rho = \eta - \rho_0 \gamma.
\end{equation}
Notwithstanding, the point of linearization for $K(\rho)$ also shifts to $\rho$, and both effects compensate to yield the same dynamics. The freedom to pick a reference can be exploited to make $\eta$ disappear, which happens when the origin is planted in the barycenter of the function $E_P(\rho)$. In practice, this strategy comes down to selecting an appropriate voltage threshold to fix wave front position.

\subsection{Variational principle \label{sec:var}}

We have noted that the dynamics of moving fronts governed by the EOM \eqref{eomfr_lowest1} may be derived from a variational principle, in spite of the dissipative nature of the system. To this aim, we write the velocity-curvature equation as a gradient system for the variables $X^i$, ($i=1,2,3$):
\begin{equation}\label{gradientsys}
    \dd_t X^i(\s_1,\s_2,t) = - \frac{\delta G }{\delta X^i}\left(\vec X, \partial_\alpha \vec{X} \right).
\end{equation}
The potential $G$ in this expression is given by:
\begin{equation}\label{potG}
    G = G_V + G_S =
    - \frac{c_0}{6} \int\mathrm{d}^2\s \epsilon^{\alpha\beta} \epsilon_{mnl} X^m \dd_\alpha X^n \dd_\beta X^l + \gamma D_0 \int  \mathrm{d}^2\s \sqrt{h}.
\end{equation}
The first term in Eq. \eqref{potG} resembles the Polyakov action term in string theory, in which the scalar product has been substituted for a cross product \cite{Zwiebach:2004}. Alternatively, this part is recognized as the differential volume occupied by the wave front during its motion. Variation of this volume term indeed delivers:
\begin{eqnarray}\label{varGV}
    - \frac{\delta G_V}{\delta X^i} &=& \frac{c}{6} \epsilon^{\alpha\beta} \epsilon_{inl} \dd_\alpha X^n    \dd_\beta X^l
    +  \frac{c}{3} \frac{\delta}{\delta X^i} \int  \mathrm{d}^2\s \epsilon^{\alpha\beta} \epsilon_{mnl} X^m \dd_\alpha X^n \dd_\beta \delta X^l \nonumber\\
    &=& \frac{c}{6}\epsilon^{\alpha\beta} \left( \dd_\alpha\vec{X} \times  \dd_\beta \vec{X} \right)^i + \frac{c}{3} \epsilon^{\alpha\beta} \epsilon_{mni} \dd_\alpha X^n \dd_\beta X^m \nonumber
    = c n^i.
\end{eqnarray}
During derivation, we have made use of Neumann boundary conditions \eqref{boundary} at the medium boundaries.

The second term in Eq. \eqref{potG} equals the Nambu-Goto term for string action. This term amounts to the surface area of the wave front; its variation brings
\begin{eqnarray}\label{varGS}
    \delta G_S &=& D_0 \int d^2 \s \frac{1}{2\sqrt{h}} \delta h = \frac{D_0}{2} \int d^2 \s \sqrt{h} h^{ij} \delta h_{ij} \nn \\
    &=&  - D_0 \int d^2 \s \delta \vec{X} \cdot \dd_i \left( \sqrt{h} h^{ij} \dd_j \vec{X} \right)
\end{eqnarray}
so that
\begin{equation}\label{varGS2}
   - \frac{\delta G_S}{\delta \vec{X}} = D_0 \DD_i \DD^i \vec{X}.
\end{equation}
The latter expression may be manipulated after going to a local Gaussian normal frame:
   $\DD_i \DD^i \vec{X} = h^{AB} \DD_A \vec{e}_B = \TRK \vec{n},$
which concludes the proof of Eq. \eqref{gradientsys}.



\subsection{Numerical evaluation of the surface tension coefficient \label{sec:numeric_fronts}}

We have looked for numerical confirmation of the surface tension coefficient $\gamma$ as given by \eqref{defgamma}. For a given model, the parameter $\gamma$ can be predicted once the one-dimensional left- and right eigenvectors to the linearized operator $\HL$ are known. To this purpose we have adapted the technique for spiral waves exposed in \cite{Biktasheva:2006} for use in the one-dimensional case, as we now outline.

First, an explicit Euler method was used to generate the profile of a traveling wave in one spatial dimension. After measurement of the wave's velocity $c_0$, temporal evolution was continued in a co-moving frame. Here, a finer spatial grid was used to improve on the wave front shape. Meanwhile, small residual drift due to imperfect determination in $c_0$ was detected and used to update $c_0$. This iterative procedure was borrowed from \cite{Biktasheva:2006}; it apparently performs better for wave fronts than for spiral waves, as only one kinematic parameter (i.e. $c_0$) needs to converge, opposed to the drift components $v_x$, $v_y$ and rotation frequency $\omega_0$ for spirals.

Once the steady state solution had converged, the translational zero mode $\ket{\bpsi_\rho}$ for the problem was obtained as the spatial derivative of $\uu_0$. For excitation models with differentiable reaction kinetics, the Jacobian matrix $F'(\uu_0)$ can be formed explicitly, from which the adjoint operator $\HL^\dagger$ may be constructed. The adjoint mode $\bra{\bY^\rho}$ was then obtained from forward evolution of $\dd_t \uu = \HL^\dagger \uu$. Since $\bra{\bY^\rho}$ is the eigenmode of this linear equation with the largest real part, an arbitrary initial condition converges to $\bra{\bY^\rho}$. During the iterative process, the solution was normalized such that $\langle \bY^\rho \mid \bpsi_\rho \rangle = 1$. The outcome of this procedure was shown in Fig. \ref{fig:loc_fronts} for the Barkley model \cite{Barkley:1990}.

Next, the surface tension coefficient could be calculated from the modes $\ket{\bpsi_\rho}$ and $\bra{\bY^\rho}$, with the spatial integration implemented using the trapezoid rule. \\

For reference, the surface tension coefficient $\gamma$ was also estimated by linear regression of the velocity-curvature relation, as obtained from forward numerical simulation. We chose to simulate a cylindrical wave, which reduces to a numerical simulation in one spatial dimension. An \textit{inwardly} traveling wave was selected because the simulation of expanding waves is limited to initial radii larger than the critical radius for wave propagation $R_c$.

With the methods described we have determined $\gamma$ using both the theoretical result, and the outcome from forward numerical simulation. This method was conducted for three different reaction kinetics in the RDE: the FitzHugh-Nagumo model , the Barkley model \cite{Barkley:1990} and the Aliev-Panfilov model \cite{Aliev:1996}. These are all models with two state variables $\uu = (u,v)$ and differentiable reaction functions $\bF = (f(u,v), g(u,v)$. The Barkley and Aliev-Panfilov models have the respective reaction functions
\begin{align}
    f(u,v) &= \frac{1}{\eps} u(1-u) \left(u- \frac{v+b}{a}\right),& g(u,v) &= u-v, \\
    f(u,v) &= - u(v+ k(u-a)(u-1)), & g(u,v) &= \eps(u,v) (v+ ku(u-a-1)), \nn
\end{align}
 where $\eps(u,v) = \eps_0 \mu_1 v (u + \mu_2)^{-1}$. We have yet cited the prescription for the FitzHugh-Nagumo system in Eq. \eqref{RDE:FHN}. For our numerical simulations the parameter set $\{a=0.75, b=0.02, \eps = 0.05\}$ was taken for the Barkley model and $\{ a=0.15, k=8,\eps_0=0.62, \mu_1 = 0.2, \mu_2 = 0.3 \}$ for the Aliev-Panfilov model. For the FitzHugh-Nagumo system, $\{ \beta = 0.75, \gamma = 0.5, \eps = 0.3 \}$ was used. The (isotropic) electrical diffusion coefficient $D_u$ was kept equal to $1.0$, whereas $D_v$ was varied from $0.01$ to the value at which no traveling wave solution existed.

\begin{figure}[h!b] \centering
  \includegraphics[width=0.75 \textwidth]{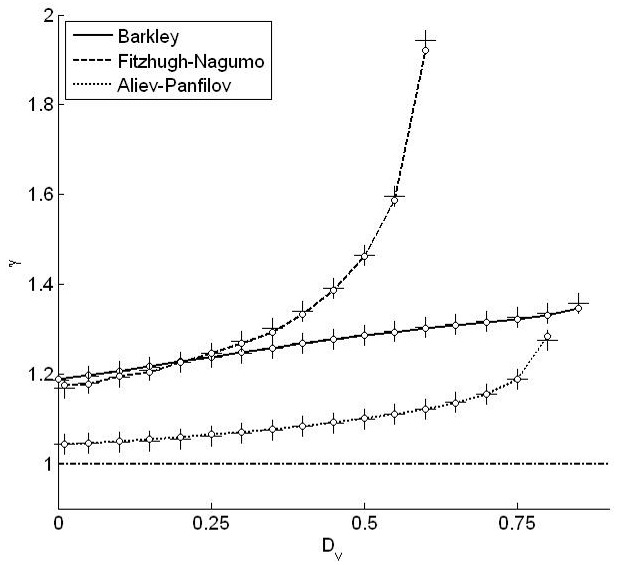}\\
  \caption[Surface tension coefficient: theory vs. experiment]{Surface tension coefficient: theory vs. experiment, for three types of reaction kinetics. Theoretical predictions ($+$) agree well with linear regression on the $c(K)$ relation obtained from forward numerical simulation.  }\label{fig:gamma3models}
\end{figure}

The final result of our numerical check is presented in Fig. \ref{fig:gamma3models}, with predicted theoretical values based on Eq. \eqref{defgamma} denoted by crosses, whereas circles indicate the outcome from fitting to an numerical $c(K)$ relationship. A close match is found between theory and numerical verification.

\section{Velocity-curvature relation for periodic waves}

\subsection{Curvature effects in the high-frequency limit}

An interesting scenario takes place when activation waves repeatedly excite a piece of tissue. If the temporal spacing is narrow enough, the incomplete recovery of the medium is likely to affect the traveling wave properties, such as wave form and propagation speed. In particular, all dynamical coefficients that depend on the underlying medium properties are subject to change. We will address the lowest order dependence $c(T)$ and $\gamma(T)$ here, for periodic stimulation with stimulus interval $T$. In the context of myocardial depolarization waves the $c(T)$ dependency is known as the dispersion relation. This correlation is of fundamental importance with respect to arrhythmogenesis, for dispersion curves which have $|\dd_T c(T)| >1$ lead to conduction block \cite{Panfilov:CBH} and may therefore initiate cardiac arrhythmias.

The situation under study may serve to represent high-frequency periodic activation of myocardium, e.g. due to an arrhythmia or artificial stimulation. Moreover, the present analysis extends the validity range of our modified velocity-curvature relations to oscillatory media, which occur in given chemical reactions, cyclic biological processes and autorhythmic tissues.

\subsection{Derivation of the dispersive velocity-curvature relation}

\begin{figure}[h!t] \centering
  \includegraphics[width=0.8 \textwidth]{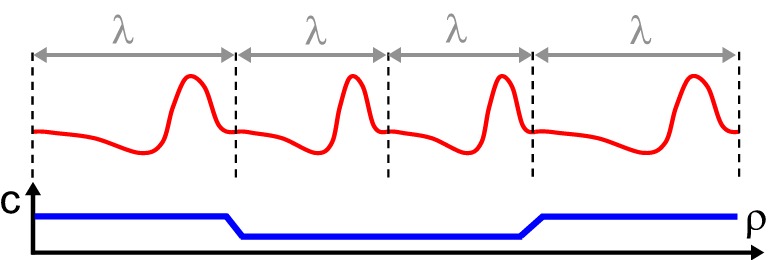}\\
  \caption[Local stretching of wave trains due to varying propagation velocity]{Local stretching of wave trains to accommodate varying propagation velocity.}\label{fig:wavetrains}
\end{figure}

As before, we take the plane wave solution as a reference; we will provide a derivation for an isotropic medium. The main difference with our treatment of isolated pulses is that periodic spatial boundary conditions will be used, which fix the wavelength to $\ell(T) = c_0(T) T$. As pointed out in \cite{Wellner:1997}, one needs to take into account that the plane wave profile $\uu_0$ locally stretches or contracts to accommodate  the local wave velocity $c(K(r))$; see also Fig. \ref{fig:wavetrains}. For periodic stimulation with fixed frequency $\Omega=2\pi/T$, the wave number and wavelength generally differ due to local curvature:
\begin{align}
\mathcal{K} &= \Omega / c(T), & \ell(T) &= c(T) T.
\end{align}
Following \cite{Wellner:1997}, the points which are a distance $r$ away from the wave front may be attributed a relative phase $\phi$ (which is termed `$\rho$' in the original work):
\begin{equation}\label{phase}
    \phi(r) = \int \mathcal{K}(r) \mathrm{d}r - \Omega t.
\end{equation}
We proceed by introducing locally stretched coordinates $\rho = \phi(r)/k_0$ perpendicular to the wave front, which is situated at the position $X(t)$. Here, $k_0$ equals $\Omega/c_0$. The new reference frame is given in terms of the exact but yet unknown wave front velocity $c$:
\begin{eqnarray}
    \rho(r,t) &=& \int\limits_{R(t)}^r \frac{\mathcal{K}(r)}{k_0} \mathrm{d}r - c t, \label{defrho1}\\
    \tau(r,t) &=& t.
\end{eqnarray}
The rescaling may be understood as follows: in regions with significantly higher (lower) front velocity (due to wave front curvature), the standard profile $\uu_0$ is wider (narrower) relative to the reference situation. While this effect is in lowest order negligible in the limit of isolated pulses, we need to take it into account for periodic wave trains. An important distinction with our derivation for isolated pulses is that we now let the coordinate frame exactly coincide with the wave front, i.e. $c_M = c$ instead of $c_M = c_0$, which we have used before.

In the next step, definition \eqref{defrho1} is reexpressed in terms of the wave front velocity $c$:
\begin{equation}
    \rho(r,t) = \int\limits_{R(t)}^r \frac{c_0}{c\left(K(r)\right)} \mathrm{d}r - c t. \label{defrho2}
\end{equation}
Note that, while $\vec{e_r} = \vec{e^r}$ in the laboratory frame, scale factors apply for the vectors $\vec{e_\rho},\ \vec{e^\rho}$, which also follow from \eqref{defrho2}:
\begin{equation}\label{erho}
    \vec{e^\rho} = \dd_r \rho\  \vec{e_r}= \frac{c_0}{c} \vec{e_r}.
\end{equation}
The metric tensor retains its block diagonal shape
\begin{equation}\label{metric}
    g_{\mu\nu} = \left(
                 \begin{array}{cc}
                   g_{\rho\rho} & 0\ 0 \\
                   \begin{array}{c}
                     0 \\
                     0
                   \end{array}
                    & h_{\alpha \beta} \\
                 \end{array}
               \right)
\end{equation}
with
\begin{align}
   g^{\rho \rho} = \left(\frac{c_0}{c}\right)^2, && g_{\rho \rho} = \left(\frac{c}{c_0}\right)^2.
\end{align}
Thus follows also
\begin{equation}\label{erhodown}
    \vec{e_\rho} = \frac{c}{c_0} \vec{e_r}.
\end{equation}
We proceed by the familiar expansion for position coordinates:
\begin{align} \label{taylorx}
    x^i &= X^i(\sigma, \tau) + \rho \vec{e_\rho} + \OO(\rho^2), &
    t &= \tau.
\end{align}
The chain rule for the coordinate transition $(x^i, t)\rightarrow (\rho, \sigma^\alpha, \tau)$ now yields
\begin{align}
  \dd_r &= \frac{c_0}{c} \dd_r, &
  \dd_t &= \dd_\tau - \dot{\vec{X}}\cdot \vec{e^\rho} \dd_\rho \label{ddt}.
\end{align}
Since we have formally appealed to the exact propagation speed $c$ we can write without any approximation
\begin{equation}\label{defR}
    \vec{X} =  c \tau \vec{e_r} = c_0 \tau \vec{e_\rho},
\end{equation}
which turns Eq. \eqref{ddt} into
\begin{equation}\label{ddt2}
    \dd_t = \dd_\tau - c_0 \dd_\rho.
\end{equation}
For future use, we notice that terms involving $\rho$-derivatives of $g_{\rho\rho}$ are small, since $\dd_\rho K$ is assumed to be smaller than $\OO(\la)$:
\begin{equation}\label{ddrgrr}
    \dd_\rho g_{\rho\rho} = 2 \left(\frac{c}{c_0^2}\right)\dd_\rho c(K) = 2 \left(\frac{c}{c_0^2}\right) \gamma \dd_\rho K(\rho).
\end{equation}
Having refined the way in which we lay down our unperturbed solution in space through the choice of coordinates, we start reformulating the RDE \eqref{RDE_chapt_fronts} using the same ansatz \eqref{fronts_ansatz} as for isolated pulses. Up to leading order, this procedure now leads to
\begin{multline}\label{RDE2}
    \la \dd_\tau \tuu - c_0 \dd_\rho \uu = D_0 \mathbf{P} \left(\frac{c_0}{c}\right)^2 \dd_\rho^2 \uu + D_0 \dd_i g^{i\rho} \mathbf{P} \dd_\rho \uu + \bPhi(\uu_0)\\ + \bPhi'(\uu_0) \dd_\rho \tuu + \OO(\la^2),
\end{multline}
from which is inferred that
\begin{equation}\label{RDE3}
    \mathbf{0} =  \bra{\bY^\rho} \HP \left[\left(\frac{c_0}{c}\right)^2 -1 \right] \dd_\rho \ket{\bpsi_\rho} + \bra{\bY^\rho} \dd_i g^{i\rho} \mathbf{P} \ket{\bpsi_\rho} +\OO(\la^2).
\end{equation}
For a medium with constant metric determinant one finds
\begin{equation}\label{digir}
    \dd_i g^{i\rho} = \dd_\mu g^{\mu \rho} + \Gamma^{\mu}_{\mu \nu} g^{\nu \rho} = \dd_\rho g^{\rho \rho} + \Gamma^{\mu}_{\mu \rho} g^{\rho \rho}.
\end{equation}
The first term can be omitted due to \eqref{ddrgrr}, while the second term involves
\begin{equation}\label{digir2}
     \Gamma^{\mu}_{\mu \rho} = \Gamma^{\rho}_{\rho \rho} + \vec{e^\alpha} \DD_\alpha \vec{e_\rho} = - K \frac{c}{c_0}.
\end{equation}
The extra scaling factor stems from the fact that the normal vector that is used for measuring extrinsic curvature needs be normalized. We conclude that $\dd_i g^{i\rho} = K \left(\frac{c_0}{c}\right)$, and insert this finding into Eq. \eqref{RDE3}:
\begin{equation}\label{RDE4}
    \mathbf{0} = \mathbf{P} \left[\left(\frac{c_0}{c}\right)^2 -1 \right] \dd_\rho^2 \uu - K \left(\frac{c_0}{c}\right) \HP \dd_\rho \uu_0.
\end{equation}
Evidently, the bracket product that is used here spans only an interval of length $\ell$, i.e. the wavelength of a plane wave with the same stimulus interval $T$:
\begin{equation}\label{newbracket}
  \langle \mathbf{f} , \mathbf{g} \rangle = \int_{a-\ell}^a \mathbf{f}^H \mathbf{g} \dd \rho.
\end{equation}
We further specify two matrix elements:
\bsub \label{def_P0P1} \begin{eqnarray}
  P_0 &=& \bra{\bY^\rho} \HP \ket{\bpsi_\rho}, \\
  P_1 &=& \bra{\bY^\rho} \HP \ket{\dd_\rho \bpsi_\rho}.
\end{eqnarray} \esub
Inserting the ansatz $c(K)=c_0 - \gamma K + \OO(\la^2)$ in \eqref{RDE4} then demands that
\begin{equation}\label{eik1}
    0 = 2 \gamma K P_1 + P_0 K,
\end{equation}
from which is concluded
\begin{equation}\label{def_gamma_disp}
    \gamma = P_0 \left( -\frac{c_0}{2 P_1}\right).
\end{equation}
So we have found an extra correction factor $\zeta$ to the leading order coefficient in the velocity-curvature relation:
\begin{equation}\label{gamma_disp}
    \gamma = - \frac{c_0}{2} \frac{\bra{\bY^\rho} \HP \ket{\bpsi_\rho} }{\bra{\bY^\rho} \HP \ket{\dd_\rho \bpsi_\rho}}.
\end{equation}
The dispersive surface tension coefficient may be rewritten
\begin{equation}\label{result}
    \gamma = \zeta \bra{\bY^\rho} \HP \ket{\psi_\rho}
\end{equation}
with
\begin{equation}\label{zeta}
      \zeta = -\frac{c_0}{2 \bra{\bY^\rho} \HP \ket{\dd_\rho \psi_\rho}}.
\end{equation}
In the low-frequency limit, we have demonstrated before that $\gamma = P_0$, a limiting case which can only hold if
\begin{equation}\label{lim}
    \lim\limits_{T\rightarrow\infty} \zeta(T) = 1.
\end{equation}
This low frequency limit can be proven from the original RDE, as we now progress to show.

\subsection{Explicit influence of incomplete recovery \label{sec:proofzeta}}

To not overload notation in this proof of statement \eqref{lim}, we will omit the index $\rho$ when writing $\bY^\rho$, $\bpsi_\rho$; moreover the prime denotes the ordinary spatial derivative with respect to $\rho$.

From the definition of the Goldstone modes
\begin{subequations}
\begin{eqnarray}
  \HL \bpsi &=& \mathbf{P} \bpsi '' + c_0 \bpsi' + \bPhi'(\uu_0) \psi = 0,  \label{Leq1}\\
  \HL^\dagger \bY &=& \mathbf{P}^T \bY'' - c_0 \bY' + \bPhi'^T(\uu_0) \bY = 0, \label{Leq2}
\end{eqnarray}
\end{subequations}
 we eliminate the reaction term by left-multiplication of \eqref{Leq1} with $\bY$, right-multiplication of \eqref{Leq2} with $\bpsi$ and subtraction:
\begin{equation}\label{subLeqs}
    \bY(\rho) \mathbf{P} \bpsi''(\rho) - \bY''(\rho) \bP \bpsi(\rho)+ c_0 (\bY(\rho) \bpsi(\rho))' = 0.
\end{equation}
Note that this expression holds for all $\rho$; no integral (bracket) has been taken yet. To capture both isolated pulses and wave trains, we assume an integration domain $]a^-, a^+[$, where $a^- = a-\ell/2,$ $a^+ = a+\ell/2$. The limit of isolated pulses is then retrieved when $\ell$ takes arbitrarily high values. Now, integrate between $a^- = a-\ell/2$ and $\rho$ and use the partial integration theorem to find
\begin{equation}\label{intsubLeqs}
     \bY(\rho)\mathbf{P} \bpsi'(\rho) -  \bY'(\rho) \mathbf{P} \bpsi(\rho) + c_0  \bY(\rho) \bpsi(\rho) = \mathcal{B}(a^-).
\end{equation}
The boundary term
\begin{equation}\label{boundary_disp}
 \mathcal{B}(a^-) = \bY(a^-) \mathbf{P} \bpsi'(a^-) -  \bY'(a^-) \mathbf{P} \bpsi(a^-) + c_0  \bY(a^-) \bpsi(a^-)
\end{equation}
vanishes in the limit of isolated pulses ($\ell \rightarrow \infty$), since the plane wave solution is asymptotically flat at large distances. From \eqref{intsubLeqs}, we infer that the boundary term does not depend on the arbitrarily chosen coordinate $a^-$, and indeed $\dd_{a^-} \mathcal{B} = 0$ by virtue of \eqref{subLeqs}. Performing one more integration of \eqref{intsubLeqs} over the entire interval $]a^-, a^+[$ yields:
\begin{equation}\label{intsubLeqs2}
    \int\limits_{a^-}^{a^+} \bY \mathbf{P} \bpsi' \mathrm{d}\rho - \int\limits_{a^-}^{a^+} \bY' \mathbf{P} \bpsi \mathrm{d}\rho + c_0 =\ell \mathcal{B},
\end{equation}
from which follows that
\begin{equation} \label{intsubLeqs3}
    2 \int \limits_{a^-}^{a^+} \bY \mathbf{P} \bpsi' \mathrm{d}\rho + c_0 = \ell \mathcal{B}.
\end{equation}
In the case of isolated pulses, the boundary terms \eqref{boundary_disp} tends to zero. Therefore, \eqref{intsubLeqs3} proves in the low frequency limit that
\begin{equation}\label{intsubLeqs4}
    T \rightarrow \infty \quad \Rightarrow \mathcal{B} \rightarrow 0  \Rightarrow  \quad  2 P_1 + c_0 \rightarrow 0 \quad \Rightarrow  \quad \zeta \rightarrow 1,
\end{equation}
justifying Eq. \eqref{lim}.

\subsection{Localization property for wave trains}

\begin{figure}[h!t] \centering
  \includegraphics[width=0.9 \textwidth]{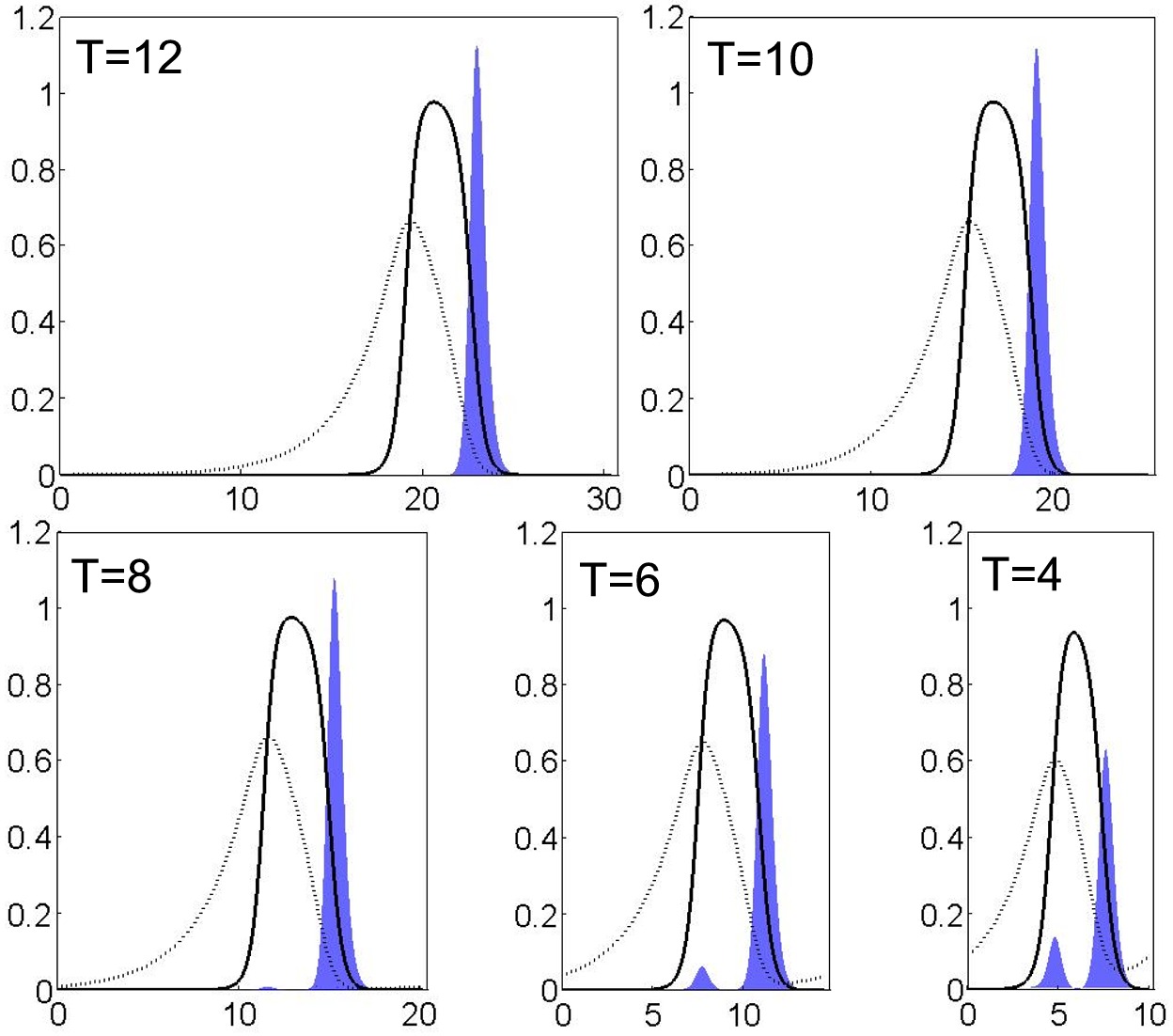}
  \caption[Localization property for wave trains]{Localization property for wave trains in the Barkley model for decreasing cycle length $T$. This effect was realized by imposing periodic spatial boundary conditions with box length $c(T) T$.}
  \label{fig:loc_periodic}
\end{figure}

The numerical technique that we have used for isolated pulses in paragraph \ref{sec:numeric_fronts} is readily generalized to the case of wave trains by imposing periodical boundary conditions to the one-dimensional domain. The effect of decreasing the period $T$ is shown in Fig. \ref{fig:loc_periodic} for the Barkley model with the same parameters as before. Interestingly, for short activation cycles, the density function $E_P$ (indicated by the shaded area) develops a secondary bulge near its wave back.

\subsection{Discussion: dispersion and restitution}

For periodic waves, we have gained a general formula \eqref{gamma_disp} for the dispersive surface tension coefficient $\gamma$, which was proven consistent with the result for isolated pulses in section \ref{sec:proofzeta}. Like our other dynamical coefficients in the EOM for wave fronts, $\gamma(T)$ was expressed using matrix elements that contain the zero modes of the one-dimensional solution to the RD system.

The boundary term $\mathcal{B}(T)$ is directly related to the restitution curve c(T) via Eq. \eqref{boundary_disp}, which can be rearranged to:
\begin{equation}\label{cTcurve}
  c_0(T) = \frac{-2 P_1(T)}{1- T \mathcal{B}(T)}.
\end{equation}
One observes that $c_0(T)$ tends asymptotically to $c_0(\infty)$ for large stimulus interval $T$.

Combining Eqs. \eqref{cTcurve} and \eqref{gamma_disp}, we calculate the dispersive surface tension coefficient $\gamma$ as
\begin{equation}\label{gamTcurve}
  \gamma(T) = \frac{P_0(T)}{1- T \mathcal{B}(T)}.
\end{equation}

\subsection{Critical radius and ratio for conduction block}

The critical radius $R_c$ to initiate propagation as given by Eq. \eqref{Rc_modified} scales proportional to $\gamma$. As a consequence of \eqref{gamTcurve}, the critical radius to initiate an action potential is also expected to grow unbounded as the pacing interval is shortened.

\begin{figure}[h!b] \centering
  \includegraphics[width=1.0\textwidth]{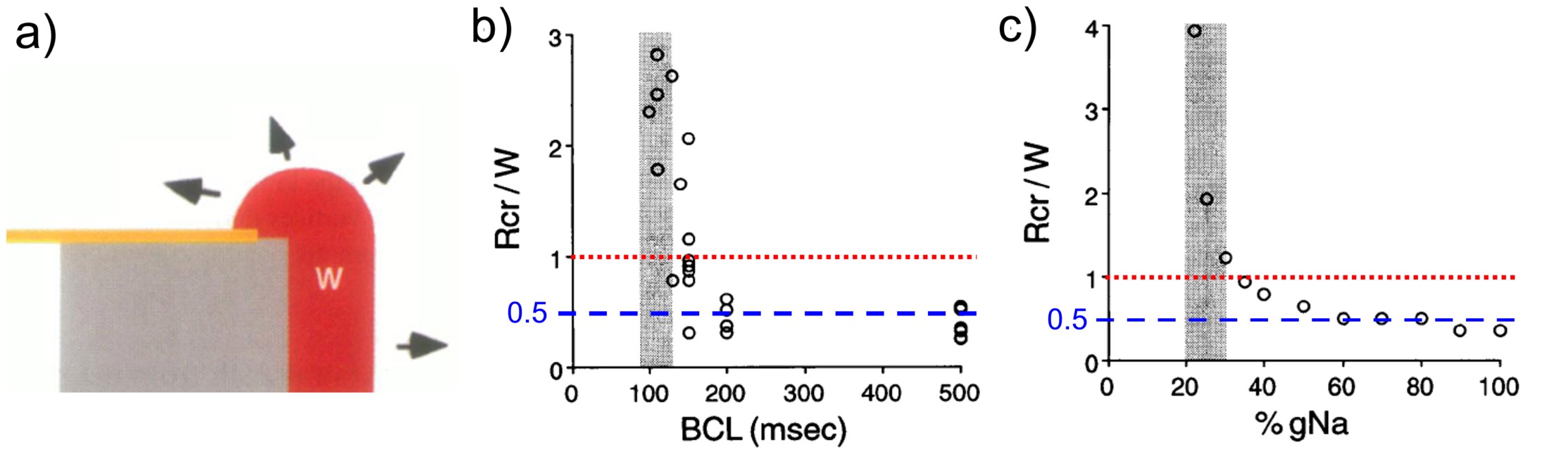}\\
  \caption[Critical ratio for the onset of electrical turbulence by Cabo \etal]{Critical ratio for the onset of electrical turbulence by Cabo \etal Panel (a) displays the mechanism of spiral wave generation when $d_f > R_c$, with $W \equiv d_f$. In panel (b), experiments in sheep heart deliver a critical radius close to 1/2 for large basic cycle lengths ($BCL \equiv T$). The third panel exemplifies the effect of decreasing sodium conductance in a numerical simulation, which also causes an upward deviation of $r_c$. These three figures have been adapted from \cite{Cabo:1996}.} \label{fig:Cabo}
\end{figure}

In \cite{Pertsov:1983} and \cite{Cabo:1996}, a critical ratio was discussed that estimates the risk of conduction block under the circumstance of rapid pacing. Following situation was considered in these papers: when a wave front propagates along an inexcitable boundary that comes to a sudden end, a spiral wave will emerge if the lateral part of the wave break (of length $d_f$) is larger than the critical radius for stimulation $R_c$. Thus the conceptually simple dimensionless ratio
\begin{equation}\label{crit_cabo0}
  r_c = \frac{R_c}{d_f}
\end{equation}
indicates higher risk for the development of arrhythmias if $r_c>1$. By means of Eqs. \eqref{Rc_modified} and \eqref{gamma_disp}, $r_c$ may for a cylindrical wave be expressed in terms of the matrix elements $P_0$ and $P_1$:
\begin{equation}\label{crit_cabo2}
  r_c = - \frac{P_0}{ 2 P_1 d_f}.
\end{equation}
In the limit where action potential upstroke is steep, one could approximate (after partial integration):
\begin{equation}\label{P0P1approx}
  P_1 = - \bra{\dd_\rho \bY^\rho} \HP \ket{\psi_\rho} \approx + \frac{1}{d_f} \bra{\dd_\rho \bY^\rho} \HP \ket{\uu_0} = - \frac{P_0}{d_f}.
\end{equation}
Here, we have used that a traveling wave of depolarization is most sensitive to perturbations that take place near its front, leading to a dominant contribution of $\dd_\rho \bY^\rho$ in this area. In the light of Eq. \eqref{P0P1approx}, the critical radius of Cabo \etal is now rated to
\begin{equation}\label{crit_cabo3}
  r_c \approx \frac{1}{2}
\end{equation}
for steep wave fronts. From Fig. \ref{fig:Cabo}b, which was taken from the original paper \cite{Cabo:1996}, it can be appreciated that $r_c$ indeed tends to a numerical value of $0.5$ in the limit of long $T$, and in the limit of steep wave fronts displayed in panel (c).

The behavior of $r_c(T)$ at high frequencies may be further addressed based on our observation that the density function in a sequence of traveling waves exhibits peaks not only near the wave front $W_f$, but also at its back $W_b$. We may therefore state that for periodic stimulation
\begin{equation}\label{mass_fb}
  P_0 = \int_{a^-}^{a^+} E_P(\rho) \mathrm{d} \rho = \int_{W_f} E_P(\rho) \mathrm{d} \rho + \int_{W_b} E_P(\rho) \mathrm{d} \rho = m_f + m_b,
\end{equation}
where the integrated densities near wave front and back have been suggestively denoted as `masses' $m_f$, $m_b$. If one writes $d_b$ for the typical width of wave back, one may redo \eqref{P0P1approx} to obtain
\begin{equation}\label{crit_caboT}
  r_c \approx \frac{1}{2} \left(1 - \frac{m_b}{m_f+m_b} \frac{d_f}{d_b}\right)^{-1}.
\end{equation}
Therefore, the deviation of the critical ratio from its asymptotical value $1/2$ may be explained for high-frequency pacing in terms two effects, namely the wave back starting to contribute to the total density $P_0$ and the increased width of the wave front due to a loss of excitability under high-frequency pacing.

\clearpage{\pagestyle{empty}\cleardoublepage}

\graphicspath{{fig/}{fig/fig_filiso/}}

\renewcommand\evenpagerightmark{{\scshape\small Chapter 7}}
\renewcommand\oddpageleftmark{{\scshape\small Filament dynamics in isotropic media}}

\renewcommand{\bibname}{References}

\hyphenation{}

\chapter[Filament dynamics in an isotropic medium]{Dynamics of scroll wave filaments\\ in an isotropic medium}
\label{chapt:filiso}



The content of this chapter and the subsequent one constitutes the very heart of this thesis. For, the striking resemblance of equilibrating scroll wave filaments to cosmic strings motivated the original development of our geometrical analysis.\\

In the course of this chapter, we restrict our analysis to scroll waves in an isotropic medium. Although this regime is inadequate to reproduce cardiac tissue, it is instructive for two important reasons. First, scroll wave activity is known to occur in oscillatory excitable media, which are typically isotropic. In fact, studying these systems parallels the historical development in the study of autowave activity, in which the scientific focuss also shifted from chemical systems to electrically active myocardium. Secondly, we can draw from the local equivalence principle, which enables to claim that the laws governing filament dynamics remain the same in anisotropic media, apart from the supplementary tidal forces (i.e. Ricci tensor terms) that will appear in an anisotropic context.

We shall in this part of the text first review the milestones that have led to the present-day description of cardiac arrhythmias in terms of scroll waves and filaments. Remarkably, the most advanced description of filament dynamics in literature is still phenomenological \cite{Echebarria:2006}.

Next, we provide a rigorous derivation for the filament EOM up to third order in curvature effects, and discuss its implications for filament instability. In particular, twist and translational degrees of freedom are identified to mutually interact; these processes might be relevant to the development of scroll wave turbulence.\\

The general approach adopted here is nearly identical to our treatment of wave fronts in the previous chapter. Still working in three spatial dimensions, we shall now divide space in 1+2 dimensions (i.e. one direction tangential to the filament and two transverse to it) instead of the 2+1dimensions that we have taken for wave fronts.

In a way, our approach to filaments therefore exhibits similar geometric duality as in the development of dual q-ball imaging exposed in Chapters \ref{chapt:imag} and \ref{chapt:QBI}.

\section{Spiral waves and scroll waves}

Before getting involved with three-dimensional scroll wave dynamics, it is instructive to first deal with the two-dimensional spiral wave solutions to the RDE \eqref{RDE_mono}.

\subsection{Spatial trajectories of the spiral tip}

\begin{figure}[h!t] \centering
  \includegraphics[width=0.8 \textwidth]{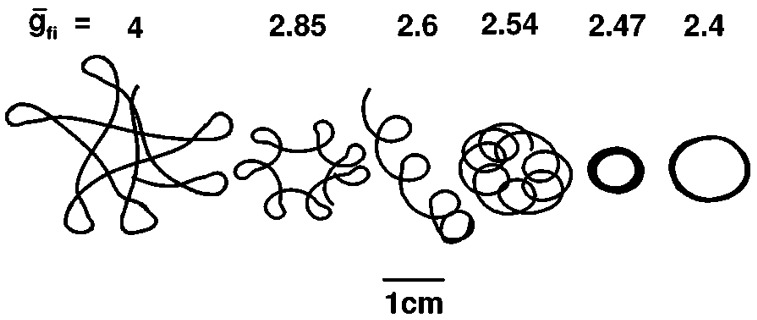}\\
  \caption[Meandering tip trajectories of spiral waves]{Meandering tip trajectories of spiral waves in different parameter regimes, corresponding to numerical modification of fast inward sodium current. Figure as appeared in \cite{Fenton:1998}. }\label{fig:tip_meander}
\end{figure}

Most cross-sections through the arm of a spiral wave closely match the plane wave solution as in Eq. \eqref{fronts_ansatz}; for that reason the outer zones of the spiral wave obey the velocity-curvature relation discussed in the previous chapter. In the central region of the spiral wave, however, this approximation breaks down as the active cells not necessarily fulfill a complete excitation cycle. Therefore, the diffusion term in the RDE manifestly comes into play near the spiral's tip and affects its motion. Figure \ref{fig:tip_meander} exemplifies some typical trajectories for spiral tips.

Disregarding the effects of incomplete recovery (i.e. refractoriness), the tip's motion is identical for all time frames. This simplest case results in a circular movement of the spiral tip; the region enclosed by the tip trajectory is known as the spiral's \textit{core}. In a different parameter regime, an oscillatory movement may add to the spiral tip motion, leading to an epicycle trajectory. This regime is called \textit{meander} and has been related to the symmetries of the Euclidean plane in \cite{Barkley:1994, Biktashev:1996}. In some detailed models of cardiac excitation, the diffusive properties around the spiral tip are even more pronounced, such that eventually only the refractoriness of previously excited tissue governs the tip trajectory; this regime delivers so-called linear cores.
In this work, we operate in the parameter regime where spiral waves have circular cores; the origin of our coordinate system will lie in the instantaneous rotation center of the activation pattern; consequently the distance between the spiral tip and our origin equals the radius of the spiral core.

\subsection{Response functions for spiral waves}

Our intention to track only the spiral tip instead of the entire spiral wave pattern is motivated by the historical observation that the motion of a spiral is only affected by an external perturbation if this stimulus takes place near the center of the spiral, i.e. in the neighborhood of its core. This remarkable fact has been checked in experiments and through numerical simulations; an analytical description was put down in \cite{Biktashev:1995b}. Outlining their proof here serves us to introduce appropriate notations.

Similar to our analysis of wave fronts, Biktashev \etal started from a known, unperturbed solution to the reaction diffusion equation \eqref{RDE_mono}, formulated for a two-dimensional isotropic medium:
\begin{equation}\label{RDE_chapt_filiso}
  \dd_t \uu = D_0 \Delta \mathbf{P} \uu + \mathbf{\bPhi}(\uu).
\end{equation}
From here on, we recycle the notation $\uu_0$ to denote the unperturbed spiral wave solution to \eqref{RDE_chapt_filiso} in an unbounded, isotropic, two-dimensional medium. It is convenient to work in a rotating frame with Cartesian coordinates $(\rho^1, \rho^2)$ or polar coordinates $(\rho,\theta)$. Rotation occurs together with the unperturbed spiral at its natural frequency $\omega_0$. As before, time in the co-rotating frame is designated $\tau$.

Due to the perturbative nature of the problem considered, the system equations were linearized around the rigidly rotating solution $\uu_0(\rho^A, \tau)$, with $A \in \{ 1,2 \}$:
\begin{equation}\label{def_Lfil}
  \HL = D_0  \Delta \mathbf{P} + \omega_0 \dd_\theta + \bPhi'(\uu_0).
\end{equation}
Compared to wave fronts, spiral wave dynamics is intricately more involved, as more degrees of freedom apply to the two-dimensional Euclidean plane. Since the RDE is invariant under translation and rotation, spiral wave solutions that are infinitesimally shifted or rotated still form a solution; in physics such modes are known as Goldstone modes (GM). Working in the co-rotating frame only preserves the rotational GM $\ket{\bpsi_\theta} = \ket{\dd_\theta \uu_0}$ as a true zero mode
\begin{equation}\label{GM_rot}
  \HL \ket{\bpsi_\theta} = \ket{\boldsymbol{0}},
\end{equation}
whereas the translational Goldstone modes $\ket{\bpsi_A} = \ket{\dd_A \uu_0}$ obey
\begin{equation}\label{GM_trans}
  \HL \ket{\bpsi_A} = - \omega_0 \eps_A^{\hs B} \ket{\bpsi_B}.
\end{equation}
When the translational GMs are diagonalized with respect to $\HL$,
\begin{align}\label{GM_pm}
\ket{\bpsi_\pm} &= \frac{1}{\sqrt{2}} \left( \ket{\bpsi_1}  \pm \ket{\bpsi_2} \right), &
   \HL \ket{\bpsi_\pm} &= \pm i \omega_0 \ket{\bpsi_\pm},
\end{align}
it is apparent that the combinations $\ket{\bpsi_+}, \ket{\bpsi_-}$ have eigenvalues $+  i \omega_0, -i \omega_0$ that lie on the imaginary axis. \\

In the treatment of spiral drift due to external perturbations, it is furthermore assumed that the adjoint operator $\HL^\dagger$ also has exactly three eigenfunctions for which
\begin{align}\label{RF_pm}
   \HL^\dagger \bY^- &= i \omega_0 \bY^-, & \HL^\dagger \bY^\theta  &=0, & \HL^\dagger \bY^+ &= -i \omega_0 \bY^+.
\end{align}
Furthermore, based on the natural inner product for the Euclidean plane
\begin{equation}\label{def_inner2D}
 \langle \mathbf{f} \mid \mathbf{g} \rangle = \langle \mathbf{f} , \mathbf{g} \rangle = \int \limits_{-\infty}^\infty \int \limits_{-\infty}^\infty  \mathbf{f}^H \mathbf{g}\  \mathrm{d} \rho^1  \mathrm{d} \rho^2,
\end{equation}
the adjoint modes in Eq. \eqref{RF_pm} can be made biorthogonal to the Goldstone modes by a Gramm-Schmidt process, such that
\begin{equation}\label{RF_GM_biortho}
 \langle \bY^{\hat{\mu}} \mid \bpsi_{\hat{\nu}} \rangle = \delta^{\hat{\mu}}_{\hs \hat{\nu}}, \qquad (\hat{\mu}, \hat{\nu} \in \{ 1, 2, \theta \}).
\end{equation}

Using $\vec{X}(\tau)$ to denote the position of the instantaneous rotation center of a spiral wave under an small external perturbation $\mathbf{h}(\vec{\rho}, \tau)$, it was shown that the spiral wave drifts at a rate
\begin{equation}\label{spiraldrift_tr}
  \vec{e}^A \cdot \dot{\vec{X}} = - \langle \bY^A \mid \mathbf{h}(\vec{\rho}, \tau) \rangle
\end{equation}
while its rotation frequency changes to
\begin{equation}\label{spiraldrift_rot}
  \omega = \omega_0 +  \langle \bY^\theta \mid \mathbf{h}(\vec{\rho}, \tau) \rangle.
\end{equation}
Since the adjoint modes $\bY^\theta,\, \bY^A$ fully account for the spiral waves rotational and translational drift in the presence of a small external perturbation, these functions were termed \textit{response functions} (RFs) by Biktashev and co-workers. \\

Note that a comparable rationale may be given to treat the response of a wave front subjected to a small external perturbation. Therefore, the one-dimensional translational zero mode $\bra{\bY^\rho}$ that we have encountered in Chapter \ref{chapt:fronts} with the same right deserves to be called a response function.

\subsection{Wave-particle duality of spiral waves}

\begin{figure}[h!t] \centering
\includegraphics[width = 1.0 \textwidth]{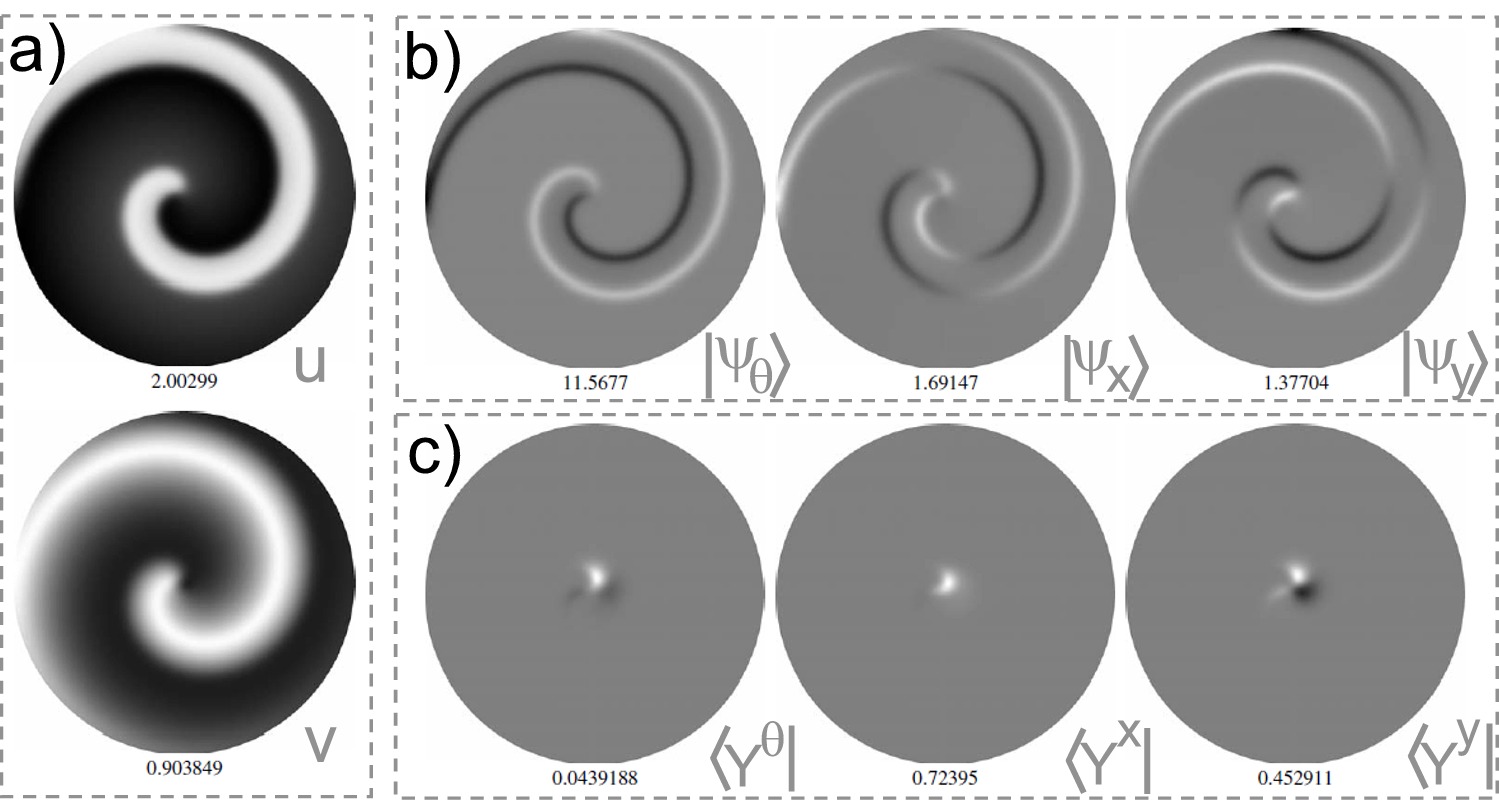}
  \caption[Particle-wave duality of spiral-shaped activation patterns]{Particle-wave duality of spiral-shaped activation patterns from numerical simulations using the two variable FitzHugh-Nagumo model, taken from \cite{Biktasheva:2009}. Panel (a) displays the spiral solution $\uu_0$ for the $u$ and $v$ variables. While the rotational and translational Goldstone modes (b) also exhibit wave nature, the response functions (c) are clearly localized. For GMs and RFs, only the $u$ component is given here; numerical values indicate the absolute scaling factors with respect to shading.}\label{fig:pw_duality}
\end{figure}

Strikingly, spiral waves only drift if the external stimulus $\mathbf{h}(\vec{\rho}, \tau)$ is applied close to its center. This would imply that the three RFs have a tempered nature, which was verified analytically and numerically by Biktasheva and co-workers in the context of spiral waves in the complex Ginzburg-Landau equation \cite{Biktasheva:2003}. It turns out that a spiral wave takes two distinct forms in its mathematical description: its Goldstone modes exhibit the typical spiral shape that extends to the edges of the medium, whereas its response function is a strongly localized function. This dualistic character was interpreted as a fundamental wave-particle nature by Biktasheva \etal and is also apparent in Fig. \ref{fig:pw_duality}.\\

The particle nature of the spiral wave is essential to our elaborations on scroll wave filaments: the local sensitivity of spiral waves close to their core is inherited by scroll waves, which are most susceptible to disturbances near their filaments.

\subsection{Geometrical properties of filaments}

Strictly speaking, the filament is the locus of the scroll wave that contains its phase singularities, i.e. the collection of spiral tips from the spiral waves that together form the scroll wave. Hence, the meandering pattern discussed above is expected to affect filament motion as well. As long as the tip trajectory has finite width, one may however perform a temporal moving average over the (pseudo)period of the wave in order to obtain a stationary filament even for meandering patterns. Note that the critical value $g_{fi}=2.6$ in Fig. \ref{fig:tip_meander} offers an example of a drifting spiral which cannot be averaged to a remain stationary.

As we shall rigorously consider only scroll waves with circular cores, the filament curve denotes the central axis of the core. In this regime, this notion of the filament curve is therefore unambiguously defined. \\

Being a continuous curve in three-dimensional Euclidean space, the filament exhibits geometric curvature $k$ and torsion $\tau_g$, as is commonly obtain from its description in a Frenet-Serret frame. Denoting the filament at a given instance of time as $\vec{X}(\s, t)$ and assuming sufficient smoothness of the curve, subsequent derivative action with respect to arc length $\s$ (notation $\dd_\s f =f '$) brings about a right-handed orthonormal triad given by the tangent vector $\vec{T} = \vec{X}'$, normal vector $\vec{N}$ and binormal vector $\vec{B}$:
\begin{equation}\label{Frenetframe}
  \left(
    \begin{array}{c}
      \vec{T}' \\
      \vec{N}' \\
      \vec{B}' \\
    \end{array}
  \right) =
  \left(
    \begin{array}{ccc}
      0 & k & 0 \\
      -k & 0 & \tau_g \\
      0 & -\tau_g & 0 \\
    \end{array}
  \right).
    \left(
    \begin{array}{c}
      \vec{T} \\
      \vec{N} \\
      \vec{B} \\
    \end{array}
  \right).
\end{equation}
The quantity $k$ here stands for the extrinsic curvature of the filament; remark that a one-dimensional curve cannot exhibit intrinsic curvature (in contrast to wave fronts). 
Noteworthily, the Frenet-Serret frame is degenerate in points where the filament curvature $k$ vanishes.\\

Apart from the filament shape, the phase of the scroll wave attached to the filament yet comprises another geometric degree of freedom. As an example, one could imaging building a scroll wave from two-dimensional spiral wave solutions stacked parallel to the XY plane, which are slightly rotated when moving along the Z-axis. The phase angle $\phi$ of the scroll wave in a plane transverse to the filament can be quantified by considering a vector $\vec{p}$ connecting the filament to a landmark point in the activation pattern, e.g. the spiral tip. Phase evolution of the scroll wave may then be tracked by following the motion of $\vec{p}$ in its turning motion around the filament. Evidently, one has that
\begin{equation}\label{def_omega}
  \dd_t \phi(\s, t) = \omega(\s, t),
\end{equation}
with $\omega$ the local rotation velocity of the scroll wave.

Even more important than the scroll's phase angle $\phi$ is its spatial derivative along the filament, as a varying phase angle $\phi$ induces additional diffusion of state variables along the filament. The quantity
\begin{equation}\label{def_w}
  \dd_\s \phi(\s, t) = w(\s, t)
\end{equation}
is therefore called the \textit{twist} of the associated scroll wave. Note that the twist may be equivalently expressed as \cite{Echebarria:2006}
\begin{equation}\label{def_wp}
  w = \left( \vec{p} \times \dd_\s \vec{p} \right) \cdot \vec{T}.
\end{equation}

\begin{figure}[h!t] \centering
\mbox{
  \raisebox{3.25cm}{a)\ }\includegraphics[height= 3.5 cm]{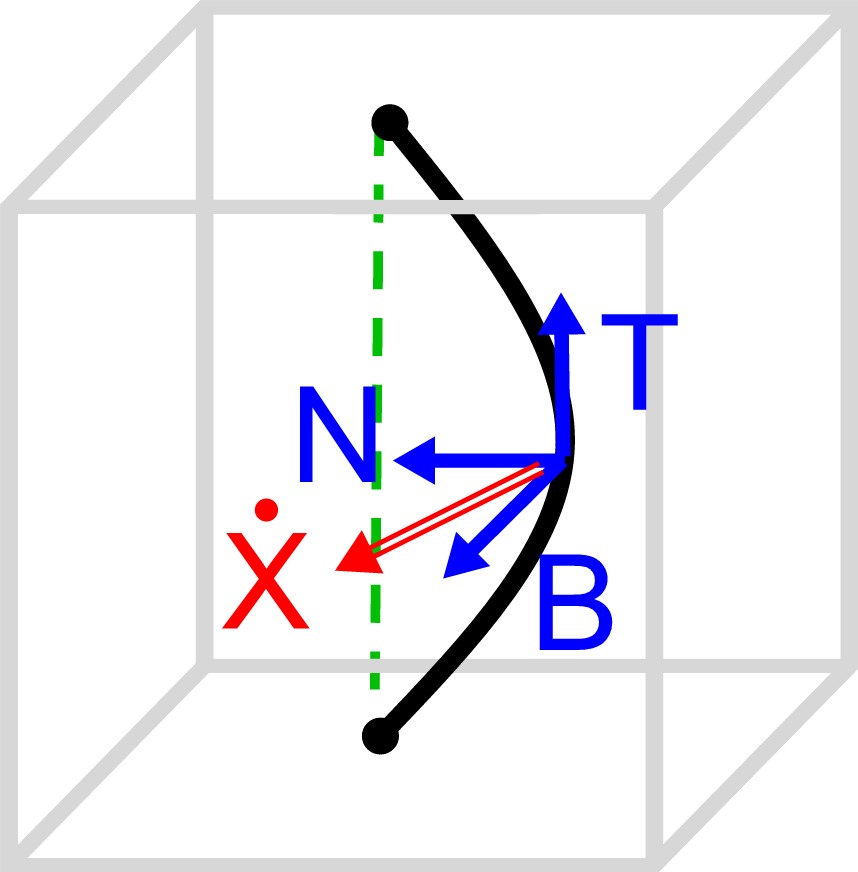}\ \
  \raisebox{3.25cm}{b)\ }\includegraphics[height= 3.5 cm]{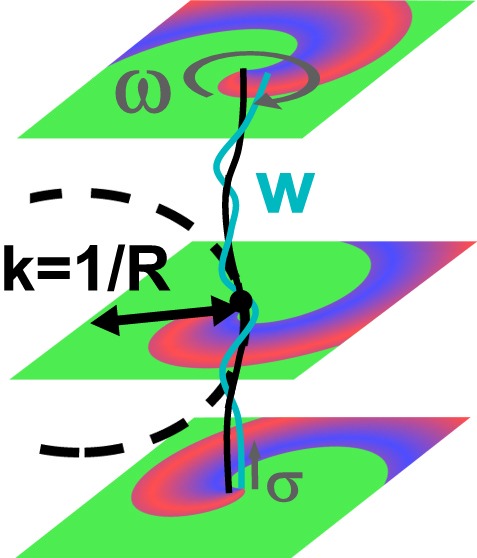}\ \
  \raisebox{3.25cm}{c)\ }\includegraphics[height= 3.5 cm]{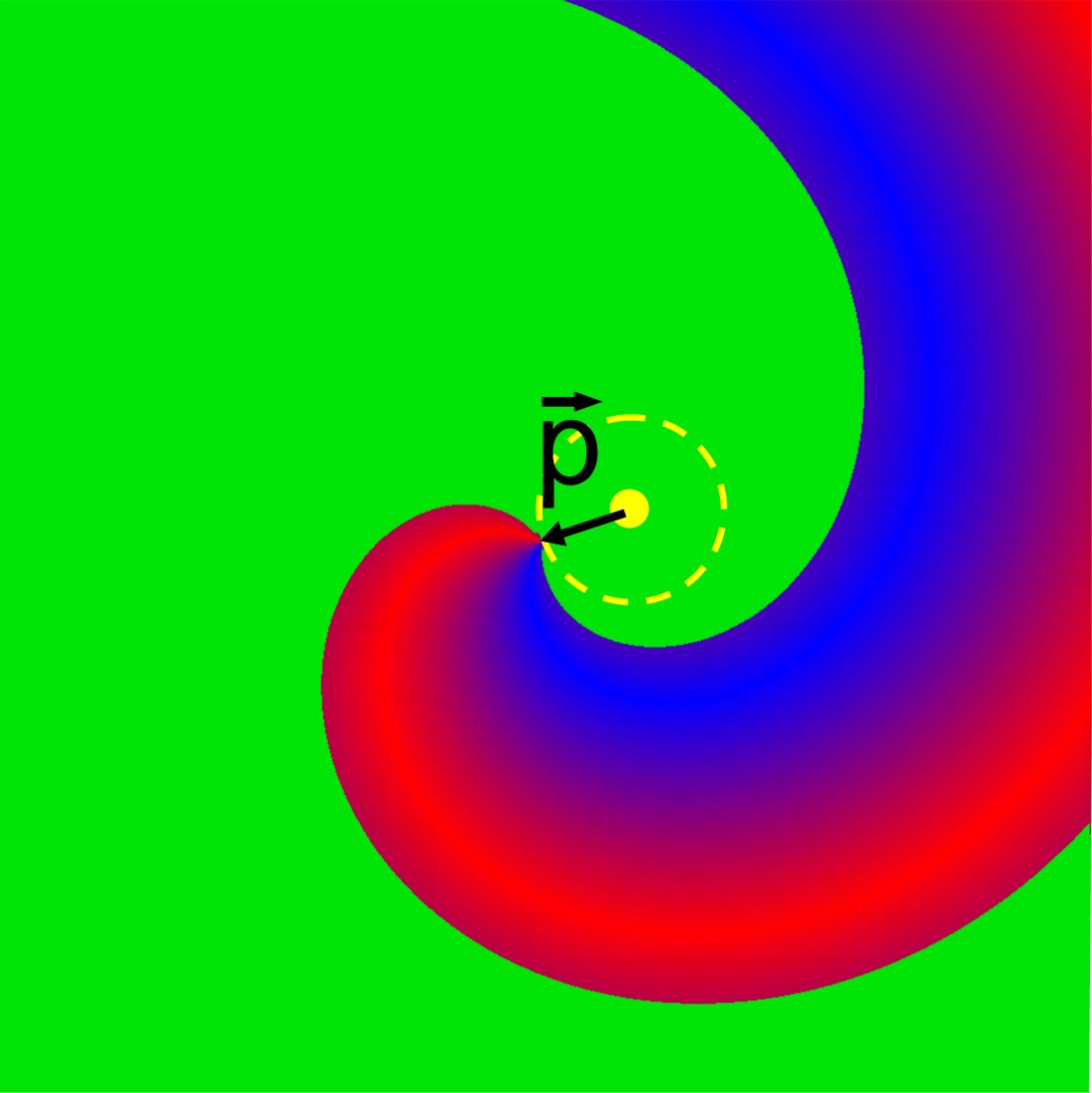}
  }
  \caption[Geometrical properties of a scroll wave filament]{Geometrical properties of a scroll wave filament: Frenet-Serret frame (a), geometric properties (b) and definition of the vector $\vec{p}$ from which the scroll wave's phase may be inferred (c).}\label{fig:fil_geom}
\end{figure}

\subsection{Filament motion due to tension}

Using the Frenet-Serret frame to define coordinates adapted to the filament shape, Keener and Tyson \cite{Keener:1992} managed to derive the effective EOM for filaments in lowest order in curvature and twist effects. In this work, the Fredholm alternative theorem was used, which comes down to projection onto a set of RFs as we have conducted above for wave fronts.

Keener's original EOM contained initially nine dynamical coefficients, but in \cite{Biktashev:1994} it was proven that only four of them do not vanish after averaging over a spiral period. As a result, a scroll wave filament with small curvature and twist was established to obey following laws of motion (adapted from \cite{Keener:1986, Biktashev:1994} to our notation):
\bsub \label{EOMfil_KB} \begin{eqnarray}
\dot{ \vec{X}} \cdot \vec{N} &=& \gamma_1 k,  \label{EOMfil_KB_N} \\
\dot{ \vec{X}} \cdot \vec{B} &=& \gamma_2 k,  \label{EOMfil_KB_B} \\
\omega &=& \omega_0 + a_0 w^2 + d_0 \dd_s w.  \label{EOMfil_KB_OM}
\end{eqnarray} \esub
We have yet imposed the transverse gauge condition here that $\dot{ \vec{X}} \cdot \vec{T}=0$; see also Fig. \ref{fig:fil_geom}a. The emergent dynamical coefficients $\gamma_1$, $\gamma_2$, $a_0$, $d_0$ are akin to the parameters $\gamma, \eta$ for wave fronts, in the sense that they depend on underlying electrophysiology, and they can be calculated by taking overlap integrals between GMs and RFs of a lower dimensional solution to the RDE. Explicit prescriptions were given in \cite{Keener:1992, Biktashev:1994} and will be derived again below. Notably, the translational degrees of freedom ($\dot{\vec{X}}, k$) fully decouple from the rotational motion ($\omega, w$) in Eqs. \eqref{EOMfil_KB}.

From the phase evolution of the scroll wave \eqref{EOMfil_KB_OM} the temporal evolution of twist may be deduced, since $\dd_\s \omega = \dd^2_{\s t} \phi = \dot{w}$:
\begin{equation}\label{EOMfil_KB_tw}
  \dot{w} = \dd_\s \omega_0 + 2 a_0 w \dd_\s w + d_0 \dd_\s^2 w.
\end{equation}
From this expression, twist is seen to be generated by differential rotation frequencies along the filament, e.g. due to varying medium properties. The coefficient $d_0$ acts as a twist diffusion coefficient, and therefore needs be positive in a medium where untwisted filaments are stable. The $w \dd_\s w$ term turns Eq. \eqref{EOMfil_KB_tw} into the Burgers' equation, which is known to support shock waves.

Subsequently, one may appreciate an elegant geometrical interpretation of Eqs. \eqref{EOMfil_KB_N}-\eqref{EOMfil_KB_B}. While the sign of $\gamma_2$ depends on the rotation direction of the spiral wave, the sign of $\gamma_1$ determines whether a bent filament locally drifts towards its normal vector, or in the opposite half plane. In the first case ($\gamma_1 >0$), a moving filament thereby reduces its length, as was explicitly proven in \cite{Biktashev:1994}. This led the authors to interpret $\gamma_1$ as the filament tension, which later inspired us to denote $\gamma$ the surface tension for wave fronts. In the regime of positive filament tension, untwisted scroll rings will contract in this regime until they annihilate, whereas transmural filaments are expected to straighten up if twist effects are small.

The full EOM \eqref{EOMfil_KB} thus has a straight untwisted scroll wave as an equilibrium state; this state is stable only if $\gamma_1 >0$ and $d_0 >0$. In the regime of negative filament tension, however, the filament will tend to increase its total length and generally take a more complex shape during this evolution. Filament instability of such type has been named a possible pathway to spark fibrillation events in the heart under given physiological tissue parameters; see also \cite{Fenton:2002}.
\subsection{The ribbon model for filament dynamics}

To describe the onset and nonlinear evolution of twist-induced filament instability in isotropic media, Echebarr\'ia \etal proposed a simple phenomenological model \cite{Echebarria:2006}. Using the notation $[\ \cdot\ ]_\perp$ to denote that part of a vector which is orthogonal to the filament tangent vector, their `ribbon model' prescribes:
\begin{eqnarray}
  \dot{\vec{X}} &=& \verb"a"_1\ \dd^2_\s \vec{X} + \verb"a"_2\ \dd_\s \vec{X} \times \dd^2_\s \vec{X}   \label{ribbon1} \\
&&  + \verb"d"_1\ w \ \dd_\s \vec{X} \times \dd^3_\s \vec{X}  - \verb"d"_2\  [\dd^3_\s \vec{X}]_\perp
  - \verb"b"_1\ [\dd^4_\s \vec{X}]_\perp  - \verb"b"_2\  \dd_\s \vec{X} \times \dd^4_\s \vec{X}. \nn
\end{eqnarray}
This equation was combined with Keener's phase evolution equation \eqref{EOMfil_KB_OM} to capture time evolution of the twist as well. Note that the proposed form of \eqref{ribbon1} is purely phenomenological, and the coefficients can only be estimated from forward numerical simulation of filaments for a given model. In section \ref{sec:extEOMfiliso}, we derive the proper equation of motion using a gradient expansion, and discuss its relation to the ribbon model.

For reference, one may expand the equations \eqref{ribbon1} using the Frenet-Serret relations \eqref{Frenetframe} to obtain
\bsub \label{EOMfil_ribbon} \begin{eqnarray}
\dot{ \vec{X}} \cdot \vec{N} &=& \verb"a"_1 k - \verb"d"_1 k w \tau_g - \verb"d"_2 w \dd_\s k \label{EOMfil_ribbon_N}\\
&& \qquad + \verb"b"_1 \left(k^3 - \dd_\s^2 k + k \tau_g^2 \right) + \verb"b"_2 \left(\dd_\s k + \dd_\s(k \tau_g) \right), \nn  \\
\dot{ \vec{X}} \cdot \vec{B} &=& \verb"a"_2 k - \verb"d"_2 k w \tau_g + \verb"d"_1 w \dd_\s k  \label{EOMfil_ribbon_B}\\
&& \qquad + \verb"b"_2 \left(k^3 - \dd_\s^2 k + k \tau_g^2 \right) - \verb"b"_1 \left(\dd_\s k + \dd_\s(k \tau_g) \right). \nn
\end{eqnarray} \esub

\section[The lowest order EOM for filaments in an isotropic medium]{The lowest order equation of motion for filaments \\in an isotropic medium \label{sec:lowEOMfiliso}}

\subsection{Reference frames adapted to a curve}
Being the most common choice, the Frenet frame \eqref{Frenetframe} as drawn in Fig. \ref{fig:frames_fil}a is not the only possibility to construct an orthonormal frame that is adapted to the filament shape. Due to orthonormality of the basis vectors, any such frame will also have a $3\times3$ coefficient matrix which is skew-symmetric, leaving three non-vanishing coefficients. Among these frames, the Frenet frame fulfills a special role since one of the coefficients always equals zero. We will give up this particularity here to adapt the frame to the rotation phase along the filament as well. The resulting reference system classifies as a `relatively parallel adapted frame' in the terminology used by R.~Bishop in \cite{Bishop:1975}.

Consider a normal vector field $\vec{M}_1(\s)$ along the $C^2$ curve $\mathcal{C(\s)}$ in Euclidean 3-space that is relatively parallel, i.e. its derivative is tangential. Obviously, a relatively parallel normal field has constant length, which we can choose equal to unity. Given a unit vector in $\mathcal{C}(\s_0)$ that is orthogonal to both $\vec{M}_1(\s_0)$ and the local unit tangent to the filament $\vec{T}(\s_0)$, there exists a unique relatively parallel field $\vec{M}_2(\s)$, which is orthogonal to $\vec{M}_1(\s)$ and $\vec{T}(\s)$ along the entire filament (see theorem 1 in \cite{Bishop:1975}). The oriented orthonormal triad $(\vec{T},\vec{M_1},\vec{M_2})$ is known as a \textit{relatively parallel adapted frame} and such frames are determined up to a rotation around the filament over a constant angle. The corresponding coefficient matrix has only two non-vanishing entries $k_1(\s), k_2(\s)$, which completely determine the shape of the filament (relying on theorem 3 from \cite{Bishop:1975}):
\begin{equation}\label{RPAFframe}
  \left(
    \begin{array}{c}
      \vec{T}' \\
      \vec{M}_1' \\
      \vec{M}_2' \\
    \end{array}
  \right) =
  \left(
    \begin{array}{ccc}
      0 & k_1 & k_2 \\
      -k_1 & 0 & 0 \\
      -k_2 & 0 & 0 \\
    \end{array}
  \right).
    \left(
    \begin{array}{c}
      \vec{T} \\
      \vec{M}_1 \\
      \vec{M}_2 \\
    \end{array}
  \right).
\end{equation}
Since relatively parallel adapted frames are minimally twisted to accommodate the shape of the filament, \eqref{RPAFframe} is particularly suited to frame untwisted scroll waves, as an alternative to the Frenet frame. Figure \ref{fig:frames_fil}b displays an example of such relatively parallel frame; in physicist terms, one would say that the orthonormal triad is being Fermi-Walker transported along the filament \cite{MTW}.

\subsection[A frame that translates, rotates and twists with the filament]{A frame that translates, rotates and twists\\ with the filament}

\begin{figure}[h!t] \centering
\mbox{
  \raisebox{4.5cm}{a)\ }\includegraphics[height= 4.75 cm]{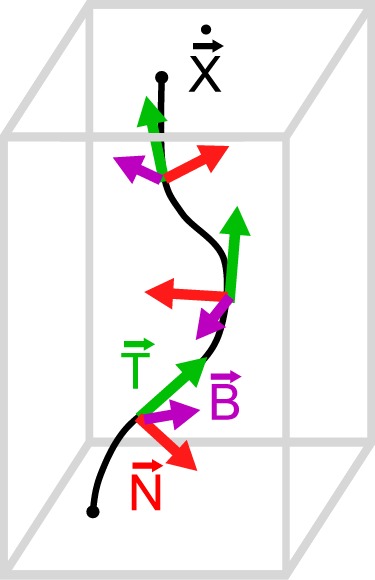}\ \
  \raisebox{4.5cm}{b)\ }\includegraphics[height= 4.75 cm]{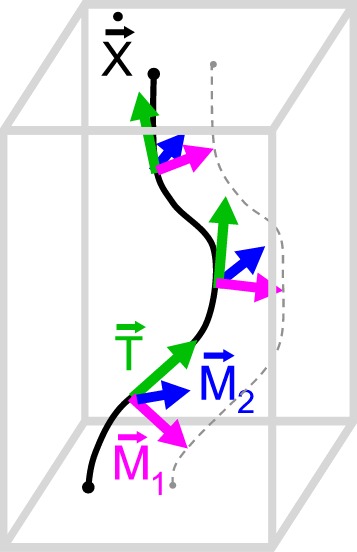}\ \
  \raisebox{4.5cm}{c)\ }\includegraphics[height= 4.75 cm]{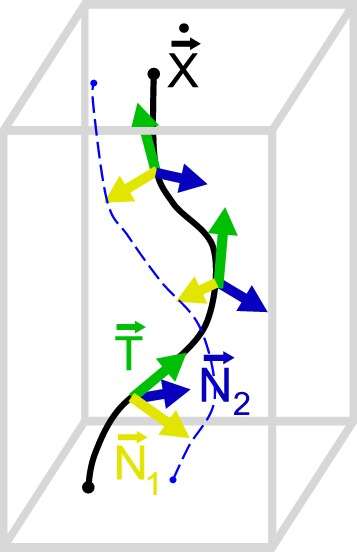}
  }
  \caption[Orthonormal triads adapted to the filament]{Orthonormal triads adapted to the filament: Frenet-Serret frame (a), relatively parallel adapted frame (b), as in \cite{Bishop:1975} and our custom frame (c) that accommodates twist as well. Note that in panel (c), $\vec{M}_1$ can be chosen equal to the fiducial vector $\vec{p}$, after which the filament together with $\vec{M}_1$ constitute the `ribbon' from \cite{Echebarria:2006}.}\label{fig:frames_fil}
\end{figure}

With twisted scroll waves, the normalized reference vector $\vec{p}$ from Eq. \eqref{def_wp} will enclose an oriented phase angle $\phi(\s,t)$ with $\vec{M}_1(\s,t)$ in each plane transverse to the filament. Hence we define at time $t=t_0$ a new orthonormal triad $(\vec{T}(\s,t_0),\vec{N_1}(\s,t_0),\vec{N_2}(\s,t_0))$ (with $\phi_0(\s) = \phi(\s, t_0)$), and let the triad rotate around the filament with the scroll wave's angular velocity $\omega(\s, t) = \dd_t \phi(\s, t)$.
In other words, the reference vectors $\vec{N}_1, \vec{N}_2$ relate to the reference vector fields $\vec{M}_1, \vec{M}_2$ from the previous paragraph as
\bsub  \begin{eqnarray}
 \vec{N}_1 = \vec{p} &=& \ \cos \phi  \vec{M}_1 + \sin \phi \vec{M}_2,  \\
 \vec{N}_2 = \vec{T} \times \vec{N}_1 &=& -\sin \phi \vec{M}_1 + \cos \phi \vec{M}_2.
\end{eqnarray}
\esub
The filament twist $w(\s,t)$ may now be introduced as the derivative of phase angle $\phi(\s,t)$ (measured in the relatively parallel adapted frame) with respect to arc length along the filament:
\begin{equation}\label{def_twistiso}
  w(\s,t) = \dd_\s \phi(\s,t).
\end{equation}
One can check that this definition is consistent with twist as defined in Eq. \eqref{def_wp} and Biktashev's quantity $\dd_\s \phi - \tau_g$ in the Frenet frame \cite{Biktashev:1994}. Decomposing $\vec{T}' = k\vec{N}$ in the basis $\vec{N}_A$, $A \in \{1,2\}$ delivers
\begin{equation}\label{def_LAa}
  \LA_A =  \vec{T}' \cdot \vec{N}_A = - \vec{N}_A' \cdot \vec{T}, \qquad (A=1,2).
\end{equation}
Consequently, the coefficient matrix for the frame adapted to the filament shape and phase (depicted in Fig. \ref{fig:frames_fil}c) reads
\begin{equation}\label{twistframe}
  \left(
    \begin{array}{c}
      \vec{T}' \\
      \vec{N}_1' \\
      \vec{N}_2' \\
    \end{array}
  \right) =
  \left(
    \begin{array}{ccc}
      0 & \LA_1 & \LA_2 \\
      -\LA_1 & 0 & w \\
      -\LA_2 & - w & 0 \\
    \end{array}
  \right).
    \left(
    \begin{array}{c}
      \vec{T} \\
      \vec{N}_1 \\
      \vec{N}_2 \\
    \end{array}
  \right).
\end{equation}
Note that twist is readily obtained from the coefficient matrix:
\begin{equation}\label{def_twist_iso2}
  w = \frac{1}{2} \eps^{AB} \vec{N}'_A \cdot \vec{N}_B.
\end{equation}
Furthermore, the extrinsic curvature $k$ of the filament is found as
\begin{equation}\label{def_curv_iso2}
        k = ||\vec{T}'|| = \sqrt{\LA_1^2+\LA_2^2}.
\end{equation}
If needed, the extrinsic curvature components can be expressed as a function of phase angle using $\LA_1 = k_1 \cos \phi+ k_2 \sin \phi$, $\LA_2 = -k_1 \sin \phi+ k_2 \cos \phi$.

\subsection{Parameterization of the space around the filament \label{subsec:curvilin1}}
In the next step, we make use of the triad $(\vec{T},\vec{N_1},\vec{N_2})$ to parameterize the neighborhood of the filament.
Whenever a point $\mathcal{P}$ is close enough to the filament, it possesses a unique perpendicular line to the filament, with intersection $\mathcal{Q}$. A curvilinear variant of cylindrical coordinates can be chosen to consist of the triplet $(\rho, \theta, \s)$, with $\rho = |\mathcal{PQ}|$ (the distance along this line to the filament), $\theta$ being the oriented angle between $\vec{N}_1$ at the point of intersection and $\s$ the arc length parameter of the intersection point $\mathcal{Q}$. For a straight filament, the thus constructed coordinates coincide with standard cylindrical coordinates. After the transformation $\rho^1 = \rho \cos \theta, \rho^2 = \rho \sin \theta$, nearly Cartesian coordinates are established such that
\begin{equation}\label{coord_iso}
 \begin{cases}
 x^i(\rho^A,\s,\tau) &=  X^i (\s,\tau) + \rho^A N_A^i(\s), \\
 t  &= \tau,
 \end{cases}
\end{equation}
with $\tau$ the time parameter in the co-moving frame. Clearly, the proper assignment of these coordinates can only be carried out for distances smaller than the local radius of curvature $R = 1/k$ of the filament. The breakdown of the coordinate system used eventually bounds the validity range of our analytical treatment.

As from here, we adopt an index convention to distinguish the different coordinate frames, similar to our conventions regarding wave fronts. The Latin indices $i,j,k \in \{1,2,3\}$ refer to the original Cartesian coordinates, as measured in a laboratory frame by heart section or imaging;  the new curvilinear coordinate system is denoted $\mu,\nu,... \in \{1,2,\s \}$ and the indices $A,B,C,... \in \{1,2\}$ refer to the components transverse to the filament in this system.

The base vectors with lower label away from the filament simply follow from
\bsub \begin{eqnarray}
 \vec{e}_A (\rho^B, \s, \tau) &=& \frac{\dd \vec{x}}{\dd \rho^A}= \vec{N}_A(\s, \tau), \\
 \vec{e}_\s (\rho^B, \s, \tau) &=& \frac{\dd \vec{x}}{\dd \s} = (1 - \LA_A \rho^A) \vec{T}(\s, \tau) + w \eps_{A}^{\ B} \rho^A \vec{N}_B(\s, \tau).\qquad
\end{eqnarray} \esub
Calculation of the metric tensor $g_{\mu\nu} = \vec{e}_\mu \cdot \vec{e}_\nu$ illustrates that the reference frame is exactly orthonormalized only on the filament and in each transverse plane:
\begin{equation}\label{gcov_iso}
  g_{\mu \nu} = \left(
                  \begin{array}{cc}
                    \delta_{MN} &  \rho^A \eps_{AM} w \\
                    \rho^A \eps_{AN} w  & 1- 2 \rho^A \LA_A  + \rho^A \rho^B\left(w^2 \delta_{AB} + \LA_A \LA_B \right) \\
                  \end{array}
                \right).
\end{equation}
For points that do not lie on the filament, orthogonality of the basis vectors is therefore lost in the case of twisted filaments. Hence, we resort to a biorthogonal system with contravariant base vectors
\begin{align}\label{contrabase}
  e^A_i  =& \frac{\dd \rho^A}{\dd x^i}, &
  e^\s_i  =& \frac{\dd \s}{\dd x^i},
\end{align}
which automatically fulfill the biorthonormality condition
\begin{equation}\label{biorth}
  \vec{e}_\mu \cdot \vec{e}^{\,\nu} = \delta_\mu^\nu.
\end{equation}
For practical calculations, the metric tensor $\tens{g}$ is also needed in contravariant shape (upper indices) , with $g_{\mu\nu} g^{\nu \kappa} = \delta_\mu^\kappa$. The matrix inversion, which involves formal expansion in $\rho$, can be performed using a Neumann series, since $(1-\tens{g})^n = \OO(\rho^n)$:
\begin{equation}\label{neumannexp}
  \tens{g}^{-1} = \sum\limits_{n=0}^\infty (1-\tens{g})^n = \sum\limits_{n=0}^3 (1-\tens{g})^n +\OO(\rho^4).
\end{equation}
Applying Eq. \eqref{neumannexp} to expression \eqref{gcov_iso} yields, at least for an isotropic medium,
\bsub \begin{eqnarray}
    g^{AB} &=& \delta^{AB} + \rho^C \rho^D \eps^A_{\hs C} \eps^B_{\hs D}  w^2 + 2 \rho^C \rho^D \rho^E \eps^A_{\hs C} \eps^B_{\hs D} w^2 \LA_E  \\
    && \qquad \qquad \qquad + 3 \rho^C \rho^D \rho^E \rho^F w^2 \LA_E \LA_F \eps_{AC} \eps_{BD} + \OO(\rho^5), \nn \\
    g^{A \s} &=& \rho^C \eps^A_{\hs C}  w + 2 \rho^C \rho^D \eps^A_{\hs D} w \LA_C + 3 \rho^C \rho^D \rho^E \eps^A_{\hs C} w \LA_D\LA_E  +  \OO(\rho^4),\quad  \\
    g^{\s \s} &=& 1 + 2 \rho^C \LA_C + 3 \rho^C \rho^D  \LA_C \LA_D + 4 \rho^C \rho^D \rho^E \LA_C \LA_D \LA_E +  \OO(\rho^4). \qquad \qquad \label{gcon_iso}
\end{eqnarray} \esub
Here, the $g^{AB}$ components have been calculated up to fourth order in $\rho$, as terms up to this order are needed in the subsequent calculations.


\subsection{RDE expressed in coordinates adapted to the filament}

The present analysis will be valid for a general reaction-diffusion system in an unbound homogeneous and isotropic medium with three spatial dimensions. The general RDE in three spatial dimensions
\begin{equation}\label{RDE}
  \dd_t \uu =  \delta^{ij} \dd_i  \left(D_0\dd_j \HP \uu\right) + \mathbf{F}(\uu).
\end{equation}
is assumed to support rigidly rotating spiral wave solutions when acting in a two-dimensional plane; this condition is obviously not restrictive for the study of filaments. In standard polar coordinates $(\rho, \theta)$, such exact solution has the form $\uu_{0}(\rho, \theta-\omega_0 t)$, with $\omega_0$ depicting the spiral's natural rotation speed\footnote{$\omega_0$ is chosen positive for spirals that rotate counter-clockwise.}. In a co-rotating frame, $\uu_{0}$ satisfies following time-independent equation:
\begin{equation}\label{RDEcomov}
 \Delta_2 \HP \uu + \omega_0 \dd_\theta \uu + \mathbf{F}(\uu) =0.
\end{equation}

We now assert that the unknown exact scroll wave solution in three dimensions $\uu$ is well approximated by the spiral wave solution $\uu_{0}$ in each plane transverse to the filament; this ansatz will be checked afterwards to hold for a finite time interval. As a straight, untwisted filament is exactly matched by the spiral solutions, the corrections for a filament that is bent or twisted will grow with filament curvature $k$ and twist $w$ of the scroll wave. In analogy to the dual theory for wave fronts, we introduce a different perturbation parameter $\la$, which keeps track of curvature and twist corrections and is defined relatively to a characteristic length scale for filament thickness, denoted $d$:
\begin{equation}\label{def_lam}
  \la = \text{max}(kd, wd).
\end{equation}
At this point, we bring up our well-prepared curvilinear coordinate system $(\rho^A, \s)$ from paragraph \ref{subsec:curvilin1}. Regarding the scroll wave solution as a stack of spiral waves can now be stated mathematically as
\begin{equation}\label{Ansatz}
  \uu(\rho^A, \s, \tau) = \uu_{0}(\rho^A) + \la \tuu(\rho^A, \s, \tau) + \OO(\la^2).
\end{equation}
To remove ambiguity in the definition of $\tuu$ due to translational or rotational invariance, a gauge condition is imposed:
\begin{equation}\label{gaugetuu}
  \langle \bY^{\hat{\mu}} \mid \tuu \rangle = 0, \qquad(\hat{\mu} = 1,2, \theta).
\end{equation}
We additionally assume that the derivatives of the perturbation $\tuu$ along the filament are order $\la$ smaller than its spatial derivatives in transverse plane:
\begin{equation}\label{dsdlam}
  \dd_\s \tuu = \dd_A \tuu \, . \,\OO(\la).
\end{equation}\\

To proceed, we reexpress the system equations \eqref{RDE} in the new coordinates \eqref{coord_iso}. The chain rule for derivatives here implies, with $\dd_\theta = \eps_A^{\ \ B} \dd_B$,
\bsub \begin{eqnarray}
\dd_\tau &=& \dd_\tau x^i \dd_i + \dd_\tau t \dd_t = \dd_t + \dot{\vec{X}} \cdot \nabla + \omega \dd_\theta, \label{timediff} \\
\dd_A &=& \dd_A x^i \dd_i + \dd_\tau t \dd_t = \vec{N}_A \cdot \nabla, \\
\dd_\s &=& \dd_\s x^i \dd_i + \dd_\s t \dd_t = (1-\rho^A \LA_A) \vec{T} \cdot \nabla + w \dd_\theta.
\end{eqnarray} \esub
From the first of these expressions $\dd_t \uu$ can be linked to $\dd_\tau \uu$. As with wave fronts, we expand the reaction term as
\begin{equation}\label{reacterm}
 \mathbf{F}(\uu)  = \mathbf{F}(\uu_0) + \la \mathbf{F}'(\uu_{0}) \tuu + \frac{\la^2}{2} \tuu \bF(\uu_0) \tuu + \OO(\la^3).
\end{equation}
For a spatially varying electrical diffusion coefficient $D_0$, the diffusion term may be written
\begin{equation}\label{diff_term_iso0}
  \delta^{ij} \dd_i  \left(D_0\dd_j \HP \uu\right) = \delta^{ij} \left(\dd_i D_0 \right) \HP \dd_j \uu + D_0 \Delta \HP \uu.
\end{equation}
For a homogeneous isotropic medium, the first term is dropped, and the diffusion process can be expressed in the new coordinate system using the general expression for the Laplacian in curvilinear coordinates, with $g$ the determinant of the covariant metric tensor:
\begin{eqnarray}
\Delta &=&  \frac{1}{\sqrt{g}} \dd_\mu \left( \sqrt{g} g^{\mu\nu} \dd_\nu \right) =  \left(\frac{1}{\sqrt{g}} \dd_\mu \sqrt{g} \right) g^{\mu\nu} \dd_\nu + \dd_\mu g^{\mu\nu} \dd_\nu + g^{\mu\nu} \dd^2_{\mu\nu} \nn \\
&=& \Gamma^\alpha_{\mu \alpha} \dd_\nu + \dd_\mu g^{\mu\nu} \dd_\nu + g^{\mu\nu} \dd^2_{\mu\nu} = - \Gamma^\nu_{\alpha \beta} g^{\alpha \beta} \dd_\nu  + g^{\mu\nu} \dd^2_{\mu\nu}. \label{diffterm}
\end{eqnarray}
The last equality is justified by the Ricci identity \eqref{Ricci_id}; the coefficients $\Gamma$ represent the Christoffel symbol of the second kind (see appendix \ref{app:diff_geom}). Putting Eqs. \eqref{timediff}, \eqref{reacterm}, \eqref{diffterm} in the time-dependent RD equation using ansatz Eq. \eqref{Ansatz}, we obtain following expansion:
\begin{eqnarray}
&&\hspace{-0.5cm} \left(\dd_\tau - \HL\right) \tuu - (\omega-\omega_0)\left( \dd_\theta \uu_{0} + \la \dd_\theta \tuu\right)  - \dot{\vec{X}}\cdot \vec{N}^A \left( \dd_A \uu_{0} + \la \dd_A \tuu \right) \nn\\
&=& - \Gamma^A_{\alpha\beta} g^{\alpha\beta} \HP \dd_A \uu_{0} + \left( g^{AB} - \delta^{AB} \right) \HP \dd^2_{AB} \uu_{0}   \label{RDE_alg_iso} \\
&&  - \Gamma^\nu_{\alpha\beta} g^{\alpha\beta} \HP  \la \dd_\nu \tuu  +  g^{\mu\nu} \dd^2_{\mu\nu}  \HP \la \tuu - \delta^{AB} \dd^2_{AB}\HP  \la \tuu  + \frac{\la^2}{2} \tuu \bF(\uu_0) \tuu.  \nn
\end{eqnarray}
Here, we have absorbed the constant factor $D_0$ in the diffusion operator $\HP$ to relieve notations. The next step is to integrate these field evolution equations over the spatial coordinates transverse to the filament in order to obtain an vector evolution equation.  Projection on the translational response functions $\bra{\bY^A}$ will procure the EOM for the filament, whereas projection onto the rotational mode $\bra{\bY^\theta}$ brings about the corrections on the scroll wave's angular velocity $\omega$.

Before carrying out the explicit projections, we expand relevant terms in Eq. \eqref{RDE_alg_iso} in orders of $\rho$. In gradient expansion around the filament, expressions \eqref{gcon_iso} deliver for the Christoffel symbols of the first kind:
\begin{equation}\label{chr1_iso}
  \begin{array}{r@{\,=\,}l@{\ \ }r@{\,=\,}l}
    \Gamma_{A,BC} & 0, &
    \Gamma_{A, \s \s} & \LA_A + \dd_\s w \rho^B \eps_{BA} - \rho^B(w^2 \delta_{AB} + \LA_A \LA_B), \\
    \Gamma_{A,B\s} & \eps_{BA} w, &
    \Gamma_{\s, A \s} & - \LA_A + \rho^B (w^2 \delta_{AB} + \LA_A \LA_B),  \\
    \Gamma_{\s,BC} & 0, &
       \Gamma_{\s,\s\s} & -\rho^A \dd_\s \LA_A + \rho^A \rho^B \left( w \dd_\s w \delta_{AB} + \LA_A \dd_\s \LA_B \right).
  \end{array}\quad
\end{equation}
With these formulae we calculate, up to third order in $\rho$, following quantity in Eq. \eqref{RDE_alg_iso}:
\begin{eqnarray}
  \Gamma^A_{\alpha\beta} g^{\alpha\beta} &=& \LA^A + \rho^B \left( w^2 \delta^A_{B} + \dd_\s w \eps_B^{\hs A}+ \LA^A \LA_B \right)
  + \rho^C \rho^D \left(-w^2 \LA_B \eps_C^{\hs A} \eps_{D}^{\ \ B} \right. \nn \\ && \left. + 2 w^2 \LA_C \delta^A_{D} +2 \dd_\s w \LA_C \eps_D^{\hs A} + \dd_\s \LA_D \eps_C^{\hs A} w + \LA^A \LA_C \LA_D  \right) \nn \\
  && + \rho^C \rho^D \rho^E \left(3 \dd_\s w \eps_{E}^{\hs A} \LA_C \LA_D+ 3 w^2 \delta_C^{\hs A} \LA_D \LA_E + 3 \eps_D^{\hs A} w \dd_\s \LA_E \LA_C  \right. \nn\\
  && \left.  -3 w^2 \eps_E^{\ \ B}\LA_B \eps_D^{\hs A} \LA_C   + \LA^A \LA_C \LA_D \LA_E  \right) + \OO(\rho^4). \label{gGamg}
\end{eqnarray}
The equivalent term with free index $\s$ is needed for stability analysis of the perturbative field correction $\tuu$. We evaluate up to first order in $\rho$:
\begin{equation}\label{gGamgs}
   \Gamma^\s_{\alpha\beta} g^{\alpha\beta} =  \rho^B w \LA_A  \eps_B^{\ \ A} - \rho^A  \dd_\s \LA_A.
\end{equation}
Now we are ready to derive the filament equation of motion by projecting Eq. \eqref{RDE_alg_iso} on the translational and rotational RFs.

\subsection{Projection on the translational mode}
After projection onto the translational response function $\bra{\bY^B}$, the first term in Eq. \eqref{RDE_alg_iso} disappears due to the gauge condition \eqref{gaugetuu} and time invariance of the eigenmodes in the rotating reference frame:
\begin{equation}\label{RDEterm1}
 \langle \bY^B \mid \left(\dd_\tau - \HL\right)\mid \tuu \rangle = \dd_\tau \langle \bY^B \mid \tuu \rangle - \eps_A^{\hspace{7pt}B} \langle \bY^A \mid \tuu \rangle = 0.
\end{equation}
This relation brings the projection of Eq. \eqref{RDE_alg_iso} to
\begin{equation}\label{EOMtr1}
  \dot{\vec{X}}\cdot \vec{N}^B =  \bra{\bY^B}  \Gamma^A_{\alpha\beta} g^{\alpha\beta} \HP  \ket{\bpsi_A} -  \bra{\bY^B} \left( g^{AC} - \delta^{AC} \right) \HP \dd_C  \ket{\bpsi_A} + (\text{Rem}_T)^B,
\end{equation}
with the translational remainder term encompassing the surviving contributions of the field perturbation $\tuu$, which are at least $\OO(\la^2)$:
\begin{eqnarray}
(\text{Rem}_T)^B &=& - \la \dot{\vec{X}}\cdot \vec{N}^A  \langle \bY^B \mid \dd_A \tuu \rangle +  \la \bra{\bY^B} \Gamma^\nu_{\alpha\beta} g^{\alpha\beta} \HP   \ket{\dd_\nu \tuu}\nn \\
&& - \la \bra{\bY^B}  g^{\mu\nu} \HP \ket{\dd^2_{\mu\nu} \tuu} + \la \bra{\bY^B} \delta^{AC} \HP \ket{\dd^2_{AC} \tuu} \quad \label{RemT} \\
&& - (\omega-\omega_0) \la \langle \bY^B \mid \dd_\theta \tuu \rangle -   \frac{1}{2} \la^2 \langle \bY^B \mid \tuu \mathbf{F}''(\uu_0) \tuu \rangle. \nn
\end{eqnarray}
In zeroth order in $\rho$ one has $\Gamma_{A,\s\s} (\rho=0) = \LA_A$. Since $\vec{e}_A$ coincides with the vector $\vec{N}$ on the filament, we obtain 
\begin{equation}\label{EOMtr2}
  \dot{\vec{X}}\cdot \vec{N}^B =  \LA^A \bra{\bY^B} \HP  \ket{\bpsi_A} + \OO(\la^2).
\end{equation}

\subsection{Convention for denoting matrix elements}
For bookkeeping the different inner products of the type $\bra{{\bY}^{\cdot}} \ldots \ket{{\bpsi}_{\cdot}}$, we henceforth use a generic notation, which is detailed in the first section of appendix \ref{app:isotropic_tensors}. Inspired by overlap integrals in quantum mechanics, we shall also refer to these overlap integrals as `matrix elements'. In our custom notation, a matrix element is indicated by the operator included between the brackets (e.g. $\HP$ or $\hat{I}$), to which a superscript is added that indicates either projection on a translational $(P^T)$ or rotational $(P^R)$ response function. The superscript also contains a numerical value, labeling the number of coordinate functions $\rho^A \rho^B$ that are included before integration. For the matrix element in expression \eqref{EOMtr2}, our rules generate the notation
\begin{equation}\label{matrix_el_PT0}
  \bra{\bY^B} \HP  \ket{\bpsi_A} = (P^{T0})^B_{\hs A}.
\end{equation}
Taking also into account the gauge condition $\dot{\vec{X}} \cdot \vec{T} =0$
enables to deduce from Eq. \eqref{EOMtr2} that
\begin{equation}\label{EOMtr3}
  \dot{\vec{X}} =  \LA_A (P^{T0})^A_{\hspace{7pt}B}\vec{N}^B + \OO(\la^2).
\end{equation}
Reverting to an inertial frame of reference, we have with $\LA_A = k\vec{N}\cdot \vec{N}_A$,
\begin{equation}\label{EOMtr4}
  \dd_t X^j =  k N^i \left(N^A_i \LA_A (P^{T0})^{AB} N_B^j \right) + \OO(\la^2).
\end{equation}

\subsection{Temporal or rotational averaging of the EOM}
Since the vectors $N^A_i(\tau)$ are tied to the scroll wave's rotation phase, these functions oscillate in time with the scroll wave's local rotation frequency $\omega(\s,t)$ when viewed in the laboratory frame of reference. The angle between the instantaneous filament velocity vector $\dot{\vec{X}}$ and the normal vector $\vec{N}$ therefore also changes quasi-periodically, such the intersection of the instantaneous filament with a transverse plane yields a cycloidal trajectory. We can remove this epicycle movement by averaging the EOM \eqref{EOMtr4} over the instantaneous spiral period $T = 2\pi/\omega$, akin to the approach by Biktashev \etal., who included the time-averaging in their definition of the inner product \cite{Biktashev:1994}; compare to our definition \eqref{def_inner2D}.

We will adopt an alternative approach here, which decomposes the occurrent tensor in its isotropic components to generate automatically the EOM averaged over an entire spiral wave period. Referring to Appendix \ref{app:isotropic_tensors} for a procedure to acquire rotationally invariant tensor components, we may thereafter state
\begin{equation} \label{decPT0}
 \bra{\bY^B} \HP  \ket{\bpsi_A} = (P^{T0})^B_{\hs A} =  P^{T0}_s \delta_A^{\hs B} + P^{T0}_a \eps_A^{\hs B}.
\end{equation}
With this decomposition, EOM \eqref{EOMtr3}  reduces to
\begin{equation}
\dot{\vec{X}} =  P^{T0}_s \LA_A \vec{N}^A + P^{T0}_a \LA_B \eps_A^{\hs B} \vec{N}^A  + \OO(\la^2).
\end{equation}
In an inertial frame of reference, Eq. \eqref{EOMtr4} becomes, with respect to a Frenet frame,
\begin{equation} \label{EOMtr_la1}
\dd_t \vec{X} =  \gamma_1 k \vec{N} +  \gamma_2 k \vec{B}  + \OO(\la^2).
\end{equation}
where we have defined
\begin{align} \label{defgamma12}
\gamma_1 =& P^{T0}_s,& \gamma_2 =& P^{T0}_a.
\end{align}
This lowest order result in $\la$, which we have published in \cite{Verschelde:2007}, matches the equations obtained by Keener \cite{Keener:1988} and Biktashev et al. \cite{Biktashev:1994}, with identifications $(P^{T0}_s, P^{T0}_a) = (b_2, c_3)$. The sign convention for $\gamma_2 $ used throughout this work differs from the one previously adopted in \cite{Verschelde:2007}, here delivering in only plus signs in the EOM \eqref{EOMtr_la1}.

The leading terms in the filament EOM may be written without reference to the Frenet frame using
\begin{equation} \label{EOMtr_la1int}
\dd_t \vec{X} =  \gamma_1\ \dd_\s^2 \vec{X} + \gamma_2\ \dd_\s \vec{X} \times \dd_\s^2 \vec{X}  + \OO(\la^2).
\end{equation}
Identification with the ribbon model prescription \eqref{ribbon1} brings that
\begin{align}
  \verb"a"_1 &= \gamma_1 = P^{T0}_s,&          \verb"a"_2 &= \gamma_2 = P^{T0}_a.
\end{align}

\subsection{Projection on the rotational mode}

Taking up our calculation in Eq. \eqref{RDE_alg_iso}, we now project the obtained field equations onto the rotational response function
$\bra{\bY^\theta}$. Relying once more on the gauge condition \eqref{gaugetuu} and time invariance of the angular response function, the first term in \eqref{RDE_alg_iso} vanishes here too:
\begin{equation}\label{RDEterm1_rot}
  \langle \bY^\theta \mid \left(\dd_\tau - \HL\right) \mid \tuu \rangle = \dd_\tau \langle \bY^{\theta} \mid \tuu \rangle - \langle \mathbf{0} \mid \tuu \rangle = 0.
\end{equation}
This simplifies the result of projection to
\begin{equation}\label{EOMth1}
  \omega-\omega_0 =   \bra{\bY^\theta}  \Gamma^A_{\alpha\beta} g^{\alpha\beta} \HP  \ket{\bpsi_A} -  \bra{\bY^\theta}   \left( g^{AB} - \delta^{AB} \right) \HP \dd_B  \ket{\bpsi_A} + \text{Rem}_\theta,
\end{equation}
where the remainder term groups all non-vanishing contributions of the field perturbation $\tuu$, which are at least $\OO(\la^2)$:
\begin{eqnarray}
\text{Rem}_\theta &=& - \la \dot{\vec{X}}\cdot \vec{N}^A  \langle \bY^{\theta} \mid \dd_A \tuu \rangle - \la (\omega-\omega_0) \dd_\theta \tuu
 \la \bra{\bY^\theta} \Gamma^\nu_{\alpha\beta} g^{\alpha\beta} \HP   \ket{\dd_\nu \tuu}\nn \\
&& - \la \bra{\bY^\theta}  g^{\mu\nu} \HP  \ket{\dd^2_{\mu\nu} \tuu} + \la \bra{\bY^\theta} \delta^{AB} \HP \ket{\dd^2_{AB} \tuu}.  \label{remth}
\end{eqnarray}
To extract the true phase evolution equation, we re-use expressions \eqref{gGamg}, and decompose the resulting matrix elements:
\bsub \label{dec_rot_lowest}\begin{eqnarray}
(P^{R1})_A^{\hs B} \left( w^2 \delta^A_{\hs B}  + \LA^A \LA_B  \right) &=& (2 w^2 +k^2) P^{R1}_s, \qquad \\
(P^{R1})_A^{\hs B} \dd_\s w \eps_B^{\hs A}  &=&  - 2 \dd_\s w P^{R1}_a, \qquad \qquad \\
(P^{R2 \delta })_{AB}^{\hs\hs CD} w^2 \eps^A_{\hs C} \eps^B_{\hs D} &=& 2 w^2 \left(P^{R2 \delta }_{00} - P^{R2 \delta }_{s} \right).  \qquad
\end{eqnarray} \esub
Forging all elements together delivers following form for the time and position-dependent angular velocity
\begin{eqnarray}\label{EOMrot_iso_lowest}
    \omega &=& \omega_0 + a_0 w^2 + b_0 k^2 +d_0 \dd_\s w,
\end{eqnarray}
with coefficients given by linear combinations of matrix elements:
\bsub \label{coeffEOMrot_iso_lowest}
\begin{eqnarray}
          a_0 &=& 2 \left( P^{R1}_s + P^{R2\delta}_{00} - P^{R2\delta}_{s} \right), \\
          b_0 &=& 2 P^{R1}_s, \\
          d_0 &=& - 2 P^{R1}_{a}.
\end{eqnarray} \esub
We observe that, up to order $\la^2$, three dynamical coefficients determine how much a scroll wave's rotation speed $\omega$ deviates from its natural spiral frequency $\omega_0$. Our parameters $a_0$ and $d_0$ are known from Keener's and Biktashev's works, where they were denoted $a_1$ and $b_1$, respectively. However, the $k^2$ term fell beyond the scope of their derivation.

The evolution equation for twist along a filament in the given order directly follows from Eq. \eqref{EOMrot_iso_lowest}:
\begin{eqnarray}\label{EOMtw_iso1}
   \dot{w} &=&  \dd_\s \omega_0 + 2 a_0\,  w \dd_\s w + b_0\, \dd_\s{k^2} + d_0\,   \dd^2_\s w.
\end{eqnarray}
We remark that the $k^2$ term in Eqs. \eqref{EOMrot_iso_lowest}-\eqref{EOMtw_iso1} provides a first example of the coupling of the translational mode to the rotational degrees of freedom.

\section[The extended EOM for filaments in an isotropic medium]{The extended equation of motion for filaments \\in an isotropic medium \label{sec:extEOMfiliso}}

The higher order contributions in the filament EOM are of two disparate types: some corrections are obtained from evaluating the diffusion term to regions further away from the filament core, whereas other motion effects stem from local accommodation $\tuu$ of the standard spiral wave profile to the geometric perturbations.

\subsection[Higher order translation effects]{Higher order translation effects in the EOM\\ from expanding the diffusion term}

As our calculation involves symmetrized terms, we make use of underlined indices from now on to indicate that all possible index permutations need to be added, e.g.
\bsub \label{conv_uline} \begin{eqnarray}
 T_{\uA \uB} &=& T_{AB} + T_{BA}, \\
 T_{\uA \uB \uC} &=& T_{ABC} + T_{CAB} + T_{BCA} + T_{ACB} + T_{BAC} + T_{CBA}.
\end{eqnarray} \esub
Remark that no combinatorial normalization factor is included in the definition.

Using the expansion \eqref{gGamg}, the continuation to \eqref{EOMtr_la1} becomes, when neglecting $\tuu$ terms:
\begin{eqnarray}\label{EOMtr5}
  \dot{\vec{X}}\cdot \vec{N}^B &=&  \LA^A \bra{\bY^B} \HP  \ket{\bpsi_A} + \left( w^2 \delta^A_F + \dd_\s w \eps_{F}^{\hs A}+ \LA^A \LA_F \right) \bra{\bY^B}  \rho^F \HP  \ket{\bpsi_A} \nn \\
  && + \left(-w^2 \LA_E \eps_C^{\hs A} \eps_{D}^{\hs E} + 2 w^2 \LA_C \delta^A_{D} +2 \dd_\s w \LA_C \eps_{D}^{\hs A} \right. \nn \\
  && \left. + \dd_\s \LA_D \eps_{C}^{\hs A} w  + \LA^A \LA_C \LA_D \right)  \bra{\bY^B}   \rho^C \rho^D \HP \ket{\bpsi_A} \nn \\
  && - 2 \eps^A_{\hs \uC} \eps_{F \uD} w^2 \LA_{\uE}  \bra{\bY^B}   \rho^C \rho^D \rho^E \HP \dd_F \ket{\bpsi_A} + \OO(\la^5).
\end{eqnarray}
The `remainder term' that includes matrix elements with $\ket{\tuu}$ have been omitted here, as such contributions will be evaluated in subsequent paragraphs. Due to the fact that there exist no isotropic tensors of odd rank, $ \bra{\bY^B}  \rho^F  \ket{\bpsi_A} = 0$ and the terms that are one order higher in $\rho$ than given in \eqref{EOMtr5} vanish as well. Again, we can break the matrix elements down to their isotropic content:
\bsub \label{dec4all}\begin{eqnarray}
(P^{T2})_A^{\hs BCD}\left(-w^2 \LA_E \eps_{CA} \eps_D^{\hs E} \right) &=& w^2 \LA^B \left(- P^{T2}_{00} + P^{T2}_s \right) \nn \\
 &&+ w^2 \eps^{EB} \LA_E \left(- P^{T2}_{20} + P^{T2}_{a} \right), \qquad \\
(P^{T2})_A^{\hs BCD}\left( 2w^2 \LA_C \delta_{AD} \right) &=& 2 w^2 \LA^B \left(P^{T2}_{00} + P^{T2}_s \right) \nn \\
&&+ 2 w^2 \eps^{CB} \LA_C  \left(P^{T2}_{20} + P^{T2}_{a} \right), \qquad \\
(P^{T2})_A^{\hs BCD}\left(2 \dd_\s w \LA_C \eps_{DA} \right) &=& 2 \dd_\s w \LA_C \eps^{CB} \left(P^{T2}_{00} - P^{T2}_s \right) \nn \\
&&+ 2 \dd_\s w \LA^B \left(- P^{T2}_{20} + P^{T2}_{a} \right), \qquad \label{dec4c}\\
(P^{T2})_A^{\hs BCD}\left(w \dd_\s\LA_D \eps_{CA} \right) &=& w \dd_\s \LA_C \eps^{CB} \left(P^{T2}_{00} - P^{T2}_s \right) \nn \\
&&+ w \dd_\s \LA^B \left(- P^{T2}_{20} + P^{T2}_{a} \right), \qquad
\end{eqnarray}
\begin{eqnarray}
(P^{T2})_A^{\hs BCD}\left(\LA_A \LA_C \LA_D \right) &=& k^2 \LA^B \left(P^{T2}_{00} + \frac{1}{2} P^{T2}_s \right) \nn \\
&&+ k^2   \eps^{CB} \LA_C \left( P^{T2}_{20} + \frac{1}{2}  P^{T2}_{a} \right). \qquad
\end{eqnarray} \esub
For the last term in \eqref{EOMtr5}, we rely on decomposition \eqref{R6b} of a rank 6 tensor with (12)(456) symmetry, which yields
\begin{align}
 (P^{T3\delta})_{AF}^{\hs \hs BCDE} \left(  - 2 w^2 \eps^A_{\hs \uC} \eps^F_{\hs \uD} \LA_{\uE} \right) =& & \label{dec6}\\
  \frac{4}{3} w^2 \LA^B \left(  -2 P^{T3\delta}_{000} + P^{T3\delta}_{s0s} + P^{T3\delta}_{s0} \right)
   + &\frac{4}{3} w^2 \eps^{CB} \LA_C \left( 2 P^{T3\delta}_{020} +  P^{T3\delta}_{a0a} - P^{T3\delta}_{a0} \right).\nn
\end{align}


The advanced EOM \eqref{EOMtr5} can now be substantiated by plugging in Eqs. \eqref{dec4all} and \eqref{dec6}. Due to the transverse gauge $\dot{\vec{X}}\cdot \vec{N}^A = 0$, an expression of the type $\dot{\vec{X}} \cdot \vec{N}^A = T^A$ is found equivalent to $T^A \vec{N}_A = (\dot{\vec{X}} \cdot \vec{N}^A) \vec{N}_A = \dot{\vec{X}}$. Furthermore we use
\begin{align}
 \LA^A &= k \vec{N} \cdot \vec{N}^A, & \dd_\s \vec{N}^A &= w \eps^A_{\hs B} \vec{N}^B = w \vec{T} \times \vec{N}^A
\end{align}
to deduce that
\bsub \label{vecN2dX}\begin{eqnarray}
\LA^B \vec{N}_B &=& (k\vec{N} \cdot \vec{N}^B) \vec{N}_B = \dd_\s^2 \vec{X},\\
\LA^E \eps_E^{\hs B} \vec{N}_B &=& (k\vec{N} \cdot \vec{N}^B \times \vec{T}) \vec{N}_B = \dd_\s \vec{X} \times \dd_\s^2 \vec{X},\\
\dd_\s \LA^C \vec{N}_C &=& [\dd_\s (k\vec{N})]_\perp + w [ k\vec{N} \cdot (\vec{T} \times \vec{N}^A) ] \vec{N}_A \nn \\
&=& [\dd_\s^3\ \vec{X}]_\perp - w  \dd_\s \vec{X} \times \dd_\s^2 \vec{X}, \label{vecN2dXc}\\
\dd_\s \LA^C \eps_C^{\hs B} \vec{N}_B &=& \dd_\s \LA^C (\vec{T} \times \vec{N}_C)
= \dd_\s \vec{X} \times \dd_\s^3 \vec{X} + w \dd_\s^2 \vec{X}.\quad
\end{eqnarray} \esub\\

Inserting expressions \eqref{vecN2dX} into Eqs. \eqref{dec4all}-\eqref{dec6} and subsequent summation finally delivers a dynamical equation for $\dot{\vec{X}}$. Therein, following coefficients appear:
\begin{equation}
        \begin{array}{r@{\,=\,}l r@{\,=\,}l }
          \gamma_1 & P^{T0}_s & \gamma_2 & P^{T0}_a, \\
          a_1 & 2 \left( P^{T2}_{00} + P^{T2}_{s}\right)  &
          a_2 & 2 \left( P^{T2}_{20} + P^{T2}_{a}\right) \\
          \multicolumn{2}{c}{\qquad +\frac{4}{3} \left(  - 2 P^{T3\delta}_{000} +  P^{T3\delta}_{s0s} + P^{T3\delta}_{s0} \right),\quad } &
          \multicolumn{2}{c}{\qquad  +\frac{4}{3} \left( 2 P^{T3\delta}_{020} + P^{T3\delta}_{a0a} - P^{T3\delta}_{a0} \right), } \\
          b_1 &  \left(P^{T2}_{00}   + \frac{1}{2} P^{T2}_{s} \right), &
          b_2 &  \left(P^{T2}_{20}   + \frac{1}{2} P^{T2}_{a} \right), \\
          c_1 & - P^{T2}_{20} + P^{T2}_{a},  &
          c_2 & \left(P^{T2}_{00} - P^{T2}_{s} \right), \\
          d_1 & 2 \left( P^{T2}_{a} - P^{T2}_{20} \right), & d_2 & 2 \left( P^{T2}_{00} - P^{T2}_{s} \right).
        \end{array}\label{coeffEOMtr_iso} \qquad
\end{equation}
We observe that the some of these ten dynamical coefficients are related:
\begin{align} \label{eq_cd}
 d_1 &= 2 c_1,& d_2 &= 2c_2.
\end{align}
Therefore, only eight model-specific parameters determine the translational movement of the filament up to third order in curvature and twist (albeit in the limit of negligible remodeling of the spiral wave profile):
\begin{eqnarray}
  \dot{\vec{X}} &=& \left(\gamma_1  + a_1 w^2 + b_1 k^2  + d_1 \dd_\s w \right)\ \dd^2_\s \vec{X} \nn\\
&& +  \left(\gamma_2  + a_2 w^2 + b_2 k^2  + d_2 \dd_\s w \right)\ \dd_\s \vec{X} \times \dd^2_\s \vec{X}   \nn\\
&&  + c_1 w [\dd^3_\s \vec{X}]_\perp + c_2 w  \dd_\s \vec{X} \times \dd^3_\s \vec{X}.\quad \label{ribbon2}
\end{eqnarray}
Direct comparison with the ribbon model \eqref{ribbon1} indicates that their $\verb"d"_1, \verb"d"_2$ coefficients explicitly follow from the list \eqref{coeffEOMtr_iso}:
\begin{align}
 \verb"d"_1 &= c_2,&  -\verb"d"_2 &= c_1.
\end{align}
However, the curvature and twist corrections to the filament tension in our law \eqref{ribbon2} are not present in the phenomenological ribbon model.

If desired, one may also restate \eqref{ribbon2} relative to the Frenet-Serret frame; the outcome is
\bsub \label{EOMtr_iso1} \begin{eqnarray}
\dot{\vec{X}} \cdot \vec{N} &=& \left(\gamma_1  + (a_1+c_2) w^2 + b_1 k^2  - c_2 w \tau_g  + d_1 \dd_\s w \right)k  + c_1 w \dd_\s k, \qquad \quad \\
\dot{\vec{X}} \cdot \vec{B} &=&  \left(\gamma_2 + (a_2-e_1)  w^2 + b_2  k^2 + c_1 w \tau_g  + d_2 \dd_\s w\right) k + e_2 w \dd_\s k. \qquad \quad
\end{eqnarray} \esub

The discussion of Eqs. \eqref{ribbon2}, \eqref{coeffEOMtr_iso} is postponed to section \ref{sec:discuss_EOMiso}, for we will first show how to update the phase evolution equation in a similar way.

\subsection[Higher order rotation effects]{Higher order rotation effects in the EOM\\ from expanding the diffusion term}

In the RDE that has been projected on the unique rotational RF $\bra{\bY^\theta}$, i.e. Eq. \eqref{EOMth1}, the following higher order corrections emerge:
\bsub \label{dec_rot_all}\begin{eqnarray}
(P^{R3})_A^{\hs ECD} \left(3 \dd_\s w \eps_E^{\hs A} \LA_C \LA_D \right) &=& - 2 k^2 \dd_\s w P^{R3}_{20},\qquad \\
(P^{R3})_A^{\hs CDE} \left(3 w^2 \delta^A_{\hs C}  \LA_D \LA_E \right) &=& 2 k^2 w^2 P^{R3}_{00}, \qquad \\
(P^{R3})_A^{\hs DCE} \left(3 w \eps_D^{\hs A} \dd_\s \LA_E \LA_C \right) &=& - 6 w k \dd_\s k P^{R3}_{20},\qquad \\
(P^{R3})_A^{\hs DCE} \left(3 w^2 \eps_D^{\hs A} \LA_C \eps_{EB} \LA^B \right) &=& 0, \qquad \\
(P^{R3})_A^{\hs CDE} \LA_A \LA_C \LA_D \LA_E &=& 2 k^4  P^{R3}_{00},\\
(P^{R2 \delta })_{AB}^{\hs\hs CD} w^2 \eps^A_{\hs C} \eps^B_{\hs D} &=& 2 w^2 \left(P^{R2 \delta }_{00} - P^{R2 \delta }_{s} \right),  \qquad \\
(P^{R4 \delta })_{AB}^{\hs\hs CDEF} \left(- 3 w^2 \eps^A_{\hs C} \eps^B_{\hs D} \LA_E \LA_F \right) &=& 3 w^2 k^2 \left( - 2 P^{R4 \delta}_{000} + P^{R4 \delta}_{s0} \right). \qquad \quad
\end{eqnarray} \esub
As a result, the extended phase evolution equation becomes (omitting wave accommodation effects):
\begin{eqnarray}\label{EOMrot_iso}
    \omega &=& \omega_0 + a_0 w^2 + b_0 k^2 +d_0 \dd_\s w \nn \\
                  & & + k^2 \left(a_0^{(1)} w^2 + b_0^{(1)} k^2 +d_0^{(1)} \dd_\s w  \right) + e_0^{(1)} w k \dd_\s k.
\end{eqnarray}
The coefficients are given by overlap integrals between the GMs and RFs:
\begin{equation}
        \begin{array}{r@{\,=\,}l r@{\,=\,}l }
          a_0 & 2 \left( P^{R1}_s + P^{R2\delta}_{00} - P^{R2\delta}_{s} \right), &
          a_0^{(1)} & 2  P^{R3}_{00} - 6 P^{R4 \delta}_{000} + 3 P^{R4 \delta}_{s0},\\
          b_0 & 2 P^{R1}_s, &
          b_0^{(1)} & 2 P^{R3}_{00}, \\
          d_0 & - 2 P^{R1}_{a},  &
          d_0^{(1)} & -2 P^{R3}_{20},  \\
          \multicolumn{2}{c}{}  &
          e_0^{(1)}  &  -6 P^{R3}_{20}.
        \end{array}\label{coeffEOMrot_iso}
\end{equation}
With the extended phase evolution equation, following twist evolution equation is associated:
\begin{eqnarray}\label{EOMtw_iso2}
     \dd_t w =  (a_0 + k^2 a_0^{(1)}   ) 2 w \dd_\s w + (b_0+ k^2 b_0^{(1)} ) \dd_\s{k^2} +(d_0 + k^2 d_0^{(1)} ) \dd^2_\s w \quad \\
     \qquad  \quad + 2 k \dd\s k \left(a_0^{(1)} w^2 + b_0^{(1)} k^2 +( d_0^{(1)}+ e_0^{(1)} ) \dd_\s w  \right) + \frac{1}{2}e_0^{(1)} w \dd_\s (k^2). \nn
\end{eqnarray}
However, Eqs. \eqref{ribbon2}, \eqref{EOMtw_iso1}, \eqref{EOMtw_iso2} are still incomplete, as the wave accommodation effect has not yet been taken into account.

\subsection{Quantification of the correction term $\tuu$ \label{sec:breather}}

Similar to our treatment of wave fronts, we proceed by estimating the correction term $\tuu$. For the first time, we are quantitatively investigating the feedback mechanisms between the perturbed scroll wave shape and motion of the filament. For, curvature and twist of the filament not only invoke net translation or rotation, but also cause deformations that cannot be attributed to excitation of the Goldstone modes. Although these deviations from the unperturbed spiral wave profile do not affect filament motion in lowest order, they are seen to cause drift in higher order by coupling to the translational and rotational RFs, similar to the resonant drift process studied by Biktashev \etal in \cite{Biktashev:1995b}.

Our present aim is to substantiate the initial ansatz \eqref{Ansatz} to
\begin{equation}\label{uu012}
  \uu(\rho, \s, \tau) = \uu_0(\rho^A) + \la(\s, \tau) \uu_1(\rho^A) + \la^2(\s, \tau) \uu_2(\rho^A) + \OO(\la^3).
\end{equation}
The formal notation $\la(\s, \tau)$ is used to emphasize that in a reasonable approximation, the $\s$ and $\tau$ dependence of the wave adjustment may be realized through $k(\s, \tau)$ and $w(\s,\tau)$.

To start with, we recapitulate the evolution equation \eqref{RDE_alg_iso} for $\tuu$, up to second order in $\la$:
\begin{eqnarray}\label{eq_tuu_fil}
\left(\dd_\tau -  \HL\right) \tuu &=&  (\omega-\omega_0) \dd_\theta \uu_0 + \left( \dot{\vec{X}}\cdot \vec{N}^A - \LA^A \HP \right)  \dd_A (\uu_0 + \la \tuu) \\
&& - (w^2 \delta_{AB} + \dd_\s w \eps_{BA} + \LA_A \LA_B) \rho^B \bP \dd_A \uu_0 \nn \\
&& - \eps_{AC} \eps_{BD} w^2 \rho^C \rho^D \bP \dd^2_{AB} \uu_0 \nn + \frac{\la^2}{2}  \tuu \mathbf{F}''(\uu_0) \tuu + \OO(\la^3).
\end{eqnarray}
The sole term that is first order in $\la$ on the right-hand side is $- \LA^A \bP \dd_A \uu_0$. The lowest order correction $\tuu$ thus follows from
\begin{equation}\label{tuu_lowest}
   \left(\dd_\tau -  \HL\right) \ket{\tuu} = - \LA_A(\tau) \HP \ket{\bpsi_A}.
\end{equation}
It is convenient here to perform the spectral decomposition of the operator $\HL$. In analogy to Eq. \eqref{eigmodes_L_fr} we put
\begin{align} \label{eigmodes_L_fil}
 \HL  \ket{\bpsi_{[j]}} &= \zeta_{[j]}  \ket{\bpsi_{[j]}}, &   \bra{\bY^{[j]}} \HL &= \zeta^*_{[j]}  \bra{\bY^{[j]}},
\end{align}
with $\{\zeta_{[j]}\}$, $\left([j] \in \{\theta, x, y, 1, 2, \ldots\}\right)$ the eigenvalue spectrum of $\HL$. We can filter out the GMs in \eqref{tuu_lowest} by left-multiplication with $\sum_{j \geq 1} \ket{\bpsi_{[j]}} \bra{\bY^{[j]}}$ to obtain
\begin{equation}\label{tuu_lowest_L0}
   \left(\dd_\tau -  \HL_0 \right) \ket{\tuu} = \LA_A(\tau)  \bra{\bY^{[j]}} \HP \ket{\bpsi_A} \ \ket{\bpsi_{[j]}},
\end{equation}
where we have used gauge condition \eqref{gaugetuu}. We have yet implicitly brought in the invertible operator
\begin{equation} \label{def_L0fil}
\HL_0 = \sum_{j \geq 1} \zeta_{[j]} \ket{\bpsi_{[j]}} \bra{\bY^{[j]}}.
\end{equation}
In our reference frame that rotates together with the spiral, the normal vector to the filament is seen to rotate. Its rotation is opposite to the spiral wave's rotation sense, such that $\LA^A$ exhibits a dominant temporal dependency $\LA^A(\tau) =  \mathrm{Re}(\LA^A_0 e^{-i\omega \tau})$. Therefore, Eq. \eqref{tuu_lowest_L0} can be solved exactly using the eigenmode expansion
\begin{equation}\label{tuu_lowest_ansatz}
\ket{ \tuu } = c_{[j]} e^{-i \omega \tau} \ket{\bpsi_{[j]}},
\end{equation}
which delivers
\begin{equation}\label{tuu_cj}
 c_{[j]}  =  - \LA^A_0 \frac{-\zeta_{[j]}+ i \omega}{\zeta^2_{[j]} + \omega^2} \bra{\bY^{[j]}} \HP \ket{\bpsi_A}.
\end{equation}
In polar notation for the complex plane, one may put $\tan(\alpha_{[j]}) =\frac{{\rm Im} \zeta_{[j]}-\omega}{{\rm Re} \zeta_{[j]}}$ and $r_{[j]} = (\zeta_{[j]}^2 + \omega^2)^{-1/2}$ such that
\begin{equation}\label{tuu_complex}
\ket{\tuu}  = - \sum \limits_{j \geq 1} \LA^A r_{[j]} \cos \alpha_{[j]} \bra{\bY^{[j]}} \HP \ket{\bpsi_A} \ket{\bpsi_{[j]}} + \OO(\la^2).
\end{equation}
Hence the standard spiral wave profile $\uu_0(\rho^A)$ is in first order in $\la$ complemented with the periodical function \eqref{tuu_complex}. The components of this disturbance are rotationally shifted over angles $\alpha_{[j]}$ with respect to the external perturbation vector $\vec{N}$, due to the finite lifetime $- 1/\mathrm{Re}\, \zeta_{[j]}$ of the decaying modes $\ket{\bpsi_{[j]}}$. \\

From Eq. \eqref{tuu_complex} may be furthermore noted that the lowest order correction vanishes for equally diffusive systems, which have $\HP = \hat{I}$. Moreover, only the evanescent modes with eigenvalues that have a (negative) real part close to zero significantly contribute to the modification of the wave profile. In a sense, the situation is reminiscent to quantum mechanical scattering theory, in which bound states only affect the scattering properties if they have a (negative) energy eigenvalue close to zero. Translated to the actual context of spiral-shaped activation patterns, examples of bound states which have eigenvalues close to zero are the so-called meandering modes \cite{Hakim:2002}, which cause non-trivial tip trajectories (remind Fig. \ref{fig:tip_meander}) as soon as their eigenvalues cross the imaginary axis towards the positive half-plane. Just before the meandering sets in, these modes exhibit eigenvalues for which $|\zeta| \not \ll |\omega_0|)$, and thereby substantially contribute to the wave shape via Eq. \eqref{tuu_complex}.

The modulation \eqref{tuu_complex} of the spiral profile in the transverse plane occurs at the scroll wave's rotation frequency $\omega$ and is therefore in good approximation periodic. This inspires us to call the emergent oscillating deformation a \textit{breather} mode; such modes typically appear in non-linear dynamical systems, e.g. in the sine-Gordon model \cite{Ablowitz:1973}. Strictly speaking, oscillating modes need to be localized in space to qualify as a breather; we have not yet proven this for the mode considered here.\\

The dominant breathing modes can be expected to interact with the translational and rotational zero modes. Fortunately, the awkward time dependency which they are expected to introduce in the dynamical coefficients for the filament EOM is likely to simplify by introducing additional internal degrees of freedom for the scroll wave (i.e. amplitude and phase for the dominant breather modes). However, this task falls outside the scope of our present study.

Not taking into account the breathing modes comes down to assuming quasi-stationarity. Otherwise put, the evanescent modes $\ket{\bpsi_{[j]}}$ are henceforth assumed to decay fast enough relative to the undisturbed spiral period $2\pi/\omega_0$. In comparison with our treatise of wave fronts, one could say that here too, we are disregarding explicit memory effects. \\

The quasi-stationary approximation implies that $||\dd_\tau \tuu|| \ll ||\HL \tuu||$, whence we may omit the time-derivative term in Eq. \eqref{eq_tuu_fil}. After filtering out the GMs, an immediate solution for $\tuu$ is obtained by spectral inversion of $\HL_0$,
relying on the fact that $\HL_0^{-1}$ acting on GMs or RFs always yields zero. The lowest order result \eqref{tuu_lowest_L0} becomes
\begin{equation}\label{def_uu1}
   \veps \ket{ \uu_1 } = \veps \LA^A \HL^{-1}_0 \HP \ket{\bpsi_A} =  \veps \LA^A  \sum \limits_{j \geq 1} \zeta^{-1}_{[j]} \bra{\bY^{[j]}} \HP \ket{\bpsi_A} \ \ket{\bpsi_{[j]}}.
\end{equation}
The formal time-scale parameter $\veps$ was included here since $\HL^{-1}_0 \HP \ket{\bpsi_A} = \OO(\veps)$.
Writing $\uu_1 = \veps \LA^A (\uu_1)_A$, the outcome may be used iteratively to obtain the solution in the next-to-leading order:
\begin{eqnarray}\label{def_uu2}
  \ket{\uu_2} &=&  \veps \LA^A \LA^B\   \HL_0^{-1} \HP \ket{\dd_B (\uu_1)_A} - \veps^2 \frac{1}{2} \LA^A \LA^B\  \HL^{-1}_0 \ket{(\uu_1)_A \mathbf{F}''(\uu_0) (\uu_1)_B} \nn \\
  &&\ + \left(w^2 \delta^A_{B} + \dd_\s w \eps_{B}^{\hs A} + \LA^A \LA_B + (P^{T0})^A_{\hs C} \LA^C \LA^B \right)\   \HL_0^{-1}  \rho^B \HP \ket{\bpsi_A} \nn \\
 &&\  - \eps^A_{\hs C} \eps^B_{\hs D} w^2 \rho^C \rho^D\  \HL_0^{-1} \HP \dd_B \ket{\bpsi_A}  + \OO(\la^3).
\end{eqnarray}
To not overload future notations, we sort all terms in $\ket{\uu_2}$ according to geometrical meaning:
\begin{eqnarray}\label{def_uu2_short}
\ket{\uu_2} =  \LA^A \LA^B \ket{(\uu_2^k)_{AB} } + \veps \dd_\s w \ket{\uu_2^w } +  w^2 \ket{\uu_2^{ww} }.
\end{eqnarray}
Remark that $ \ket{(\uu_2^k)_{AB} } $ on itself is not symmetrical with respect to interchanging the indices. Noteworthily, only the  $\ket{\uu_2^w }$ term is of first order in the time-scale parameter $\veps$, since $\eps_B^{\hs A} \rho^B \bpsi_A = \bpsi_\theta$. On the other hand, left-multiplication of $\ket{\uu_1}$ or $\ket{\uu_2}$ with $\bra{\bY^A} \HP$ or $\bra{\bY^\theta} \HP $ will increase the order in $\veps$ by one.

Expressions \eqref{def_uu1}-\eqref{def_uu2} are of the proposed shape from Eq. \eqref{uu012}, which can in hindsight only be achieved in the quasi-stationary approximation. Our next step will be to insert $\uu_1, \uu_2$ in the transformed RDE \eqref{RDE_alg_iso}, to see how these effects affect filament motion.

\subsection{Terms in the translational EOM due to wave accommodation}

Here, we collect the contributions in the `remainder term' \eqref{RemT} that are $\OO(\la^3)$ and therefore comparable to the next-to-leading order translational corrections in the filament EOM. In the process, we additionally impose rotational invariance to eliminate terms that do not affect the time-averaged motion of the filament.

First, we discern in the left-hand side of \eqref{RDE_alg_iso} that the filament velocity gets multiplied by a factor
\begin{equation}\label{mfac}
  1 + \langle \bY^B \mid \dd_A \tuu \rangle = 1 + \langle \bY^B \mid \dd_A \uu_2 \rangle + \OO(\la^3).
\end{equation}
As the right hand side of the EOM contains no zeroth order terms in $\la$, this correction term may simply be brought to the remainder term \eqref{RemT} by means of the expansion $(1+x^2)^{-1} = 1 - x^2 + \OO(x^4)$, which leads to
\bsub \label{RemTall}
\begin{equation} \label{RemT1}
 -\dot{\vec{X}}\cdot \vec{N}^A \langle \bY^B \mid \dd_A \tuu \rangle = - (P^{T0})^A_{\hs C} \LA^C \langle \bY^B \mid \dd_A \uu_2  \rangle + \OO(\la)^4.
\end{equation}
The factor $\langle \bY^B \mid \dd_A \uu_2  \rangle$ is subsequently expanded with Eq. \eqref{def_uu2_short}.

Second, we treat the contribution  $\bra{\bY^B} \Gamma^\nu_{\alpha\beta} g^{\alpha\beta} \HP   \ket{\dd_\nu \tuu}$ to expression \eqref{RemT}. Choosing $\nu = A$ brings about a term that expands to
\begin{eqnarray}\label{RemT2}
  \LA^A \bra{\bY^B} \HP \ket{\dd_A \uu_2} &=&  \LA^A \LA^C \LA^D \bra{\bY^B} \HP \ket{\dd_A (\uu_2^k)_{CD}} \\
  &&  + \veps \LA^A \dd_\s w \bra{\bY^B} \HP \ket{\dd_A \uu_2^w} +  w^2 \LA^A \bra{\bY^B} \HP \ket{\dd_A \uu_2^{ww}}.\nn
\end{eqnarray}
One could also have taken $\nu = \s$, but using Eq. \eqref{gGamgs}, the resulting term is seen to be $\OO(\la^4)$.

Third, we take into consideration $- \la \bra{\bY^B}  \left(g^{\mu\nu} \dd^2_{\mu\nu} - g^{AB} \dd^2_{AB}\right) \HP \ket{\tuu}$. Making $(\mu, \nu) = (\s, \s)$ produces
\begin{equation}\label{RemT3}
  - \bra{\bY^B} \HP \ket{\dd_\s^2 \tuu} = - \veps^2 \dd^2_\s \LA^A \bra{\bY^B} \HP \ket{(\uu_1)_A} ,
\end{equation}
whereas the choice $(\mu, \nu) = (A,B)$ grants
\begin{multline}\label{RemT4}
  - \bra{\bY^B} \HP \left(g^{CD} - \delta^{CD} \right) \dd^2_{CD} \ket{\uu_1} =\\ -\veps  w^2 \LA^A \eps_{CE} \eps_{DF} \bra{\bY^B} \HP \rho^E \rho^F \ket{(\dd^2_{CD} \uu_1)_A} + \OO(\la^4).
\end{multline}
The alternate possibility $\{\mu, \nu\} = \{A,\s\}$ (counted twice) brings forth
\begin{equation}\label{RemT5}
  - 2 \bra{\bY^B} \HP g^{A\s}\ket{\dd_{A \s}^2 \tuu} = - 2 \veps w \dd_\s \LA^D \bra{\bY^B} \HP  \rho^C \eps_{AC} \ket{\dd_A(\uu_1)_D}.\quad
\end{equation}

The fourth term in \eqref{RemT} reads $-(\omega-\omega_0) \langle \bY^B \mid \dd_\theta \tuu \rangle$ and has non-vanishing contribution too:
\begin{equation}\label{RemT6}
 - (\omega-\omega_0) \langle \bY^B \mid \dd_\theta \tuu \rangle = - \veps \left( a_0 w^2 + b_0 k^2 +d_0 \dd_\s w \right) \LA^A \langle \bY^B \mid \dd_\theta (\uu_1)_A \rangle.
\end{equation}

Finally, the fifth and last term implements the modified reaction process due to wave pattern disturbance:
\begin{eqnarray}
\hspace{-0.5cm}- \frac{1}{2} \langle \bY^B \mid \tuu \mathbf{F}''(\uu_0) \tuu \rangle &=& - \veps \LA^A \langle \bY^B \mid (\uu_1)_A \mathbf{F}''(\uu_0) \uu_2 \rangle \nn \\
&=& - \veps \LA^A \LA^C \LA^D \langle \bY^B \mid (\uu_1)_A \mathbf{F}''(\uu_0) (\uu_2^k)_{CD}\rangle \nn \\
&& - \veps^2 \LA^A \dd_\s w \langle \bY^B \mid (\uu_1)_A \mathbf{F}''(\uu_0) \uu_2^w \rangle \nn \\
&&- \veps \LA^A w^2 \langle \bY^B \mid (\uu_1)_A \mathbf{F}''(\uu_0) \uu_2^{ww} \rangle. \label{RemT7}
\end{eqnarray}
\esub

\subsection[Translational EOM with adjustment of the wave profile]{Extended filament EOM for translation with adjustment of the wave profile}

With the preceding calculations on wave deformation, we have derived the translational EOM \eqref{ribbon2} up to third order in filament twist and curvature (in the quasi-stationary approximation). We shall annotate the additional dynamical coefficients owing to wave profile accommodation $\tuu$ with tildes; a matrix element expression for these coefficients can be recovered from the findings from the preceding section.

Six out of the seven contributions \eqref{RemTall} do not invoke novel terms in the extended EOM \eqref{ribbon2}, but merely cause a shift of the coefficients:
\begin{align}
a_1 \rightarrow \bar{a}_1 &= a_1 + \tilde{a}_1, & a_2 \rightarrow \bar{a}_2 &= a_2 + \tilde{a}_2,\nn \\
b_1 \rightarrow \bar{b}_1 &= b_1 + \tilde{b}_1, & b_2 \rightarrow \bar{b}_2 &= b_2 + \tilde{b}_2, \nn \\
c_1 \rightarrow \bar{c}_1 &= c_1 + \veps \tilde{c}_1, & c_2 \rightarrow \bar{c}_2 &= c_2 + \veps \tilde{c}_2, \nn \\
d_1 \rightarrow \bar{d}_1 &= d_1 + \veps \tilde{d}_1, & d_2 \rightarrow \bar{d}_2 &= d_2 + \veps \tilde{d}_2.  \label{shift_filco}
\end{align}
Remark that the modification coefficients $a_1,a_2,b_1,b_2$ that capture corrections to the spiral profile may also contribute substantially for small or vanishing values of the time scale parameter $\veps$.

The additional term that needs be incorporated in the filament EOM is \eqref{RemT3}. This term of $\OO(\eps^2)$ may be restated using
\begin{equation}\label{ds2LA}
  \dd_\s \LA^A \vec{N}_A = [\dd^4_\s \vec{X}]_\perp - 2 w \vec{T} \times \dd_\s^3 \vec{X} - \dd_\s w \vec{T} \times \dd_\s^2 \vec{X} - w^2 \dd_\s^2 \vec{X}.
\end{equation}
For that reason, the term \eqref{RemT3} not only causes $\OO(\veps^2)$ corrections to the twist-related coefficients $a_1,a_2, c_1, c_2, d_1, d_2$, but also establishes the supplementary contribution of the form $\dd^4_\s \vec{X}$. We decompose the novel term
\begin{equation}\label{decf1f2}
  \bra{\bY^B} \HP \ket{(\uu_1)_A} = e_1 \delta_A^{\ B} + e_2 \eps_A^{\hs B}
\end{equation}
in analogy to Eq. \eqref{decPT0}. Our most complete EOM for filament translational motion in an isotropic medium now becomes:
\begin{eqnarray}
  \dot{\vec{X}} &=& \left(\gamma_1  + \bar{a}_1 w^2 + \bar{b}_1 k^2  + \bar{d}_1 \dd_\s w \right)\ \dd^2_\s \vec{X} \nn\\
&& +  \left(\gamma_2  + \bar{a}_2 w^2 + \bar{b}_2 k^2  + \bar{d}_2 \dd_\s w \right)\ \dd_\s \vec{X} \times \dd^2_\s \vec{X}   \nn\\
&&  + \bar{c}_1 w [\dd^3_\s \vec{X}]_\perp + \bar{c}_2 w  \dd_\s \vec{X} \times \dd^3_\s \vec{X}\nn \\
&&  + \veps^2 e_1 [\dd^4_\s \vec{X}]_\perp + \veps^2 e_2 \dd_\s \vec{X} \times \dd^4_\s \vec{X}.
\quad \label{ribbon3}
\end{eqnarray}

One must not conclude from \eqref{ribbon3} that the coefficients $\verb"b"_1$, $\verb"b"_2$ in the ribbon model formulated in \cite{Echebarria:2006} need to be of order $\veps^2$ since they confer to our parameters $-e_1, -e_2$. Rather, because the $a_1$, $a_2$ terms were not included in the law of motion postulated by Echabarr\'ia \etal, the $\verb"b"_1$, $\verb"b"_2$ contributions in the ribbon model need to account for the twist induced weakening of the filament tension, so a correspondence
\begin{align}
 \verb"b"_1 &\approx a_1, &  \verb"b"_2 &\approx a_2
\end{align}
is more appropriate. In lowest order in twist and curvature, our observation does not affect the conclusion of the paper \cite{Echebarria:2006}.

\subsection[Rotational EOM with wave modification]{Extended filament EOM for rotation in the presence of wave profile modification}

A rigorous treatment of the rotational corrections up to order $\OO(\la^4)$ requires determining the deformation correction $\tuu$ of the wave profile up to $\OO(\la^3)$, which falls outside our present scope. We shall therefore only quantify these corrections up to second order in curvature and twist. From the `remainder term' \eqref{remth}, it is seen that the deviation $\omega-\omega_0$ is augmented with
\begin{multline}\label{term_omega}
  - \veps (P^{T0})^A_{\hs B} \LA^B \LA^C \langle \bY^\theta \mid \dd_A (\uu_1)_C \rangle + \veps \LA^A \LA^B \bra{\bY^\theta} \HP \ket{\dd_A (\uu_1)_B} \\- \frac{\veps^2}{2} \LA^A \LA^B \langle \bY^\theta \mid (\uu_1)_A \bF''(\uu_0) (\uu_1)_B \rangle +  \OO( \la^4).
\end{multline}
Because these contributions are all proportional to $k^2$, they cause in lowest order the substitution
\begin{equation}\label{subs_deform}
  b_0 \rightarrow \bar{b}_0 = b_0 + \veps^2 \tilde{b}_0
\end{equation}
in the phase and twist equations \eqref{EOMrot_iso_lowest} and \eqref{EOMtw_iso1}, with
\begin{multline}\label{def_b0tilde}
  \tilde{b}_0 = \veps^2 \frac{(\gamma_1-1)}{2} \delta^{AB} \langle \bY^\theta \mid \dd_A (\uu_1)_B \rangle
  + \frac{\veps^2}{2} \gamma_2  \eps^{AB} \langle \bY^\theta \mid \dd_A (\uu_1)_B \rangle \\
- \frac{\veps^2}{4} \delta^{AB} \langle \bY^\theta \mid (\uu_1)_A \bF''(\uu_0) (\uu_1)_B \rangle.
\end{multline}

\section[Discussion of the extended EOM for filaments in isotropic media]{Discussion of the extended EOM for filaments \\in isotropic media \label{sec:discuss_EOMiso}}

\subsection{Equilibrium states}

In this section, we look for stationary solutions for the filament shape, while allowing rigid translation or rotation. The stability analysis of such equilibrium states is postponed to the next section.

Straight and untwisted filaments still make out an equilibrium solution to our equations, since from Eq. \eqref{ribbon3} follows
\begin{equation}\label{straight=stable}
  \dd_\s^2 \vec{X} \equiv 0 \Rightarrow \dd_t \vec{X} = 0.
\end{equation}
In such case the twist evolution equation \eqref{EOMtw_iso2} simplifies to
\begin{eqnarray}\label{straight_tw}
    && \dd_t w =  2 a_0 w \dd_\s w + d_0 \dd^2_\s w.
\end{eqnarray}
For $d_0>0$, the untwisted filament is a stable solution to this equation. In the limit of low twist, \eqref{straight_tw} reduces to a diffusion equation, and for that reason $d_0$ may be called the twist diffusion coefficient. As such, this physiology-dependent parameter plays a similar role as surface tension coefficient $\gamma$ and the filament tension $\gamma_1$. It also shares the property with these other tension coefficients that it reduces to unity in the limit of small time scale ratio $\veps$.

Remarkably, Eq. \eqref{straight_tw} evenly allows a non-trivial steady state solution for twist around a straight filament: separation of variables leads to
\begin{equation}\label{straight_wss}
  w(\s) = \sqrt{\frac{A}{a_0}} \mathrm{tanh}\left( \frac{\s \sqrt{A a_0}}{d_0} \right)
\end{equation}
with $A$ a constant of integration of the same sign as $a_0$ and bearing the dimension of $time^{-1}$. In particular, \eqref{straight_wss} provides a boundary layer description between a region of constant twist $w_0$ and a medium edge where twist must vanish. In such case, the twist profile will be shape as a hyperbolic tangent, with thickness given by $s = d_0 / (a_0 w) $.

Relaxing the condition of vanishing curvature, we can seek other stable solutions with a high degree of symmetry. For instance, an untwisted circular scroll ring of radius $R$ obeys
\begin{equation}\label{circ_scroll}
  \dd_t R = \frac{\gamma_1}{R}  + \frac{\bar{b}_1-e_1}{R}
\end{equation}
and therefore possesses a critical radius at which contraction or expansion is stopped, if only $\gamma_1$ and $\bar{b}_1-e_1$ have opposite sign. In such case, the final radius of the scroll ring equals
\begin{equation}\label{circ_scroll2}
  R(t \rightarrow \infty) = \sqrt{ -\,\frac{\gamma_1}{\bar{b}_1-e_1}}.
\end{equation}
Note that such scroll will still translate along its symmetry axis with constant  velocity.

Circular scroll waves with constant twist $w = m / (2 \pi R)$, $m \in \mathbb{Z}$ may also form an equilibrium solution; for their description it suffices to augment the term $\bar{b}_1-e_1$ in Eqs. \eqref{circ_scroll}-\eqref{circ_scroll2} with $(\bar{a}_1 + \bar{c}_1)(2 \pi n)^2$. \\

Lastly, we consider helices of constant pitch $2\pi / p $ and radius $r$, which have a Cartesian representation $(x,y,z) = (r \cos pz, r \sin pz, z)$. These curves are calculated to yield
\begin{align}\label{prop_helix}
 d\s &= \sqrt{1+p^2r^2} dz, & k &= \frac{r p^2}{1+p^2 r^2}, & \tau &= \frac{1}{1+p^2 r^2}.
\end{align}
If the helix has constant twist, we may use the parallel adapted frame around the curve to define its twist, given that the scroll's phase changes over $\Delta \phi$ in the laboratory frame over one winding of the helix:
\begin{equation}\label{tw_helix}
 w = \frac{\Delta \phi - 2\pi }{\Delta \s}  = \frac{(\Delta \phi - 2 \pi) p }{\sqrt{1+ p^2 r^2}}.
\end{equation}
The stationary radius ($r>0$) of a helix in an unbounded medium is obtained by inserting \eqref{prop_helix}-\eqref{tw_helix} in the filament EOM. As this condition results in a fourth order polynomial in $\sqrt{1+p^2 r^2}$ we do not present an explicit solution here. Owing to inequality of the coefficients for filament motion along the normal and binormal directions, the helicoidal filament with stable radius will still rotate as a whole around its central axis.

\subsection[Stability of a straight and twisted filament]{Linear stability analysis of a straight filament \\with homogenous twist \label{sec:sproing}}

Let us perform linear stability analysis around a straight filament that has homogeneously distributed twist $w_0$ in the laboratory frame of reference. This situation can be realized by either considering an infinitely long filament or applying periodic boundary conditions to the filament parameterization. It is known from numerical experience that an initially stable filament will destabilize and take a helicoidal shape as soon as the twist $w_0$ exceeds a threshold value; this phenomenon is referred to as `sproing' in literature \cite{Henze:1990, Henry:2002, Echebarria:2006}.\\

After laying down the Z axis of our Cartesian coordinate frame along the filament, the position of the perturbed filament in a transverse plane can be captured by the parameterization $\left(X_0 + x(z,t), Y_0 + y(z,t)\right)$; the twist equals $w_0+w(z,t)$. Since it can be proven that the set of helices with (small) radius $R$ and pitch $b$ provide a complete basis for the linear perturbations $x(z,t), y(z,t)$, we take on such dependency here.

Adopting complex notation for position in the transverse plane, we write $X_c = x+iy = A e^{ipz}$. Inserting this ansatz in the filament EOM delivers an expression of the form $\dd_t X_c = \Omega_c X_c$. the filament is linearly stable only if the real part of the growth rate $\Omega = \mathrm{Re} \Omega_c$ is negative. We shall write $\Omega = Re(\Omega_c)$. During the derivation one needs to take into account that Eq. \eqref{tw_helix} implies $w = (w_0 - p) / \sqrt{1 + p^2 r^2}$, i.e. the initial twist is changed by the natural winding of the helix.\\

\begin{figure}[h!b] \centering
  \includegraphics[width=0.9\textwidth]{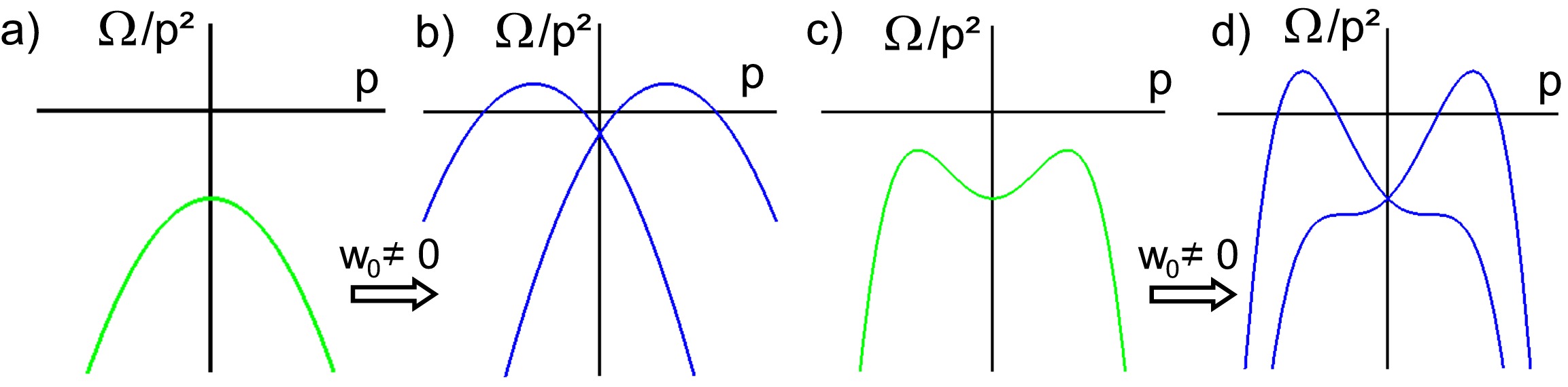}\\
  \caption[Growth rate for the sproing instability]{Growth rate for the sproing instability as a function of the wave number $p$ of the instability. We compare the cases without (a,c) and with (b,d) twist, for two qualitatively different curves $\Omega(p)$.}\label{fig:sproing}
 \end{figure}

Up to the order of our gradient expansion, we infer that the growth rate $\Omega$ for the helix radius $r$ depends on $p$ as a fourth order polynomial:
\begin{equation}\label{ReOM}
 \Omega(p, w_0) = p^2 \left[C_2(w_0^2) + C_3 w_0 p + C_4 p^2 \right].
\end{equation}
Part of this leading order dependence was also obtained from the ribbon model by Echebarr\'ia \etal However, we have obtain the coefficients explicitly as linear combinations of matrix elements:
\begin{align}
C_2 &= -  \gamma_1 - \bar{a}_1 w_0^2 ,
& C_3&= (\bar{c}_2 + 2 \bar{a}_1),
& C_4&= - \bar{a}_1 - \bar{c}_2 + e_1,
\end{align}
with an additional coupling term $w_0^2 p^2$ that was absent in the ribbon model treatment. The requirement that an untwisted filament is linearly stable requires that $\Omega(w_0=0)$ is negative for all $p$. Thereto, the unperturbed filament tension coefficient $\gamma_1$ needs to be positive. The coefficient $C_4$ is allowed to take small positive values, if only the higher order polynomial terms in $p$ manage to keep $\Omega/p^2$ negative for all $p$; compare panels (a) and (c) in Fig. \ref{fig:sproing}.

In the presence of homogenous twist, the curve $\Omega/p^2$ shifts vertically due to the $\bar{a}_1 w_0^2$ term; a negative $\bar{a}_1$ causes positive values for $\Omega$ for high enough twist rate $w_0$, even for $p=0$. Additionally, the linear term in $w_0$ brings an asymmetry in the profile, and therefore displaces the wave numbers first reach a positive growth rate. This process is shown in panels (b) and (d) of Fig. \ref{fig:sproing}.

For the case where higher order terms can be neglected, the critical twist at which the filament destabilizes is easily deduced due to the factorization in Eq. \eqref{ReOM}; this situation necessitates that $C_4<0$. From Eq. \eqref{ReOM}, it is then seen that a secondary maximum in the growth rate curve develops for twist rate higher than
\begin{equation}\label{crit_twist}
  w_{\rm sec} = \frac{4}{3} \sqrt{\frac{2 \gamma_1 }{ -\frac{C_3^2}{C_4} - \frac{32}{9} \bar{a}_1}}.
\end{equation}
Whether positive growth rate values are attained may be inferred from the parabolic dependence in $\Omega/p^2$: the top of this quadratic function takes positive values only for twist larger than
\begin{equation}
 w_{\rm crit} = 2 \sqrt{\frac{\gamma_1 }{ -\frac{C_3^2}{C_4} - 4 \bar{a}_1}}.
\end{equation}
If the domain is large enough such that all helix wave numbers are admitted, helix with the largest growth rate has pitch
\begin{equation}
 p_{\rm cr} = \sqrt{\frac{\gamma_1}{ - C4 (1+ 4 \frac{\bar{a}_1 C_4}{C_3^2})}}.
\end{equation}

\subsection{Total filament length \label{sec:filtension}}

Knowing that in lowest order in curvature and twist the filament tension $\gamma_1$ determines the overall evolution of the filament shape, we attempt in this section to recover the concept of filament tension from the extended EOM for filaments.


As in Biktashev \etal \cite{Biktashev:1994}, we write $s$ here to denote the arc length parameter for all times $t$. The filament $\s$ was conceived so that the local filament velocity is always orthogonal to the filament and therefore cannot keep up with $s$ when the filament locally expands or contracts. The time derivative of the total filament length $S(t) = \int \dd_\s \vec{X} \cdot \vec{T} \mathrm{d} \s$ is found as
\begin{eqnarray}\label{totalfillength}
    \frac{dS}{dt} &=& \left. \dd_\s \vec{X} \cdot \vec{T}\  \frac{ds}{dt} \right|^{s_2(t)}_{s_1(t)} + \int \dd_\s \dd_t \vec{X} \cdot \vec{T}  \mathrm{d} \s = -  \int \dd_t \vec{X} \cdot \dd^2_\s \vec{X} \mathrm{d} \s,\qquad
\end{eqnarray}
where the boundary terms disappear when the filament is closed or when no-flux boundary conditions are imposed on the medium edges. A filament that ends on a no-flux boundary can be treated by the method of mirror sources, from which follows that in the endpoints $s_i$, $(i=1,2)$:
\begin{align} \label{BC_filament}
w(s_i) &= 0,  &           \dd_\s k(s_i) &= 0,  & \tau_g(\s_1) &= 0.
\end{align}

We now insert our EOM for the filament that is valid up to third order in curvature and twist into Eq. \eqref{totalfillength}. Meanwhile, we denote the coefficient that precedes $\dd_\s^2 \vec{X}$ in Eq. \eqref{ribbon3} as
\begin{equation}\label{def_G1}
  G_1 = \gamma_1 + \bar{a}_1 w^2 + \bar{b}_1 k^2 + \bar{d}_1 \dd_\s w,\qquad
\end{equation}
and similarly for $G_2$. Since triple products that contain a parallel pair of vectors disappear, the $G_2$ term vanishes and the $e_2$ term may be dropped after partial integration:
\begin{eqnarray}
  \frac{dS}{dt}  &=& - \int G_1 k^2 \mathrm{d} \s - \bar{c}_1 \int  w \dd^3_\s \vec{X} \cdot \dd^2_\s \vec{X} \mathrm{d} \s   \label{fille2}\\
&&  - \bar{c}_2 \int w  \left(\dd_\s \vec{X} \times \dd^3_\s \vec{X} \right) \cdot  \dd^2_\s \vec{X} \mathrm{d} \s
   - \veps^2  e_1 \int \dd^4_\s \vec{X}  \cdot  \dd^2_\s \vec{X} \mathrm{d} \s. \nn
\end{eqnarray}
Using partial integration, a factor $k^2$ can be separated in most of the terms, eventually yielding:
\begin{equation}
  \frac{dS}{dt}  = - \int \left[G_1 - \frac{\bar{c}_1}{2} \dd_\s w - \bar{c}_2 w \tau_g - e_1(\tau_g^2 + k^2) \right] k^2 \mathrm{d} \s
- \veps^2 e_1 \int (\dd_\s k)^2 \mathrm{d} \s.
\quad \qquad\label{fille3}
\end{equation}
Note that the last term only contributes for heavily curved filaments; moreover its pre-factor $e_1$ is of order $\veps^2$. When smallness of the $e_1$ term is assumed, we infer that that the concept of filament tension can be given a local character. As the quantity between brackets in \eqref{fille3} replaces the filament tension $\gamma_1$ that was encountered in \cite{Biktashev:1994}, we generalize the concept of filament tension to
\begin{equation}\label{eff_filtens}
  \gamma_1^{\rm eff}(\s) =  \gamma_1 + \bar{a}_1 w^2 + (\bar{b}_1- e_1) k^2 + \veps \left(\tilde{d}_1 - \frac{\tilde{c}_1}{2}\right) \dd_\s w - \bar{c}_2 w \tau_g - e_1 \tau_g^2.
\end{equation}
The difference $(d_1 - \frac{c_1}{2})$ has dropped from the expression due to their linear dependence \eqref{eq_cd}. As a result, variations in scroll wave twist only contribute to the effective filament tension through modification of the wave profile.\\

The perhaps dramatic implications of this equation are immediately evident: any of the geometrical terms that appear in the effective filament tension given by Eq. \eqref{fille3} may cause the filament to increase its length, even when the medium parameters guarantee that straight and untwisted filaments are stable, i.e. $\gamma_1 = P^{T0}_s > 0$. This issue is relevant to the development of cardiac arrhythmias, as lengthening filaments give rise to filament multiplication; the latter process is thought a tentative pathway to cardiac fibrillation.

\subsection{Twist evolution equation}

The twist evolution equations \eqref{EOMtw_iso1}-\eqref{EOMtw_iso2} may be recognized as the inhomogeneous Burgers' equation, which supports both shock waves and rarefaction waves; see also the discussion in \cite{Setayeshgar:2002}. In isotropic media, the source term to the twist evolution equation is formed either by a differential natural rotation frequency $\omega_0$ due to varying reaction kinetics or by curvature of the filament, with coupling coefficient $\bar{b}_0$.

Stability of the twist against small perturbations is granted by the model-specific parameter $d_0$, which is expected to lie close to one for healthy myocardium. From our preliminary inquiries on the higher order terms, this `twist diffusion coefficient' is also seen to be altered by curvature:
\begin{equation}\label{d0eff}
  d_0^{eff} = d_0 + d_0^{(1)} k^2.
\end{equation}
Hence, if a filament has a region of intense curvature, the spreading of twist along the filament can be expected to be lowered if $d_0^{(1)}<0$. Therefore, cusps in the filament shape may facilitate twist accumulation and eventually destabilize the filament due to weakening of its tension, as discussed in the previous paragraphs. The tentative relation of this phenomenon to the nature of \textit{twistons} as described in \cite{Fenton:2002} merits further investigation.

For high filament curvature, the effective twist diffusion coefficient may be even rendered negative when $d_0^{(1)}<0$, enabling another pathway to scroll wave instability. An estimate for the critical curvature in analogy to Eq. \eqref{crit_twist} is here identified as
\begin{equation}\label{crit_curv}
  k^{\rm{crit}} = \sqrt{\frac{d_0}{-d_0^{(1)}}}.
\end{equation}

Furthermore, it can be expected that wave profile accommodation effects (which are not explicitly listed in the present work) will cause additional contributions to the twist diffusion coefficient \eqref{d0eff} that are proportional to $w^2$ and $\dd_\s w$.

\subsection{Localization of the spiral wave}

Using our matrix element formalism, an estimate can be given here to the filament effective thickness $d$, as arose in the definition of our perturbation parameter $\la$ (Eq. \eqref{def_lam}). For diffusive-coupled processes we therefore infer
\begin{equation}\label{fil_thickness}
  d^2 = \bra{\bY^A} \HP \rho^2 \ket{\bpsi_A} = 4 P^{T2}_s.
\end{equation}
Evidently, this quantification fits into the wave-particle duality of spiral waves as described in \cite{Biktasheva:2003} and Fig. \ref{fig:pw_duality}.

\subsection{Metric drift corrections \label{sec:metricdriftfiliso}}

In our derivation of the filament EOM \eqref{ribbon3}, we had assumed that the diffusion coefficient $D_0$ is constant throughout space. If one relaxes this condition in Eq. \eqref{diff_term_iso0}, an additional drift term is generated:
\begin{equation}
 \vec{e}^B \cdot \dot{\vec{X}} = - \bra{\bY^B} \dd_\mu(D_0) g^{\mu \nu} \HP \ket{\dd_\nu(\uu_0 + \la \tuu)}.
\end{equation}
In lowest order, the supplementary drift of the filament is given by
\begin{equation}
 \vec{e}^B \cdot \dot{\vec{X}} = - \dd_C D_0 \delta^{CB} \bra{\bY^B} \HP \ket{\bpsi_A}
\end{equation}
such that
\begin{equation} \label{metric_drift_iso}
 (\dot{\vec{X}})_{m.d.} = - \gamma_1\nabla D_0  - \gamma_2 \vec{T} \times \nabla D_0.
\end{equation}
This additional drift component is termed `metric drift' here and in \cite{Dierckx:2009}, as it is related to the determinant of the metric tensor in the curved space formalism. Our present treatment is restricted to the leading order contribution here; it may be noted that the diffusivity gradient $\nabla D_0$ couples to filament curvature and twist in higher order. The metric drift correction also affects the wave profile.

Not coincidentally, the filament tension coefficients also determine the metric drift properties. For, in lowest order, $\gamma_1$ and $\gamma_2$ comprise the only rotationally invariant tensor components that exclusively couple to the diffusing variables. As a consequence to Eq. \eqref{metric_drift_iso}, filaments with positive tension $\gamma_1$ migrate towards zones with a lower scalar diffusion coefficient.

\subsection{Fixed angle principle for filament drift}

From our formalism follows that the leading order response of the filament to a small vectorial perturbation is to drift at a fixed angle relative to that perturbation. We have seen this instance yet exemplified in Eq. \eqref{ribbon3} and in the case of metric drift, i.e. \eqref{metric_drift_iso}. Similarly, we have found for an inhomogeneous distribution of a parameter $p$ in the reaction kinetics \cite{Dierckx:2009}:
\begin{equation}\label{par_drift}
\vec{e}^B \cdot \dot{\vec{X}} = - \dd_A p \bra{\bY^B} \rho^A \ket{\bF(\uu_0)} + \OO(\dd^3 p).
\end{equation}
which also leads to drift under a fixed angle with respect to the (small) vectorial perturbation $\vec{v} = \nabla p$. The rationale behind this universal scroll wave behavior is the following: coordinate invariance demands that $\vec{e}^B \cdot \dot{\vec{X}}$ is of the form $v_A T^{AB}$. In lowest order, $T^{AB}$ is a constant matrix element and therefore the drift occurs at an oriented angle $\Gamma$ with
\begin{equation}
 \tan \Gamma = \frac{\eps_{AB} T^{AB}}{\delta_{AB} T^{AB}}
\end{equation}
relative to the perturbation vector $\vec{v}$. Note that in the higher order corrections as well as in anisotropic media, composite tensors may be formed, e.g. $v_A \LA^A \LA^B$, and therefore only a weaker version of the fixed angle rule will hold.

\subsection{Numerical determination of the dynamical parameters}

Through our analytical investigations, we have provided explicit expressions to the dynamical parameters that determine filament motion. In order to obtain numerical values for these coefficients for a given RD model, one needs to know the GMs and RFs to high accuracy. Numerical procedures to estimate these functions with high numerical precision have been developed; see \cite{Barkley:1992, Hakim:2002, Biktasheva:2009}. Since these methods rely on an direct estimation of the Jacobian matrix for the reaction functions, only RD models with continuously differentiable reaction kinetics can be treated at present. Also, calculating this set of coefficients for system with a large number of state variable may prove difficult.

An alternative way to assess the dynamical parameters for a given numerical model, is to perform a series of numerical simulations of filaments of simple geometry, from which a fitting procedure allows to quantify the coefficients in the EOM. Note that such methodology applies to an unrestricted class of monodomain RD models that produce filaments.\\

Of course, the combination of both strategies outlined here is most interesting, as a comparison of their outcome can serve to validate our newly obtained filament equations. Obviously, such validity check is recommendable, as it not only allows to detect human error in the calculations but also provides sound numerical evaluation and an estimate to the scope of our theory. We have scheduled to perform such numerical validation in the near future.

\clearpage


\clearpage{\pagestyle{empty}\cleardoublepage}

\graphicspath{{fig/}{fig/fig_filaniso/}}

\renewcommand\evenpagerightmark{{\scshape\small Chapter 8}}
\renewcommand\oddpageleftmark{{\scshape\small Filament dynamics in anisotropic media}}

\hyphenation{}

\chapter[Filament dynamics in anisotropic media]{The dynamics of scroll wave filaments in generic anisotropic media}
\label{chapt:filaniso}

In this section we derive for the first time the extended equation of motion for scroll wave filaments in an excitable medium with arbitrary anisotropy. To this purpose, we adopt the formalism put forward in Chapter \ref{chapt:activ}, which transforms generic anisotropy to a curved background space.

The lowest order result for filament motion remains unchanged relative to isotropic media, if only one replaces ordinary by covariant spatial derivatives. This outcome could have been predicted based on the local equivalence principle, and at the same time proves the minimal principle that was raised by Wellner, Pertsov and co-workers \cite{Wellner:2002}. This part of our derivation has been published in \cite{Verschelde:2007, Dierckx:2009}.

We then continue to derive the EOM for filaments up to third order in curvature; in this case the intrinsic curvature of the background space explicitly enters the analytical description. The tidal effects that arise from tissue anisotropy are for the first time recorded in this dissertation.

As an important application, we discuss the decrease of effective filament tension in a medium with rotational anisotropy, which might be responsible for the loss of stability when the transmural fiber rotation rate is increased. Also, the drift of intramural filaments is elegantly covered by our dynamical equations.

Our derivations appeal to a Taylor expansion of the metric tensor around the filament curve in so-called Fermi coordinates. We borrow the expansion from works on quantum gravity \cite{Nesterov:1999, Brewin:2009}. Ironically, the author of the latter paper valued establishing this expansion in the following way: \textit{``This latter calculation involves the solution of a two-point boundary value problem -- not a job for the faint hearted!''}.


%
%
%
%

\section[Fermi coordinates around the filament]{Fermi coordinates around the filament}


\subsection{Fermi-Walker transport of a reference triad along a curve}

In a first step, we need a counterpart to the reference triad ($\vec{T}, \vec{N}_1, \vec{N}_2$) from Chapter \ref{chapt:filiso}, that was adapted to extrinsic filament curvature $k$ and twist $w$. We now show how a suitable coordinate frame can be constructed along the filament, which is Euclidean up to second order in the distance in each plane transverse to the filament. Inspiration is drawn from a similar problem in the theory of general relativity: how can a traveler who steers a spacecraft at relativistic speed, keep track of the events that take place close to him?



\begin{figure}[h!b]
\begin{center}
  \includegraphics[width=0.45\textwidth]{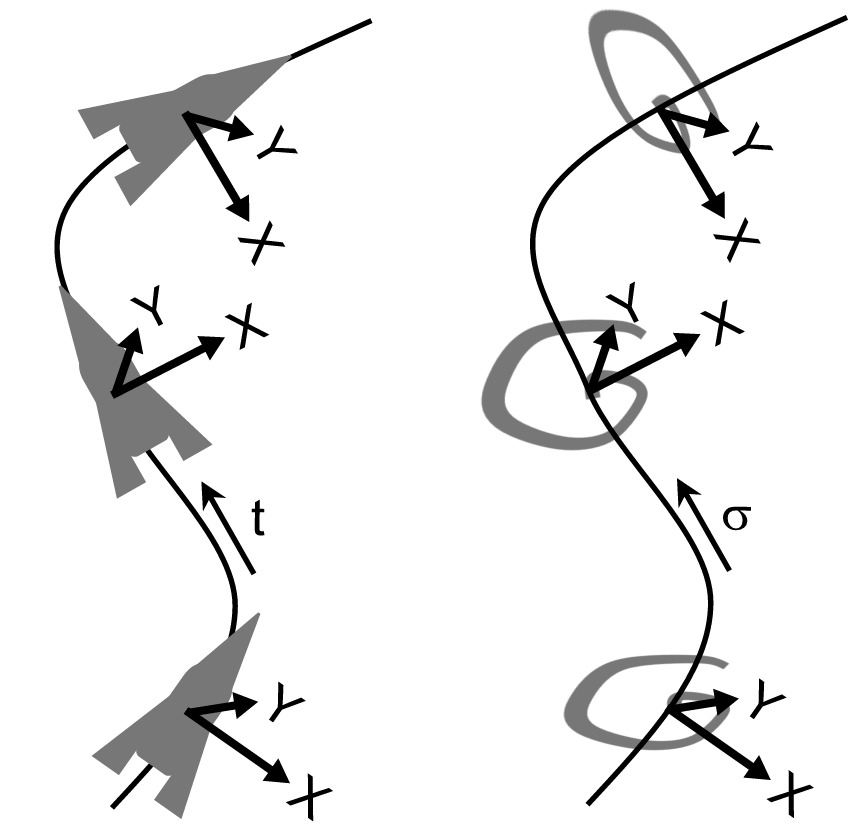}
  \caption[Spacecraft analogy]{The spacecraft analogy: finding a simple coordinate system to describe the surroundings of each point on the scroll wave filament is equivalent to setting up a permanent frame around an accelerated spacecraft in (2+1) Minkowskian space-time. The parameter $t$ denotes the proper time of the spacecraft pilot. The space-like surfaces are not parallel due to the mixing of space and time in a relativistic context.}\label{fig:spacecraft}
\end{center}
\end{figure}

A powerful analogy between the space-time trajectory of a relativistic spacecraft and an instantaneous scroll wave filament is illustrated in Figure \ref{fig:spacecraft}: both are one-dimensional curves in a curved space background. The spacecraft pilot could choose a local inertial frame for every instance of time, but then he/she would have to redefine space coordinates over and over again. If the trajectory is known a priori, however, a much more advantageous choice is to conceive a single (curvilinear) coordinate frame for the entire flight. These coordinates were put forward by Fermi as soon as 1922 \cite{Fermi:1922}, and are now known as Fermi coordinates. These should not be confounded with Fermi normal coordinates, which deal with the special case of a filament that lies along a geodesic or, in space traveler terms, a spacecraft with its engine turned off. The formal introduction of Fermi (normal) coordinates in relativistic context can be found in e.g. \cite{Fermi:1922, Manasse:1963, MTW}.

The procedure for scroll wave filaments can be readily translated,  when omitting one spatial dimension and changing space-time signature to Euclidean (+,+,+):
\begin{enumerate}
  \item Parameterize the filament using its arc length $\s$ as a coordinate, measured in in the space with metric tensor  $g^{ij} = D_0^{-1} D^{ij}$. The unit vector tangent to the filament (previously $\vec{T}$) is now denoted $\vec{e}_\s$.
  \item Construct an orthonormal triad at an arbitrary point on the filament, and put $\s=0$ in that point. Thereto, introduce two other unity vectors $\vec{e}_1,\,\vec{e}_2$; this construction can always be done since every Riemannian manifold can be made locally Euclidean in a single point. Now transport the orthonormal triad $(\vec{e}_1,\ \vec{e}_2,\  \vec{e}_\s)$ along the filament, with rotation around $\vec{e}_\s(\s)$ to follow the scroll wave's phase angle if the scroll is twisted. In the absence of twist, this procedure is known as `Fermi-Walker transport' and the resulting frame is the curvilinear generalization of the twist-adapted frame in the isotropic case from Chapter \ref{chapt:filiso}. The inclusion of twist is here seen to be equivalent to spinning of the spacecraft cabin around its instantaneous velocity vector.
  \item So far we have not taken into account the indispensable rotating character of scroll waves yet. To this purpose we make the triad $\vec{e}_\mu$ rotate around the filament with the instantaneous angular velocity $\omega(\s, \tau)$. Time $t$ is treated as an external parameter, having no equivalent in the spacecraft analogy. As in the isotropic case, $\tau$ denotes the time coordinate in the co-moving frame.
  \item For each $\s$, fill up the immediate vicinity of the filament with geodesics bristling up from the filament, perpendicularly to $\vec{e}_\s(\s)$. Thereupon, we will assign coordinates to a given point in the neighborhood of the filament where the geodesics do not yet intersect. In this region, only one of the laid out geodesics will connect the given point to the filament. This unique geodesic can be fully characterized by telling where and in which direction it leaves the filament. That information is contained in the parameter $\s$ and the tangent vector to the geodesic $\vec{n} = n^1 \vec{e}_1(\s) + n^2 \vec{e}_2(\s)$, respectively. The arc length $\rho$ along the geodesic provides the third parameter. Hence each point not far from the filament can be labeled unambiguously using the triple $(\rho^1, \rho^2, \s) = (\rho n^1, \rho n^2, \s)$.
  \item The three coordinates $(\rho^1, \rho^2, \s)$ make up a mesh that covers the neighborhood of the filament, which allows to extend the triad $\vec{e}_\mu$ around the filament:
      \begin{equation}\label{deftriad1}
          \vec{e}_\mu(\rho^1, \rho^2, \s) = \frac{\dd \vec{x}}{\dd \rho^\mu}(\rho^1, \rho^2, \s).
      \end{equation}
  \item Like in any non-orthogonal coordinate system, introduce a dual triad of contravariant vectors
      \begin{equation}\label{deftriad2}
          \vec{e}^\mu(\rho^1, \rho^2, \s) = \nabla \rho^\mu.
      \end{equation}
    Both triads are mutually orthogonal, and normalized due to using arc length for measuring distances:
    \begin{equation}\label{norme}
        \vec{e}_\mu \cdot \vec{e}^\nu = e_\mu^i e_i ^\nu = \delta_\mu^{\hs \nu}.
    \end{equation}
\end{enumerate}

\begin{figure}[h!t] \centering
  \mbox{
 \raisebox{4.5cm}{a)}\includegraphics[height=5cm]{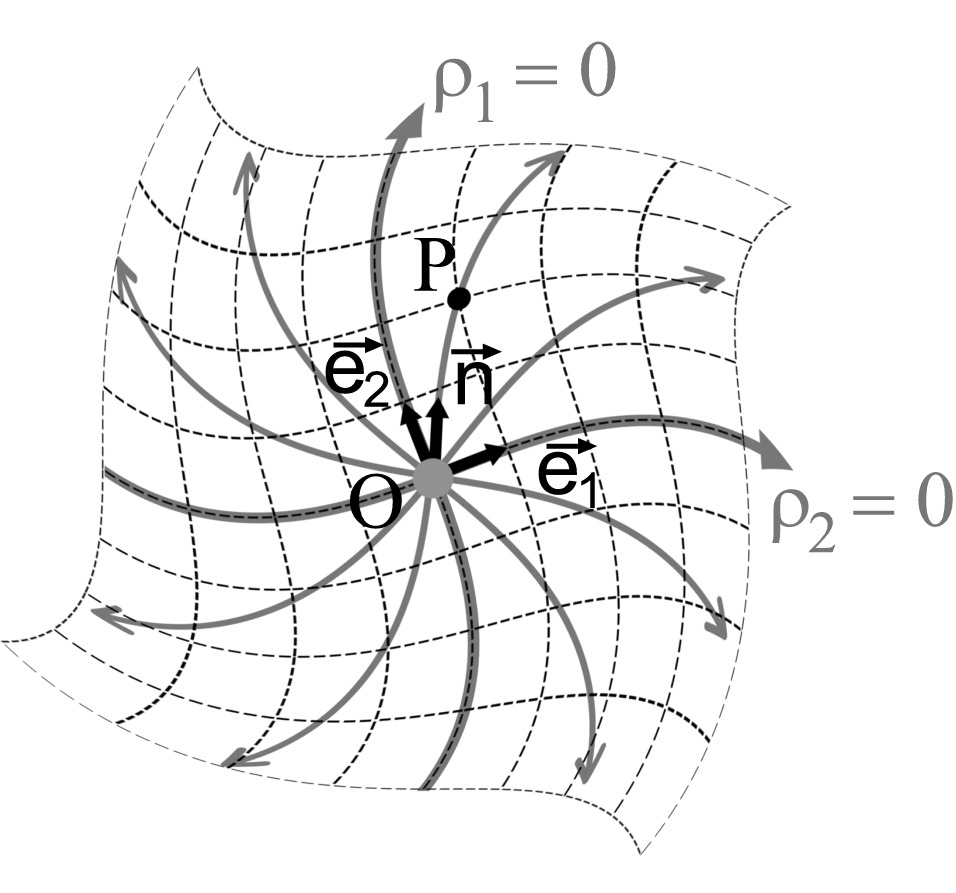}
 \raisebox{4.5cm}{b)}\includegraphics[height=5cm]{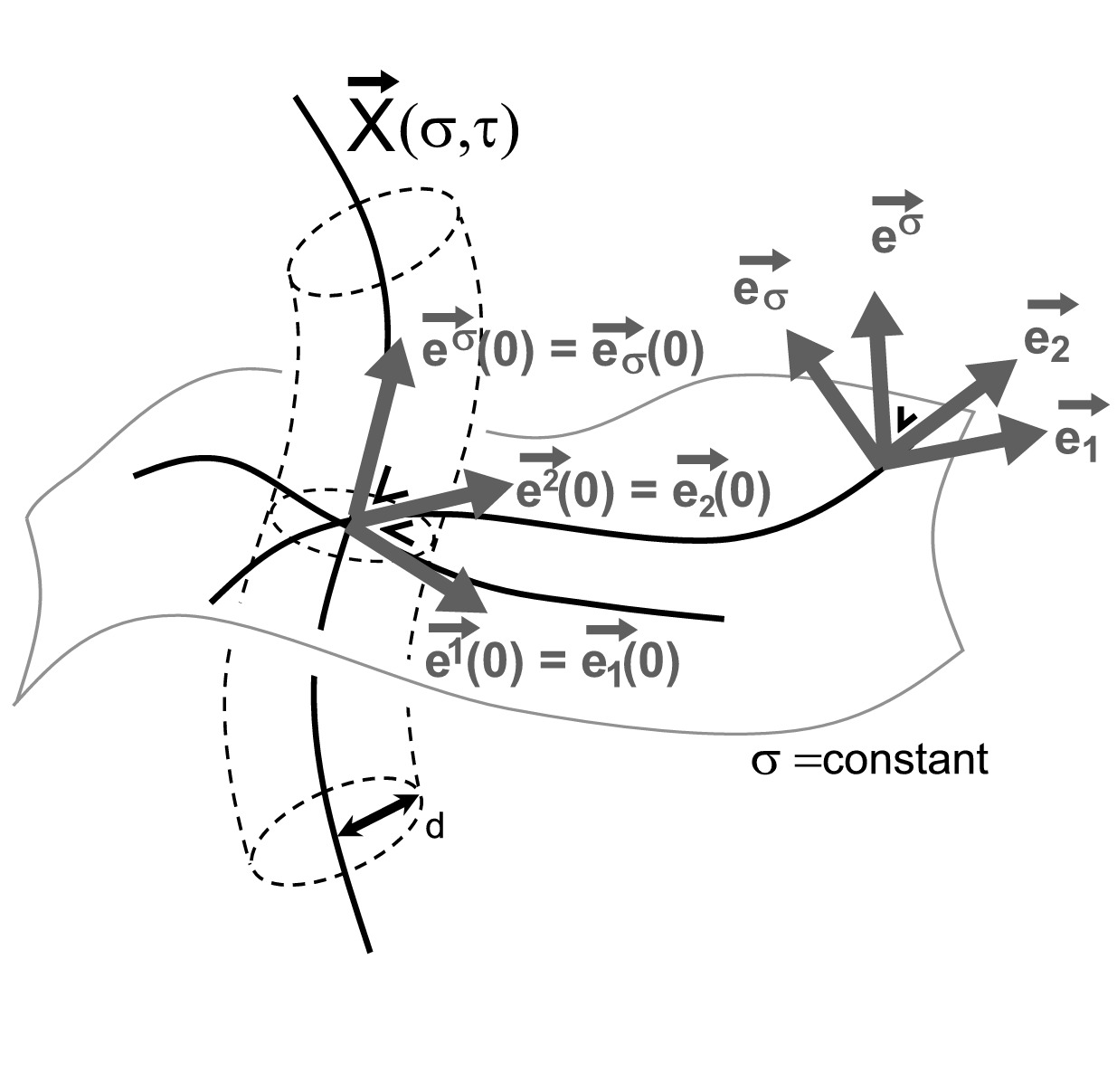}
  }
  \caption[Fermi coordinates adapted to the filament]{Fermi Coordinates adapted to the filament. a) Coordinates in a slice with constant $\s$, defined using the geodesics that originate from the filament in point $O$. Points $P$ on this family of geodesics are assigned polar coordinates. Using the $\vec{e}_1, \vec{e}_2$ defined along the filament by twist-accommodated Fermi-Walker transport, one can draw a complete coordinate mesh (dashed lines). The point P on this drawing would receive the coordinates $\rho^1 = 1$, $\rho^2 = 2$. b) Illustration of the reference triads on the filament and close to it. Whilst $e^\mu_{\ i} = e_{\mu i}$ on the filament, surfaces with $\s$ constant are curved and therefore the triad $e^\mu_{\ i}$ cannot be mapped onto $e_{\mu i}$ away from the filament.}\label{fig:coordpic}
\end{figure}

We will stick to the same index convention  as in the isotropic case: $i,j,k \in \{1,2,3\}$ refer to Cartesian coordinates, with metric $g^{ij} = D_0^{-1} D^{ij}$ and determinant $det(g_{ij}) = |g_{cart}|$. The co-moving Fermi coordinates are denoted with $\mu,\nu,... \in \{1,2,\s \}$ and the indices $A,B,C,... \in \{1,2\}$ refer to the $\rho^A$ components, which are transverse to the filament.

\subsection{Metric tensor in co-moving Fermi coordinates}
From the transformation law of the metric tensor \eqref{transf_g} follows that
   \begin{equation}\label{transg}
        g_{\mu\nu} = e^i_\mu g_{ij} e^j_\nu = \vec{e}_\mu \cdot \vec{e}_\nu.
    \end{equation}
On the filament, Fermi-Walker transport of the triad ensures orthonormality for all $\s$. The metric tensor reduces to the identity matrix in all points of the filament:
\begin{equation}
\eval{g_{\mu\nu}}{\rho=0} = \delta_{\mu \nu}, \qquad \forall \s,\, \tau.
\end{equation}
We now demonstrate that the constructed frame also makes the Christoffel symbols $\Gamma^\mu_{AB}$ vanish when evaluated on the filament. This fact arises from the geodesic equation \eqref{geodesic_eq}, which states that parallel transport along a geodesic preserves its tangent vector \cite{MTW}:
\begin{equation}\label{geod_eq}
    \DD_s^2 x^\mu = \dd^2_s x^\mu + \Gamma^\mu_{\alpha\beta} \dd_s x^\alpha  \dd_s x^\beta = 0.
\end{equation}
In this, $s$ stands for the arc length parameter along the geodesic. In our context, the geodesics originating perpendicularly from the filament have parametrization $x^\mu (\rho) = (\rho n^1, \rho n^2, \s)$, so the geodesic equation simplifies to
\begin{equation}\label{geod_eq2}
    \Gamma^\mu_{A B} n^A n^B = 0.
\end{equation}
Since this equation is by construction fulfilled for all vectors $\vec{n}$ perpendicular to the filament, we have verified that
\begin{equation}
    \eval{\Gamma^\mu_{A B}}{\rho=0}  = 0, \qquad \forall \s,\, \tau
\end{equation}
on the filament. This result can be stated equivalently as $\eval{ \DD_A \vec{e}_B}{\rho=0} = \vec{0}$ or $\eval{ \dd_A g_{BC}}{\rho=0}  = 0$.\\

Establishing the Taylor expansion of the metric tensor in powers of $\rho^A$ is rather involved because the parallel transport of base vectors along radial geodesics needs be carried out explicitly. Fortunately, the task has already been performed in the general relativity community: see \cite{Nesterov:1999} and references therein. Evidently, one first needs to prescribe the transport of the reference triad on the observer's worldline, i.e. , the filament:
\begin{equation}\label{transport}
   \eval{ \DD_\s \vec{e}_\mu }{\rho=0} = - \boldsymbol{\Omega} \cdot \eval{ \vec{e}_\mu }{\rho=0},
\end{equation}
from which immediately follows that $\boldsymbol{\Omega}$ is an antisymmetric tensor that lives on the filament, obeying
\begin{equation}\label{Omega}
  \Omega^\nu_{\hs \mu} =  \eval{ \Gamma^\nu_{ \s \mu} }{\rho=0}.
\end{equation}
For a curved and twisted filament, we derive
\bsub \label{defOM} \begin{eqnarray}
        \Omega_{A\s} &=& \eval{ \vec{e}_A \cdot \DD_\s \vec{e}_\s}{\rho=0} =  \LA_A,\\
        \Omega_{AB} &=& \eval{ \vec{e}_A \cdot \DD_\s \vec{e}_B}{\rho=0} = w \eps_{BA}.
\end{eqnarray} \esub
The expansion of the metric tensor in Fermi coordinates as a function of the transport coefficients $\Omega$ and Riemann tensor up to third order in $\rho$ is quoted from \cite{Nesterov:1999}:
\bsub \label{Nest} \begin{eqnarray}
 g_{\s\s} &=& 1 + 2 \OM_{\s A} \rho^A + \left(\OM_{\mu A} \OM^\mu_{\ B} + R_{\s A B \s} \right) \rho^A \rho^B  \\ \nn
                 &&+ \left( R_{D AB\s} \OM^D_{\hs C} + \frac{4}{3} R_{\s AB \s} \OM_{\s C} + \frac{1}{3} R^{\s}_{\ AB\s , C} \right)\rho^A \rho^B \rho^C, \\
  g_{\s F} &=&  \OM_{F A} \rho^A + \frac{2}{3} R_{F A B \s} \rho^A \rho^B + \left( \frac{1}{6} R_{FABD} \OM^D_{\ C}  \right. \\   \nn
                  &&  \left. + \frac{1}{4} R_{FAB\s} \OM_{\s C}  + \frac{1}{6} R_{\s AB \s} \OM_{F C} + \frac{1}{4} R^F_{\ AB\s , C} \right)\rho^A \rho^B \rho^C, \qquad \\
 g_{EF} &=& \delta_{EF} +  \frac{1}{3} R_{E A B F}  \rho^A \rho^B  + \frac{1}{12} \left(R_{F AB\s} \OM_{EC}  \right. \\
               &&  \left. + R_{E AB\s} \OM_{FC} +  R^E_{\ ABF, C} + R^F_{\ ABE, C} \right)\rho^A \rho^B \rho^C. \qquad \nn
\end{eqnarray} \esub
Here, the Riemann tensor components and their derivatives are implicitly evaluated on the filament. The simpler case of an isotropic medium is embedded in this result as the flat space outcome where all $R_{\alpha\beta \mu \nu}$ equal zero. Before going any further, we reexpress all ordinary derivatives in Eqs. \eqref{Nest} as covariant derivatives plus correction terms: (all quantities are evaluated on the filament)
\bsub\begin{eqnarray}
     R^E_{\hs ABF, C}  &=&  R^E_{\hs ABF; C} - \OM^E_{\hs C} R^\s_{\hs ABF}, \\
     R^\s_{\hs AB\s, C}  &=& R^\s_{\hs AB\s; C} + \OM^D_{\hs C} R^\s_{\hs ABD}, \\
     R^F_{\hs AB\s, C}  &=&  R^F_{\hs AB\s; C} - \OM^F_{\hs C} R_{\s AB\s} + \OM_{\s C} R_{FAB\s} + \OM^D_{\hs C} R^F_{\hs ABD}. \qquad\quad
\end{eqnarray} \esub
The components of the covariant metric tensor now read
\bsub \label{gcov_aniso} \begin{eqnarray}
 g_{\s\s} &=& 1 + 2 \OM_{\s A} \rho^A + \left(\OM_{\mu A} \OM^\mu_{\ B} + R_{\s A B \s} \right) \rho^A \rho^B  \\ \nn
             &&+ \left( \frac{4}{3} R_{DAB\s} \OM^D_{\hs C} + R_{\s AB \s} \OM_{\s C} + \frac{1}{3} R_{\s AB\s ; C} \right)\rho^A \rho^B \rho^C, \\
  g_{\s F} &=&  \OM_{F A} \rho^A + \frac{2}{3} R_{F A B \s} \rho^A \rho^B + \left( \frac{5}{12} R_{FABD} \OM^D_{\ C}  \right. \\   \nn
                 &&  \left. + \frac{1}{2} R_{FAB\s} \OM_{\s C}  + \frac{1}{12} R_{\s AB \s} \OM_{CF} + \frac{1}{4} R_{FAB\s ; C} \right)\rho^A \rho^B \rho^C, \qquad\\
 g_{EF} &=& \delta_{EF} +  \frac{1}{3} R_{E A B F}  \rho^A \rho^B  + \frac{1}{6} R_{EABF; C}  \rho^A \rho^B \rho^C. \qquad
\end{eqnarray} \esub
We proceed by calculating the contravariant metric tensor, as in our derivation for the isotropic case.
The formal Neumann series \eqref{neumannexp} for $\mathbf{g}$ results in:
\bsub \label{gcon_aniso} \begin{eqnarray}
 g^{\s \s} &=&  1 + 2 \OM_{A\s} \rho^A + \left(3 \OM_{\s A} \OM_{\s B}  - R_{\s A B \s} \right) \rho^A \rho^B  \quad \qquad\\
 && + \left( 4 \OM_{A \s}   \OM_{ B\s}  \OM_{C\s} + \frac{8}{3}  R_{\s AB \s } \OM_{\s C} - \frac{1}{3} R_{\s AB \s; C} \right) \rho^A \rho^B \rho^C, \nn \\
g^{\s F} &=&  \OM_A^{\hs F} \rho^A + \left( 2 \OM^F_{\hs A} \OM_{\s B} - \frac{2}{3} R^F_{\hs AB \s} \right) \rho^A \rho^B \\
               &&+ \left( 3 \OM_C^{\hs F}  \OM_{A\s} \OM_{B\s}  -\frac{13}{12} R_{\s AB\s} \OM_C^{\hs F}  + \frac{5}{6} R^F_{\hs ABD} \OM^D_{\hs C} \right. \nn \\
               && \left. - \frac{1}{2} R^F_{\hs AB\s} \OM_{\s C} - \frac{1}{4} R^F_{\hs AB\s; C} \right) \rho^A \rho^B \rho^C, \nn \\
g^{EF} &=&  \delta^{EF} + \left( \OM^E_{\hs A}  \OM^F_{\hs B} - \frac{1}{3} R^{E\hs \hs F}_{\hs AB} \right) \rho^A \rho^B  \\
               && + \left( 2 \OM_{A \s} \OM^E_{\hs B} \OM^F_{\hs C} + \frac{2}{3} R^\uE_{\hs AB\s} \OM^\uF_{\hs C} - \frac{1}{6} R^{E\hs \hs F}_{\hs AB \hs ; C}  \right) \rho^A \rho^B \rho^C. \nn
\end{eqnarray} \esub

\section[Derivation of the lowest order EOM in anisotropic media]{Derivation of the lowest order EOM for filaments\\in anisotropic media}

\subsection{Ansatz to the gradient expansion}

As a direct generalization of our derivation in the isotropic case, we introduce the dimensionless perturbation $\la$, which now bounds the effect of anisotropy as well:
\begin{equation}\label{def_lam_aniso}
  \la = \text{max}(kd, wd, \sqrt{R_{\alpha\beta \mu \nu}}\, d ).
\end{equation}
From dimensional arguments it may be remarked that the Riemann tensor corrections only arise in second order in $\la$.
Our ansatz is now that the two-dimensional spiral solution $\uu_{0}(\rho^A)$ approximates well the true three-dimensional solution, when laid out in the space around the filament using the natural Fermi coordinate frame, i.e.
\begin{equation}\label{Ansatz_aniso}
  \uu(\rho^A, \s, \tau) = \uu_0 (\rho^A) + \la \tuu(\rho^A, \s, \tau) + \OO(\la^2).
\end{equation}
As before, we furthermore assume that
\bsub \begin{eqnarray}
\dd_\s \tuu &=& \dd_A \tuu . \OO(\la),  \label{dstuu_aniso} \\
 \dot{\vec{X}} &=& \OO(\la). \label{vecXsmall}
\end{eqnarray} \esub
supplemented with the gauge conditions
\begin{align} \label{gauges_aniso}
 \langle \bY^B \mid \tuu \rangle &= 0,&  \langle \bY^\theta \mid \tuu \rangle &= 0,& \dot{\vec{X}} \cdot \vec{e}^{\,\s} &= 0.
\end{align}
 We are now ready to restate the RDE in the co-moving Fermi coordinates, for which we write following expansion around the filament:
\begin{equation}\label{xexp_Fermi}
  x^i = X^i(\s, \tau) + e^i_A (\s, \tau) \rho^A + \frac{1}{2} C^i_{AB} \rho^A \rho^B + \OO(\rho^3).
\end{equation}

The overall work flow to handle the anisotropic case is precisely the same as in chapters \ref{chapt:fronts} and \ref{chapt:filiso}:
we substitute the approximated solution \eqref{Ansatz_aniso} in the RDE
\begin{equation}\label{RDE_aniso}
  \dd_t \uu = \dd_i  \left(D^{ij} \dd_j \bP \uu\right) + \mathbf{F}(\uu),
\end{equation}
supplemented with \eqref{gauges_aniso} and \eqref{xexp_Fermi}. The expansion of the reaction term in the RDE using ansatz \eqref{Ansatz_aniso} occurs fully analogous to the isotropic case; see Eq. \eqref{reacterm}. However, expanding the time-derivative and reaction terms from the RDE is a more challenging task in generic anisotropic setting.

\subsection{Time derivative term in Fermi coordinates}
Using the chain rule for going from coordinates $(x^i, t)$ to $(\rho^A, \s, \tau)$, we find here too that $\dd_t = \dd_\tau - (\vec{e}^{\, \mu} \cdot \dd_\tau \vec{x}) \dd_\mu$. Application up to third order in $\la$ yields, with Eqs. \eqref{dstuu_aniso}-\eqref{vecXsmall},
\begin{multline} \label{dtu3D}
    \dd_t \uu = \la \dd_\tau \tuu - \left(\vec{e}^A(\rho) \cdot \dot{\vec{X}}\right) \dd_A (\uu_0 + \la \tuu) - \omega \dd_\theta (\uu_0 + \la \tuu)
    \\ - \dd_\tau (C^i_{BC})  e^A_i  \rho^B \rho^C \dd_A(\uu_0 + \la \tuu). \qquad
\end{multline}
For further expansion of the terms $\vec{e}^A(\rho)$ and $\dd_\tau (C^i_{BC})$, we borrow following result from \cite{Nesterov:1999}:
\begin{eqnarray}\label{eni}
    \vec{e}^{\,\mu}(0) \cdot \vec{e}_\s(\rho) &=& \delta^\mu_\s + \Omega^\mu_{\hs A} \rho^A + \frac{1}{2} R^\mu_{\hs AB\s} \rho^A \rho^B \\
    &&\qquad+ \frac{1}{6} \left(R^\mu_{\hs AB \nu} \OM^\nu_{\hs C} + R^\mu_{\hs AB \s ; C} \right) \rho^A \rho^B \rho^C + \OO(\rho^4), \nn\\
    \vec{e}^{\,\mu}(0) \cdot \vec{e}_F(\rho) &=& \delta^\mu_F + \frac{1}{6} R^\mu_{\hs ABF} \rho^A \rho^B
   + \frac{1}{12} R^\mu_{\hs ABF;C} \rho^A \rho^B \rho^C + \OO(\rho^4). \nn
   \end{eqnarray}
Based on Eqs. \eqref{eni} we can evaluate the $\dd_\tau (C^i_{BC})$ term in Eq. \eqref{dtu3D}. Since the basis vectors $\vec{e}_A$ do not rotate in the co-moving frame, only the Cartesian index holds rotation effects relative to time coordinate $\tau$. Secondly, spiral drift also entails $\tau$-dependence through a convection term, such that
\begin{eqnarray}\label{dtauCi}
  e^A_i \dd_\tau (C^i_{BC}) &=& - \omega \eps_D^{\hs A} \left(\frac{1}{6} R^D_{\hs \uB \uE C} \rho^E + \frac{1}{24} R^D_{\hs \uE \uF C; \uB} \rho^E \rho^F  \right)\\
  && \quad + (\dot{\vec{X}} \cdot \vec{e}^D) \left[ \frac{1}{6} R^A_{\hs \uB \uD C} + \frac{1}{12} R^A_{\uB \uD C; \uE} \rho^E \right]. \nn
\end{eqnarray}
As in the previous chapter, the underlined indices indicate that symmetric terms are implicitly added.

The series for $\vec{e}^A(\rho)$ in Eq. \eqref{dtu3D} can be obtained from formal matrix inversion of Eqs. \eqref{eni} owing to the biorthogonality relation $e^\mu_i e_\nu^i = \delta^\mu_{\hs \nu}$. We do the exercise up to second order in $\rho$:
\begin{eqnarray}\label{eArho}
    \vec{e}_\mu(0) \cdot \vec{e}^A(\rho) &=& \delta^A_\mu - \delta^\s_\mu \OM^A_{\hs B} \rho^B \\
    &&\ + \left( - \delta_\mu^F \frac{1}{6} R^A_{\hs CDF} - \delta_\mu^\s \frac{1}{2} R^A_{\hs CD\s} + \delta^\s_\mu \OM^A_{\hs C} \OM^\s_{\hs D} \right) \rho^C \rho^D. \nn
\end{eqnarray}
Altogether we have obtained for the time-derivative term in the RDE:
\begin{eqnarray} \label{tterm_aniso}
    \dd_t \uu &=& \la \dd_\tau \tuu - \left(\vec{e}^A(0) \cdot \dot{\vec{X}}\right) \dd_A (\uu_0 + \la \tuu) - \omega \dd_\theta (\uu_0 + \la \tuu) \\
&&+  \left(\vec{e}^D(0) \cdot \dot{\vec{X}}\right) \left(\frac{1}{3} R^A_{\hs BCD} + \frac{1}{6} R^A_{\hs BCD; E} \rho^E \right) \rho^B \rho^C \dd_A \uu_0 + \OO(\la^4) \nn.
\end{eqnarray}
Note that the order corrections \eqref{dtauCi} completely disappear after multiplication with $\rho^B \rho^C$ due to the symmetries of the Riemann tensor.

\subsection{Series expansion for the diffusion term}

In order to describe anisotropic excitable media with spatially varying principal diffusivities, we now further specify how the separation $D^{ij} = D_0 g^{ij}$ should be performed. In the following is convenient to impose that the metric tensor $\mathbf{g}$ has constant determinant, so that we impose:
\begin{align} \label{cond_metricdrift}
 D^{ij} &= D_0(\vec{r}) g^{ij}(\vec{r}), & \dd_i |g_{Cart}| &= 0.
\end{align}
Remark that the metric determinant in Cartesian coordinates equals the product of the principal electrical diffusivities. Therefore, if one would consider a medium in which the anisotropy ratio changes while the product of the diffusivities remains constant, no additional metric drift term proportional to $\nabla D_0$ will be generated. Referring to Eq. \eqref{diffterm} for intermediate steps, we end up with
\begin{equation}\label{diffterm_aniso}
  \dd_i(D_0 g^{ij} \dd_j \bP \uu) = \dd_\mu D_0 g^{\mu\nu} \dd_\nu \bP \uu - D_0 \Gamma^\nu_{\alpha \beta} g^{\alpha \beta} \dd_\nu \bP \uu + D_0 g^{\mu\nu} \dd^2_{\mu\nu} \bP \uu.
\end{equation}
The metric drift term can be treated as in section \ref{sec:metricdriftfiliso}. Therefore, we do not treat again the metric drift term here. In the assumption that $D_0$ is a medium constant, we shall also absorb this factor in the diffusion projection operator $\bP$. Plugging in the ansatz \eqref{Ansatz_aniso} then delivers
\begin{eqnarray}
  \dd_i(g^{ij} \dd_j \bP \uu) &=&  - \Gamma^A_{\alpha \beta} g^{\alpha \beta} \dd_A \bP \uu + g^{AB} \dd^2_{AB} \bP \uu \nn \\
 &&  - \Gamma^\nu_{\alpha \beta} g^{\alpha \beta} \dd_\nu \bP \tuu + g^{\mu\nu} \dd^2_{\mu\nu} \bP \tuu. \qquad \label{diffterm_aniso2}
\end{eqnarray}
The task at hand is to expand these terms involving the metric tensor around the filament using the Fermi coordinates. As we only consider corrections to the rotation frequency up to order $\la^3$, the expansion \eqref{gcon_aniso} suffices for the $g^{AB}$, $g^{\mu \nu}$ terms. Notwithstanding, the quantity $\Gamma^A_{\alpha \beta} g^{\alpha \beta}$ needs to be known up to $\OO(\la^3)$ to describe the tidal forces that act on the filament.\\

We now explicitly calculate the quantities
\bsub \label{digia_aniso} \begin{eqnarray}
  \dd_i g^{iA} &=& - \Gamma^A_{\mu\nu} g^{\mu\nu} = \dd_\mu g^{\mu A} + \frac{1}{2} g^{\alpha A} g^{\mu \nu} \dd_\alpha g_{\mu \nu}, \\
  \dd_i g^{i\s} &=& - \Gamma^\s_{\mu\nu} g^{\mu\nu} = \dd_\mu g^{\mu \s} + \frac{1}{2} g^{\alpha \s} g^{\mu \nu} \dd_\alpha g_{\mu \nu},
\end{eqnarray} \esub
which we for the ease of calculation expand as
\begin{equation}\label{digia_aniso_exp}
- \Gamma^\mu_{\alpha\beta} g^{\alpha\beta} =  \dd_i g^{i\mu} = \GG^\mu + \GG^\mu_B \rho^B + \frac{1}{2} \GG^\mu_{BC} \rho^B \rho^C + \OO(\rho^3).
\end{equation}
When inserting Eqs. \eqref{gcov_aniso}, \eqref{gcon_aniso} in \eqref{digia_aniso}, we obtain after some arithmetics that
\bsub \label{GS} \begin{eqnarray}
    \GG^\s &=& 0,  \\
    \GG^\s_B &=& \DD_\s \LA_B - \frac{2}{3} R_{B\s}.
\end{eqnarray} \esub\\

Similarly, the lowest order expansion to $\dd_i g^{iA}$ is calculated to be
\bsub \label{GAB}\begin{eqnarray}
 \GG^A &=&  - \LA^A,\\
 \GG^A_B &=& \DD_\s w \eps^A_{\hs B} - w^2 \delta^A_B - \LA^A \LA_B + \frac{1}{3} \kappa \delta^A_B - R^A_{\hs B}.
\end{eqnarray} \esub
Here, we have used that $ \DD_\s \OM^A_{\hs B} =  \dd_\s \OM^A_{\hs B}$; it can be checked with the definition of covariant differentiation \eqref{def_covderiv} that both transport terms indeed cancel. Also, the intrinsic curvature of the plane transverse to the filament has been denoted $R^1_{\hs 212} = \kappa$. All quantities involved in Eqs. \eqref{GS}-\eqref{GAB} are implicitly evaluated at the filament, i.e. at $\rho^A=0$.

The elaboration of the $\GG^A_{BC}$ is more tedious and is postponed to section \ref{sec:sectionGABC}.

\subsection{Translational EOM in the presence of anisotropy}

Reassembling Eqs.  \eqref{tterm_aniso} and \eqref{diffterm_aniso} into the the RDE \eqref{RDE_aniso} delivers
\begin{eqnarray}
\left(\dd_\tau - \HL\right) \tuu &=& \left(\vec{e}^A(0) \cdot \dot{\vec{X}}\right) \dd_A(\uu_0 + \la \tuu) +  (\omega-\omega_0)\left( \dd_\theta \uu_{0} + \la \dd_\theta \tuu\right) \nn \\
&& - \Gamma^A_{\alpha \beta} g^{\alpha \beta} \bP \dd_A \uu_0 + (g^{AB}-\delta^{AB}) \bP \dd^2_{AB} \uu_0 - \Gamma^\nu_{\alpha \beta} g^{\alpha \beta} \bP \dd_\nu \tuu \nn \\
&& + (g^{\mu\nu} \dd^2_{\mu\nu} - \delta^{AB} \dd^2_{AB} ) \bP \tuu + \frac{\la^2}{2} \tuu \bPhi''(\uu_{0}) \tuu + \frac{\la^3}{6} \bPhi'''(\uu_{0}) \tuu^3 \nn \\
&&- \frac{1}{3} \left(\vec{e}^D(0) \cdot \dot{\vec{X}}\right) R^A_{\hs BCD}  \rho^B \rho^C \dd_A \uu_0
 + \OO(\la^4).\quad
 \label{RDE_alg_aniso}
\end{eqnarray}

Proceeding as in the isotropic case, we project onto the translational adjoint Goldstone mode $\bra{\bY^B}$ to obtain in lowest order:
\begin{equation}\label{eBX}
  \vec{e}^B \cdot \dot{\vec{X}} = \bra{\bY^B} \HP \ket{\psi_A} \LA^A + \OO(\la^3).
\end{equation}
Combining with the transverse gauge $\dot{\vec{X}} \cdot \vec{e}^\s = 0$ and taking the same steps as in Eqs. \eqref{decPT0}-\eqref{defgamma12}, we are led to
\begin{eqnarray}\label{EOM_filaniso_lowest}
  \dot{\vec{X}} &=& \gamma_1 \DD_\s \vec{e}_\s + \gamma_2 \vec{e}_\s \times \DD_\s \vec{e}_\s + \OO(\la^3).\\
                        &=& \gamma_1 \DD^2_\s \vec{X} + \gamma_2 \DD_\s \vec{X} \times \DD^2_\s \vec{X} + \OO(\la^3). \nn
\end{eqnarray}
Clearly, this lowest order result is nothing else but the EOM  \eqref{EOMtr_la1int} obtained by Keener and Biktashev, generalized to a curved background space that captures the functional anisotropy of the medium. We have communicated this particular finding yet in \cite{Verschelde:2007} and \cite{Dierckx:2009}.\\

Judging on Eq. \eqref{EOM_filaniso_lowest}, the condition for a filament to remain stationary is
\begin{equation} \label{filaniso_equil}
  \dd_t \vec{X}=0 \Leftrightarrow \DD^2_\s \vec{X} = 0.
\end{equation}
The stationarity condition precisely matches the geodesic equation \eqref{geodesic_eq0}; therefrom we learn that the equilibrium positions for scroll wave filaments need to be geodesic curves. In other words, our law of motion \eqref{EOM_filaniso_lowest} provides a general proof to the minimal principle by Wellner and Pertsov.

Notwithstanding, the higher order terms in the expansion (i.e. the tidal forces) could entail equilibrium positions that do not exactly match geodesic curves.

\subsection{Rotational EOM in the presence of anisotropy}

If we had projected Eq. \eqref{RDE_alg_aniso} on the rotational adjoint mode $\bra{\bY^\rho}$ we end up with
\begin{multline}
(\omega- \omega_0) \left( 1+ \langle\bY^\theta \mid \dd_\theta \tuu \rangle \right) = - \GG^A_B \bra{\bY^\theta} \HP \rho^B \ket{\bpsi_A} - (\vec{e}^A \cdot \vec{X})  \bra{\bY^\theta} \HP \ket{\dd_A \tuu} \\
- \frac{1}{2} \dd^2_{CD} g^{AB}(0) \bra{\bY^\theta} \HP \rho^C \rho^D \dd_B \ket{\bpsi_A} - \frac{\la^2}{2} \langle \bY^\theta \mid \tuu \bF''(\uu_0) \tuu \rangle. \quad \label{RDE_rot_aniso}
\end{multline}
Without considering adjustments to the wave profile, this equation would imply
\begin{eqnarray}
\omega = \omega_0 - \GG^A_B \bra{\bY^\theta} \HP \rho^B \ket{\bpsi_A} - \frac{1}{2} \dd_{CD} g^{AB}(0) \bra{\bY^\theta} \HP \rho^C \rho^D \dd_B \ket{\bpsi_A}.
\end{eqnarray}
Appealing to the tensor decomposition \eqref{dec_rot_lowest} performed in the isotropic case, and expression \eqref{GAB} for $\GG^A_B$, we arrive at an expression of the form
\begin{eqnarray} \label{EOM_filaniso_rot0}
\omega = \omega_0 + a_0 w^2 + b_0 k^2 + d_0 \DD_\s w + F_0 \RR + G_0 \kappa.
\end{eqnarray}
To reexpress the quantity $\kappa$ in other coordinate frames, it is convenient to rewrite the curvature of the hypersurface transverse to the filament in terms of
\begin{equation}\label{defRss}
  \DD_\s \vec{X} \cdot \bR \cdot   \DD_\s \vec{X} = R_{\s \s} = \frac{\RR}{2} - \kappa.
\end{equation}
This step transforms Eq. \eqref{EOM_filaniso_rot0} to
\begin{eqnarray} \label{EOM_filaniso_rot}
\omega = \omega_0 + a_0 w^2 + b_0 k^2 + d_0 \DD_\s w + f_0 \RR + g_0   \DD_\s \vec{X} \cdot \bR \cdot   \DD_\s \vec{X}
\end{eqnarray}
with dynamical coefficients given by
\bsub \label{coeffEOMrot_aniso_lowest}
\begin{eqnarray}
          a_0 &=& 2 \left( P^{R1}_s + P^{R2\delta}_{00} - P^{R2\delta}_{s} \right), \\
          b_0 &=& 2 P^{R1}_s, \\
          d_0 &=& - 2 P^{R1}_{a} = \bra{\bY^\theta} \HP \ket{\psi_\theta},\\
          f_0 &=&  \frac{4}{3} P^{R1}_s + \frac{2}{3} \left( P^{R2\delta}_{00} - P^{R2\delta}_{s} \right),\\
          g_0 &=& - \frac{1}{3} P^{R1}_s - \frac{1}{3} \left( P^{R2\delta}_{00} - P^{R2\delta}_{s} \right).
\end{eqnarray} \esub
Apparently, the newly introduced coefficients are fully determined by the coefficients $a_0, b_0$ that had already risen in an isotropic medium.

In the quasi-stationary approximation, we can also investigate the effect of the wave profile correction $\tuu$ on the scroll wave's phase evolution. In lowest order in curvature we obtain from Eq. \eqref{RDE_aniso} that $\la \tuu = \la \uu_1 + \la^2 \tuu^{(1)}$, with
\begin{equation}\label{def_u1_aniso}
  \uu_1 = \veps \LA^A \HL_0^{-1} \HP \ket{\psi_A} = \veps \LA^A (\uu_1)_A.
\end{equation}
Considering only terms up to $\OO(\la^2)$, the same contributions are identified in \eqref{RDE_rot_aniso} as in the anisotropic case, such that Eq. \eqref{EOM_filaniso_rot} is amended to
\begin{eqnarray} \label{EOM_filaniso_rot2}
\omega = \omega_0 + a_0 w^2 + \bar{b}_0 k^2 + d_0 \DD_\s w + f_0 \RR + g_0 \DD_\s \vec{X} \cdot \bR \cdot   \DD_\s \vec{X}
\end{eqnarray}
with $\bar{b}_0 = b_0 + \veps^2 \tilde{b}_0$; the coefficient $\tilde{b}_0$ is still given by Eq. \eqref{def_b0tilde}. Taking this correction into account, there is only one linear relation left that ties $f_0$ and $g_0$ to the other dynamical coefficients in \eqref{EOM_filaniso_rot2}.

The anisotropy-induced terms in Eq. \eqref{EOM_filaniso_rot2} are also inherited by the twist evolution equation:
\begin{multline} \label{EOM_filaniso_tw}
\dd_\tau w -  d_0 \DD^2_\s w -  a_0 \DD_\s (w^2) =\\ \bar{b}_0 \DD_\s (k^2) + \DD_\s \omega_0 + f_0 \DD_\s \RR + g_0 \DD_\s \left(   \DD_\s \vec{X} \cdot \bR \cdot   \DD_\s \vec{X} \right).
\end{multline}
This expression was stated here as the inhomogeneous Burgers' equation. Importantly, variations in the intrinsic curvature of the background space around the filament are found to act as sources of twist. The special case of twist evolution around a straight transmural filament in the presence of rotational anisotropy has been treated by Setayeshgar and Bernoff in \cite{Setayeshgar:2002}, where the same type of equation was gained and discussed.

\section[Quantification of tidal forces resulting from medium anisotropy]{Quantification of the tidal forces that act on filaments in anisotropic media}

Ultimately, we here calculate the third order corrections in curvature and twist that influence filament motion; in this context \textit{curvature} applies to the extrinsic filament curvature ($k$) as well as to the intrinsic curvature of space ($\RR, \RR_{\mu \nu}, \kappa$). After collecting all relevant terms, the major task involved is to calculate the term $\GG^A_{BC}$ in \eqref{digia_aniso_exp}, i.e. to evaluate $ \dd_i g^{iA} = - \Gamma^A_{\mu\nu} g^{\mu\nu}$ up to second order in $\rho^A$.

\subsection{The full RDE projected on the translational mode}

Retaining all terms up to $\OO(\la^3)$ after projecting Eq. \eqref{RDE_alg_aniso} on the translational RF $\bra{\bY^F}$, we obtain following EOM:
\begin{eqnarray}
&& \hspace{-0.7cm} - \vec{e}^F(0) \cdot \dot{\vec{X}} =  \GG^A \bra{\bY^F} \HP \ket{\bpsi_A} + \dd^3_{CDE} g^{AB}(0) \bra{\bY^F} \frac{1}{6} \rho^C \rho^D \rho^E \HP \ket{\dd^2_{AB} \uu_0} \nn\\
&& + \frac{1}{2} \GG^A_{BC} \bra{\bY^F} \rho^B \rho^C \HP \ket{\bpsi_A} - \frac{1}{3} (\vec{e}^D(0) \cdot \dot{\vec{X}}) R^A_{\hs BCD}  \bra{\bY^F} \rho^B \rho^C \HP \ket{\bpsi_A} \nn \\
&& + (\vec{e}^A(0) \cdot \dot{\vec{X}}) \langle \bY^F \mid \dd_A \tuu \rangle + (\omega-\omega_0) \langle \bY^F \mid \dd_\theta \tuu \rangle
+ \GG^A  \bra{\bY^F}  \HP \ket{\dd_A \tuu} \nn 
\end{eqnarray}
\begin{eqnarray}
&&+ 2 \dd_B g^{A\s} \bra{\bY^F} \rho^B \HP \ket{\dd^2_{A\s} \tuu} + \bra{\bY^F} \HP \ket{\dd^2_\s \tuu} \nn \\
&& + \frac{1}{2} \langle \bY^F \mid \tuu \bF''(\uu_0) \tuu \rangle + \frac{1}{6} \langle \bY^F \mid \bF'''(\uu_0) \tuu^3 \rangle + \OO(\la^4).  \label{RDE_alg_aniso2}
\end{eqnarray}
We shall first consider the anisotropy-induced correction which is found as the fourth term on the right-hand side of \eqref{RDE_alg_aniso2}, where we use the lowest order approximation \eqref{EOM_filaniso_lowest} to $\vec{e}^D(0) \cdot \dot{\vec{X}}$. Decomposition of the matrix element of type \eqref{case34} delivers
\begin{multline}
 - \frac{1}{3} (\vec{e}^D(0) \cdot \dot{\vec{X}}) R^A_{\hs BCD}  \bra{\bY^F} \rho^B \rho^C \HP \ket{\bpsi_A} =  \\
 -\frac{\kappa}{3} \LA^F \left[ \gamma_1 (I^{T2}_{s}- I^{T2}_{00}) - \gamma_2  (I^{T2}_{a}- I^{T2}_{20})  \right]\\
 -\frac{\kappa}{3} \LA_A \eps^{AF} \left[ \gamma_1 (I^{T2}_{a}- I^{T2}_{20}) + \gamma_2  (I^{T2}_{s}- I^{T2}_{00})  \right]. \label{kapppa_I}
\end{multline}

The second term on the right-hand side of Eq. \eqref{RDE_alg_aniso2} can be evaluated using expansion \eqref{gcon_aniso} and the decomposition \eqref{R6b}. After some calculations we find
\begin{eqnarray}
  \dd^3_{CDE} g^{AB}(0) \bra{\bY^F} \frac{1}{6} \rho^C \rho^D \rho^E \HP \ket{\dd^2_{AB} \uu_0} =\hspace{4cm} \label{d3gAB} \\
 - w^2 \LA^F \left(-\frac{8}{3} P^{T3\delta}_{000} + \frac{4}{3} P^{T3\delta}_{s0s} + \frac{4}{3} P^{T3\delta}_{s0}  \right) \nn \\
- w^2 \LA_A \eps^{AF} \left(\frac{8}{3} P^{T3\delta}_{020} + \frac{4}{3} P^{T3\delta}_{a0a} - \frac{4}{3} P^{T3\delta}_{a0}  \right) \nn \\
 - \DD^F \kappa \left(-\frac{1}{6} P^{T3\delta}_{000} + \frac{1}{18} P^{T3\delta}_{s0s} + \frac{1}{18} P^{T3\delta}_{s0}  \right) \nn \\
- \DD_A \kappa \eps^{AF} \left(\frac{1}{18} P^{T3\delta}_{020} + \frac{1}{18} P^{T3\delta}_{a0a} + \frac{1}{18} P^{T3\delta}_{a0}  \right) \nn\\
 - w R_{\s F}  \left(-\frac{4}{9} P^{T3\delta}_{020} - \frac{4}{9} P^{T3\delta}_{a0a} + \frac{2}{9} P^{T3\delta}_{a0}  \right) \nn \\
- w R_{\s A} \eps^{AF} \left(-\frac{4}{9} P^{T3\delta}_{000} + \frac{4}{9} P^{T3\delta}_{s0s} + \frac{2}{9} P^{T3\delta}_{s0}  \right). \nn
\end{eqnarray}
Separate paragraphs are dedicated to the computation of the $\GG^A_{BC}$ term and the contributions $\tuu$ that describe remodeling of the wave profile.

\subsection{Inventory of terms in $\GG^A_{BC}$ \label{sec:sectionGABC}}

From its definition \eqref{digia_aniso_exp} follows that
\begin{eqnarray}
\GG^A_{BC} &=& \left. \dd^2_{BC} \left( \dd_\mu g^{\mu A} + \frac{1}{2} g^{A\alpha} \dd_\alpha g_{\mu \nu} g^{\mu \nu} \right) \right|_{\rho=0} \label{calc_digia1}\\
&=& \dd^3_{BCD} g^{DA} + \dd^3_{BC\s} g^{A\s} + \frac{1}{2} \dd^2_{BC} g^{AD} \dd_D g_{\s\s} + \frac{1}{2} \dd^3_{ABC} g_{\mu \nu} \delta^{\mu \nu} \nn \\
&&+ \frac{1}{2} \dd_A g_{\mu \nu} \dd^2_{BC} g^{\mu\nu} + \frac{1}{2} \dd_\uB g^{A\s} \dd^2_{\s \uC} g_{\s\s} + \frac{1}{2} \dd^2_{A\uB} g_{\mu \nu} \dd_\uC g^{\mu \nu}.\nn
\end{eqnarray}
As before (see Eq. \eqref{conv_uline}), the underlined indices indicate that the terms with symmetrized indices should be added.
We have manually performed the elaboration of \eqref{calc_digia1} based on the expansions \eqref{gcov_aniso}-\eqref{gcon_aniso} for the metric tensor. Altogether, over forty terms were generated, which we do not cite here. During calculation, we have substituted $\OM^\mu_{\hs \nu}$ for $\LA^A$ and $w$ terms, according to Eqs. \eqref{defOM}. Furthermore, we have rephrased Riemann tensor components in terms of the Ricci tensor using expressions in \eqref{Riemann2Ricci} and \eqref{Rprop3}. Upon introduction, ordinary derivatives with respect to $\s$ were converted to covariant ones; most notably, we have
\begin{eqnarray}
    \dd_\s \LA_C = \DD_\s \LA_C + w \eps_C^{\hs A} \LA_A.
\end{eqnarray}
Here, the powerful nature of the covariant derivative of $\LA_C$ is clearly felt: whereas the quantity $ \dd_\s \LA_C$ is merely the change of a numerical value defined relative to the coordinate frame used, $ \DD_\s \LA_C$ naturally accounts for the rate of change which is due to the twist in the coordinate frame. Compare also with Eq. \eqref{vecN2dXc}. \\

Eventually, our pen and paper calculations for \eqref{calc_digia1} lead to
\begin{eqnarray}
 \GG^A_{BC} &=& - \LA^A \LA_\uB \LA_\uC - 2 w^2 \LA_\uB \delta^A_\uC + 2 \DD_\s w \LA_\uC \eps^A_{\hs \uB}  + w \DD_\s \LA_\uC \eps^A_{\hs \uB} \nn \\
 && + \frac{4}{3} w R_{D\s} \delta_{\uB\uC} \eps^A_{\hs D} - 2 w R_{\uB\s} \eps^A_{\hs \uC} - \frac{2}{3} \LA^E R_{\uB E} \delta^A_\uC  +\frac{2}{3} \LA^E R^A_{\hs E} \delta_{\uB\uC}  \nn \\
 && - \frac{1}{3} \LA^A \RR \delta_{\uB\uC} - \frac{7}{6} \LA_A \kappa \delta_{\uB\uC}
 + \LA_\uC \kappa \delta^A_\uB -\frac{1}{2} \LA^A R_{\uB\uC}  -2 \LA_\uB R^A_{\hs \uC} \nn \\
 &&+\frac{1}{3} \LA_D \kappa \eps^A_{\hs \uB} \eps_{\uC D} - \frac{2}{3} \DD_\s R_{\s \uB} \delta^A_\uC + \frac{2}{3} \DD_\s R^A_{\hs \s} \delta_{\uB\uC} + \frac{1}{12} \DD_A \kappa \delta_{\uB\uC} \nn \\
 &&- \frac{1}{6} \DD^A R_{\uB\uC}-\frac{1}{3} \DD_\uC R^A_{\hs \uB} - \frac{1}{6} \DD_D \kappa \eps^A_{\hs \uB} \eps_\uC^{\hs D}. \quad \label{calc_digia2}
\end{eqnarray}
As a consistency check, one can return to the isotropic case by putting all Riemann tensor components equal to zero. Indeed, the first four terms in \eqref{calc_digia2} are evenly identified in expression \eqref{gGamg}.

When the RDE \eqref{RDE_alg_aniso} is projected onto the translational adjoint mode $\bra{\bY^F}$, the listed terms appear in $\frac{1}{2} \GG^A_{BC} \bra{\bY^F} \rho^B \rho^C \HP \ket{\psi_A} = \frac{1}{2} \GG^A_{BC} (P^{T2})_A^{\hs FBC}$. Decomposition in rotationally invariant components for all terms involved can be systematized by observing that the terms in Eq. \eqref{calc_digia2} appear in three distinct symmetry types. Therefore, the decomposition needs only be performed for those prototypical cases. In the following, $T_{EF}$ is taken a symmetric tensor and $V_D$ has vector character:
\bsub \label{symmGABC}
\begin{eqnarray}
V^A T_{BC} (P^{T2})_A^{\hs F BC} &=&
\left( P^{T2}_{00} - \frac{P^{T2}_{s}}{2}\right) V^F \mathrm{Tr}(T) + P^{T2}_{s} V^A T_A^{\hs F}  \\
&&+ \left( P^{T2}_{20} - \frac{P^{T2}_{a}}{2}\right) V^A \eps_A^{\hs F} \mathrm{Tr}(T) + P^{T2}_{a} V^A T_{AB} \eps^{BF}, \nn \\
V_B T^A_{\hs C} (P^{T2})_A^{\hs F BC} &=& P^{T2}_{00} V_C T^{CF} + \frac{P^{T2}_{s} }{2} V_F \mathrm{Tr}(T) \nn \\
&&+ P^{T2}_{20} V_C T^{CA} \eps_A^{\hs F} + \frac{P^{T2}_{a} }{2} V_A \eps^{AF} \mathrm{Tr}(T),
\end{eqnarray}
\begin{eqnarray}
\eps^A_{\hs B} V_C (P^{T2})_A^{\hs F BC} &=& \left(P^{T2}_{s} - P^{T2}_{00} \right)  V_C \eps^{CF} + \left( P^{T2}_{20} - P^{T2}_{a} \right)  V^F.\qquad
\end{eqnarray} \esub
Note that these three cases also appeared in our treatment of filaments in isotropic media, although with Eq. \eqref{dec4all} a systematic approach was not really required as only five terms were involved.

The relations \eqref{symmGABC} allow to contract expression \eqref{calc_digia2} to obtain the matrix element $\frac{1}{2} \GG^A_{BC} (P^{T2})_A^{\hs FBC}$. In the process, the factor $1/2$ cancels with the symmetrized indices $BC$ in \eqref{calc_digia2}. After some calculations we arrive at
\begin{eqnarray}
&&\hspace{-0.6cm} \frac{1}{2} \GG^A_{BC} (P^{T2})_A^{\hs FBC} = \label{calc_digia3} \\
&& - \left(P^{T2}_{00} + \frac{P^{T2}_{s}}{2} \right) k^2 \LA^F - \left(P^{T2}_{20} + \frac{P^{T2}_{a}}{2} \right) k^2 \LA_A \eps^{AF} \nn \\
&& - 2 \left(P^{T2}_{00} + P^{T2}_{s} \right) w^2 \LA^F - 2 \left(P^{T2}_{20} + P^{T2}_{a} \right) w^2 \LA_A \eps^{AF} \nn \\
&& - 2 \DD_\s w \left(P^{T2}_{00} - P^{T2}_{s} \right) \LA_A \eps^{AF}  - 2 \DD_\s w \left(P^{T2}_{a} - P^{T2}_{20} \right) \LA^F\nn \\
&& -  w \left(P^{T2}_{00} - P^{T2}_{s} \right) \DD_\s \LA_A \eps^{AF}  - w \left(P^{T2}_{a} - P^{T2}_{20} \right) \DD_\s \LA^F\nn \\
&& - w \left(- \frac{8}{3} P^{T2}_{00} + 2 P^{T2}_{s} \right) R_{D\s} \eps^{DF}  - w \left(\frac{8}{3} P^{T2}_{20} - 2 P^{T2}_{a} \right) R_\s^{\hs F}\nn \\
&& - \RR \left(\frac{7}{12} P^{T2}_{00} + \frac{3}{8} P^{T2}_{s} \right) \LA^F  - \RR \left(\frac{7}{12} P^{T2}_{20} + \frac{3}{8} P^{T2}_{a} \right) \LA_A \eps^{AF} \nn \\
&& - \kappa \left( P^{T2}_{00} - \frac{7}{12} P^{T2}_{s} \right) \LA^F  - \kappa \left(P^{T2}_{20} - \frac{7}{12} P^{T2}_{a} \right) \LA_A \eps^{AF} \nn \\
&& - \LA_A R^{AF} \left( 2 P^{T2}_{00} + \frac{7}{6} P^{T2}_{s} \right)  - \LA_A R^{AB} \eps_B^{\hs F} \left( 2 P^{T2}_{20} + \frac{7}{6} P^{T2}_{a} \right) \nn \\
&&  - \DD^F \RR \left(\frac{1}{4} P^{T2}_{00} + \frac{1}{8} P^{T2}_{s} \right)  - \DD_A \RR \eps^{AF} \left(\frac{1}{4} P^{T2}_{20} + \frac{1}{8} P^{T2}_{a} \right) \nn \\
&& - \DD^F \kappa \left(- \frac{1}{12} P^{T2}_{00} + \frac{1}{4} P^{T2}_{s} \right)  - \DD_A \kappa \eps^{AF}  \left( - \frac{1}{12} P^{T2}_{20} + \frac{1}{4} P^{T2}_{a} \right) \nn \\
&&  - \DD_\s R^{\s F} \left(-\frac{1}{3} P^{T2}_{00} + \frac{1}{2} P^{T2}_{s} \right)  - \DD_\s R^{\s B} \eps_B^{\hs F} \left(- \frac{1}{3} P^{T2}_{20} + \frac{1}{2} P^{T2}_{a} \right). \nn
 \end{eqnarray}
During calculation, the contributions of the type $\DD_A R^{AF}$ were redistributed over the $\DD^A \RR$ and $\DD_\s R^{\s A}$ terms. For, contracting the Bianchi identity \eqref{Bianchi} twice delivers that
\begin{equation}\label{Einsteintensor}
  \DD_\mu \left(R^{\mu \nu} - \frac{\RR}{2} g^{\mu \nu} \right) = 0.
\end{equation}
The divergenceless quantity in brackets is usually called the Einstein tensor and plays an important role in the theory of general relativity. For our application, we inferred from \eqref{Einsteintensor} that
\begin{equation}\label{DBRAB}
  \DD_B R^{BA} = \frac{1}{2} \DD^A \RR - \DD_\s R^{\s A}.
\end{equation}

\subsection{Filament EOM without wave modification effects}

Inserting Eqs. \eqref{d3gAB}, \eqref{kapppa_I} and \eqref{calc_digia3} into the projected RDE \eqref{RDE_alg_aniso2} already generates the filament EOM in the limit of negligible wave adaptation. As a preparation to properly formulating the EOM, we calculate
\bsub \begin{eqnarray}
\DD_\s \vec{X} &=& \vec{e}_\s, \\
\DD^2_\s \vec{X} &=& \LA_A \vec{e}^A, \\
\DD^3_\s \vec{X} &=& (\DD_\s \LA_A) \vec{e}^A + k^2 \DD_\s^2 \vec{X} - w \DD_\s \vec{X} \times \DD_\s \vec{X}.
\end{eqnarray} \esub
Moreover, one may write for points on the filament:
\bsub \begin{eqnarray}
R_{\s F} \vec{e}^F &=& [\DD_\s \vec{X} \cdot \bR]_\perp,\\
R_{\s A} \eps^{AF} \vec{e}_F &=& \DD_\s \vec{X} \times (\bR \cdot \DD_\s \vec{X}),\\
\DD_F S \vec{e}^F &=& [\nabla S]_\perp,\\
\DD_A S \eps^{AF} \vec{e}_F &=& \DD_\s \vec{X} \times  \nabla S,\\
\kappa &=& \frac{\RR}{2} - \DD_\s \vec{X} \cdot \bR \cdot \DD_\s \vec{X}.
\end{eqnarray} \esub
Additionally, we will need that
\begin{eqnarray}
 (\DD_\s R_{\s A}) \vec{e}^A &=& \DD_\s(\DD_\s \vec{X} \cdot \bR) - R_{\s A} \Gamma^A_{\s \mu} \vec{e}^\mu \\
 &=& [(\DD_\s \bR) \cdot \DD_\s \vec{X}]_\perp + [\bR \cdot \DD^2_\s \vec{X}]_\perp \nn \\
 && - \left(  \DD_\s \vec{X} \cdot \bR \cdot \DD_\s \vec{X}  \right) \DD_\s^2 \vec{X} + w \DD_\s \vec{X} \times (\bR \cdot \DD_\s \vec{X}). \nn
\end{eqnarray}

By gathering all terms in Eq. \eqref{RDE_alg_aniso2} that are not due to the field corrections $\tuu$, we prove following form for the filament EOM in generic anisotropic media:
\begin{eqnarray}
 &&\hspace{-0.8cm} \dot{\vec{X}} = \left( \gamma_1  + a_1 w^2 + b_1 k^2  + d_1 \DD_\s w + f_1 \RR + g_1 \DD_\s \vec{X} \cdot \bR \cdot \DD_\s \vec{X}  \right)\ \DD^2_\s \vec{X} \nn\\
&+&  \left( \gamma_2  + a_2 w^2 + b_2 k^2  + d_2 \DD_\s w + f_2 \RR + g_2  \DD_\s \vec{X} \cdot \bR \cdot \DD_\s \vec{X} \right)\ \DD_\s \vec{X} \times \DD^2_\s \vec{X}   \nn\\
&+& c_1 w [\DD^3_\s \vec{X}]_\perp + c_2 w  \DD_\s \vec{X} \times \DD^3_\s \vec{X}\nn \\
&+& h_1 [(\DD_\s \bR) \cdot \DD_\s \vec{X}]_\perp + h_2 \DD_\s \vec{X} \times  (\DD_\s \bR) \cdot \DD_\s \vec{X}   \label{ribbon_aniso}\\
&+& i_1 w [\bR \cdot \DD_\s \vec{X}]_\perp + i_2 w \DD_\s \vec{X} \times  (\bR \cdot \DD_\s \vec{X}) \nn \\
&+& j_1 [\bR \cdot \DD^2_\s \vec{X}]_\perp + j_2 \DD_\s \vec{X} \times  (\bR \cdot \DD^2_\s \vec{X}) \nn \\
&+& m_1 [\nabla \RR]_\perp + m_2 \DD_\s \vec{X} \times \nabla \RR \nn \\
&+& n_1 [\nabla  \left( \DD_\s \vec{X} \cdot \bR \cdot \DD_\s \vec{X} \right) ]_\perp + n_2 \DD_\s \vec{X} \times \nabla \left( \DD_\s \vec{X} \cdot \bR \cdot \DD_\s \vec{X} \right) + \OO(\la^5). \nn
\end{eqnarray}
Some immediate conclusions may yet be drawn from Eq. \eqref{ribbon_aniso}. First, the effective filament tension is altered by $\RR$ and $\kappa$ terms, and is therefore affected by myocardial fiber rotation rate. Secondly, untwisted filaments that have aligned with a geodesic are nevertheless subject to gradients in $ \RR$ and $\kappa$, which may therefore pull the filament away from the geodesic solution.\\

One could also have predicted the same overall shape of the equation \eqref{ribbon_aniso} solely based on dimensional and symmetry arguments. For that reason, it is not unimportant that our constructive proof of \eqref{ribbon_aniso} simultaneously offers analytical expressions for the dynamical coefficients that appear in the EOM.

We have identified following parameters that govern filament behavior in an anisotropic media -- up to (rather likely) human error in the arithmetics:
\bsub \begin{eqnarray}
\gamma_1 &=& P^{T0}_{s},\\
\gamma_2 &=& P^{T0}_{a},\\
a_1 &=& 2 \left(P^{T2}_{00} + P^{T2}_{s} \right) + \left(-\frac{8}{3} P^{T3\delta}_{000} + \frac{4}{3} P^{T3\delta}_{s0s} + \frac{4}{3} P^{T3\delta}_{s0}  \right),\\
a_2 &=& 2 \left(P^{T2}_{20} + P^{T2}_{a} \right)+ \left(\frac{8}{3} P^{T3\delta}_{020} + \frac{4}{3} P^{T3\delta}_{a0a} - \frac{4}{3} P^{T3\delta}_{a0}  \right), \\
b_1 &=&\left(P^{T2}_{00} + \frac{P^{T2}_{s}}{2} \right),\\
b_2 &=& \left(P^{T2}_{20} + \frac{P^{T2}_{a}}{2} \right), \\
c_1 &=& \left(P^{T2}_{a} - P^{T2}_{20} \right),\\
c_2 &=& \left(P^{T2}_{00} - P^{T2}_{s} \right), \\
d_1 &=& 2 \left(P^{T2}_{a} - P^{T2}_{20} \right),\\
d_2 &=& 2 \left(P^{T2}_{00} - P^{T2}_{s} \right),\\
f_1 &=& \left(\frac{13}{12} P^{T2}_{00} + \frac{1}{12} P^{T2}_{s} \right) + \frac{1}{6}  \left[ \gamma_1 (I^{T2}_{s}- I^{T2}_{00}) - \gamma_2  (I^{T2}_{a}- I^{T2}_{20})  \right],\\
f_2 &=& \left(\frac{13}{12} P^{T2}_{20} + \frac{1}{12} P^{T2}_{a} \right) + \frac{1}{6}  \left[ \gamma_1 (I^{T2}_{a}- I^{T2}_{20}) + \gamma_2  (I^{T2}_{s}- I^{T2}_{00})  \right],\\
g_1 &=&\left( - \frac{2}{3} P^{T2}_{00} + \frac{1}{12} P^{T2}_{s} \right)- \frac{1}{3}  \left[ \gamma_1 (I^{T2}_{s}- I^{T2}_{00}) - \gamma_2  (I^{T2}_{a}- I^{T2}_{20})  \right], \qquad \quad \\
g_2 &=& \left(-\frac{2}{3} P^{T2}_{20} + \frac{1}{12} P^{T2}_{a} \right)- \frac{1}{3}  \left[ \gamma_1 (I^{T2}_{a}- I^{T2}_{20}) + \gamma_2  (I^{T2}_{s}- I^{T2}_{00})  \right], \qquad \\
h_1 &=&  \left( \frac{1}{2} P^{T2}_{s} - \frac{1}{3} P^{T2}_{00}  \right),\\
h_2 &=& \left( \frac{1}{2} P^{T2}_{a} - \frac{1}{3} P^{T2}_{20}  \right),\\
i_1 &=& \left( 3 P^{T2}_{20} - \frac{5}{2} P^{T2}_{a}  \right) + \left(-\frac{4}{9} P^{T3\delta}_{020} - \frac{4}{9} P^{T3\delta}_{a0a} + \frac{2}{9} P^{T3\delta}_{a0}  \right),
\end{eqnarray}
\begin{eqnarray}
i_2 &=& \left( -3 P^{T2}_{00} - \frac{5}{2} P^{T2}_{00}  \right)  +  \left(-\frac{4}{9} P^{T3\delta}_{000} + \frac{4}{9} P^{T3\delta}_{s0s} + \frac{2}{9} P^{T3\delta}_{s0}  \right),\\
j_1 &=& \frac{5}{3} \left( P^{T2}_{00} + P^{T2}_{s} \right),\\
j_2 &=& \frac{5}{3} \left(  P^{T2}_{20} +P^{T2}_{a} \right),\\
m_1 &=&\left( \frac{1}{4} P^{T2}_{00} + \frac{1}{8} P^{T2}_{s}  \right),\\
m_2 &=& \left( \frac{1}{4} P^{T2}_{20} + \frac{1}{8} P^{T2}_{a}  \right),\\
n_1 &=&\left(- \frac{1}{12} P^{T2}_{00} + \frac{1}{4} P^{T2}_{s} \right) +  \left(-\frac{1}{6} P^{T3\delta}_{000} + \frac{1}{18} P^{T3\delta}_{s0s} + \frac{1}{18} P^{T3\delta}_{s0}  \right), \qquad \quad\\
n_2 &=& \left( - \frac{1}{12} P^{T2}_{20} + \frac{1}{4} P^{T2}_{a} \right) + \left(\frac{1}{18} P^{T3\delta}_{020} + \frac{1}{18} P^{T3\delta}_{a0a} + \frac{1}{18} P^{T3\delta}_{a0}  \right). \qquad
\end{eqnarray} \esub
It can be noted here that
\begin{align}
d_1 &= 2c_1,& d_2 &= 2 c_2,& 4 m_1 &= b_1,& 4 m_2 &= b_2;
\end{align}
in total the coefficients in the filament EOM as given here contain 18 independent components. For systems with equally diffusion state variables, the coefficients $g_1,g_2$ simplify to
\begin{align}
 f_1 &= \frac{11}{12} I^{T2}_{00} + \frac{1}{4} I^{T2}_{s}, & f_2 &= \frac{11}{12} I^{T2}_{20} + \frac{1}{4} I^{T2}_{a}.\\
 g_1 &=  I^{T2}_{00} - \frac{1}{4} I^{T2}_{s}, & g_2 &=  I^{T2}_{20} - \frac{1}{4} I^{T2}_{a}.
\end{align}\\

\subsection{Contributions due to the perturbation term $\tuu$}

To see how the perturbative corrections $\tuu$ affect the filament EOM, we determine the correction term $\tuu$ up to $\OO(\la^2)$ (see Chapter \ref{chapt:filiso}) in the quasi-stationary approximation. While the first order estimate $\uu_1 = \LA^A (\uu_1)_A$ remains unaltered, anisotropy manifests in second order. Compared to Eqs. \eqref{def_uu2}- \eqref{def_uu2_short} we obtain:
\begin{multline}\label{def_uu2_aniso}
\ket{\uu_2} =  \LA^A \LA^B \ket{(\uu_2^k)_{AB} } + \veps \dd_\s w \ket{\uu_2^w } \\ + \left(w^2- \frac{\kappa}{3} \right) \ket{\uu_2^{ww} } + R^A_{\hs B} \ket{(\uu_2^R)^B_{\hs A}},\quad
\end{multline}
with $\ket{(\uu_2^R)^B_{\hs A}} = \HL^{-1}_0 \HP \ket{ \rho^B \bpsi_A}$.\\

Akin to the isotropic case, these corrections cause the coefficients $a_i,b_i,c_i,d_i$ $i=1,2$ to shift as in Eq. \eqref{shift_filco}. Moreover, the additional term in the EOM \eqref{ribbon3} with coefficients $\veps^2 e_1, \veps^2 e_2$ is also generated in the anisotropic case. One can check from \eqref{def_uu2_aniso} and \eqref{RDE_alg_aniso2} that some other coefficients in the EOM \eqref{ribbon_aniso} are also affected by the modification of the spiral wave profile:
\begin{align}
f_1 \rightarrow \bar{f}_1 &= f_1 + \tilde{f}_1, & f_2 \rightarrow \bar{f}_2 &= f_2 + \tilde{f}_2,\nn \\
g_1 \rightarrow \bar{g}_1 &= g_1 + \tilde{g}_1, & g_2 \rightarrow \bar{g}_2 &= g_2 + \tilde{g}_2, \nn \\
j_1 \rightarrow \bar{j}_1 &= j_1 + \veps \tilde{j}_1, & j_2 \rightarrow \bar{j}_2 &= j_2 + \veps \tilde{j}_2.
\label{shift_filco_aniso}
\end{align}

As a consequence, the filament EOM \eqref{ribbon_aniso} needs be amended to
\begin{eqnarray}
 &&\hspace{-0.8cm}  \dot{\vec{X}} = \left(\gamma_1  + \bar{a}_1 w^2 + \bar{b}_1 k^2  + \bar{d}_1 \DD_\s w + \bar{f}_1 \RR + \bar{g}_1\DD_\s \vec{X} \cdot \bR \cdot \DD_\s \vec{X} \right)\ \DD^2_\s \vec{X} \nn\\
&+&  \left(\gamma_2  + \bar{a}_2 w^2 + \bar{b}_2 k^2  + \bar{d}_2 \DD_\s w + \bar{f}_2 \RR + \bar{g}_2 \DD_\s \vec{X} \cdot \bR \cdot \DD_\s \vec{X}\right)\ \DD_\s \vec{X} \times \DD^2_\s \vec{X}   \nn\\
&+& \bar{c}_1 w [\DD^3_\s \vec{X}]_\perp + \bar{c}_2 w  \DD_\s \vec{X} \times \DD^3_\s \vec{X}\nn \\
&+& \veps^2 e_1 [\DD^4_\s \vec{X}]_\perp + \veps^2 e_2 \DD_\s \vec{X} \times \DD^4_\s \vec{X}\nn \\
&+& h_1 [(\DD_\s \bR) \cdot \DD_\s \vec{X}]_\perp + h_2 \DD_\s \vec{X} \times  (\DD_\s \bR) \cdot \DD_\s \vec{X} \qquad \label{ribbon_aniso2} \\
&+& i_1 w [\bR \cdot \DD_\s \vec{X}]_\perp + i_2 w \DD_\s \vec{X} \times  (\bR \cdot \DD_\s \vec{X}) \nn \\
&+& \bar{j}_1 [\bR \cdot \DD^2_\s \vec{X}]_\perp + \bar{j}_2 \DD_\s \vec{X} \times  (\bR \cdot \DD^2_\s \vec{X}) \nn \\
&+& m_1 [\nabla \RR]_\perp + m_2 \DD_\s \vec{X} \times \nabla \RR \nn \\
&+& n_1 [\nabla \left( \DD_\s \vec{X} \cdot \bR \cdot \DD_\s \vec{X} \right) ]_\perp + n_2 \DD_\s \vec{X} \times \nabla \left( \DD_\s \vec{X} \cdot \bR \cdot \DD_\s \vec{X} \right) + \OO(\la^5). \nn
\end{eqnarray}

\section[Discussion of filament motion in the anisotropic case]{Discussion of filament motion \\in the anisotropic case}

\subsection{Tidal forces due to anisotropy}

With our dynamical equations \eqref{EOM_filaniso_rot2} and \eqref{ribbon_aniso}, we have been able to mathematically capture the effect of non-trivial anisotropy of the myocardium with respect to the propagation of electrical activity. The same equations naturally hold for isotropic media; such `flat space' scenario is easily recovered by equating all Ricci tensor components to zero.

The terminology \textit{tidal forces} that we have been using to denote the dynamical effects of locally varying anisotropy deserves more attention here. In classical physics terms, the ocean tides on earth are seen to result from non-homogeneity in the gravitational field due to the sun and moon. More specifically, the differential gravitational attraction that each of these two bodies exert on the earth, causes a tidal force when considered relative to the earth's midpoint. Consequently, such tidal force can only be felt by an extended body, and generally tends to change the body's shape.

Tidal forces may evenly be described using geodesic divergence: due to the finite distance between earth and sun, the radial geodesics that start in the sun's center are not perfectly parallel near the earth. In the perpetual falling motion of the earth around the sun, the constituents of the earth can therefore not follow the geodesic paths as freely falling particles: this effect is felt as the tidal force. From this simple reasoning can be seen that in a curved space formalism, tidal forces originate from the relative spatial acceleration of initially parallel geodesic curves. As we encounter a mathematically equivalent situation in our description of generic anisotropy, the dynamical effects have also been referred to as `tidal'.

\subsection{Residual force on a straight filament}

Interestingly, from our law of filament motion \eqref{ribbon_aniso2} no simple stationary solutions can be read for generic types of tissue anisotropy. For a filament that has aligned with a geodesic one has $\DD_\s^2 \vec{X} =0$, whence
\begin{eqnarray}
  &&\hspace{-0.8cm}  \dot{\vec{X}} = i_1 w [\bR \cdot \DD_\s \vec{X}]_\perp + i_2 w \DD_\s \vec{X} \times  (\bR \cdot \DD_\s \vec{X}) \label{force_geod} \\
&+& h_1 [(\DD_\s \bR) \cdot \DD_\s \vec{X}]_\perp + h_2 \DD_\s \vec{X} \times  (\DD_\s \bR) \cdot \DD_\s \vec{X}\nn \\
&+& m_1 [\nabla \RR]_\perp + m_2 \DD_\s \vec{X} \times \nabla \RR \nn \\
&+& n_1 [\nabla \left(  \DD_\s \vec{X} \cdot \bR \cdot \DD_\s \vec{X} \right) ]_\perp + n_2 \DD_\s \vec{X} \times \nabla\left( \DD_\s \vec{X} \cdot \bR \cdot \DD_\s \vec{X}\right) + \OO(\la^5).\nn
\end{eqnarray}
Even in the case of vanishing twist along the filament, the gradient of the intrinsic curvature measures $\RR$ and $R_{\s\s}$ as well as off-diagonal components of $\DD_\s \bR$ may thus exert residual forces on the filament. As such, we have identified (possibly small) deviations from the minimal principle presented in \cite{Wellner:2002}.

\subsection{Evolution of total filament length \label{sec:filtension_aniso}}


In Eq. \eqref{totalfillength}, the change in filament length was found equal to \cite{Biktashev:1994}:
\begin{eqnarray}\label{totalfillength2}
    \frac{dS}{dt} &=&  -  \int \dd_t \vec{X} \cdot \dd^2_\s \vec{X} \mathrm{d} \s
\end{eqnarray}
under no-flux or periodic boundary conditions. Generally, the Ricci curvature terms in the filament EOM therefore do not sustain monotonous growth of filament length.

For the special case where the filament aligns with one of the principal axes of the Ricci tensor, one has $\DD_\s^2 \vec{X} \cdot \bR \cdot \DD_\s \vec{X} = 0$; in such case we may write:
\begin{equation}
    \frac{dS}{dt} =  -  \int  \left(\gamma_1^{\rm iso} + \bar{f}_1 \RR + \bar{g}_1\vec{T} \cdot \bR \cdot \DD_\s \vec{T} + \bar{j}_1 \vec{N} \cdot \bR \cdot \vec{N} + \bar{j}_2 \vec{N} \cdot \bR^\ast \cdot \vec{B} \right) k^2\mathrm{d} \s.
\end{equation}
Here, $\gamma_1^{\rm iso}$ is shorthand for the quantity \eqref{eff_filtens} in an isotropic medium, and $\vec{N}$ equals $k^{-1} \DD^2_\s \vec{X}$. Moreover, the $j_2$ term has been reformulated using the dualized Ricci tensor:
\begin{align}
 (R^\ast)_{AD} = \eps_{AB} R^{BC} \eps_{CD}.
\end{align}

\subsection{Filament motion as a result of geodesic divergence}

\begin{figure}[h!b] \centering
  \includegraphics[width=0.9\textwidth]{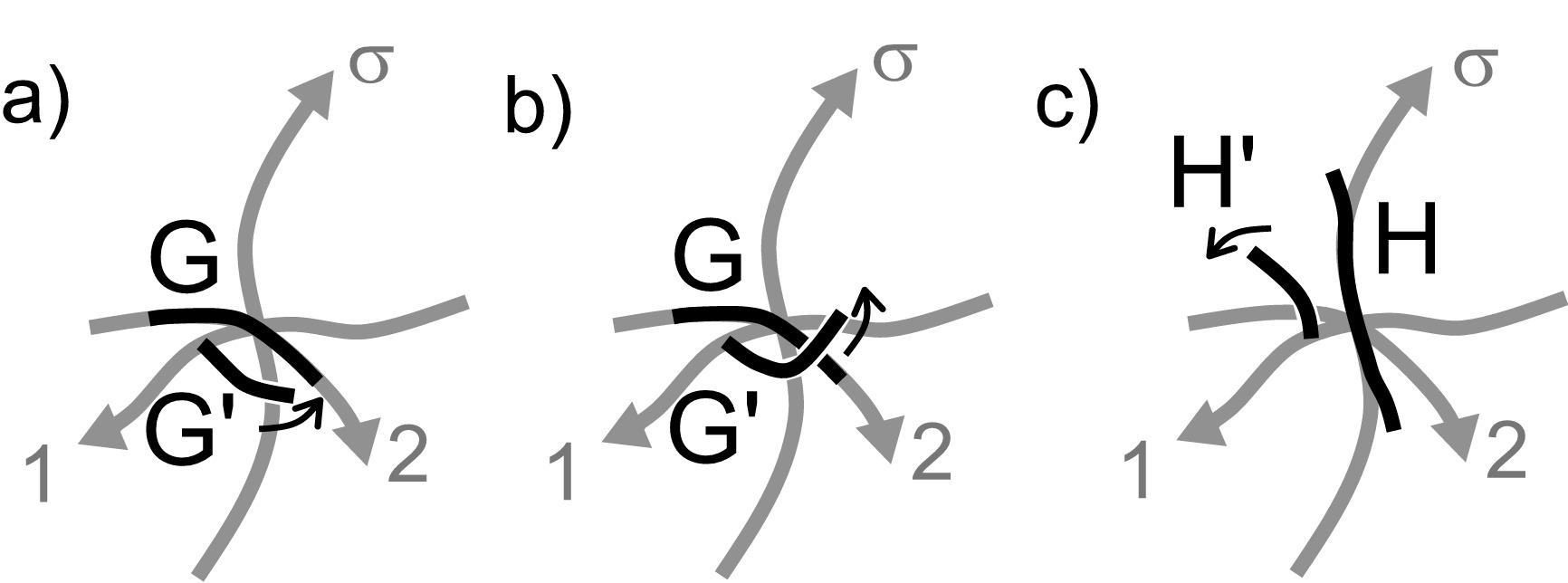}\\
  \caption[Geodesic deviation around the filament]{Types of geodesic deviation along the filament. The first two panels show intrinsic curvature (a) and shear (b) of the transverse plane, assessed by the relative acceleration of the test geodesic $G'$ relative to the fiducial geodesic $G$. The fiducial geodesic $H$ in panel (c) is chosen locally tangent to the filament, and relative deviation of the test geodesic $H'$ indicates divergence (c)of  nearby geodesics that run locally parallel to the filament.
  }\label{fig:geod_dev}
\end{figure}

The curved space formalism exposed in Chapter \ref{chapt:activ} not only offers a mathematical description, but also provides a way to reason in physical terms on filament behavior. For, in the end, filament motion is governed by their local geometry with respect to the surrounding space; therefore one may interpret the Ricci curvature terms that have emerged in our EOM in terms of the divergence of geodesics, as depicted in Fig. \eqref{fig:geod_dev}. \\

We shall take the 3-direction of the local coordinate system here tangential to the filament, so that it can be parameterized by the arc length $\s$. The different components of Ricci and Riemann tensor can then be thought of in terms of divergence and intertwining of geodesics near the filament. The starting point for this interpretation is the geodesic deviation equation \eqref{geod_dev_app}:
\begin{equation}\label{geod_dev}
  \left( \frac{\DD^2\vec{n}}{\DD s^2} \right)^\alpha + R^\alpha_{\hs \beta \mu \nu} u^\beta n^\mu u^\nu = 0.
\end{equation}
The present discussion is restricted to the three dimensional case.

\begin{enumerate}
    \item \textbf{Intrinsic curvature of the transverse plane}\\
            The single component of the Riemann tensor that has only ties to the surface transverse to the filament in a given point is $R_{1212}$. A geodesic $G'$ in the transverse plane that runs initially parallel to the geodesic $G$ that leaves the filament in the 2-direction, will accelerate towards it at a rate $R^1_{\hs 212}$, according to Eq. \eqref{geod_dev}. For notational simplicity, we will henceforth denote
            \begin{equation}\label{def_kappa}
                    R^1_{\hs 212} =   \kappa,
            \end{equation}
            and the above reasoning tells that $\kappa < 0$ correlates to divergence of geodesics in the transverse plane (i.e. it becomes saddle-like), whilst $\kappa > 0$ indicates sphere-like curvature of the hypersurface transverse to the filament.
    \item \textbf{Shear of the transverse plane}\\
            Next, consider the quantity $R^\s_{\hs 212}$, which also equals $R_{22}$ when evaluated on the filament. The quantity considered indicates how much the test geodesic $G'$ from the previous case accelerates in the direction of the filament (with positive $R^\s_{\hs 212}$ denoting deviation towards the negative $\s$-axis). Therefore, $R_{22}$ measures shear of the transverse surface in the 1-direction.
    \item \textbf{Convergence of geodesics along the filament}\\
            When choosing the fiducial geodesic $H$ locally along the filament, the test geodesic, say $H'$, that is slightly displaced in the 1-direction, bends towards the filament at a rate $R^1_{\hs \s 1\s}$. Therefore, one could say that the $R^1_{\hs \s 1\s}$ element probes geodesic convergence in the plane spanned by the filament tangent and the 1-direction, being positive in case of convergence, and negative if geodesics locally diverge.
            This quantity can be made rotationally invariant by summing $R^1_{\hs \s 1\s}$ and $R^2_{\hs \s 2\s}$ to equal $R_{\s\s}$. The element $R^1_{\hs \s 2 \s}$ indicates how the geodesic $H'$ turns in the 2-direction. An appropriate rotation of the $1,2$ axes around the filament can make this term vanish.
    \item \textbf{Local scalar curvature of space}\\
            When considering the divergence of geodesics in all three coordinate planes, the summation $ R^{\mu\nu}_{\hs \hs\mu \nu} $ needs to be performed, which eventually delivers the Ricci scalar curvature $\RR$.
\end{enumerate}
A more complete discussion of sets of initially parallel geodesics can be found in relativity textbooks, e.g. \cite{Caroll:2004}.

\subsection{Spiral wave dynamics on a curved, anisotropic surface}

Our investigations also allow to derive the rotation velocity of a spiral wave on a curved surface with anisotropy. In such case, the filament degenerates to a phase singularity point and therefore no twist or filament curvature effects arise. The driving force for spiral drift and rotational corrections is the surface's intrinsic curvature $\kappa=R_{1212}$; for isotropic two-dimensional media, $\kappa$ equals the Gaussian curvature $K_G$. In this section, the formal expansion parameter should therefore be interpreted as $\la = \sqrt{|\kappa d^2|}$.

Next, we locally endow the surface with a nearly Euclidean coordinate system $(\rho^1, \rho^2)$, using radial geodesics that emanate from the instantaneous rotation center of the spiral. Evidently, these geodesics are constrained to the surface under study. It may be noted here that the thus constructed coordinates are Riemann normal coordinates in two dimensions \cite{MTW}.

In two dimensions, the metric tensor in Eqs. \eqref{gcov_aniso}-\eqref{gcon_aniso} reduces to
\bsub \label{g_aniso_dim2} \begin{eqnarray}
g_{EF} &=& \delta_{EF} +  \frac{1}{3} R_{E A B F}  \rho^A \rho^B  + \frac{1}{6} R_{EABF; C}  \rho^A \rho^B \rho^C, \qquad\\
g^{EF} &=&  \delta^{EF}  - \frac{1}{3} R^{E\hs \hs F}_{\hs AB} \rho^A \rho^B - \frac{1}{6} R^{E\hs \hs F}_{\hs AB \hs ; C} \rho^A \rho^B \rho^C.
\end{eqnarray} \esub

The lowest order drift of the spiral wave turns out to be entirely due to metric drift
\begin{equation}\label{spiral_drift_0}
  \vec{e}^F \cdot \dot{ \vec{X}} = \frac{1}{2} \frac{\dd_A|\mathbf{D}|}{|\mathbf{D}|} \delta^{AB} \bra{\bY^F} \HP \ket{\bpsi_B} + \OO(\la^3),
\end{equation}
from which follows
\begin{equation}\label{spiral_drift}
 \dot{\vec{X}} = \gamma_1 \frac{\nabla |\mathbf{D}|}{2|\mathbf{D}|} + \gamma_2 \vec{n} \times \frac{\nabla  |\mathbf{D}|}{ 2|\mathbf{D}|} + \OO(\la^3).
 \end{equation}
with $n$ a normal vector to the medium oriented in the same sense as $\omega_0$. This lowest order result is identical to the metric drift for filaments in a three-dimensional excitable medium. If metric drift corrections are present (i.e. the local properties of the medium viewed in a laboratory frame are not homogeneous), the correction $\tuu$ reads, in the quasi-stationary approximation,
\begin{equation}\label{uu1_spiral}
  \uu_1 = \delta^{AB} \dd_A \frac{\nabla |\mathbf{D}|}{2|\mathbf{D}|} (\uu_1)_B
\end{equation}
with $(\uu_1)_A = \HL^{-1}_0 \HP \ket{\bpsi_A}$ as before in Eq. \eqref{def_uu1}.

Allowing metric drift corrections, the spiral wave's rotational frequency up to $\OO(\la^3)$ is readily obtained from \eqref{RDE_alg_aniso2}:
\begin{eqnarray}
    \omega-\omega_0 &=& - \bra{\bY^\theta} \frac{1}{\sqrt{|g_c|}} \dd_A(D_0 \sqrt{|g_c|}) \HP g^{AB} \dd_B \ket{\psi_A} + \bra{\bY^\theta} \Gamma^A_{BC} g^{BC} \HP \ket{\psi_A} \nn\\
     && - \frac{1}{2} \bra{\bY^\theta} \left(g^{AB}-\delta^{AB}\right) \HP \dd_B \ket{\psi_A} - \frac{\la^2}{2} \langle \bY^\theta \mid \tuu \bF''(\uu_0) \tuu \rangle.
\end{eqnarray}

With expressions \eqref{g_aniso_dim2}, one computes that $\Gamma^A_{BC} g^{BC} = \frac{2}{3} \kappa \rho^A + \OO(\la^3)$ and $\frac{1}{2} \dd^2_{CD} g^{AB}(0) = \frac{\kappa}{3} \eps^A_{\hs C} \eps^B_{\hs D}  + \OO(\la^3)$, whence
\begin{multline}
    \omega-\omega_0 = s_0 \kappa +  P^{R1}_s \frac{\Delta |\mathbf{D}|}{2|\mathbf{D}|}
    + \langle \bY^\theta \mid (\uu_1)_A\  \bF''(\uu_0)\  (\uu_1)^A \rangle \left(\frac{\nabla |\mathbf{D}|}{2|\mathbf{D}|}\right)^2 +   \OO(\la^4).
\end{multline}
with
\begin{equation}\label{def_s0}
s_0 = \frac{2}{3}  \left(2 P^{R1}_s + P^{R2\delta}_{00} - 2 P^{R2\delta}_s \right).
\end{equation}
For the particular case of a spiral wave rotating on a sphere with large enough radius $R$ to avoid self-interaction of the spiral wave, we obtain with $\kappa = K_G = 1/R^2$ that
\begin{equation}
    \omega = \omega_0 + \frac{s_0}{R^2}.
\end{equation}
In \cite{Zykov:1996}, the parabolical dependence of $\omega(R^2)$ was obtained phenomenologically; we have obtained the coefficient of the quadratic term here in an explicit way. Due to the circumferential deficit that arises for positive $\kappa$, it may be expected \textit{a priori} that $s_0$ typically takes positive values. \\

For the higher order curvature corrections to the spiral drift, we restrict ourselves curved surfaces that are isotropic and homogenous. This choice makes the metric drift corrections vanish and moreover $\tuu = \OO(\kappa) = \OO(\la^2)$. For that reason, modifications of the spiral wave profile effects can only induce spiral drift from $\OO(\la^5)$ onwards. In the footsteps of the three-dimensional case, we are led to
\begin{eqnarray}
 -\vec{e}^F \cdot \dot{\vec{X}} &=& - \frac{1}{2} \GG^A_{BC} \bra{\bY^F} \rho^B \rho^C \HP \ket{\bpsi_A} \nn\\ &&\ + \frac{1}{6} \dd^3_{CDE} g^{AB}(0) \bra{\bY^F} \rho^C \rho^D \rho^E \HP \dd_B \ket{\bpsi_A} + \OO(\la^5).\qquad
 \end{eqnarray}
Finally, we reach following formula for spiral drift on an inhomogeneously curved two-dimensional surface:
\begin{equation} \label{EOM_spiral}
\dot{ \vec{X}} = s_1\  \nabla \kappa + s_2\  \vec{n} \times \nabla \kappa + \OO(\la^5),
\end{equation}
with dynamical coefficients
\bsub \label{coeff_s12} \begin{eqnarray}
s_1 &=& \left(\frac{1}{4}P^{T2}_{00} + \frac{1}{2} P^{T2}_{s} \right) + \frac{1}{9}\left(-2 P^{T3\delta}_{000} + P^{T3\delta}_{s0s} +  P^{T3\delta}_{s0}  \right),\\
s_2 &=& \left(\frac{1}{4}P^{T2}_{20} + \frac{1}{2}P^{T2}_{a} \right)+ \frac{1}{9} \left(2 P^{T3\delta}_{020} +  P^{T3\delta}_{a0a} - P^{T3\delta}_{a0}  \right).
\end{eqnarray} \esub
Here too, the dynamics of the excitation pattern is significantly altered by the sign of a dynamical coefficient. For positive $s_1$, the spiral wave will drift towards the points with largest Gaussian curvature (i.e. bulges in the surface), whilst negative $s_1$ attracts spiral cores to those saddle points of the surface that have the most negative value for $\kappa$.

Note that the present treatment also holds for a surface that is locally composed of the same prototypical fiber, and could therefore to some extent represent a patch of atrial tissue. Also, thin tissue slabs that exhibit transmural myofiber rotation can be approached with the same formalism, using an effective electrical diffusion tensor.

\section{Filament motion in rotational anisotropy}

\subsection{Finding geodesics from a symmetry principle \label{sec:symm_geod}}

An elegant way to assure that a curve in a medium with local anisotropy may be found in turning to a symmetry principle. Hereto, we formally rely on a theorem in Chapter 11 of \cite{Frankel:1997}; it is the answer to the question: ``If you fold a sheet of paper once, why is the crease a straight line?''\\

The theorem states that \textit{``the fixed set of an isometry consists of connected components, each of which is a totally geodesic submanifold''}. Otherwise put, if one can identify a symmetry of the medium that preserves length, an invariant set $V$ of points is guaranteed to be a totally geodesic submanifold, i.e. if one initiate a geodesic curve tangent to $V$, the entire geodesic will be included in $V$. As a corollary, the symmetry axes of an anisotropic medium are always geodesic curves. Additionally, symmetry planes are totally geodesic submanifolds.

Applied to an excitable medium with rotational anisotropy, the transmural axis is immediately proven to be a geodesic; the symmetry operation involved is a rotation of $180^\circ$ around the transmural axis. Note that this condition is also obeyed in the case when the myofiber rotation rate is not constant throughout the wall. Under the condition of constant myofiber rotation rate, one can moreover deduce that all local material axes in the medium (which are translationally invariant) are geodesic curves.

\subsection{Drift of a straight intramural filament}

As a particular case to our filament EOM, we consider a straight untwisted filament that is oriented intramurally in a medium with rotational anisotropy. This problem was considered by Wellner \etal in \cite{Wellner:2000}.\\

First, we remark that a line orthogonal to the axis of fiber orientation is left invariant under point reflection through a point in that line. Provided that myocardial fiber rotation rate is constant, such operation preserves distance in the anisotropic medium, whence the line is a geodesic. Combined with the fact that we investigate an untwisted filament here, most of the terms in \eqref{ribbon_aniso2} vanish. Given that rotational anisotropy with fixed fiber rotation rate yields a space with constant Ricci scalar, only two terms in the filament EOM survive at all:
\begin{eqnarray}
  \dot{\vec{X}} &=& n_1 \nabla \left(  \DD_\s \vec{X} \cdot \bR \cdot \DD_\s \vec{X} \right)  + n_2 \DD_\s \vec{X} \times \nabla\left( \DD_\s \vec{X} \cdot \bR \cdot \DD_\s \vec{X}\right).\quad \label{EOM_intramural}
\end{eqnarray}
Following section \ref{sec:rot_aniso}, we lay the Z-axis of our standard Cartesian frame in the transmural direction. As in \cite{Wellner:2000}, we take the infinitely long filament parallel to the Y-axis, and the fiber direction aligned with the X-axis in the plane with $z=0$. From Eq. \eqref{EOM_intramural}, the filament is seen to respond to the gradient of
\begin{equation}\label{intramural_pert}
  T =  \DD_\s \vec{X} \cdot \bR \cdot \DD_\s \vec{X}
\end{equation}
by drifting at an angle $\Gamma$ with
\begin{equation}\label{intramural_psi}
 \tan  \Gamma = \frac{n_2}{n_1}
\end{equation}
relative to $\nabla T$; herein $\Gamma$ applies to the rescaled (curved) space. As the anisotropy considered exhibits only dependence on the transmural coordinate, $\nabla T$ points in the transmural direction. In our view, the problem has turned equivalent to the drift of a filament in the presence of a gradient in the medium properties, such as temperature or a reaction parameter \cite{Vinson:1999, Dierckx:2009}.

Using accents to denote the rescaled (curved) space, the drift \eqref{EOM_intramural} may be written
\begin{align} \label{intra_dzdt}
 \frac{dz'}{dt} &= n_1 \DD_{z'} T,&  \frac{dx'}{dt} &= n_2 \DD_{z'} T.
\end{align}
Given the local rescaling laws
\begin{align} \label{RA_resc}
dx &= \sqrt{\frac{D^{xx}}{D_0}} dx',& dz &= \sqrt{\frac{D^{zz}}{D_0}} dz',
\end{align}
it follows that
\begin{equation}\label{intramural_traj}
  \frac{dx}{dz} = \frac{n_2}{n_1} \sqrt{\frac{D^{xx}}{D^{zz}}} = \frac{n_2}{n_1} \sqrt{ \frac{D_L \cos^2 pz + D_P \sin^2 k z}{D_T} }.
\end{equation}
From this, the filament trajectory in the XZ plane is obtained as an incomplete elliptic integral of the second kind; our result is identical to the findings in \cite{Wellner:2000}.

Hereafter we determine the filament's transmural coordinate as a function of time, based on Eq. \eqref{intra_dzdt}. Aided by the rescaling properties \eqref{RA_resc} we obtain
\begin{equation}\label{intra_dzdt}
  \frac{dz}{dt} = n_1 \frac{D^{zz}}{ D_0} \DD_{z} T.
\end{equation}
Next, it is noted that
\begin{equation}\label{TRT1}
T =  \DD_\s \vec{X} \cdot \bR \cdot \DD_\s \vec{X} =  R_{yy} \left( \frac{\dd y}{\dd y'}\right)^2 = \frac{D^{yy}}{D_0} R_{yy}.
\end{equation}
When the covariant derivative acts on $T$, the factor $D^{yy}$ can be brought in front of the derivative action, as it is proportional to the metric tensor. After computation of the appropriate transport coefficients one is led to:
\begin{align}\label{DDT}
  \DD_{z} R_{yy} &= - \mu^3 \chi \sin \mu z \cos \mu z,
\end{align}
with $\mu$ the (constant) myofiber rotation rate and $\chi$ a dimensionless parameter that relates to the principal diffusivities in the medium:
\begin{equation}
 \chi = \frac{D_T (D_L-D_P)(D_L+D_P)^2}{D_L^2 D_P^2}.
 \end{equation}
This way we arrive at
\begin{equation}\label{intra_dzdt2}
  \frac{dz}{dt} = - n_1 \frac{D_T}{ D_0} \frac{D^{yy}(\mu z)}{D_0} \chi \mu^3 \sin \mu z \cos \mu z.
\end{equation}
The resulting one-dimensional dynamical system has equilibrium states for positions with $\mu z = m\pi$, $m \in \mathbb{Z}$. These loci correspond to places where the filament is either aligned or orthogonal to the local myofiber direction. The stability of the preferred positions depends on the sign of the coefficient $n_1$: if $n_1<0$, the drifting filament stabilizes in layers where myofibers run parallel to it. For $n_1>0$, however, the myofiber direction will be orthogonal to the filament in the final state. Once more, we observe qualitatively different filament dynamics if the sign of a dynamical coefficient is altered.\\

\begin{figure}[h!t] \centering
  \mbox{
  \raisebox{3cm}{a)}\includegraphics[height=4cm]{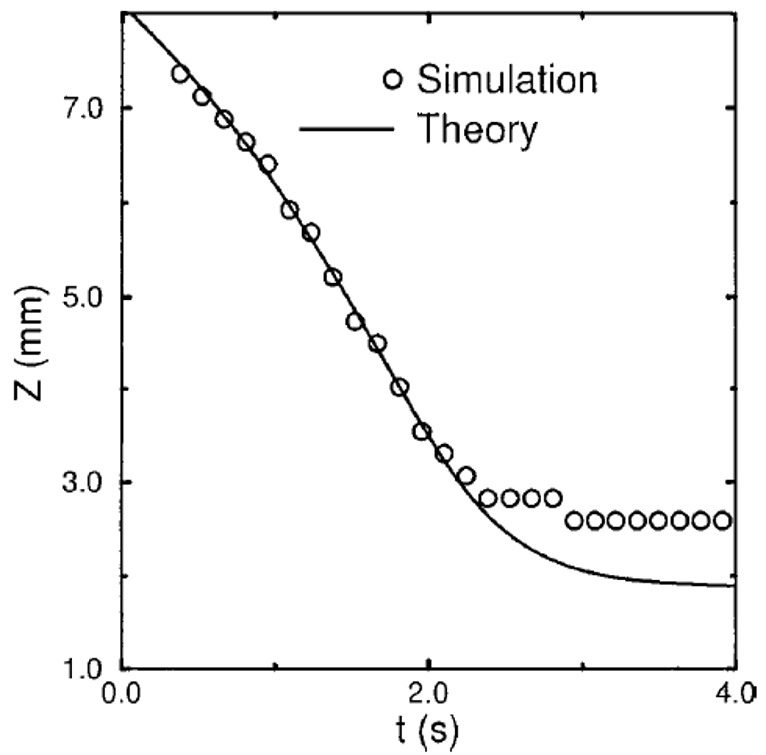}
  \raisebox{3cm}{b)}\includegraphics[height=4cm]{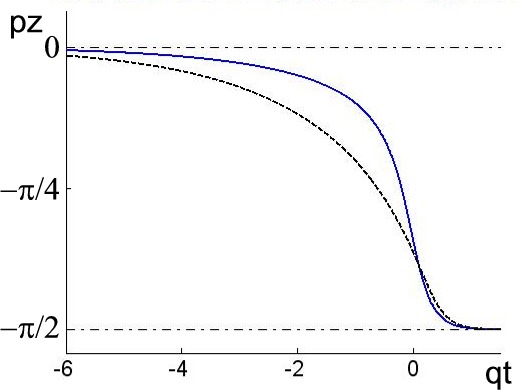}
  }
  \caption[Transmural coordinate for a drifting intramural filament]{Transmural coordinate $z$ for a drifting intramural filament for rotational anisotropy with constant pitch $p$. Panel (a) presents the theoretical and numerical outcome from \cite{Wellner:2000}. The solid line in panel (b) displays the predicted behavior according to our Eq. \eqref{sol_intramural} with $n_1<0$, compared to Wellner \etal (dashed line).
   }\label{fig:Wellner_intramural}
\end{figure}

Upon inserting $D^{yy} = D_P \cos^2 \mu z + D_L \sin^2 \mu z$, expression \eqref{intra_dzdt2} may be integrated to yield the time course of the filament. After a substitution $u=\sin^2 \mu z $, we obtain the solution
\begin{equation}\label{sol_intramural}
  q(t-t_0) = \ln (\sin^2 \mu z) - \frac{D_P}{D_L} \ln (\cos^2 \mu z) - \frac{D_L-D_P}{D_L} \ln\left(\sin^2 \mu z + \frac{D_P}{D_L-D_P} \right).
\end{equation}
Herein, the time scale parameter for the drift of the filament towards a plane of stability reads
\begin{equation}\label{timesc_intramural}
  q = - 2 n_1 \mu^4 \chi \frac{D_P D_T}{D_0^2}.
\end{equation}
Notably, the characteristic time for drift scales proportional to the fiber rotation rate to the power four. The third term in \eqref{sol_intramural} is not encountered in the treatment by Wellner \etal , who also had $q \propto \mu^2$. A visual comparison between their result and our outcome is offered in Fig. \ref{fig:Wellner_intramural}. The time scales between both solutions can not be related quantitatively here, as the time scale identified in \cite{Wellner:2000} was defined using the matrix element $P^{T0}$ instead of $P^{T2}$. Note that the relatively sharp transition to the equilibrium position from numerical experiments as depicted in Fig. \ref{fig:Wellner_intramural}a is in qualitative agreement with our result \eqref{sol_intramural}, denoted by the solid line in panel (b) of the same figure.

\subsection{Stability of a straight transmural filament}

To conclude our treatment of filaments in anisotropic media, we lay the scroll wave filament in the transmural direction in a medium with rotational anisotropy. It is known from numerical simulations that such filament may destabilize when fiber rotation is taken into account, even when the straight filament is stable in an isotropic medium \cite{Fenton:2002}. A quantitative explanation to this phenomenon has not yet been presented in literature.

From the symmetry argument in \ref{sec:symm_geod}, we already know that curves parallel to the $Z$-axis are geodesics. To simplify further calculations, we take a coordinate frame that twists with \textit{myofiber rotation}, i.e. $\vec{e}_1 || \vec{e}_f$ for all $z$. Note that these are not the twist-adapted Fermi coordinates from our theory. Rather, the reference system taken here corresponds to the relatively parallel adapted frame that was sketched in Fig. \ref{fig:frames_fil}b.

We have yet calculated the components of the Ricci tensor for rotational anisotropy in paragraph \ref{sec:curv_RA}. However, as these were referenced to the laboratory (Euclidean) reference system, we need the Jacobian matrix for the coordinate transformation $x^i \rightarrow x'^\mu$ to find the Ricci tensor components in the current frame, using Eq. \eqref{def_tensor}. Luckily, we have
\begin{equation}
 J^i_{\hs \mu} = \frac{\dd x^i}{\dd x'^\mu} = e^i_\mu
\end{equation}
so that the prescribed rotation of the reference triad immediately yields:
\begin{equation} \label{jacob_RA}
 J^i_{\hs \mu} = e^i_\mu = \left(
                   \begin{array}{ccc}
                     \cos \alpha \sqrt{\frac{D_L}{D_0}} & -\sin \alpha \sqrt{\frac{D_P}{D_0}} & 0 \\
                     \sin \alpha \sqrt{\frac{D_L}{D_0}} & \cos \alpha \sqrt{\frac{D_P}{D_0}} & 0 \\
                     0 & 0 & \sqrt{\frac{D_T}{D_0}} \\
                   \end{array}
                 \right).
\end{equation}
It can be checked with \eqref{jacob_RA} that the metric tensor $g_{\mu \nu} = e_\mu^i e_\nu^j g_{ij}$ reduced to the identity matrix, and the Ricci tensor gains a constant diagonal shape with entries $\RR \frac{(DL+DP)}{(DL-DP)}$, $-\RR \frac{(DL+DP)}{(DL-DP)}$ and $\RR$. From the diagonal shape of the Ricci tensor and $\RR$ and $\RR_{\s\s}$ being constant, it can be concluded that a filament parallel to the Z-axis exhibits translational equilibrium, irrespective of its twist properties. To investigate whether such equilibrium is stable or unstable is the task at hand.\\

Turning to the twist equation, the equilibrium state $w=0$ is unaffected by $\RR$ and $\RR_{\s \s}$ as they are constant along the filament. Due to our particular coordinate system, an untwisted filament in the natural reference frame around it is twisted when viewed in the laboratory frame, with twist equal to the myocardial rotation rate $\mu$ (with a different sign). For practical application, we will nevertheless linearize the filament around a non-zero twist $w'_0$, measured in the twisted frame. A filament that is untwisted in the laboratory frame would therefore gain
\begin{equation}
  w'_0  = \mu ' = \sqrt{\frac{D_0}{D_T}} \mu,
\end{equation}
where explicit rescaling along the Z-direction has been carried out.

We make use of the same linear basis for perturbations around the filament as in our treatment of the sproing instability in paragraph \ref{sec:sproing}. In the twisted frame around the filament, we introduce small helicoidal perturbations of radius $r$ and winding period $2\pi / p$:
\begin{align} \label{pos_initial}
 x' &= r \cos p z = r \cos p' z',&  y' &= r \sin p z = r \sin p' z'.
\end{align}
The stability analysis may now be continued by investigating under which conditions the growth rate for $r$ could becomes positive.

We shall furthermore assume that the anisotropy-induced corrections to the twist are not strong enough to cause instability at the level of the twist evolution equation. However, a helical trajectory also brings in an twist contribution, since parallel transport along the helix necessitates a phase evolution of $2\pi$ per winding of the helix. Taking into account the arc length parameter from \eqref{prop_helix}, we obtain following twist in the adapted frame:
\begin{equation}\label{twist_initial}
  w' = - w'_0 - \frac{p'}{\sqrt{1+p'^2 r^2}}.
\end{equation}

\begin{figure}[h!b] \centering
  \includegraphics[width=0.55\textwidth]{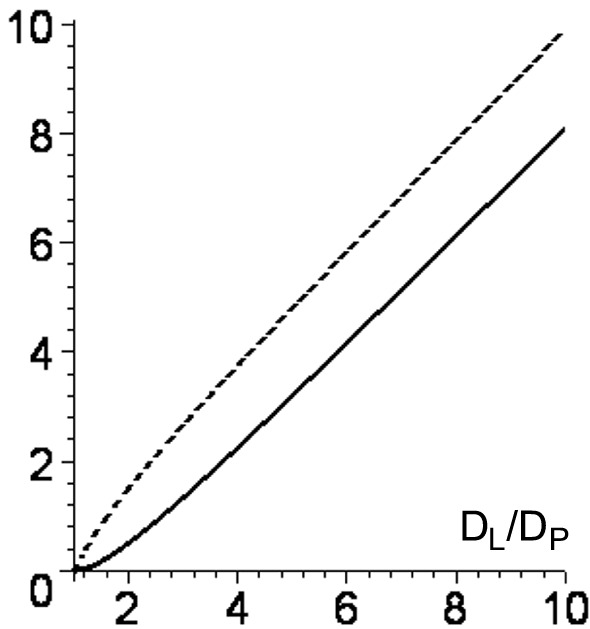}\\
  \caption[Anisotropy indices in stability analysis of a transmural filament]{Anisotropy indices in stability analysis of a transmural filament: comparison of $\mathcal{A}_f$ (solid line) and $\mathcal{A}_v$ (dashed line) for varying anisotropy ratio.}\label{fig:RAs}
\end{figure}

Further analysis can be restricted to expanding the full filament EOM in first order in $r$ for the initial position and twist given by Eqs. \eqref{pos_initial}-\eqref{twist_initial}; this procedure is easily implemented in a symbolic algebra toolbox.  Up to $\OO(\la^4)$ in our gradient expansion, the growth rate $\Omega$ turns out to be of following form:
\begin{equation} \label{growth_RA}
\dot{r} = \Omega r =  \left( \mathcal{C}_1(p' z') w_0 p' + \mathcal{C}_2(p' z') p'{}^2 + \mathcal{C}_3 w_0 p'{}^3 + \mathcal{C}_4 p'{}^4 \right) r
\end{equation}
with explicit coefficients:
\bsub \label{def_coeffs_C} \begin{eqnarray}
 \mathcal{C}_1 &=&  i_2 \frac{{\mu'}^2}{2} \mathcal{A}_f + \frac{{\mu'}^2}{2} \mathcal{A}_v \left[i_1 \sin (2 p'z') + i_2 \cos (2 p'z') \right], \\
 \mathcal{C}_2 &=& - \left[ \gamma_1 + \bar{a}_1 {w_0'}^2 - \left(\bar{f}_1 + \bar{g}_1 + i_2 \right) \frac{{\mu'}^2}{2} \mathcal{A}_f \right] \nn \\
 && + \frac{{\mu'}^2}{2} \mathcal{A}_v \left[ (\bar{j}_2- i_1) \sin (2 p'z') + (\bar{j}_1-i_2) \cos (2 p'z') \right], \\
 \mathcal{C}_3 &=& 2 \bar{a}_1 + \bar{c}_2, \\
 \mathcal{C}_4 &=& e_1 - \bar{a}_1 - \bar{c}_2
\end{eqnarray} \esub
Interestingly, the anisotropy properties of the medium gather into two distinct dimensionless ratios in the description of filament instability:
\begin{align}
\mathcal{A}_f &= \frac{(D_L-D_P)^2}{D_L D_P}, & \mathcal{A}_v &= \frac{D_L^2-D_P^2}{D_L D_P}.
\end{align}
The ratio $\mathcal{A}_f$ precedes terms that are fixed for all transmural coordinates, whereas $\mathcal{A}_v$ comes with terms that vary with transmural position. From Fig. \ref{fig:RAs} it can be seen that the $\mathcal{A}_v > \mathcal{A}_f$ if $D_L >D_P$. As the relative difference is largest in media with small anisotropy ratios $D_L / D_P$, the terms that vary with transmural coordinates are likely to play a bigger role in such media. For large anisotropy ratio, both indices tend to $D_L/D_P$.\\

\begin{figure}[h!b] \centering
  \includegraphics[width=1.0 \textwidth]{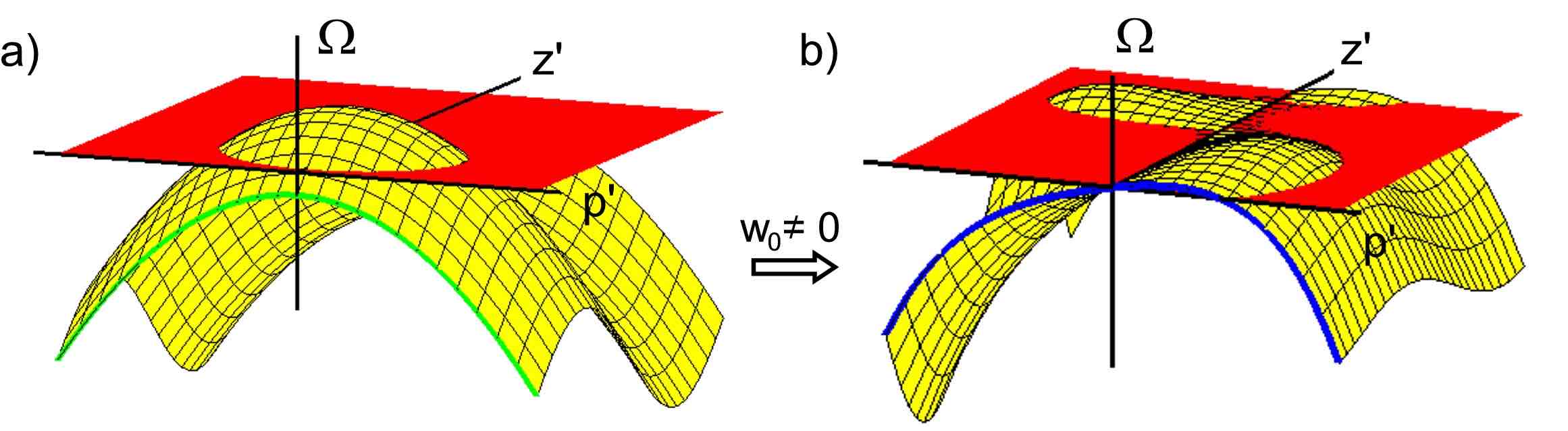}\\
  \caption[Instability of a transmural filament in rotational anisotropy]{Growth rate for instability of a transmural filament in rotational anisotropy, for a scroll wave that is twisted with the same pitch as the medium (a) and for a scroll wave that appears untwisted in the laboratory frame of reference (b). Since the growth rate for instability depends on the transmural coordinate localized instabilities develop along the filament.} \label{fig:inst_RA}
\end{figure}

Let us continue our stability analysis by considering the simplest case with $w_0' = 0$, which corresponds to a scroll wave that is initiated under equilibrium twist conditions. This makes the contributions from $\mathcal{C}_1$ and $\mathcal{C}_3$ vanish; thereby the growth rate may be written
\begin{equation}
 \Omega(p',z') =  {p'}^2 \left( \mathcal{C}_2(p' z') + \mathcal{C}_4 p'{}^2 \right).
\end{equation}
Compared to the sproing instability in the isotropic case that was displayed in panels (a) and (c) from Fig. \ref{fig:sproing}, we observe that the rotational anisotropy acts on two levels. First, the nominal filament tension $\gamma_1$ is effectively shifted over a constant value proportional to $- {\mu'}^2 \mathcal{A}_f /2$; this effect may be either a stabilizing or destabilizing, depending on the sign of $\left(\bar{f}_1 + \bar{g}_1 + i_2 \right)$. Secondly, the leading order contribution $\Omega/p^2$ is seen to exhibit cyclic variation with the transmural coordinate, with amplitude
\begin{equation} \label{def_A_RA}
 A = \frac{{\mu'}^2}{2} \mathcal{A}_v \sqrt{ (\bar{j}_2- i_1)^2 + (\bar{j}_1-i_2)^2}.
\end{equation}
Hence, when fiber rotation rate is increased, the growth rate $\Omega(p',z')$ can reach positive values only if
\begin{equation}\label{cond_growth_RA}
  \mathcal{A}_f  \left(\bar{f}_1 + \bar{g}_1 + i_2 \right) + \mathcal{A}_v  \sqrt{ (\bar{j}_2- i_1)^2 + (\bar{j}_1-i_2)^2} < 0.
\end{equation}
Moreover, the conditions for growth of an instability will first only be fulfilled in a small transmural zone at the same time. Right after this onset of translational instability, the filament tension prevents further growth of the instability, although bounded time-dependent perturbations of the instantaneous filament trajectory may be observed in this regime. The spatial dependence of the growth rate has been visualized in Fig. \ref{fig:inst_RA} by adding the transmural position in as a second coordinate axis.

When the myofiber rotation rate is further increased, the instability will more strongly develop as a local bulge in the filament and lead to an effective translational instability. The subsequent evolution of the instability falls outside the scope of our present effort. Nevertheless, in the foregoing we have shown why the growth of full helical instabilities as in the case of sproing does not occur. To summarize, if the scroll wave is initiated under twist equilibrium, the straight filament can only remain stable for high fiber rotation rate if the condition \eqref{cond_growth_RA} is not fulfilled.\\ 

We now turn to the more complicated case of a scroll wave that is initially untwisted in the laboratory frame, which has therefore $w'_0 = - \mu'$ in the reference frame adapted to the transmural myofiber rotation. In the long-wavelength limit (i.e. small $p'$), the growth rate is always found to take positive values since
\begin{equation}\label{growth_C1}
    \Omega(p', z') =  - p' \frac{{\mu'}^3}{2}  i_2 \mathcal{A}_f - p' \frac{{\mu'}^3}{2}  \mathcal{A}_v \left[i_1 \sin (2 p'z') + i_2 \cos (2 p'z') \right)] + \OO({p'}^2).
\end{equation}
As in the previous case, the benign effect of filament tension $\gamma_1$ may overcome this destabilizing effect by allowing only a narrow interval in which the growth rate is positive. Because the instability described here situates in the low-wavelength regime, the growing instability affects the part of the filament for which $ \Omega$ is positive as a whole. Short-wavelength perturbations are also a sustained by the full equation \eqref{growth_RA}. Here too, the instability first sets in only in a localized transmural region. The resulting instability will therefore also manifest as a local protrusion in the filament shape.\\

The effect of reversing the scroll wave's rotation sense is most easily be studied by changing the sign of the fiber rotation pitch $\mu$, as all dynamical coefficients in the filament EOM remain unchanged with this approach. Due to the appearance of only even powers of $p$ and $\mu$, the growth rate $\Omega$ does not change under such mirror symmetry of the medium. However, without the time averaging over one rotation period, parity breaking terms do appear and may therefore discriminate between the evolution of scroll waves of opposite chirality.\\

The outcome of our perturbative analysis is in qualitative agreement with existing numerical simulations of filament dynamics in the presence of rotational anisotropy. Evidently, a direct comparison between theory and numerical experiment would be appropriate here.

\clearpage{\pagestyle{empty}\cleardoublepage}

\graphicspath{{fig/}{fig/fig_synth/}}

\hyphenation{an-iso-tro-pic}

\renewcommand\evenpagerightmark{{\scshape\small Chapter 9}}
\renewcommand\oddpageleftmark{{\scshape\small Assessment of intrinsic curvature using RTI}}

\chapter[Synthesis: Assessment of intrinsic curvature using RTI]{Synthesis: Assessment of \\intrinsic curvature using RTI}
\label{chapt:synth}

In the previous chapter, the temporal evolution of a scroll wave filament in anisotropic cardiac tissue was shown to strongly depend on the associated Ricci curvature tensor. Here, we expose an integrative approach to realistically quantify the intrinsic curvature effects for individual hearts by fusing the curved space formalism with the diffusion tensor imaging that was discussed in chapter \ref{chapt:imag} of this thesis.  The basic observation is that the principal material axes of a piece of myocardium directly correlate to both the electrical diffusion tensor $\mathbf{D}_{\rm el}$ and the proton diffusion tensor $\mathbf{D}_{\rm pr}$.

If one manages to estimate the typical dynamical coefficients in the filament EOM, the methodology outlined here could serve for patient-specific risk stratification on the development of structure-induced cardiac arrhythmias.

\section{Methods}
Our present treatment is valid under the approximation that a single dominant laminar orientation is present in the myocardial tissue. In that circumstance, the electrical and proton diffusion tensor share the same set of eigenvectors, and possess the same order in eigenvalues. Using the direct product of vectors we may therefore write
\bsub \begin{eqnarray}
 \mathbf{D}_{\rm el} &=& D_1^{\rm el} \vec{e}_f  \otimes \vec{e}_f + D_2^{\rm el} \vec{e}_s  \otimes  \vec{e}_s + D_3^{\rm el} \vec{e}_n \otimes  \vec{e}_n,  \label{Dpr}\\
 \mathbf{D}_{\rm pr} &=& D_1^{\rm pr} \vec{e}_f   \otimes \vec{e}_f + D_2^{\rm pr} \vec{e}_s   \otimes \vec{e}_s + D_3^{\rm pr} \vec{e}_n \otimes \vec{e}_n.
\end{eqnarray} \esub
However, the eigenvalue ratios differ considerably: detailed measures of conduction velocity in \cite{Caldwell:2009} yielded an anisotropy ratio of $18.5: 3.2 :1$ for the electrical diffusion tensor, whereas ex-vivo DTI on canine hearts resulted in absolute proton diffusivities of  $\unit{(1.0, 0.55, 0.45)}{\micro \meter}^2$/ms.
Nevertheless, $\mathbf{D}_{\rm el}$ can be reconstructed given a measurement of $\mathbf{D}_{\rm pr}$ by DW-MRI means using expression \eqref{Dpr}: it suffices to plug experimentally obtained eigenvalues in Eq. \eqref{Dpr} together with the eigenvectors found by DTI. This approach is commonly chosen in present-day numerical simulations that aim to fiducially represent anisotropic cardiac tissue.

Our addition is to use the estimate of Eq. \eqref{Dpr} to compute the Ricci curvature tensor and Ricci scalar in each point of an DTI data set. Spatial derivatives in Eqs. \eqref{defRiemann},\eqref{christ} are for now evaluated by simple finite differences on the lattice formed by the DTI data.

\section{Ricci scalar map of rabbit ventricle}

For simple visualization, we have plotted the Ricci curvature scalar in an axial slice and a coronal slice for a rabbit heart\footnote{DTI data set courtesy to Dr. Flavio Fenton.}. In representing the Ricci scalar, blue color encodes negative scalar curvature, while red stands for positively curved space. As anticipated with the example of rotational anisotropy, we observe that most parts of the myocardium exhibit a negative curvature scalar. Moreover, regions with most intense intrinsic curvature relate to zones where myofiber diverge (e.g. LV-RV fusion sites) or rotate abruptly (e.g. border zone with papillary muscle); these zones have yet been suspected to play an important role in the development of cardiac arrhythmias.

\begin{figure}[h!b] \centering
  \includegraphics[width=1.0\textwidth]{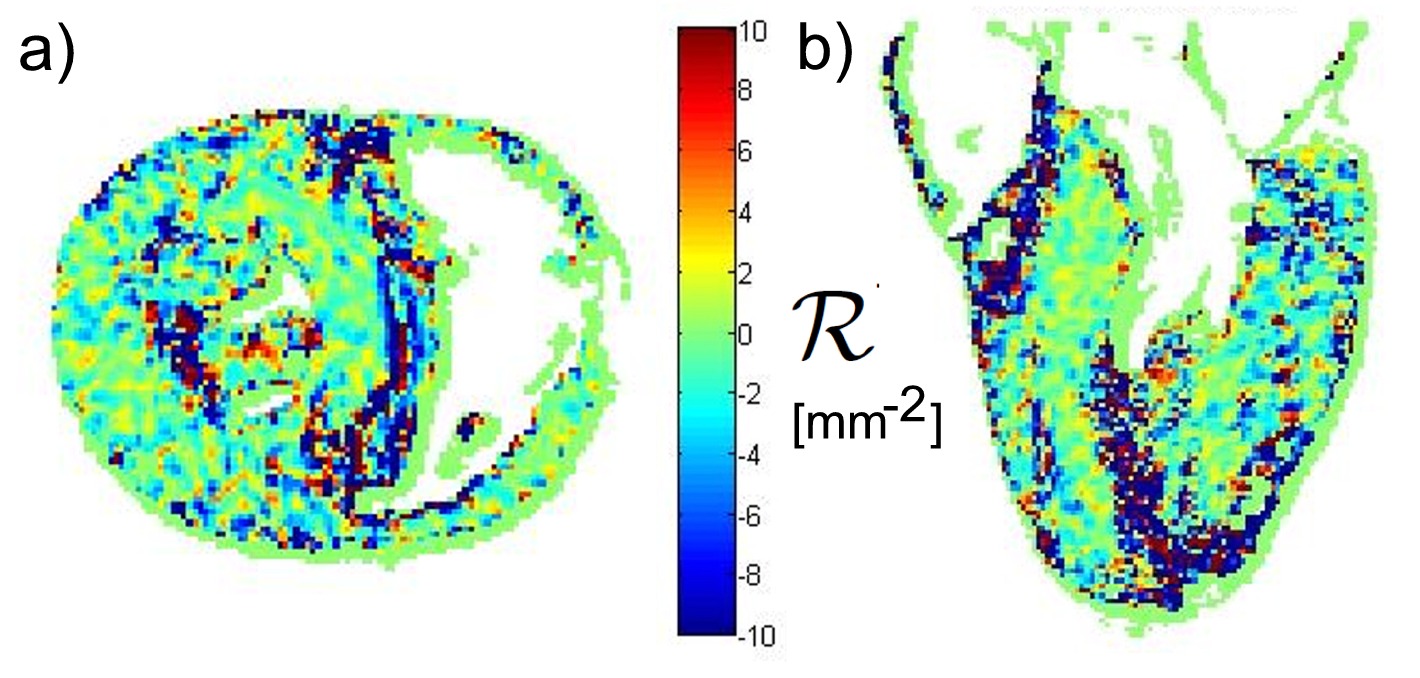}\\
  \caption[Ricci scalar map for rabbit ventricle]{Ricci scalar map for rabbit ventricle in axial (a) and coronal (b) view.} \label{fig:ricci_rabbit}
\end{figure}

\section{Discussion of RTI}

The RTI methodology presented here is still preliminary. To start with, one will need to assess reproducibility and the effects of noise in estimating the curvature tensors, as the method involves taking the second spatial derivatives of fitted tensors on a grid of modest spatial resolution. Moreover, the correspondence between the electrical and proton diffusion tensor requires further quantitative validation; perhaps the anisotropy ratios need to be adapted for different anatomical sites. In the light of the coexisting laminar populations that were investigated in Chapter \ref{chapt:QBI}, the simple strategy outlined here will require additionally amendments to deal with such complex laminar structure.

\clearpage{\pagestyle{empty}\cleardoublepage}


\renewcommand\evenpagerightmark{{\scshape\small Chapter 10}}
\renewcommand\oddpageleftmark{{\scshape\small Conclusions and outlook}}

\hyphenation{an-i-so-tro-py }

\chapter[Conclusions]{Conclusions and outlook}
\label{chapt:concl}

\section[Anisotropy, geometry and pathways to instability]{Anisotropy, geometry and \\pathways to instability.}

\subsection{Anisotropy revisited}

In the strictest sense of the word, anisotropy indicates anything that breaches local isotropy in a material. In more abstract terms, anisotropy breaks local rotational symmetry of a medium, and thereby affects the previously isotropic processes taking place within it. Throughout this work, we have engaged two different strategies to deal with this particular circumstance.\\

Our first tactic is a seminal one in both scientific and everyday problem-solving: \textit{if faced with a novel difficulty, reduce it to a known problem}\footnote{A particular realization of this option for electric systems is to turn the device off and on again. In a cardiac context, the resulting procedure is better known as defibrillation.}. Thus, when confronted with anisotropy, perform a local rescaling to restore isotropy. The price to pay is that one will have to operate in a non-Euclidean space, wherein tidal forces (Ricci curvature terms) enrich the dynamics of extended structures. We have concretized this approach by original application to the motion of wave fronts and filaments in anisotropic excitable tissues, which uncovered their effective equations of motion in a non-isotropic context.

A second stance towards anisotropy is a more pragmatic one: \textit{turn any complication to a potential advantage.} Specifically, observing processes which are subject to anisotropy may serve to examine the properties of the medium. This is how myocardial anisotropy is commonly quantified in the first place, namely by recording directional differences in the conduction velocity of action potentials. We have benefited from another process, i.e.  the thermal diffusion of water molecules, to assess the principal material axes in extended volumes of myocardial tissue with diffusion-MRI. Without having to resort to an underlying diffusion model, we were able to infer the orientation of crossing myofibers and cleavage planes on purely geometrical grounds using a method which we have called dual QBI. Although anisotropy with respect to electrical diffusion largely differs from the restricted diffusion of water, both sets of principal axes coincide with the tissue's local material axes, which enables the coupling of orientational information from one process to the other. This methodology has been used before to produce realistic anatomical maps that include fibrous and laminar organization based on DTI measurements.

In the `curved space interpretation' of anisotropy, we have drawn from DTI data sets to extract intrinsic curvature properties of the myocardium, which act on wave fronts and filaments through the lensing effect and tidal forces, respectively.

\subsection{Anisotropy effects are either tidal or fictitious}

The seemingly complicated motion of scroll wave filaments in media with rotational anisotropy has been the subject of many numerical and analytical studies. We have pursued the idea initiated in \cite{Wellner:2002} to treat the inverse of the electrical diffusion tensor as the metric tensor for the space under study. Realistic ventricular anatomy turns this background space into a heavily curved Riemannian manifold, which provokes the non-trivial dynamics of rotor filaments in the eyes of an external observer.

A scientist that starts to reason in the curved space formalism soon realizes that only the relative acceleration of initially parallel nearby  geodesics can invoke original anisotropy effects, which are known as tidal forces in gravity theory. All other phenomena that were perceived in the common laboratory system of reference should be regarded upon as merely fictitious forces, i.e. consequences of choosing a non-ideal reference frame. In the very same way, gravitational force is gauged away in Einstein's general relativity theory, for gravity may be considered fully due to intrinsic curvature of spacetime.

Finally, it is worth noting that not all tidal forces violate the minimal principle for rotor filaments. In particular, Ricci tensor terms that couple to extrinsic filament curvature in the filament EOM do not act on locally straight filaments unaffected.

\subsection{Universality in the dynamics of excitation patterns}

A major concern with present-day mathematical models that are used to represent cardiac activation sequences is that it is hard, if not impossible, to forecast how a particular model that was conceived at the single cell level will behave in extended pieces of tissue. In particular, varying the electrophysiological parameters is known to influence stability of wave fronts, spirals and scroll waves. Even more, models which have the same restitution properties and action potential profile may nevertheless give rise to substantially different filament behavior \cite{Rappel:2001}. Also, it turns out that some extremely detailed ionic models exhibit the same type degree of complexity in their activation patterns as simple reaction-diffusion models do.\\

Endorsed by the geometrical theories for activation fronts and rotor filaments that were advocated in this work, we may draw the conclusion that \textit{all} reaction-diffusion systems that sustain activation fronts behave remarkably similar, given that only a handful of dynamical coefficients $(c, \gamma, \omega, \gamma_1, \gamma_2)$ accounts for the basic motion of wave fronts and filaments (disregarding dispersion and meandering effects). Describing the more pronounced curvature effects requires additional coupling constants. We have found that four of them suffice to capture wave front dynamics up to second order in curvature, even in an anisotropic context. Since filaments have got more degrees of freedom (i.e. three zero modes), their description requires not less than 17 coefficients in isotropic media, and 16 to capture the additional effects of anisotropy (up to the order given here). In this listing, we have yet eliminated the components of the filament motion that average to zero during one rotation period.

The situation may be compared to classical mechanics, where the motion of a rigid body can be accurately determined from the initial conditions once the coupling factors to external forces are given. In lowest order, the unique parameter is the mass of the body, which for that reason therefore resembles the filament tension coefficient $\gamma_1$ for scroll wave filaments. In a more advanced description, the rigid body is seen to possess other dynamical coefficients such as a moment of inertia, a drag coefficient and perhaps an electrical charge, which mediate supplementary forces. Inclusion of the previously neglected effects brings a better description of the object's motion, at the cost of a more involved equation of motion. A major difference between filament motion and the classical mechanics of a flexible rod is that filaments have no inertia, because their EOM is first order in time. Therefore, filament dynamics is similar Aristotelian mechanics, where motion is determined by friction forces, rather than inertia.\\

Part of the success of modern physical theories lies in the insight gained by formulating evolution equations as an action principle. For wave fronts, we have been able to cast the lowest order velocity-curvature relation into an elegant variational principle, in spite of the dissipative nature of the system. Again, distinct RD models generate potential functions which only differ in the pre-factors that go with each term.

\subsection[Pathways to instability from the extended EOM]{Pathways to instability from the extended\\ equations of motion}

Based on our newly derived dynamical equations, we have been able to shed new light on dynamical instabilities that have been observed with scroll wave filaments and wave fronts. In particular, our analytical description pinpoints several mechanisms which may destabilize filaments and thereby lead to fibrillation. In isotropic media the effective tension of a rotor filament was proven to depend on several geometric invariants: the local twist and filament curvature, and derivatives of these with respect to arc length along the filament. For the first time, we have derived explicit coupling terms between twist and translational degrees of freedom; our extended equations of motion not only cover the sproing instability in quantitative terms, but also indicate that local filament curvature may impair the stabilizing twist diffusion process.\\

When taking into account tissue anisotropy within the operational distance framework, the Ricci scalar curvature of space is observed to shift the effective filament tension. For the particular case of rotational anisotropy, the Ricci curvature scales quadratically with the myofiber rotation rate, which suggests that we may have uncovered an important mechanism for anisotropy-induced filament instability. Nontrivial anisotropy further complicates filament dynamics, as it spoils the principle that the filament always reacts to external perturbations by drifting under a fixed angle with the direction along which the perturbation is applied. The culprit is a Ricci tensor term in the filament equation of motion that couples to local filament curvature. In a rotational anisotropy setting, the Ricci tensor varies with the local fiber rotation and thereby acts differently on other parts of a transmural filament, which we have proven to be a possible pathway to filament instability using linear stability analysis. In our analysis of a transmural filament in rotational anisotropy, we have managed to effectively unwind the medium by choosing a reference frame that rotates with the fiber rotation. Such frame facilitates calculation of the tidal forces, since the Ricci tensor is reduced in a constant diagonal matrix.

Our analysis of a transmural filament in rotational anisotropy shows that the destabilization of the filament is highly similar to the sproing instability in an isotropic medium, as an untwisted scroll wave in an Euclidean frame of reference confers to a homogenously twisted scroll wave in a locally Euclidean space around the filament. Remarkably, the local myofiber rotation makes the growth rate for translational instabilities not only sensitive to their wave number, but also to the transmural coordinate. For that reason, the filaments only locally destabilize in such medium. \\

The concept of physical tension as a criterion for stability has been broadened here to encompass curvature-induced wave front instabilities instead of filaments. The lowest order coefficient that enters the velocity-curvature relation was attributed the physical meaning of surface tension. In systems such as myocardium where not all state variables diffuse at the same speed, the obtained surface tension coefficient may take values that are different from one. In particular, ionic models with negative surface tension would exhibit unstable wave fronts, in analogy with tension-related filament instabilities. The second order corrections to the velocity-curvature relation, which were originally derived here, indicate that wave front curvature alters the effective surface tension, enabling instabilities to develop in activation fronts that are too intensely curved.

\section{Applications}

\subsection{Understanding arrhythmias}

First and foremost, we value the insights that are gained from a novel physical theory. Given the not quite restrictive regime in which our geometrical theories for wave fronts, wave trains and rotor filaments were developed, they yet cover a broad class of models of cardiac activation. So far, all reaction-diffusion models in the continuous monodomain approximation with differentiable kinetics have been demonstrated to obey the laws derived. The wide range of models covered permits to judge in general and quantitative terms on the dynamics of cardiac excitation patterns.

Evidently, the study of cardiac arrhythmias and disease is infinitely larger than writing down a set of equations that explain a only small part of the life cycle of a heart rhythm disorder. Luckily, our research turned out to be located on the trailing edge between a disordered heart and a failing one. Yet solely for that reason, any relevant tool from medicine, technology, pharmacology or statistics -- let alone differential geometry -- must be considered a welcome one.

\subsection{Time-efficient numerical computations}

The development of equations of motion for wave fronts and rotor filaments in regimes of moderate curvature paves the way for numerical studies of activation wave dynamics that are solely based on the motion of their organizing centers. For a given RD model, the set of dynamical coefficients that appears in the EOM up to given order may be calculated by either explicit evaluation of the overlap integrals of Goldstone modes, or parameter fitting from forward numerical simulation of the RDE in standard geometries. In future realizations, typical experimental values of such coefficients may be employed once they have been measured. Given the EOM, the initial configuration of a wave front or filament needs be discretized, after which forward evolution of the EOM, which is first order in time, may be explicitly implemented. With $N$ elements on each side of a cubic lattice, solving the full RDE is a process of order $N^3$, whereas tracking only wave fronts scales as $\OO(N^2)$ and a filament merely necessitates $\OO(N)$ elements.

The proposed approach is compatible with realistic fiber and laminar structure of the heart through the Ricci Tensor Imaging (RTI) correspondence. With these elements, we estimate it feasible to replicate a single transmural filament in a realistic anatomy and check where the filament equilibrates, or by which reason it loses stability to induce a state of fibrillation. Unfortunately, with the present description it is not possible to look further than the time where the filament becomes unstable, since one does not yet dispose of a theory of filament interaction. The same kind of investigation could be conducted with wave fronts initiated in a realistic geometries; here the evolution must be stopped when wave breaks are induced, as it is still unclear how to treat them in a way that is consistent with geometric filament dynamics.

\subsection{Relevance to other excitable and oscillatory systems}

The insights gained from our investigations on cardiac activation dynamics have been attained using generic reaction-diffusion models with an arbitrary number of state variables. Therefore, we expect our findings to be evenly applicable to other active media that are modeled using reaction-diffusion equations. Specifically, spiral wave structures may occur in both excitable and oscillatory media. Moreover, our dispersive velocity-curvature relation applies to both excitable and oscillatory systems. Importantly, we have found qualitatively different behavior of systems where all state-variables diffuse at the same rate, and those in which a non-trivial diffusion projection operator $\HP$ needs be included. Generally speaking, dynamics is more involved in the latter case because the inertia of the less diffusing variables induces additional curvature effects.

\subsection{Diffusion imaging of complex laminar structure}

The imaging of complex laminar structure by means of the dual QBI method holds future clinical potential by its non-invasive nature. As with other high-angular resolution diffusion MRI methods, current sampling strategies are far too time-consuming to nominate for routine medical practice. Notwithstanding, pilot studies may be performed to quantitatively broaden anatomical knowledge. Before going into that phase, however, further validation and optimization of the method is instructive.

\section{Outlook}

\subsection{Consolidation}

The analytical findings presented in this work urge to be validated. In a first stage, numerical simulations building on those that were included here will enable to detect missing or wrongfully calculated terms which may be present in our equation. The very same simulations may evenly serve to check the yet unknown range of validity for our gradient expansion series. In addition, further literature study and discussion are recommended to be taught other known phenomena which might be either fit within our geometric theory or which may exhibit its limitations.

Concerning the imaging part of this thesis, it would be instructive to hold the dual QBI method against other imaging techniques of various nature, to estimate the relative advantages and weaknesses of this diffusion MRI technique. Attention needs be paid to fixation and registration procedures, as well as reproducibility and anatomical variation between individuals and animal species. It remains also to be seen how portable the method is towards clinical MRI scanners, which operate at lower gradient strengths due to their larger field of view.

\subsection{Overcoming present limitations}

Given the present analytical treatment, several modifications that could enlarge its scope lie within close reach. Notably, only minor modifications are needed to make the bidomain equations for cardiac excitation satisfy the current gradient expansion approach. More challenging will be to tune in the different anisotropy ratios for the intra and extracellular domains, as have been observed in myocardial tissue \cite{Panfilov:CBH}. A second amendment would be to include the meandering of scroll wave filaments, i.e. examine how meandering spiral patterns are affected by adding a third spatial dimension. While the description of smooth meandering patterns is related to our description of breather modes in section \ref{sec:breather}, the description of linear cores remains more challenging. Nonetheless, as our dynamical equations of motion contain all possible terms that remain after rotational averaging, they can expected to hold for any quasi-periodic meandering trajectory. In such circumstance, however, it is less clear how the coefficients in the EOM can be obtained on theoretical grounds.

As indicated in the main text, the dual QBI method for laminar structure may benefit from developments similar to those coined for standard QBI in brain imaging applications. Amongst others, deconvolution strategies may prove useful for boosting angular resolving power. Another limitation of dual QBI in its present state is that it cannot discriminate between voxels where multiple laminar orientations are present, and those without cleavage planes. Rather than a filtering based on (the absence of) spatial correlation, a sound quantification of the diffusion properties of myocardial tissue is desirable. Also, the effects of dehydration and perfusion on medium parameters related to diffusion processes deserve further research. In particular, both cleavage plane widths and the surface-to volume ratio and orientation of perimysial collagen structures are expected to influence the resulting diffusion weighted images.

\subsection{Challenges}

An substantial challenge lies in translating the analytical and imaging results presented in this work towards more applied fields of cardiac research.\\

Importantly, the dual QBI method should be implemented on clinical scanners to enable the structural imaging of large hearts, such as human. Current and future advances in the diffusion weighted imaging of beating hearts \textit{in situ} will obviously need to be included in the imaging protocol.\\

For the theoretical part, an initial link with computational modeling may be established by determining the set of leading order dynamical coefficients for the EOM for various models of cardiac excitation. A less obvious step is how to further tie the determinants for wave front and filament motion to experimental recordings. Such outcome could serve as a check for computational models, in the same way as restitution curves are being used for nowadays for model fitting.

It can furthermore be expected that numerical and physiological experiments may directly benefit from the operational measure of distance promoted here. For, the formalism enables to separate between structural anisotropy and other dynamical effects.

Even without knowing precise values (if any) for the free parameters in the modified equations of motion for fronts and filaments, one may yet continue to analyze the equations to gain insights in the types of instability. Still further simplifications could be gained by deriving additional variational or conservational laws in the study of cardiac excitation patterns.

Although we may have established an important window for the study of rotor activity, the entire life course of arrhythmias has not been covered yet. Notably, it would be appealing from a theoretical point of view to dispose as well of a sound description of the evolution from wave breaks to scroll waves that takes into account the critical properties of diffusion near the phase singularity. The other end of our validity regime is the transition to fibrillation; it can be expected that further insight in the mechanisms of scroll wave interaction may lead to better understanding of the processes that occur during cardiac fibrillation. Since filament interaction is likely to have a non-perturbative character, understanding filament interaction remains a important future challenge.

%


\clearpage{\pagestyle{empty}\cleardoublepage}

\appendix
\renewcommand\evenpagerightmark{{\scshape\small Appendix \thechapter}}
\renewcommand\oddpageleftmark{{\scshape\small\leftmark}}

\renewcommand\evenpagerightmark{{\scshape\small Appendix A}}
\renewcommand\oddpageleftmark{{\scshape\small Elements from differential geometry}}

\hyphenation{}

\chapter[Elements from differential geometry]{Elements from differential geometry}
\label{app:diff_geom}

In this work, the Einstein summation convention is adopted, which allows to omit summation signs in front of sums that runs over repeated indices.

\section{Physics in curved spaces}

\subsection{The metric tensor}

A three-dimensional curved space -- more precisely: a Riemannian manifold-- can be characterized by a metric tensor, which prescribes the distance between neighboring points, given a particular coordinate system:
\begin{equation}\label{ds2}
  ds^2 = g_{\mu \nu} d x^\mu x^\nu.
\end{equation}
In the familiar three-dimensional Euclidean space, the metric takes the shape of the unity matrix ($g_{\mu \nu} = \delta_{\mu\nu}$), and \eqref{ds2} becomes the familiar Pythagorean theorem. When trading one coordinate system for another, invariance of $ds^2$ implies that
\begin{equation}\label{ds2change}
  (g')_{\mu \nu} = \frac{\dd x^i}{ \dd {x'}^\mu} g_{ij} \frac{\dd x^j}{ \dd {x'}^\nu}
\end{equation}
and from this transformation property it is assured that the metric $g$ indeed has tensor character. The determinant of $g$ will be denoted $|g|$, and obviously depends of the particular coordinate system used as it gets multiplied by the square of the Jacobian determinant when expressed in a different coordinate frame.

The inverse to $g_{\mu \nu}$ transforms in a contravariant way, and is hence written with upper indices:
\begin{equation}\label{gcov}
  g^{\mu \nu} g_{\nu \kappa} = \delta^\mu_{\hs \kappa}.
\end{equation}

\subsection{Covariant differentiation}

To define spatial derivatives of a scalar quantity is straightforward in curved spaces, as the usual definition for flat spaces makes sense:
\begin{equation}\label{dersc}
  \dd_u S (\vec{r}) = \lim_{h \rightarrow 0} \frac{1}{h} \left( S(\vec{r} + h \vec{u}) - S(\vec{r}) \right).
\end{equation}
However, for a vector or tensor quantity, the basis vectors differ from point to point in space, and their rate of change should be prescribed before the spatial derivative of such quantity can be meaningfully defined. The derivative operator which takes this effect into account is known as the `covariant derivative', and denoted here as $\DD_\mu$ \footnote{Alternative notations are $ \DD_\mu \vec{u} = \nabla_\mu \vec{u} = \vec{u}_{;\mu}$.}. To define a covariant derivative unambiguously, it suffices to prescribe the change of base vectors
\begin{equation} \label{affconn}
  \DD_\mu \vec{e}_{\nu}  = \Gamma^{\la}_{\mu \nu} \vec{e}_\la,
\end{equation}
where $\Gamma$ stands for the affine connection. In our application, the base vectors are defined as tangents to coordinate lines of varying $\rho^\mu$ in the studied space, i.e.
\begin{equation}\label{emu}
  \vec{e}_\mu = \frac{\dd \vec{x}}{ \dd \rho^{\mu}}
\end{equation}
and in this context $\Gamma$ is called the metric connection. The coefficients $\Gamma_{\mu, \nu \kappa}$ and $\Gamma^\mu_{\nu \kappa}$ are known as the Christoffel symbols of the first and second kind, respectively. In metric spaces, they are moreover uniquely defined in terms of the metric tensor:
\bsub \label{christ} \begin{eqnarray}
 \Gamma_{\mu, \nu \la} &=& \frac{1}{2}\left( \dd_\la g_{\mu \nu} + \dd_{\nu} g_{\mu \la} - \dd_\la g_{\mu \nu }\right) \\
  \Gamma^\alpha_{\mu\nu} &=& g^{\alpha \beta} \Gamma_{\beta, \mu \nu}.
\end{eqnarray} \esub
With this natural choice for the connection coefficients, \eqref{affconn} obtains the meaning of parallel transport: the basis vectors are transported parallel to themselves when translated between nearby points in space. Writing down the derivative of a vector $\vec{A} = A^\mu \vec{e}_\mu = A_{\mu} \vec{e}^{\mu}$ now yields following coordinate prescription for the covariant derivative of a vector:
\bsub \label{def_covderiv_app} \begin{eqnarray}
 \DD_\mu A^{\nu} &=& \dd_\mu A^\nu + \Gamma^{\nu}_{\mu \la} A^\la,\\
 \DD_\mu A_{\nu} &=& \dd_\mu A_\nu - \Gamma^{\la}_{\mu \nu} A_\la.
\end{eqnarray} \esub
With tensor quantities, all occurring indices should be subjected to parallel transport, e.g.
\begin{equation}\label{tenspartransp}
 \DD_\mu A^{\alpha \beta} = \dd_\mu A^{\alpha \beta} + \Gamma^{\alpha}_{\mu \nu} A^{\nu \beta} + \Gamma^{\beta}_{\mu \nu} A^{\alpha\nu}.
\end{equation}
From explicit calculation using \eqref{christ}, it can be proven that
\begin{equation}\label{Ricci_id}
  \DD_\mu g_{\alpha \beta} = 0, \qquad \DD_\mu g^{\alpha \beta}=0.
\end{equation}
These statements are known as the Ricci identity, and guarantee during calculations that raising and lowering indices commutes with covariant differentiation.  The proper extensions of the divergence and Laplacian operators are found as \cite{MTW}:
 \begin{eqnarray} \label{covdiv}
    div\,\vec{V} = \DD_\mu V^\mu &=& \frac{1}{\sqrt{|g|}} \dd_\mu \left( \sqrt{|g|} V^\mu  \right), \\
    \Delta f = \DD_\mu \DD^\mu f &=& \frac{1}{\sqrt{|g|}} \dd_\mu \left( \sqrt{|g|} g^{\mu\nu} \dd_\nu f \right). \label{covlap}
  \end{eqnarray}
The right-hand sides of these equations are manifestly covariant, i.e. any suitable coordinate system can be used to perform an explicit calculation. Importantly, we also observe that, for constant $|g|$, Eq. \eqref{covlap} coincides with the diffusion part of reaction-diffusion equations, with the identification $g^{ij} = D_0^{-1} D^{ij}$. This key remark allows handling anisotropy by curved space techniques, as described in Chapter \ref{chapt:activ}.

\subsection{Curvature tensors}
Owing to the local equivalence principle, a particular choice of coordinates can make the Christoffel symbols vanish at a given point. For that reason the curvature of space at any point cannot be contained in the first spatial derivatives of the metric tensor. However, curvature of space can be probed by parallel transport of a vector along an infinitesimal square in a coordinate plane of choice. Whereas the vector always comes out unchanged by parallel transport along a loop in flat spaces, it does not in curved space \cite{MTW}. The deviation serves as a measure for local curvature of the space, and can in a coordinate basis be expressed as
\begin{equation}\label{defRiemann}
  \left[ \DD_\alpha , \DD_\beta \right] V_\mu = R^\nu_{\hs \mu \alpha \beta} V_\nu.
\end{equation}
On the right-hand side, the Riemann curvature tensor $R_{\mu \nu \alpha \beta}$ appears, having following general coordinate expression:
\begin{equation}\label{Rabcd}
 R^\mu_{\ \nu \alpha \beta} = \dd_\alpha \Gamma^\mu_{\nu \beta} - \dd_\beta \Gamma^\mu_{\nu \alpha} + \Gamma^\mu_{\alpha\lambda} \Gamma^{\lambda}_{\nu \beta} - \Gamma^\mu_{\beta\lambda} \Gamma^{\lambda}_{\nu \alpha}.
\end{equation}
When using the metric connection, this tensor exhibits the following symmetries:
\bsub \label{Rsymm} \begin{eqnarray}
    R_{\mu\nu \alpha \beta} &=& R_{\alpha \beta \mu\nu },\\
    R_{\mu\nu \alpha \beta} &=& - R_{\mu\nu \beta \alpha } = -R_{\nu\mu \alpha \beta} =  R_{\nu\mu \beta \alpha },\\
    R_{\mu\nu \alpha \beta} &+& R_{\mu \alpha \beta \nu }+ R_{\mu \beta \nu \alpha}= 0.
\end{eqnarray}
Since covariant differentiation obeys the Jacobi identity
\begin{equation}\label{Jacobi}
 [A,[B,C]] + [B,[C,A]] + [C,[A,B]] =0,
\end{equation}
definition \eqref{defRiemann} leads to the Bianchi identities
\begin{eqnarray}\label{Bianchi}
        R_{\mu\nu \alpha \beta; \gamma} + R_{\mu\nu \gamma \alpha; \beta } + R_{\mu\nu \beta \gamma; \alpha} &=& 0,\\
        R_{\mu\nu \alpha \beta, \gamma} + R_{\mu\nu \gamma \alpha, \beta } + R_{\mu\nu \beta \gamma, \alpha} &=& 0.
\end{eqnarray} \esub
The second identity, involving ordinary (non-covariant) derivatives, is generated by the first due to antisymmetry of the Riemann tensor in the last two indices.

From the symmetries \eqref{Rsymm} follows that only one contraction of the Riemann tensor is meaningful; this defines the Ricci curvature tensor
\begin{equation}\label{Rab}
    R_{\mu \nu} = R^\alpha_{\hs \mu \alpha \nu},
\end{equation}
which is symmetric. A subsequent contraction yields the Ricci curvature scalar:
\begin{equation}\label{R}
   \RR = R^\mu_{\ \mu} = g^{\nu\mu} R_{\nu\mu}.
\end{equation}

The sign of the Ricci scalar can be easily interpreted, as it determines whether the space is locally sphere-like (elliptic space, $\RR>0$) or saddle-like (hyperbolic space, $\RR<0$). In quantitative terms, the volume $V$ and surface area $A$ of a n-dimensional ball of radius $r$ relate to their Euclidean values as
\bsub \begin{eqnarray}\label{surfdefect}
        V &=& V_{\rm Eucl.} \left(1 - \frac{\RR}{6(n+2)} r^2 + \OO(r^4) \right), \\
        S &=& S_{\rm Eucl.} \left(1 - \frac{\RR}{6 n} r^2 + \OO(r^4) \right).
\end{eqnarray} \esub
In particular, the circumference of a circle of radius $r$ on a two-dimensional curved surface amounts to
\begin{equation}\label{circumf}
  C = 2\pi r \left( 1 - \frac{\RR}{12} r^2  + \OO(r^4) \right).
\end{equation}

\subsection{Geodesics}

The important role that is played by straight lines in Euclidean geometry is in curved spaces taken over by geodesics, i.e. curves which are locally straight. As an example, one can think of the great circles on a sphere in three-dimensional Euclidean space, which are straight lines for someone confined to the surface of the sphere.

Geodesics are characterized by having a constant tangent vector field, i.e.
\begin{equation} \label{geodesic_eq0}
 \DD^2_s \vec{X} = 0,
\end{equation}
with $s$ an affine parameter along the curve. Written in components, the geodesic equation becomes:
\begin{equation}\label{geodesic_eq}
      \frac{d^2 x^\la }{ds^2} + \Gamma^{\lambda}_{\mu \nu }\frac{dx^\mu }{ds}\frac{dx^\nu }{ds} = 0.
\end{equation}
An important property in Riemannian manifolds is that every two nearby points are joined by a unique geodesic. Being a second order differential equation, Eq. \eqref{geodesic_eq} also implies that both the starting point and an initial direction should be specified to uniquely define a geodesic through a given point.

The interpretation of a geodesic being the curve of extremal length that runs between two given points follows from the definition of total length of a curve
\begin{equation}\label{length}
      S = \int_a^b \sqrt{ g_{\mu \nu} \frac{d X^\mu}{d s} \frac{d X^\nu}{d s} }\,ds.
\end{equation}
For, the Euler-Lagrange equation that is produced by varying arc length yields the geodesic equation \eqref{geodesic_eq}. Note that geodesics can be curves that either minimize or maximize the length between given endpoints.

Intrinsic curvature of space can be described by the extent to which initially parallel geodesics start to deviate. To this purpose, consider a geodesic with affine parameter $s$ and unit tangent $\vec{u}$. A second geodesic which is locally parallel to the first, but translated over the deviation vector $\vec{n}$ will start deviating from the first, at a rate given by
\begin{equation}\label{geod_dev_app}
  \left( \frac{\DD^2\vec{n}}{\DD s^2} \right)^\alpha + R^\alpha_{\hs \beta \mu \nu} u^\beta n^\mu u^\nu = 0.
\end{equation}
This relation is used in Chapter \ref{chapt:activ} to gain physical insight in what the different component of Riemann and Ricci tensors mean in terms of divergence of geodesics. 

\section[Specific relations for spaces of two and three dimensions]{Specific relations for spaces of \\two and three dimensions}

\subsection{Curvature tensors in two dimensions \label{sec:curv2}}

When considering curvature of a two-dimensional space, the Riemann tensor possesses only one degree of freedom. The same quantity determines the Ricci tensor and Ricci curvature scalar; in the context of differential geometry of surfaces, this single measure of intrinsic curvature of a surface is nothing else than the Gaussian curvature $K_G$, i.e. the product of the principal curvatures.

As we will encounter the relevant curvature tensors for surfaces embedded in a three-dimensional space, we add a superscript $(2)$ to discriminate them from the same quantities defined with all three dimensions.
\bsub \begin{eqnarray} \label{Riemann_n=2}
    {}^{(2)} R_{ABCD} &=& K_G \left(g_{AC} g_{BD} - g_{AD} g_{BC} \right), \label{R2ABCD}\\
    {}^{(2)} R_{AB} &=& K_G\, g_{AB}, \\
    {}^{(2)} \RR &=& 2 K_G.
\end{eqnarray} \esub
Note that the Gaussian curvature $K_G$ equals the product of the principal curvatures of the surface, while the mean curvature $H$ amounts to their arithmetic average:
\begin{align} \label{def_KGH}
   K_G & = K_1 K_2, &  H &= \frac{1}{2} \left( K_1 + K_2 \right).
\end{align}
In the nominal velocity-curvature relation for wave fronts in isotropic media, $K$ denotes $2H$.

\subsection{Curvature tensors in three dimensions}

When working in three dimensions, $R_{\mu\nu \alpha \beta}$ and $R_{\mu\nu}$ can be expressed in terms of each other, since they both contain six linearly independent components. As a result, the trace-free part of the Riemann tensor -known as the Weyl tensor- vanishes.  

The explicit relation that inverts \eqref{Rab} in three dimensions reads
\begin{multline}\label{Riemann2Ricci}
  R_{\alpha \beta \mu \nu} = \left( g_{\alpha \mu} R_{\beta \nu} + g_{\beta \nu} R_{\alpha \mu}   - g_{\alpha \nu} R_{\beta \mu}  - g_{\beta \mu} R_{\alpha \nu}  \right)  \\ - \frac{\RR}{2} \left( g_{\alpha \mu} g_{\beta \nu} - g_{\alpha \nu} R_{\beta \mu} \right).
\end{multline}
Since there are six index sets for which $R_{\mu\nu \alpha \beta}$ does not vanish, these sets precisely confer to the six independent components of the Riemann tensor.

\subsection{Three-dimensional space relative to the wave front}

The Riemann and Ricci curvature tensors for the full three-dimensional space may be broken down into extrinsic and intrinsic curvature contributions, with respect to the wave front surface:
\bsub \label{Riemann_front} \begin{eqnarray}
{}^{(3)} R^A_{\hs BCD} &=& {}^{(2)} R^A_{\hs BCD} + \left( K^A_{\hs D} K_{BC}  - K^A_{\hs C} K_{BD} \right),\qquad \label{Riemann_front1}\\
{}^{(3)} R^\rho_{\hs BCD} &=& \DD_D K_{BC}  - \DD_C K_{BD}, \label{Riemann_front2}\\
{}^{(3)} R^\rho_{\hs A \rho B} &=&  - \dd_\rho K_{AB} + K_{AC} K^C_{\hs B}.\label{Riemann_front3}
\end{eqnarray} \esub
In the main text, the superscript ${}^{(3)}$ is omitted.

\subsection{Three-dimensional space relative to the filament}

As in general relativity, we can put up specific relations after specifying a preferential direction in the non-Euclidean space. Similarly to choosing the worldline of an observer parameterized by his or her eigentime $\tau$, we take the instantaneous filament with arc length parameter $\s$ as a reference. After setting up Fermi coordinates around the filament as exposed in Chapter \ref{chapt:filaniso}, following relations hold (with $\kappa = R^{12}_{\hs \hs 12}$):
\bsub \label{Rprop3}\begin{eqnarray}
 R_{ABCD} &=& \kappa \eps_{AB} \eps_{CD} + \OO(\rho^2), \label{kappa_eps} \\
 R_{ABC\s} &=& R_{\s B} \delta_{AC} - R_{A\s} \delta_{BC} + \OO(\rho),\\
 R_{\s A \s B} &=& R_{A B} - \kappa \delta_{AB} + \OO(\rho),\\
 R_{ABCD,E} &=& R_{ABCD;E}  + \OO(\rho),\\
 R^\alpha_{\beta \mu \nu, \s} &=& R_{\alpha \beta \mu \nu, \s} + \OO(\rho).
\end{eqnarray} \esub


\clearpage{\pagestyle{empty}\cleardoublepage}


\renewcommand\evenpagerightmark{{\scshape\small Appendix B}}
\renewcommand\oddpageleftmark{{\scshape\small Isotropic tensors}}

\renewcommand{\bibname}{References}

\hyphenation{}

\chapter[Isotropic tensors and matrix element formalism]{Isotropic tensors and\\ matrix element formalism}
\label{app:isotropic_tensors}

\section{Notation for matrix element states}

\subsection{Definition of the inner product}
In our gradient expansion series for filament motion and wave front dynamics, several overlap integrals between response functions and Goldstone modes are encountered. Inspired by quantum mechanics, we take on the \textit{bra-ket} notation
\bsub  \label{def_matel} \begin{eqnarray}
\langle \mathbf{f}, \mathbf{g} \rangle = \langle \mathbf{f} \mid \mathbf{g} \rangle &=& \int \mathrm d^n V \mathbf{f}^H \mathbf{g}, \\
\bra{\mathbf{f}} \hat{O} \ket{\mathbf{g}}  &=& \int \mathrm d^n V \mathbf{f}^H \hat{O} \mathbf{g}.
\end{eqnarray} \esub
and refer to the resulting quantities as \textit{matrix elements}. In Eqs. \eqref{def_matel}, integration is taken over the number of dimensions of Euclidean space transverse to the object under study, i.e. $n=1$ for fronts and $n=2$ for filaments. In either case, no temporal averaging is understood in the inner product, in contrast to \cite{Keener:1986, Biktashev:1995}.

\subsection{Generic notation for matrix element states}
In our geometric theory for filaments exposed in chapters \ref{chapt:filiso} and \ref{chapt:filaniso}, several matrix states between response functions and Goldstone modes are encountered. As we have defined the response functions to be real-valued, the Hermitian conjugate symbol in Eq. \eqref{def_matel} causes only a matrix transposition. In this context, $\hat{O}$ represents either the identity operator $\hat{I}$ or the diffusive projection operator $\HP$, weighed with a function of coordinates and, possibly, a spatial derivative operator. We now introduce custom notation:
\bsub
\begin{eqnarray}
    \left(O^{Rn}\right)_{A}^{\hs C_1...C_n} &=& \bra{\bY^\theta} \hat{O} \rho^{C_1}... \rho^{C_n} \ket{\bpsi_A}, \label{matel1}\\
    \left(O^{Rn\delta} \right)_{AB}^{\hs \hs C_1...C_n} &=& \bra{\bY^\theta} \hat{O} \rho^{C_1}... \rho^{C_n} \dd_B \ket{\bpsi_A}, \\
    \left(O^{Tn}\right)_{A}^{\hs C_1...C_n D} &=& \bra{\bY^D} \hat{O} \rho^{C_1}... \rho^{C_n} \ket{\bpsi_A}, \\
    \left(O^{Tn\delta} \right)_{AB}^{\hs\hs C_1...C_n D} &=& \bra{\bY^D} \hat{O} \rho^{C_1}... \rho^{C_n} \dd_B \ket{\bpsi_A}. \label{matel4}
\end{eqnarray}
\esub
In the generic symbols on the left-hand side, the base letter $I$ or $P$ indicates whether the diffusive projection operator is needed, and superscripted $R$ or $T$ denotes projection onto the rotational or translational response function. The integer $n$ denotes the degree of the homogenous coordinate function, and a $\delta$ is added when an additional spatial derivative is present. Assignment of indices occurs as in Eqs. \eqref{matel1}-\eqref{matel4}. Furthermore, the physical dimension of the matrix element can be inferred from the sum of superscripts, where $T$ has value 0; $R$ and $\delta$ convey to -1 and $n$ has its natural value. The result should be multiplied with the dimension of $D_0$ ($length^2/time$), if this quantity has been absorbed in $\HP$.

Later in this appendix, subscripts will be assigned  to the notation \eqref{matel1}-\eqref{matel4} as well, denoting the independent tensor components after imposing rotational invariance.

\section{Theory of isotropic tensors}

\subsection{Definition}
A tensor is isotropic if and only if the tensor is invariant under rotations.

\subsection{Rank and dimension dependent properties}

\begin{itemize}
\item \textbf{Rank 1 or 0}\\
Tensors of rank zero are scalars; these are always isotropic. Rank 1 tensors (\textit{vectors}) are never isotropic, unless they have length zero.
\item \textbf{Tensors of odd rank}\\
Because a tensor of odd rank can be repeatedly contracted to obtain a vector, such tensor needs to be totally antisymmetric and can therefore only be given by the Levi-Civita tensor. As the Levi-Civita tensor has rank equal to the dimension of the space considered, spaces of even dimension do not possess an isotropic tensor of odd rank.
\item \textbf{Tensors of even rank in dimension $>2$}\\
In spaces with dimension higher than two, there is in general only one isotropic tensor for a given even rank $n=2m$; it is constructed by the symmetrized product of m Kronecker delta's:\begin{eqnarray}
                    T_{ij} &=& \delta_{ij}, \nn \\
                    T_{ijkl} &=& \delta_{ij} \delta_{kl} + \delta_{ik} \delta_{jl} + \delta_{il} \delta_{jk}, \\
                    T_{ijklmn} &=& \delta_{ij} \delta_{kl} \delta_{mn} + ... .\nn
                  \end{eqnarray}
In general there will be $(n-1)(n-3).\, \ldots\, .1 = \prod\limits_{j=0}^{m-1} [2(m-j)-1] $ terms.
\item \textbf{Tensors of even rank in dimension 2}\\
    This case is far more complex than the aforementioned and precisely the case needed in this work. Henceforth, we work in the space of tensors of even rank in dimension 2.
\end{itemize}

\subsection{Brute force calculation of the independent components}

A crude way to find a basis for the space of isotropic tensors with given rank and prescribed symmetries, is to average an arbitrary tensor over the entire rotation group. As the elements of the rotationally averaged tensor span the full space of isotropic tensors, a basis of the elements forms the basis of the space of isotropic tensors.
In two dimensions, this procedure can easily be performed using a symbolic algebra package, because it involves only simple trigonometric integrals over the full period $[0, 2\pi]$. As a simple example, consider the rank-2 tensor in a given base: $A=(a_{ij})$. Then follows, with $2\times2$ rotation matrices $R(\theta)$,
\begin{equation}\label{rotav_rank2}
  \langle A \rangle_{ij} = \int \limits_0^{2\pi} \mathrm{d}\theta R_i^{\hs k}(\theta) R_j^{\hs \ell}(\theta) a_{k\ell}.
\end{equation}
Explicit calculation results in
\begin{eqnarray}
   \langle A \rangle &=& \frac{1}{2} \left(
                                     \begin{array}{cc}
                                       a_{11}+a_{22} & a_{12}-a_{21} \\
                                       a_{21}-a_{12} & a_{11}+a_{22} \\
                                     \end{array}
                                   \right) \nn \\
                                   &=& \frac{a_{11}+a_{22}}{2}
                                   \left(
                                     \begin{array}{cc}
                                       1 & 0 \\
                                       0 & 1 \\
                                     \end{array}
                                   \right) +
                                   \frac{a_{12}-a_{21}}{2}
                                   \left(
                                     \begin{array}{cc}
                                       0 & 1 \\
                                       -1 & 0 \\
                                     \end{array}
                                   \right).\label{rotav_rank2_brute}
\end{eqnarray}
We conclude that any isotropic tensor of rank 2 in two dimensions can be written as
\begin{equation}\label{decrank2}
  T_{AB} = T_s \delta_{AB}+ T_a \epsilon_{AB}.
\end{equation}
In the rank-2 case, the subscripts $s,a$ denote the symmetric and antisymmetric tensor components. Within the generic notation scheme, we thus obtain for the filament tension components \cite{Verschelde:2007}, from $\bra{\bY^B} \HP \ket{\bpsi_A} = \gamma_1 \delta_{AB}+ \gamma_2 \epsilon_{AB}$:
\begin{align}
 \gamma_1 &= P^{T1}_s, & \gamma_2 &= P^{T1}_a.
\end{align}

For tensors of higher rank, the independent isotropic components can similarly be assessed by rotationally averaging the tensor with elements $a_i$; after determining the unique resulting matrix elements $a'_i = C_{ij} a_j $, the rank of the coefficient matrix $C_{ij}$ equals the number of isotropic components for the tensor.

\subsection{Independent components using the Pauli matrices}

There is, however, a more elegant way to explore the properties of isotropic tensors (see e.g. \cite{Avron:1998}). This method makes use of the generator of rotations, rather than explicit evaluation. To start with, we choose an appropriate basis to represent rank-2 tensors in two dimensions, based on the Pauli matrices $\{\bs^1, \bs^2, \bs^3\}$:
   \begin{align} \label{Pauli}
         \bs^1 &= \left(
              \begin{array}{cc}
                0 & 1 \\
                1 & 0 \\
              \end{array}
            \right), & \bs^2 &= \left(
              \begin{array}{cc}
                0 & -i \\
                i & 0 \\
              \end{array}
            \right), & \bs^3 &= \left(
              \begin{array}{cc}
                1 & 0 \\
                0 & -1 \\
              \end{array}
            \right).
    \end{align}
These Hermitian matrices are traceless and obey following equations, where $a,b,c \in \{1,2,3\}$:
\bsub \label{ess0}\begin{eqnarray}
  \bs^a \bs^b &=& \delta_{ab} \boldsymbol 1 + i \epsilon_{abc} \bs^c, \\
  \bs^a \bs^b + \bs^b \bs^a\ &=& 2 \delta_{ab} \one, \\
  \bs^a \bs^b - \bs^b \bs^a = \left[\bs^a, \bs^b \right] &=& 2 i \epsilon_{abc} \bs^c.
\end{eqnarray} \esub
We will slightly adapt the basis \eqref{Pauli} to our purpose by adding $\bs^0 = \boldsymbol 1$ and using the real-valued antisymmetric tensor $\beps = i \bs^2$ instead of $\bs^2$. Useful product relations are
\begin{align}\label{ess}
    \bs^1 \beps &= -\bs_3,  & \bs^3 \beps &= \bs^1, & \bs^1 \bs^3 &= -\beps, \nn \\
    \beps \bs^1 &= \bs^3, & \beps \bs^3 &= -\bs^1, & \bs^3 \bs^1 &= \beps.
\end{align}

Consider now the generator for rotations in two dimensions, i.e. $\beps = i\bs^2$. Hence, the condition for a generic rank 2 tensor
\begin{equation}
 \mathbf{T} = \sum \limits_{a=0}^3 \eta_a \bs^a
 \end{equation}
to be invariant under rotation requires that it commutator with the generator of rotations vanishes:
\begin{eqnarray}\label{2D}
    \left[\mathbf{T}, \bs^2\right] &=& 0.
\end{eqnarray}
This condition immediately leads to \eqref{decrank2} without performing any integration:
\begin{equation}\label{iso2}
    T^{iso}_{ij} = T_s \s^0_{ij} + \tilde{T}_a \s^2_{ij} = T_s \delta_{ij} + T_a \eps_{ij}.
\end{equation}\\

For the space of rank four tensors, a natural basis to expand in is
\begin{equation}\label{R4}
    T_{ijkl} = \sum \limits_{a=0}^3 c_{ab} \s^a_{ij} \s^b_{kl}.
\end{equation}
Recalling spin physics, we learn that the generator for rotation is not $\bs^2 \otimes \bs^2$, but $\mathbf{1} \otimes \bs^2 +  \bs^2 \otimes \mathbf{1}$. Rotational invariance of $\bs^\mu \otimes \bs^\nu$ then imposes the condition
\begin{eqnarray}
  \bs^\mu \bo \left[\bs^\nu , \bs^2\right] + \left[\bs^\mu , \bs^2\right] \bo  \bs^\nu &=& 0.
\end{eqnarray}
Solutions to this equation can either have $\left[\bs^\mu , \bs^2\right] = \left[\bs^\nu , \bs^2\right] = 0$ or exhibit $\left[\bs^\mu , \bs^2\right] = - \left[\bs^\nu , \bs^2\right] \neq 0$. The former case leads to the invariant components
  \begin{align}\label{comm}
    T_{00}& \bs^0 \bo \bs^0,& T_{02}& \bs^0 \bo i\bs^2,& T_{20}& i\bs^2 \bo \bs^0,& T_{22}& \bs^2 \bo \bs^2,
  \end{align}
whereas the latter case yields following commutator relations with $\bs^2$:
\begin{subequations}
  \begin{eqnarray}
    \left[\bs^1 \bo \bs^1 , \boldsymbol{1} \bo \bs^2 +  \bs^2 \bo \boldsymbol{1} \right] &=& i (\bs^3 \bo \bs^1 + \bs^1 \bo \bs^3),\\
     \left[\bs^3 \bo \bs^3 ,\boldsymbol{1} \bo \bs^2 +  \bs^2 \bo \boldsymbol{1} \right] &=& -i (\bs^1 \bo \bs^3 + \bs^3 \bo \bs^1),\\
     \left[\bs^1 \bo \bs^3 , \boldsymbol{1} \bo \bs^2 +  \bs^2 \bo \boldsymbol{1} \right] &=& i (\bs^3 \bo \bs^3 - \bs^1 \bo \bs^1),\\
     \left[\bs^3 \bo \bs^1 , \boldsymbol{1} \bo \bs^2 +  \bs^2 \bo \boldsymbol{1} \right] &=& i (- \bs^1 \bo \bs^1 + \bs^3 \bo \bs^3).
  \end{eqnarray}
  \end{subequations}
Hence there are two invariant combinations in the latter case
  \begin{align}\label{as}
     \frac{T_s}{2}& (\bs^1 \bo \bs^1 + \bs^3 \bo \bs^3), & \frac{T_a}{2} &(\bs^1 \bo \bs^3 - \bs^3 \bo \bs^1).
  \end{align}
The above derivation shows that the space of isotropic tensors of rank 4 in 2 dimensions is 6-dimensional:
  \begin{multline}\label{4all}
    T^{iso} =  T_{00} \bs^0 \bo \bs^0 + T_{02} \bs^0 \bo \beps + T_{20} \beps \bo \bs^0 + T_{22} \beps \bo \beps \\ + \frac{T_s}{2} (\bs^1 \bo \bs^1 + \bs^3 \bo \bs^3) + \frac{T_a}{2} (\bs^1 \bo \bs^3 - \bs^3 \bo \bs^1),
  \end{multline}
where the components have been labeled accordingly. The subscripts $s,a$ refer to (anti)symmetry with respect to interchange of the first and last pair of indices. The integer factors have been inserted for convenience, as they will disappear after subsequent index contractions. Note that imposing symmetry on the tensor indices may reduce the dimension of the tensor space.


To obtain the coefficients $c_{ab}$ in Eq. \eqref{R4} one has to compute explicitly
\begin{equation}\label{est_R4comps}
  c_{ab} = \frac{1}{4} \sum \limits_{i,j,k,l} T_{ijkl} \s^a_{ij} \s^b_{kl}.
\end{equation}
In such calculations, Eqs. \eqref{ess0}-\eqref{ess} come in handy.


\section{Application to the gradient expansion series}

\subsection{Decomposition of rank 2 tensors}

The occurring rank 2 tensors $P^{T0}$ and $P^{R1}$ can be simply decomposed as
 \begin{equation}\label{case20}
     T =  T_s \bs^0 + T_a \beps.
   \end{equation}

\subsection{Decomposition of rank 4 tensors}
The motivation for these calculations is the appearance of isotropic matrix elements in our expansion series. So far, we have encountered invariant tensors of rank four with three types of symmetry:

\begin{enumerate}
    \item \textbf{Exchange symmetry $(3\leftrightarrow 4)$}:\\
    The prescribed symmetry can be imposed by demanding $T_{ABCD} \eps^{CD}= 0$, from which follows that $T_{22} = T_{02} = 0$. This observation reduces the dimension of the tensor space to 4:
    \begin{multline}\label{case34}
     T =  T_{00} \bs^0 \bo \bs^0 +  T_{20} \beps \bo \bs^0 \\+ \frac{T_s}{2} (\bs^1 \bo \bs^1 + \bs^3 \bo \bs^3) +  \frac{T_a}{2} (\bs^1 \bo \bs^3 - \bs^3 \bo \bs^1).
   \end{multline}
  This type of symmetry is found in the matrix elements build with the translational RF: $(P^{T2})_{A}^{\hspace{7pt} BCD} = \bra{\bY^B}\hat{P} \rho^C \rho^D \ket{\bpsi_A}$. In a similar way one has $(I^{T2})_{A}^{\hspace{7pt} BCD} = \bra{\bY^B}\rho^C \rho^D \ket{\bpsi_A}$.
  \item \textbf{Exchange symmetry $(1 \leftrightarrow 2)$ and $(3\leftrightarrow 4)$}:\\
   The prescribed symmetry requires, in addition to the previous case, that $T_{ABCD} \eps^{AB}= 0$, whence $T_{20}=0$:
  \begin{multline}\label{case12_34}
     T =  T_{00} \bs^0 \bo \bs^0 \\+ \frac{T_s}{2} (\bs^1 \bo \bs^1 + \bs^3 \bo \bs^3) +  \frac{T_a}{2} (\bs^1 \bo \bs^3 - \bs^3 \bo \bs^1).
   \end{multline}
  A term of this type $(P^{R2\delta})_{AB}^{\hspace{14pt} CD} = \bra{\bY^\theta}\hat{P} \rho^C \rho^D \ket{\partial^2_{AB} \mathbf{u}_0}$
 is encountered in the evolution equation for twist.
   \item \textbf{Exchange symmetry $(2 \leftrightarrow 3 \leftrightarrow 4)$}:\\
   The generating set of $2\times2$ matrices is effectively smaller than compared to the first case due to the additional (23) index exchange symmetry. It turns out that, under the given symmetry,  $ 2\bs^0 \bo \bs^0 $ coincides with  $\bs^1 \bo \bs^1 + \bs^3 \bo \bs^3$ and $ - 2\beps \bo \bs^0 $ coincides with  $\bs^1 \bo \bs^3 - \bs^3 \bo \bs^1$. Therefore, such tensor has the decomposition
   \begin{equation}\label{case3}
     T =  T_{00} \bs^0 \bo \bs^0 +  T_{20} \beps \bo \bs^0.
   \end{equation}
   This symmetry type can be found in a higher order correction term to the twist equation: $(P^{R3})_A^{\hspace{7pt}BCD}  \bra{\bY^\theta}\hat{P} \rho^B \rho^C \rho^D \ket{\bpsi_A}$.
\end{enumerate}

\newpage
\subsection{Decomposition of rank 6 tensors}
We restrict the treatment to the symmetry types that appear in our calculations.

\begin{enumerate}
 \item \textbf{Exchange symmetry $ (1 \leftrightarrow 2) (3 \leftrightarrow 4 \leftrightarrow 5 \leftrightarrow 6)$}:\\
In the second order corrections to the angular frequency $\omega$ of a scroll wave, the rank 6 tensor $P^{R4\delta}$ appears. Fortunately, this matrix bears (12) and (3456) index symmetry, so that the decomposition basis can be restricted to
\begin{equation}\label{R6ta}
    T_{ijklmn} = \sum \limits_{a,b,c \in \{0,1,3\} } C_{abc} \s^a_{ij} \left( \s^b_{kl} \s^c_{mn}  + \s^c_{kl} \s^b_{mn}\right),
\end{equation}
where the (45) exchange symmetry is not yet enforced. Demanding that each of these 18 components commutes with the generator of rotations $\beps$, permits only four isotropic tensor components, including $\bs^0 \bo \bs^0 \bo \bs^0$ and $ \bs^0 \bo (\bo \bs^1 \bo \bs^1 + \bs^3 \bo \bs^3)$. Recalling the case for a rank 4 tensor with (234) index symmetry, we conclude here too that both quoted tensor components are identical, leaving only three independent contributions:
    \begin{eqnarray}\label{R6a}
    &&\hspace{-0.4cm} T =  T_{000} \bs^0 \bo \bs^0 \bo \bs^0 \\
    &&+ \frac{T_{s0}}{4}  \left( \bs^1 \bo (\bs^1 \bo \bs^0 + \bs^0 \bo \bs^1) +  \bs^3 \bo (\bs^3 \bo \bs^0 + \bs^0 \bo \bs^3) \right) \  \nn\\
    &&+  \frac{T_{a0}}{4}  \left( \bs^1 \bo (\bs^3 \bo \bs^0 + \bs^0 \bo \bs^3) -  \bs^3 \bo (\bs^1 \bo \bs^0 + \bs^0 \bo \bs^1) \right). \  \nn
   \end{eqnarray}
 \item \textbf{Exchange symmetry $ (1 \leftrightarrow 2) (4 \leftrightarrow 5 \leftrightarrow 6)$}:\\
   A suitable basis for this tensor space is found as
   \begin{equation}\label{R6t}
    T_{ijklmn} = \sum \limits_{a,c \neq 2} C_{abc} \s^a_{ij} \s^b_{kl} \s^c_{mn}.
\end{equation}
 Requesting that the commutator with $\beps$ vanishes allows only direct products of the type \eqref{case12_34} and $\bs^0$ or $\beps$, with cyclic permutation of the Pauli matrices. Of the 10 remaining tensors, only 6 are found to be linearly independent:
    \begin{eqnarray}\label{R6b}
     T &=&  T_{000} \bs^0 \bo \bs^0 \bo \bs^0
                 + T_{020} \bs^0 \bo \beps \bo \bs^0 \\
               && \quad + \frac{T_{s0s}}{2} (\bs^1 \bo \bs^0 \bo \bs^1 + \bs^3 \bo \bs^0 \bo \bs^3) \nn \\
                &&\quad +  \frac{T_{s0}}{2} (\bs^1 \bo \bs^1 \bo \bs^0 + \bs^3 \bo \bs^3 \bo \bs^0)  \nn \\
               &&\quad  + \frac{T_{a0a}}{2} (\bs^1 \bo \bs^0 \bo \bs^3 - \bs^3 \bo \bs^0 \bo \bs^1) \nn \\
               &&\quad  +  \frac{T_{a0}}{2} (\bs^1 \bo \bs^3 \bo \bs^0 - \bs^3 \bo \bs^1 \bo \bs^0)  \nn.
   \end{eqnarray}
\end{enumerate}

\clearpage{\pagestyle{empty}\cleardoublepage}

\backmatter
\pagenumbering{arabic}

\renewcommand\evenpagerightmark{{\scshape\small References}}
\renewcommand\oddpageleftmark{{\scshape\small References}}

\renewcommand\em{\it}
\bibliographystyle{phdbib}

\setcounter{page}{1}
\renewcommand{\thepage}{R-\arabic{page}}
\hyphenation{myo-fi-bers mecha-nics }

\end{document}